\documentclass{article}
\usepackage{graphicx}  
\usepackage{amsmath}   
\usepackage{adjustbox}
\usepackage[compress]{cite}
\usepackage{amssymb}   
\usepackage{bm} 
\usepackage{dcolumn}
\usepackage{color}
\usepackage{colortbl}
\usepackage{colordvi}
\usepackage{mathrsfs}
\usepackage{amsfonts}
\usepackage{varioref}
\usepackage{float}
\usepackage{graphicx}
\usepackage{multirow}
\usepackage{subcaption}
\usepackage[bottom=0.5in,top=0.5in,right=1.5in,left=1.5in]{geometry}
\RequirePackage[colorlinks,citecolor=blue,urlcolor=magenta,linkcolor=blue]{hyperref}
\input epsf
\usepackage{tablefootnote}
\def\gr{general relativity}

\labelformat{section}{Section #1} 
\labelformat{subsection}{Section #1} 
\labelformat{subsubsection}{Section #1}
\labelformat{subsubsubsection}{Section #1}
\labelformat{equation}{Eq.~(#1)} 
\labelformat{figure}{Fig.~#1} 
\labelformat{subfigure}{Fig.~\thefigure#1} 
\labelformat{table}{Table~#1} 
\labelformat{appendix}{Appendix #1}
\title{ Signatures of Einstein-Maxwell dilaton-axion gravity from the observed quasi-periodic oscillations in black holes
}

\author{Anirban Dasgupta \footnote{522ph1007@nitrkl.ac.in}~$^{1}$, Nishant Tiwari \footnote{nishant\_t@ph.iitr.ac.in}~$^{1,2}$ and Indrani Banerjee \footnote{banerjeein@nitrkl.ac.in}~$^{1}$\\
{\small{$^{1}$Department of Physics and Astronomy, National Institute of Technology, Rourkela, Odisha-769008, India}}\\
{\small{$^{2}$Department of Physics, Indian Institute of Technology, Roorkee, Roorkee, Uttarakhand-247667, India}}}

\date{ }  
\begin{document}
  
\maketitle

\begin{abstract}

String-inspired models are often believed to provide an interesting framework for quantum gravity and force unification with promising prospects to resolve issues like dark matter and dark energy, which cannot be satisfactorily incorporated within the framework of general relativity (GR). The goal of the present work is to investigate the role of the Einstein-Maxwell dilaton-axion (EMDA) gravity arising in the low-energy effective action of the heterotic string theory in explaining astrophysical observations, in particular, the high-frequency quasi-periodic oscillations (HFQPOs) observed in the power spectrum of black holes.
EMDA gravity has interesting cosmological implications, and hence it is worthwhile to explore the footprints of such a theory in available astrophysical observations. This requires one to study the stationary, axi-symmetric black hole solution in EMDA gravity, which corresponds to the Kerr-Sen spacetime. Such black holes are endowed with a dilatonic charge while the rotation is sourced from the axionic field. We investigate the orbital and epicyclic frequencies of matter rotating in the Kerr-Sen spacetime and consider eleven well-studied QPO models in this work. We compare the model-dependent QPO frequencies with the available observations of five BH sources, namely, XTE J1550-564, GRS 1915+105, H 143+322, GRO J1655-40 and Sgr A*. Our analysis provides constraints on the spins of the aforesaid black holes which, when compared with previous estimates, enables us to understand the observationally favored QPO models for each of these sources. Further, from the current data, the EMDA scenario cannot be ruled out in favor of \gr. We comment on the implications and limitations of our findings and how the present constrains compare with the existing literature.

\end{abstract}
\section{Introduction}\label{QPO_Intro}
Black holes are one of the most enigmatic predictions of \gr, which has received observational confirmation since early nineteen seventies \cite{1972Natur.235..271B,Webster:1972bsw}. With recent observational breakthroughs, such as the discovery of gravitational waves \cite{Abbott:2017vtc,Abbott:2016nmj,Abbott:2016blz} and the release of black hole images \cite{Akiyama:2019cqa,EventHorizonTelescope:2022xnr,EventHorizonTelescope:2022xqj}, their existence have only been further affirmed. Since black holes are cloaked by an event horizon, they are, astrophysically, among the simplest celestial objects, being characterized only by their mass and angular momentum. Their electric charges, if present, are eventually neutralized due to accretion. This is the statement of the no-hair theorem \cite{Bekenstein:1971hc,Israel:1967wq,Carter:1971zc,Robinson:1975bv,Mazur:1982db} which, however has several counterexamples \cite{Bizon:1990sr,Garfinkle:1990qj,Greene:1992fw,Lavrelashvili:1992ia,Torii:1993vm,Herdeiro:2014goa,Berti:2013gfa}. Black holes are objects of enormous density and possess the strongest gravitational field among all the astrophysical objects in the universe. The near-horizon regime of black holes is, therefore, an ideal astrophysical site to test the nature of strong gravity. This is important, because \gr\ (GR)$-$the most successful theory of gravity till date, breaks down at the black hole and Big-Bang singularities \cite{Penrose:1964wq,Hawking:1976ra,Christodoulou:1991yfa}, and also falls short in explaining the dark sector \cite{Bekenstein:1984tv,Perlmutter:1998np,Riess:1998cb}. This makes the quest for a more complete theory of gravity increasingly compelling, potentially involving modifications in the gravity sector, extensions in the matter sector, or both \cite{Shiromizu:1999wj,Dadhich:2000am,Harko:2004ui,
Carames:2012gr,Chakraborty:2015bja,Nojiri:2006gh,Lanczos:1932zz,Lovelock:1971yv,Padmanabhan:2013xyr,Horndeski:1974wa,Sotiriou:2013qea,Charmousis:2015txa}. 

Among the various alternatives to GR, string theory provides an interesting framework that  incorporates the quantum nature of gravity and provides a mechanism for force unification. These models account for the ultra-violet nature of gravity and can potentially replace GR in the high-curvature regime, where quantum gravity is expected to be relevant. In this work, we study the Einstein-Maxwell-dilaton-axion (EMDA) gravity, which arises in the low-energy effective action of superstring theories \cite{Sen:1992ua,Rogatko:2002qe}. While the gravity action remains the same as in GR, the matter sector includes additional fields like the dilaton and the axion which are inherited from string theory and are coupled to the metric and the Maxwell field. The dilaton and axion fields originating from string compactifications find interesting applications in inflationary cosmology and the late-time acceleration of the universe \cite{Sonner:2006yn,Catena:2007jf} and hence it is worthwhile to search for footprints of these fields in the available astrophysical observations. Different black hole solutions have been constructed in string-inspired low-energy effective theories \cite{Gibbons:1987ps,Garfinkle:1990qj,Horowitz:1991cd,Kallosh:1993yg} bearing non-trivial charges associated with the dilaton and anti-symmetric tensor gauge fields. Interestingly, the charge neutral rotating solution in string theory cannot be distinguised from the Kerr metric \cite{Thorne:1986iy,Campbell:1992hc,Psaltis:2007cw}; hence an observational confirmation of the Kerr metric may not conclusively validate GR. Since astrophysical black holes are typically rotating, one needs to look for stationary and axi-symmetric black hole solutions in EMDA gravity, which correspond to the Kerr-Sen spacetime \cite{Sen:1992ua}, where the electric charge and rotation arise from the dilaton and the axion fields, respectively.
In earlier works, the Kerr-Sen black hole has been studied in the context of photon motion, strong gravitational lensing and black hole shadows \cite{Gyulchev:2006zg,An:2017hby,Younsi:2016azx,Hioki:2008zw,Mizuno:2018lxz,Narang:2020bgo}. This spacetime has also been tested with the optical observations of Palomar-Green quasars \cite{Banerjee:2020qmi}, the jet power measurements of several microquasars \cite{Banerjee:2020ubc}, the reflection spectrum of the black hole binary EXO 1846-031 \cite{Tripathi:2021rwb}, and the images of M87* and Sgr A* \cite{Sahoo:2023czj}. This motivates us to test this metric with yet another available astrophysical observation, the high-frequency quasi-periodic oscillations (HFQPOs) in black holes, which is the goal of the present work.

Quasi-periodic oscillations are observed primarily in the power spectrum of some Low-Mass X-ray binaries (LMXRBs) \cite{2006csxs.book.....L,vanderKlis:2000ca} such as neutron stars (NSs) and black holes, and less often observed in active galactic nuclei \cite{Torok:2004xs}. These peaks in the power spectrum of BHs and NSs were first observed by NASA's Rossi X-Ray Timing Explorer satellite \cite{2006csxs.book.....L}. QPOs can be classified into low and high frequency types, such that the low-frequency quasi-periodic oscillations (LFQPOs) have frequencies $\sim \rm m Hz$ and the high-frequency quasi-periodic oscillations (HFQPOs) have frequencies $\sim$ of hundreds of Hz, for stellar-mass BH sources \cite{vanderKlis:2000ca,Maselli:2014fca}. 
In this work, we will be dealing with HFQPOs in BH sources \cite{Torok:2004xs,Abramowicz:2011xu,Aschenbach:2004kj,Kotrlova:2017wyq,Torok:2011qy}.
The HFQPOs in LMXRBs involve timescales $\sim 0.1–1~ \rm ms$, which are close to the dynamical time scales of accreting matter near the vicinity ($r<10~R_{\rm g}$) of these compact objects \cite{1971SvA....15..377S,1973SvA....16..941S,PhysRevLett8217}. These timescales can be directly inferred from the fact that the characteristic velocities of accreting fluids near these compact objects scale as, $v\sim \sqrt{(GM/r)}$, such that the dynamical timescale becomes, $t_{\rm d}\sim r/v \sim\sqrt{(r^{3}/GM)}$. Thus, for a $1.4~M_\odot$ NS at $r\sim 15~\textrm{km}$ the dynamical timescale turns out to be $t_{\rm d}\sim ~0.1~\rm ms$, while for a $10~M_\odot$ BH at $r\sim 100~\textrm{km}$ $t_{\rm d}\sim 1~\textrm{ms}$ \cite{2006csxs.book.....L,vanderKlis:2000ca}.
Dynamical timescales $\sim  \rm ms$ for stellar-mass BHs \cite{1971SvA....15..377S,1973SvA....16..941S} translate to hundreds of Hz in the frequency domain, in accordance with the HFQPO observations \cite{2006csxs.book.....L,vanderKlis:2000ca}. Thus, HFQPOs provide unique opportunities to investigate the physics of strong gravity near BHs \cite{vanderKlis:2000ca}.
HFQPO frequencies scale inversely with the mass of the central compact object since they can be related to the orbital and epicyclic frequencies associated with the motion of matter in the vicinity of the central object. Hence, for neutron stars, the HFQPOs are $\sim $kHz while for supermassive BHs like Sgr A*, these are $\sim \rm mHz$. 
They are interesting as they are associated with the motion of matter in the vicinity of the central compact object and hence are believed to hold a wealth of information regarding the nature of gravity in the high curvature regime \cite{PhysRevLett8217}.

In order to explain the HFQPO data in black holes, several theoretical models have been proposed \cite{Stella:1997tc,PhysRevLett8217,Stella_1999,Cadez:2008iv,Kostic:2009hp,Germana:2009ce,Kluzniak:2002bb,Abramowicz:2003xy,Rebusco:2004ba,Nowak:1996hg,Torok:2010rk,Torok:2011qy,Kotrlova:2020pqy,1980PASJ...32..377K,Perez:1996ti,Silbergleit:2000ck,Dexter:2013sxa,Rezzolla:2003zy,Rezzolla:2003zx}, which are based on the study of orbital and epicyclic frequencies of test particles around these compact objects.
By comparing the model-dependent QPO frequencies with the HFQPO data, we constrain the spin and the dilaton charge of these black holes,
which when compared with the previous spin estimates, provide an understanding of the observationally favored HFQPO models for each of these sources. We also note that the present data cannot rule out the EMDA scenario in favor of GR. 
This study therefore, provides a possible test-bed for string-inspired models. We comment on the implications and limitations of our findings and discuss how the present constraints compare with the earlier estimates. The paper is organized as follows: in the next section, we briefly discuss the framework of EMDA gravity and the Kerr-Sen black hole solution, following which we summarize the motion of massive test particles in such a spacetime in \ref{S3}, on which the different QPO models are based. \ref{S5} provides a brief description of each of these models and how these models compare with the HFQPO observations of black holes. In \ref{S6-1} we report constraints on the model parameters by using MCMC simulations and we conclude with a summary of our results in \ref{S7}. We work with metric convention (-,+,+,+) and consider geometrized units, i.e. $G=1=c$.


\section{Rotating Black Holes in Einstein-Maxwell dilaton-axion gravity}
\label{S2}
String inspired models can potentially replace general relativity (GR) in the strong gravity regime where quantum effects of gravity are expected to be operational. The Einstein-Maxwell-dilaton-axion (EMDA) gravity is one such string inspired model which arises in the low energy effective action of superstring theories \cite{Sen:1992ua}. When the 10-dimensional heterotic string theory is compactified on a six dimensional torus $T^6$, the 4-dimensional effective action consists of $N=4$ super Yang-Mills theory coupled to $N=4$, $d=4$ supergravity, which after suitable truncation leads to a pure supergravity theory exhibiting $S$ and $T$ dualities. The Einstein-Maxwell-dilaton-axion theory is the bosonic sector of this supergravity theory coupled to the $U(1)$ gauge field \cite{Rogatko:2002qe}.

The four-dimensional effective action of EMDA gravity is given by \cite{Campbell:1992hc},:

\begin{align}
S = \frac{1}{16\pi} \int \sqrt{-g} \, d^4x \left(R - 2\partial_\mu\chi\partial^\mu\chi - \frac{1}{3}H_{\mu\alpha\beta}H^{\mu\alpha\beta} + e^{-2\chi}F_{\alpha\beta}F^{\alpha\beta}\right), 
\label{S2-1}
\end{align}
Here, $R$ and $g$ respectively denote the Ricci scalar and the determinant associated with the 4-dimensional metric tensor $g_{\mu\nu}$. The dilaton field is denoted by $\chi$ while 
$F_{\mu\nu}=\nabla_{\mu}A_{\nu}-\nabla_{\nu}A_{\mu}$ is the second rank antisymmetric field strength tensor associated with the $U(1)$ gauge field $A_\mu$. Additionally,  $H_{\rho\sigma\delta}$ denotes the  field strength tensor associated with the Kalb-Ramond field $B_{\rho\sigma}$, such that,
\begin{eqnarray}
H_{\rho \sigma \delta} = \nabla_{\rho} B_{\sigma \delta} + \nabla_{\sigma} B_{\delta \rho} + \nabla_{\delta} B_{\rho \sigma} -(A_{\rho} B_{\sigma \delta} + A_{\sigma} B_{\delta \rho} + A_{\delta} B_{\rho \sigma})
\label{S2-2}
\end{eqnarray}
In \ref{S2-2} the cyclic permutation of $A_\mu$ with $B_{\mu\nu}$ is known as the Chern-Simons term.
In four dimensions, $H_{\rho\sigma\delta}$ can be written in terms of the pseudo-scalar axion field $\psi$ which is its Hodge-dual,
\begin{eqnarray}
H_{\rho \sigma \delta} = \frac{1}{2}e^{4\chi} \epsilon_{\rho \sigma \delta \gamma} \partial^\gamma \psi
\label{S2-3}
\end{eqnarray}
In terms of the axion field, the action takes the following form,
\begin{eqnarray}
S = \frac{1}{16\pi} \int \sqrt{-g} \, d^4x \left( R - 2 \partial_\nu \chi \partial^\nu \chi - e^{4\chi} \partial_\nu \psi \partial^\nu \psi + e^{-2\chi} F_{\rho\sigma} F^{\rho\sigma} + \psi F_{\rho\sigma} {F}^{\ast\rho\sigma} \right)
\label{S2-4}
\end{eqnarray}
where ${F}^{\ast\rho\sigma}$ is the dual of ${F}_{\rho\sigma}$.

The equation of motion corresponding to the axion field $\psi$, obtained by varying the action \ref{S2-4} with respect to it is given by:
\begin{equation} 
\nabla_\mu\nabla^\mu\psi + 4\nabla_\nu\psi\nabla^\nu\psi - e^{-4\chi}F_{\rho\sigma}F^{\ast\rho\sigma} = 0,
\label{S2-5}
\end{equation}
whereas the equation of motion for the dilaton field is:
\begin{equation} 
\nabla_\mu\nabla^\mu\chi - \frac{1}{2}e^{4\chi}\nabla_\mu\psi\nabla^\mu\psi + \frac{1}{2}e^{-2\chi}F_{\rho\sigma}F^{\rho\sigma} = 0,
\label{S2-6}
\end{equation}
The Maxwell equations with dilaton and axion field couplings can be expressed as follows,
\begin{equation} 
\nabla_\mu(e^{-2\chi}F^{\mu\nu} + \psi F^{\ast\mu\nu}) = 0, 
\label{S2-7}
\end{equation}
\begin{equation} 
\nabla_\mu F^{\ast\mu\nu} = 0, 
\label{S2-8}
\end{equation}
Solving the coupled equations \ref{S2-6}, \ref{S2-7} and \ref{S2-8}, the solutions of the axion, dilaton and the $U(1)$ gauge fields are obtained \cite{Ganguly:2014pwa,Sen:1992ua,Rogatko:2002qe},

\begin{align}
\psi &= \frac{q^{2}}{M}\frac{a\cos\theta}{r^{2} + a^{2}\cos^{2}\theta} 
\label{S2-9}\\
e^{2\chi} &= \frac{r^{2} + a^{2}\cos^{2}\theta}{r(r + r_{2}) + a^{2}\cos^{2}\theta}\label{S2-10}\\
A & =\frac{qr}{\Sigma}\bigg(-dt +a \mathrm{sin}^2\theta d\phi\bigg)
\label{S2-11}
\end{align}
Here, the Einstein's equations assume the form,
\begin{eqnarray}
G_{\mu\nu}=T_{\mu\nu}
\label{S2-12}
\end{eqnarray}
where, $G_{\mu\nu}$ is the Einstein tensor constructed with the metric $g_{\mu\nu}$ while $T_{\mu\nu}$ is the energy-momentum tensor associated with the fields $\psi$, $\chi$ and $A_\mu$. Here, $T_{\mu\nu} $ takes the form,
\begin{align}
T_{\mu\nu}= e^{2\chi} \left(4F_{\mu\rho} F_{\rho}^{\nu} - g_{\mu\nu} F^{\alpha\beta}F_{\alpha\beta} \right) - g_{\mu\nu} \left(2\partial_\gamma \chi \partial^\gamma \chi + \frac{1}{2}e^{4\chi} \partial_\gamma \psi \partial^\gamma \psi \right) + \partial_\mu \chi \partial_\nu \chi + e^{4\chi} \partial_\mu \psi \partial_\nu \psi
\label{S2-13}
\end{align}
\begin{figure}[t!]
    \centering
    \includegraphics[width=0.5\linewidth]{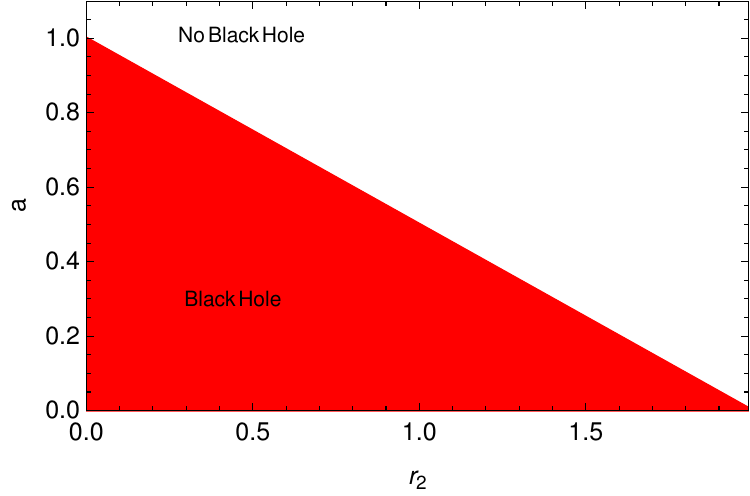}
    \caption{In the above figure, the red shaded region depicts the allowed range of spin corresponding to a given $r_2$ for the Kerr-Sen metric to represent a black hole.}
    \label{fig1}
\end{figure}
The stationary, axisymmetric black hole solution of the Einstein field equations is represented by the Kerr-Sen metric which in the Boyer-Lindquist coordinates has the following form \cite{Sen:1992ua,Garcia:1995qz}:
\begin{equation}
ds^2=-\Big(1-\frac{2Mr}{\Sigma}\Big)dt^2+\frac{\Sigma}{\Delta}(dr^2+\Delta d\theta^2)-\frac{4aMr}{\Sigma}sin^2\theta dtd\phi+sin^2\theta d\phi^2\Big[r(r+r_2)+a^2+\frac{2Mra^2sin^2\theta}{\Sigma}\Big]
\label{S2-14}
\end{equation} 
where $\Sigma = r(r + r_2) + a^2\cos^2\theta $ and $\Delta = r(r + r_2) - 2Mr+a^2$.
This metric defines the spacetime geometry surrounding a charged, rotating black hole in EMDA gravity with mass $M$. Here, $r_{2} = \frac{q^{2}}{{M}}e^{2\chi_{0}}$ is the dilaton parameter associated with the $U(1)$ gauge charge $q$ of the black hole and the asymptotic dilatonic field $\chi_{0}$. Note that, this charge stems from the coupling between the axion, dilaton  and the $U(1)$ gauge field as all the fields vanish in the event $q$ is zero (\ref{S2-9},  \ref{S2-10} and \ref{S2-11}). Further, \ref{S2-9} tells us that the rotation of the black hole is sourced by the axion field as when $q$ is non-vanishing but $a$ is vanishing the axion field strength becomes zero. The static black hole corresponding to \ref{S2-14} corresponds to a dilaton black hole characterized by its mass and dilaton charge $r_2$ \cite{Garfinkle:1990qj,Yazadjiev:1999xq}.
The event horizon $r_h$ for the Kerr-Sen black hole can be determined by solving $\Delta=0$ so that,
\begin{eqnarray}
 r_h=M-\frac{r_2}{2}\pm \sqrt{\Big(M-\frac{r_2}{2}\Big)^2-a^2}
 \label{S2-15}
 \end{eqnarray}
where we take the positive root.
From the above equation, we note that there is a real, positive value of the event horizon only if $0 \leq \frac{r_2}{M} \leq 2$. The same condition tells us that for a given $r_2$, the black hole cannot achieve an angular momentum $a>M-\frac{r_2}{2}$, in which case it becomes a naked singularity. This is clearly elucidated in \ref{fig1} which illustrates the allowed values of $r_2$ and $a$ (marked by the red shaded region) such that the Kerr-Sen metric represents a black hole.
In the rest of the paper, we consider $r_2$ and $a$ to be dimensionless, i.e., we scale $r_2 \equiv \frac{r_2}{R_g}$ and $a \equiv \frac{a}{R_g}$, where $R_g=GM/c^2$ is the gravitational radius of the black hole. 

\section{Models to explain the twin-peak High-Frequency Quasi-Periodic Oscillations in black holes}
\label{S3}
The stationary, axisymmetric spacetime described by the Kerr-Sen metric has the general form:
\begin{align}
ds^2=g_{tt}dt^2+2g_{t{\phi}}dtd{\phi}+g_{\phi\phi}d{\phi}^2+g_{rr}dr^2+g_{\theta\theta}d{\theta}^2
\label{S3-1}
\end{align}
where $g_{\nu\mu}\equiv g_{\nu\mu}(r,{\theta})$ and the metric has reflection symmetry about the equatorial plane. Since the spacetime is independent of $t$ and $\phi$, $\partial_t$ and $\partial_\phi$ are the Killing vectors and the specific energy $E$ and the specific angular momentum $L$ of test particles are conserved. 
Here, we consider the motion of massive test particles around such a spacetime. From \ref{S3-1} it is evident that the Lagrangian
associated with the motion of massive test particles is,
\begin{align}
\mathcal{L}=g_{tt}\dot{t}^2+2g_{t\phi}\dot{t}\dot{\phi}+g_{\phi\phi}\dot{\phi}^2+g_{rr}\dot{r}^2+g_{\theta\theta}\dot{\theta}^2
\label{S3-2}
\end{align}
From the constraint condition $g_{\mu\nu}u^\mu u^\nu=-1$, it can be shown that,
\begin{align}
    g_{rr} \dot{r}^2 + g_{\theta\theta} \dot{\theta}^2 + E^2 U(r, \theta) = -1
\label{S3-3}
\end{align}
where, 
\begin{align}
U(r,\theta)=g^{tt}-2\Big(\frac{L}{E}\Big)g^{t\phi}+\Big(\frac{L}{E}\Big)^{2}g^{\phi\phi}
\label{S3-4}
\end{align}

As we are considering the motion of the test particles to be confined in the equatorial plane, the above equation gives, $\dot{r}^2=V(r)$ where $V(r)=-\frac{1+E^{2}U(r_c,\frac{\pi}{2})}{g_{rr}}$.
The innermost stable circular orbit (ISCO) $r_{ms}$, can be obtained by solving $V(r)=0$, $V^{\prime}(r)=0$ and $V^{\prime\prime}(r)=0$, which for the Kerr-Sen spacetime is a function of $r_2$ and $a$. This is also the marginally stable circular orbit for the test particles.
As the particle is rotating around a black hole, it has an angular velocity $\Omega=\frac{u^{\phi}}{u^t}$ where, $u^{\mu} = \dot{x}^{\mu} = \frac{dx^{\mu}}{d\tau}$. 
From the radial Euler-Lagrangian equation and by considering circular, equatorial geodesics, it can be shown that
\begin{align}
\frac{\partial g_{tt}}{\partial r}+2\Omega\frac{\partial g_{t\phi}}{\partial r}+\Omega^{2}\frac{\partial g_{\phi\phi}}{\partial r}=0
\label{S3-5}
\end{align}
solving which we get $\Omega$,
\begin{align}
\Omega=\frac{-\partial_{r}g_{t\phi}\pm\sqrt{(\partial_{r}g_{t\phi})^{2}-(\partial _{r}g_{\phi\phi})(\partial_{r}g_{tt})}}{\partial _{r}g_{\phi\phi}}
\label{S3-6}
\end{align}
where the `+' sign corresponds to prograde and the `-' sign corresponds to retrograde orbits.
From \ref{S3-6} one can obtain the frequency of circular motion $f_\phi=\Omega/2\pi $.

\begin{figure}[t!]
    \centering
    \hspace{-2.cm}
    \begin{subfigure}{0.33\textwidth}
        \includegraphics[width=\linewidth]{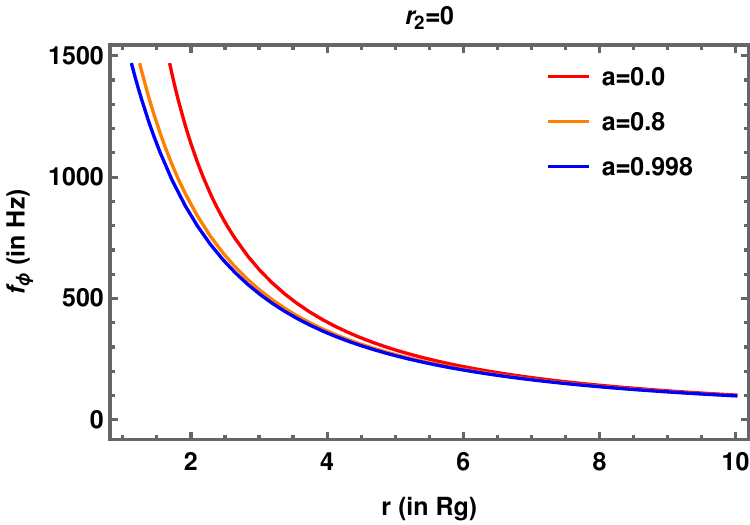}
        \caption{}
        \label{fig:subfig1}
    \end{subfigure}
    \hspace{0.02\textwidth} 
    \begin{subfigure}{0.33\textwidth}
        \includegraphics[width=\linewidth]{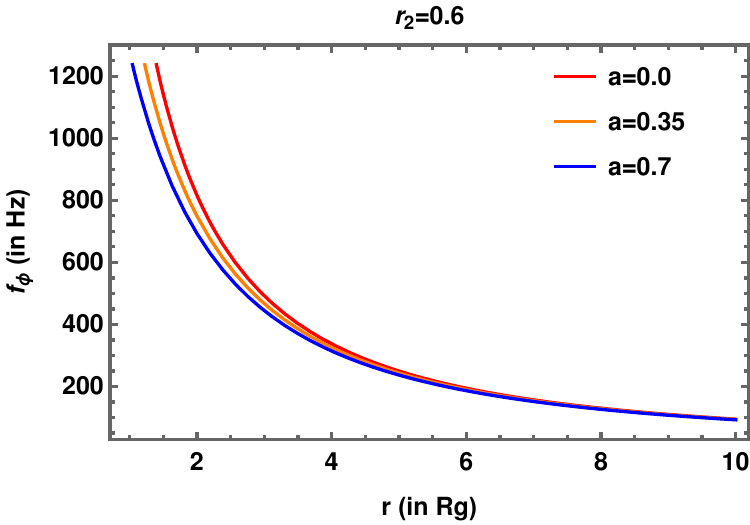}
        \caption{}
        \label{fig:subfig2}
    \end{subfigure}
    \hspace{0.02\textwidth} 
    \begin{subfigure}{0.33\textwidth}
        \includegraphics[width=\linewidth]{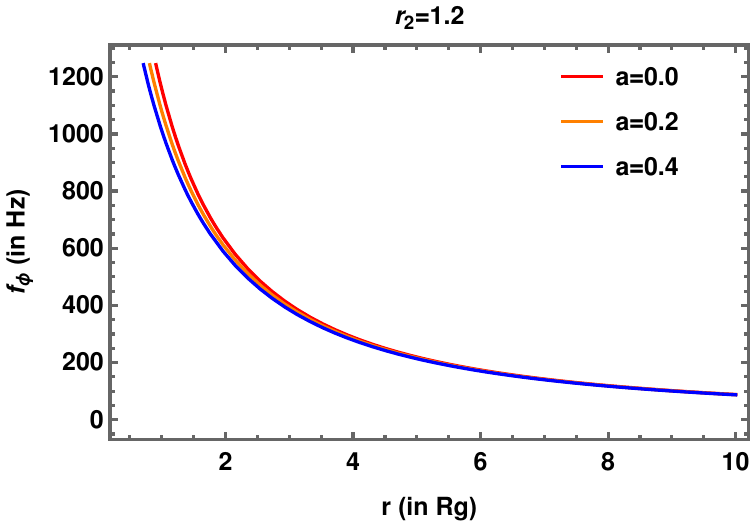}
        \caption{}
        \label{fig:subfig3}
    \end{subfigure}
    \hspace*{-2.3cm}
    \begin{subfigure}{0.33\textwidth}
        \includegraphics[width=\linewidth]{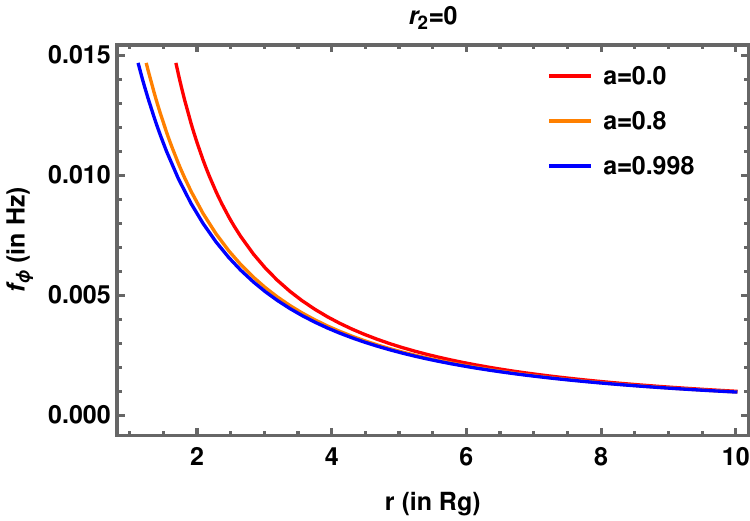}
        \caption{}
        \label{fig:subfig4}
    \end{subfigure}
    \hspace{0.02\textwidth} 
    \begin{subfigure}{0.33\textwidth}
        \includegraphics[width=\linewidth]{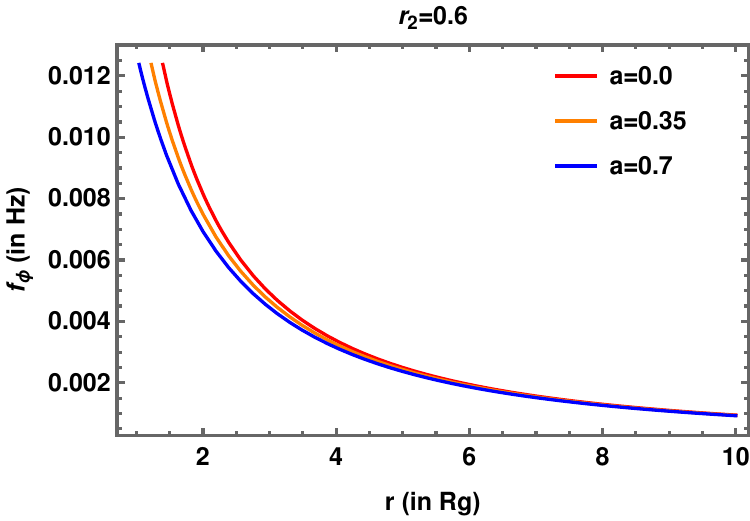}
        \caption{}
        \label{fig:subfig5}
    \end{subfigure}
    \hspace{0.02\textwidth} 
    \begin{subfigure}{0.33\textwidth}
        \includegraphics[width=\linewidth]{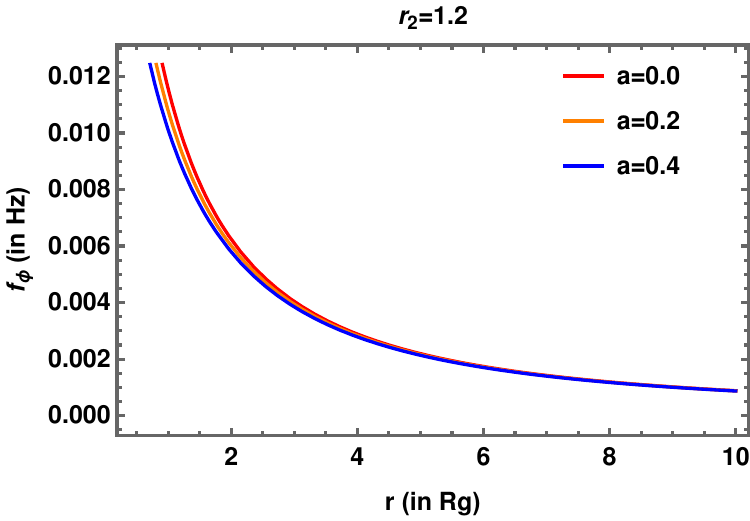}
        \caption{}
        \label{fig:subfig6}
    \end{subfigure}
    \caption{The above figure shows the radial variation of $f_{\phi}$ for (a) $r_2=0$, (b) $r_2=0.6$ and (c) $r_2= 1.2$ for a $M=10 M_\odot$ BH and (d) $r_2=0$, (e) $r_2=0.6$ and (f) $r_2= 1.2$ for a $M=10^6 M_\odot$ BH. In each sub-figure, the radial variation of $f_{\phi}$ is elucidated for three choices of spin, $a=0$ (the non-rotating case denoted by the red curve), $a=0.8$ for $r_2=0$ (denoted by the orange curve) and $a=\frac{1}{2}(1-\frac{r_2}{2})$ for non-zero $r_2$ (denoted by the orange curve), and $a \approx a_{max}=1-\frac{r_2}{2}$ (the maximal spin denoted by the blue curve). }
    \label{fig2}
\end{figure}
In \ref{fig2} we consider radial variation of $f_{\phi}$ for $r_2=0$, $r_2=0.6$ and $r_2= 1.2$ for different choices of spin, $a=0$ (red), $a=0.8$ for $r_2=0$ and $a=\frac{1}{2}(1-\frac{r_2}{2})$ for non-zero $r_2$ (orange), and $a\approx a_{max}=1-\frac{r_2}{2}$ (blue). The figures are plotted from the marginally stable circular orbit $r_{ms}$ to $10 R_g$.
The upper panel is plotted for a $M=10 M_\odot$ BH, while the lower panel is plotted for a $M=10^6 M_\odot$ BH.
The figure elucidates that for a given dilaton charge, the orbital frequency increases as we move closer to the black hole and decreases as the black hole spin increases. Also, with an increase in the dilaton charge, the orbital frequency decreases.

\begin{figure}[t!]
    \centering
    \hspace{-2.cm}
    \begin{subfigure}{0.33\textwidth}
        \includegraphics[width=\linewidth]{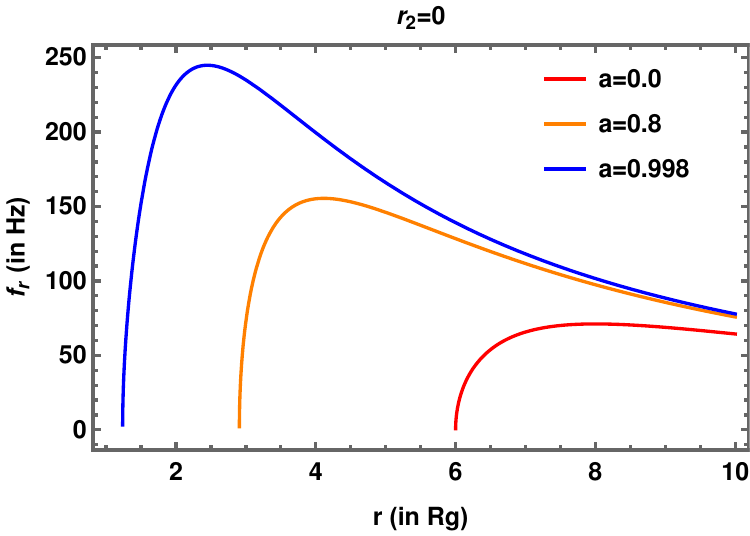}
        \caption{}
        \label{fig:subfig1}
    \end{subfigure}
    \hspace{0.02\textwidth} 
    \begin{subfigure}{0.33\textwidth}
        \includegraphics[width=\linewidth]{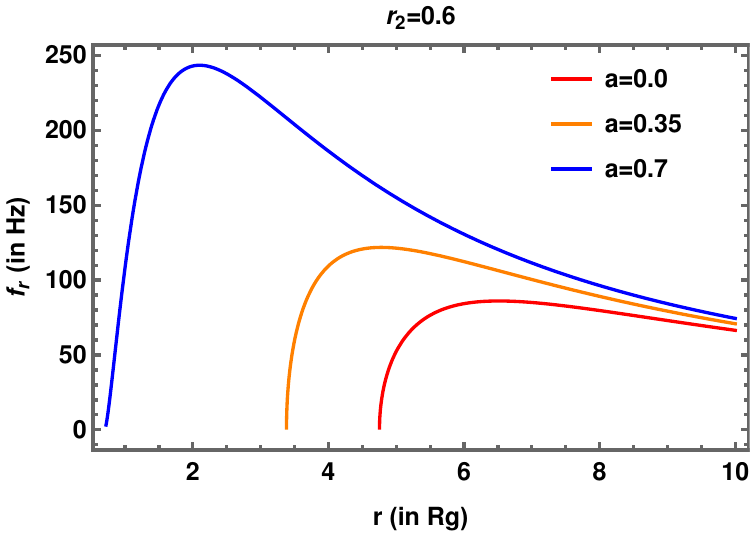}
        \caption{}
        \label{fig:subfig2}
    \end{subfigure}
    \hspace{0.02\textwidth} 
    \begin{subfigure}{0.33\textwidth}
        \includegraphics[width=\linewidth]{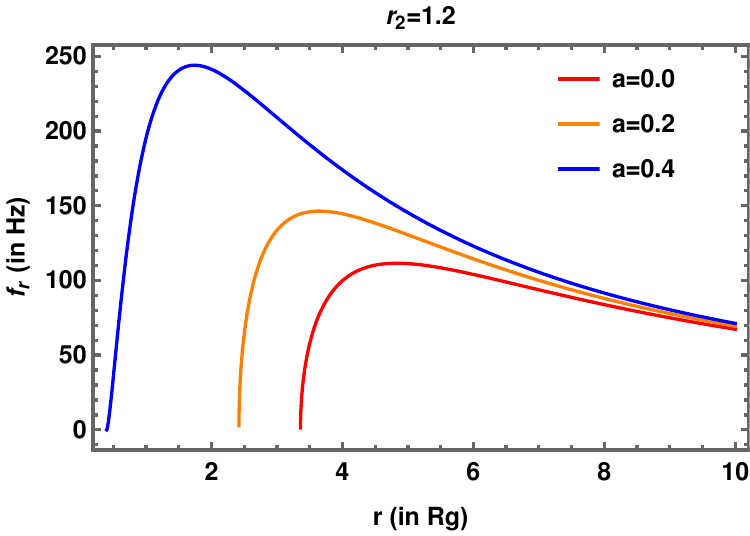}
        \caption{}
        \label{fig:subfig3}
    \end{subfigure}
    \hspace*{-2.3cm}
    \begin{subfigure}{0.33\textwidth}
        \includegraphics[width=\linewidth]{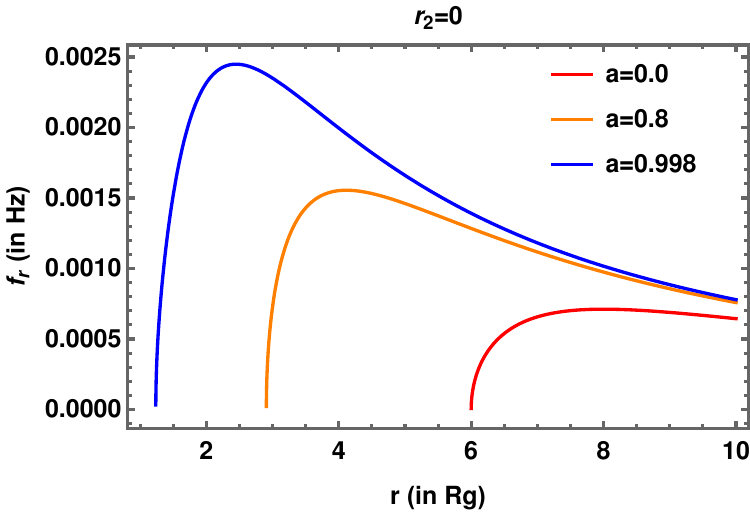}
        \caption{}
        \label{fig:subfig4}
    \end{subfigure}
    \hspace{0.02\textwidth} 
    \begin{subfigure}{0.33\textwidth}
        \includegraphics[width=\linewidth]{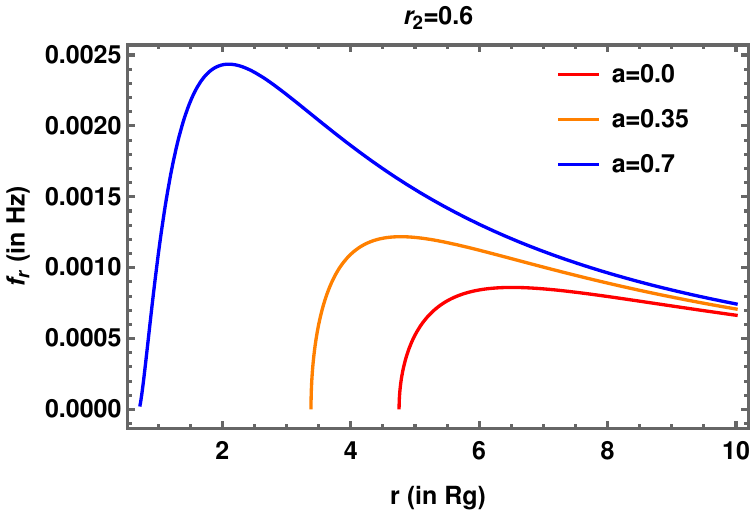}
        \caption{}
        \label{fig:subfig5}
    \end{subfigure}
    \hspace{0.02\textwidth} 
    \begin{subfigure}{0.33\textwidth}
        \includegraphics[width=\linewidth]{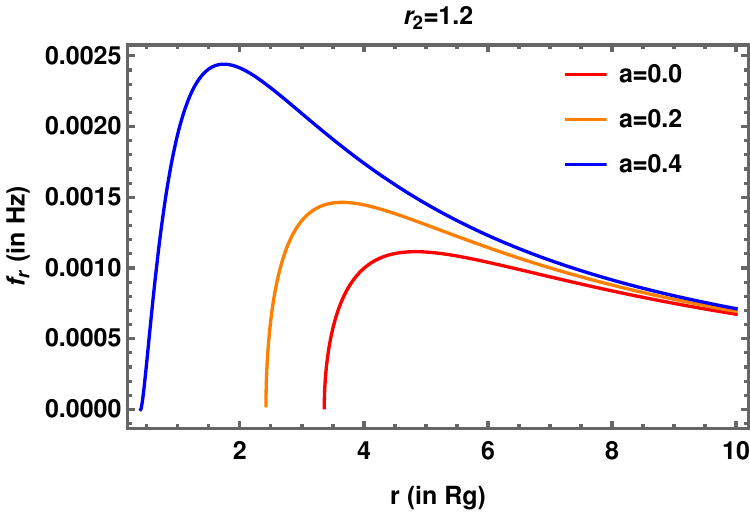}
        \caption{}
        \label{fig:subfig6}
    \end{subfigure}
    \caption{The above figure shows the radial variation of $f_{r}$ for (a) $r_2=0$, (b) $r_2=0.6$ and (c) $r_2= 1.2$ for a $M=10 M_\odot$ BH and (d) $r_2=0$, (e) $r_2=0.6$ and (f) $r_2= 1.2$ for a $M=10^6 M_\odot$ BH. In each sub-figure, the radial variation of $f_{r}$ is elucidated for three choices of spin, $a=0$ (the non-rotating case denoted by the red curve), $a=0.8$ for $r_2=0$ (denoted by the orange curve) and $a=\frac{1}{2}(1-\frac{r_2}{2})$ for non-zero $r_2$ (denoted by the orange curve), $a \approx a_{max}=1-\frac{r_2}{2}$ (the maximal spin denoted by the blue curve). }
    \label{nur}
\end{figure}

\begin{figure}[htp]
    \centering
    \hspace{-2.cm}
    \begin{subfigure}{0.33\textwidth}
        \includegraphics[width=\linewidth]{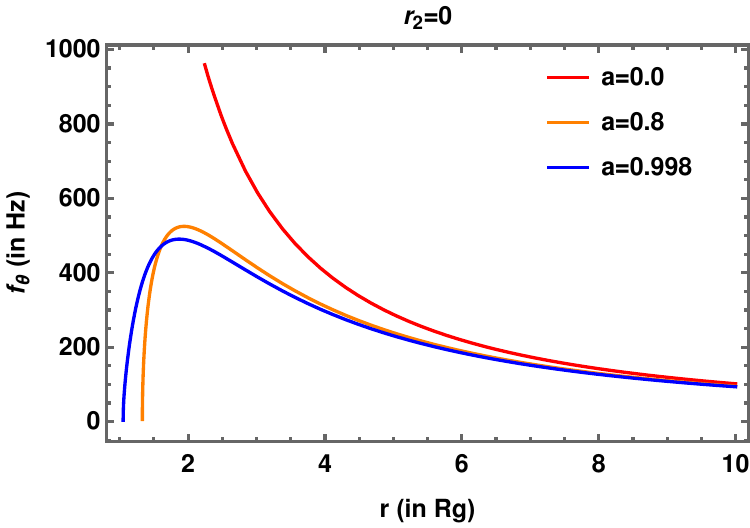}
        \caption{}
        \label{fig:subfig1}
    \end{subfigure}
    \hspace{0.02\textwidth} 
    \begin{subfigure}{0.33\textwidth}
        \includegraphics[width=\linewidth]{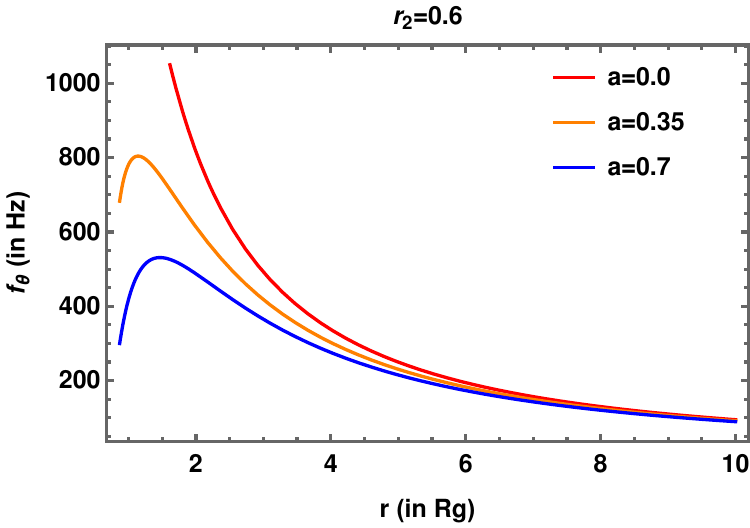}
        \caption{}
        \label{fig:subfig2}
    \end{subfigure}
    \hspace{0.02\textwidth} 
    \begin{subfigure}{0.33\textwidth}
        \includegraphics[width=\linewidth]{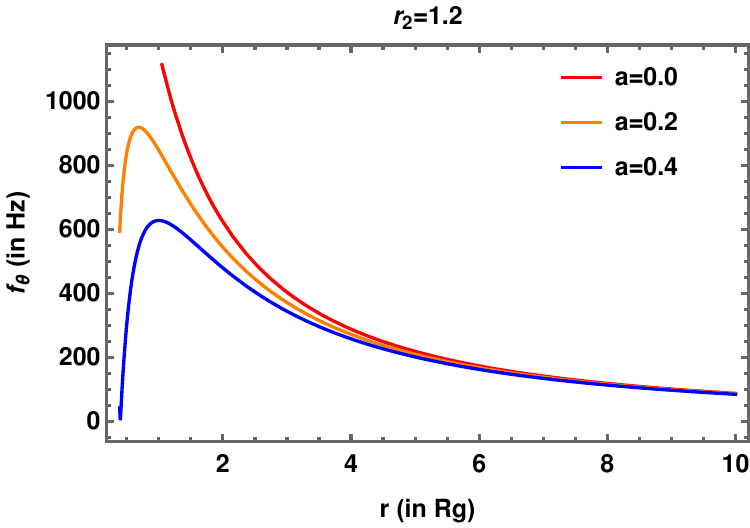}
        \caption{}
        \label{fig:subfig3}
    \end{subfigure}
    \hspace*{-2.3cm}
    \begin{subfigure}{0.33\textwidth}
        \includegraphics[width=\linewidth]{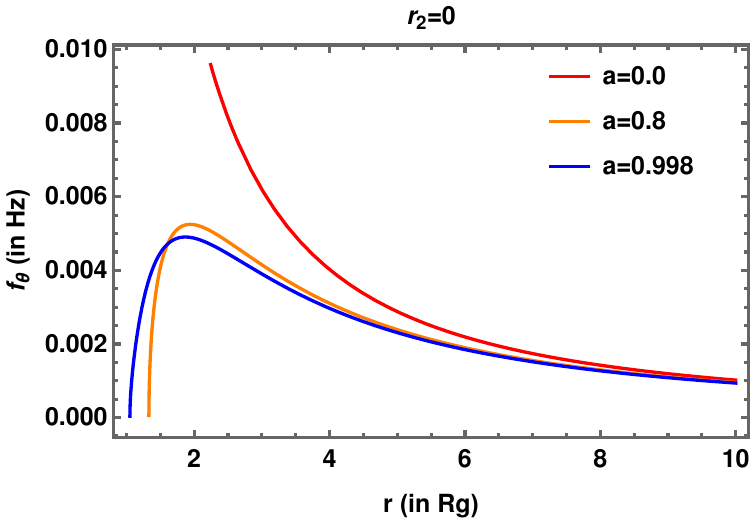}
        \caption{}
        \label{fig:subfig4}
    \end{subfigure}
    \hspace{0.02\textwidth} 
    \begin{subfigure}{0.33\textwidth}
        \includegraphics[width=\linewidth]{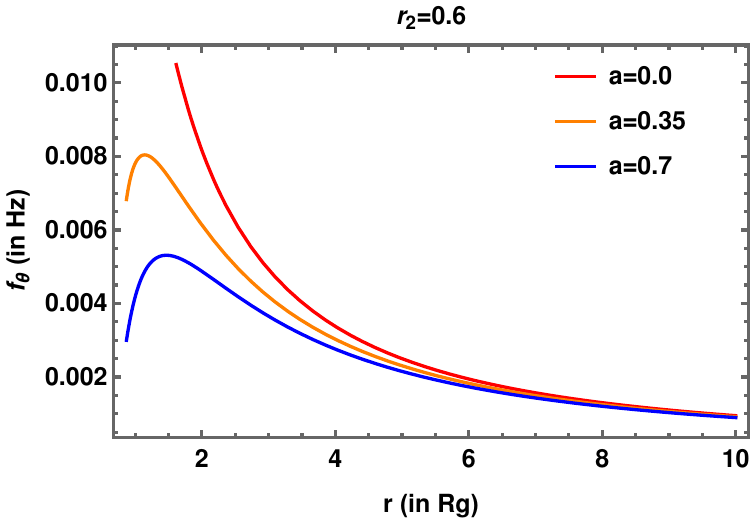}
        \caption{}
        \label{fig:subfig5}
    \end{subfigure}
    \hspace{0.02\textwidth} 
    \begin{subfigure}{0.33\textwidth}
        \includegraphics[width=\linewidth]{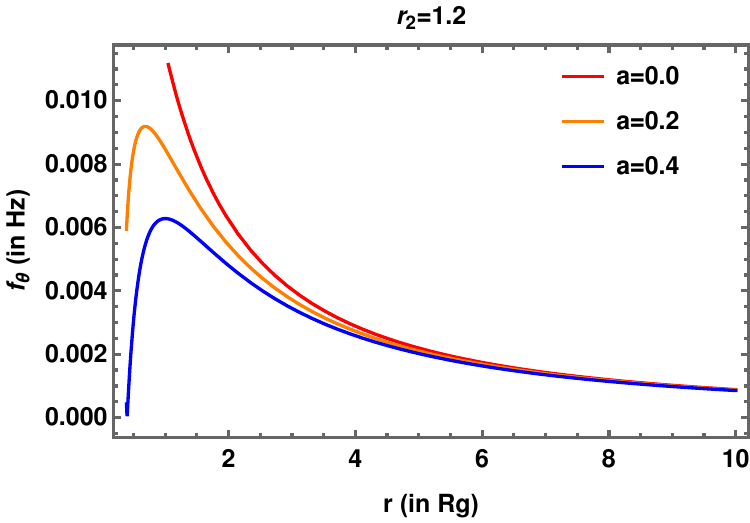}
        \caption{}
        \label{fig:subfig6}
    \end{subfigure}
    \caption{The above figure shows the radial variation of $f_{\theta}$ for (a) $r_2=0$, (b) $r_2=0.6$ and (c) $r_2= 1.2$ for a $M=10 M_\odot$ BH and (d) $r_2=0$, (e) $r_2=0.6$ and (f) $r_2= 1.2$ for a $M=10^6 M_\odot$ BH. In each sub-figure, the radial variation of $f_{\theta}$ is elucidated for three choices of spin, $a=0$ (the non-rotating case denoted by the red curve), $a=0.8$ for $r_2=0$ (denoted by the orange curve) and $a=\frac{1}{2}(1-\frac{r_2}{2})$ for non-zero $r_2$ (denoted by the orange curve), $a \approx a_{max}=1-\frac{r_2}{2}$ (the maximal spin denoted by the blue curve). }
    \label{nutheta}
\end{figure}

We now consider slightly perturbing the motion of particles in a stable, circular, equatorial orbit both radially and vertically.
We can show the perturbation by the following equations
\begin{align}
r(t)= r_c+\delta r ;\hspace{2mm}  \delta r \sim e^{i\omega_r t} ; \hspace{2mm}  \delta r<<r  \nonumber \\
\theta(t)=\frac{\pi}{2}+\delta\theta; \hspace{2mm}    \delta\theta \sim e^{i\omega_{\theta}t};\hspace{2mm}         \delta\theta<< \frac{\pi}{2}
\label{S3-7}
\end{align}
where $f_r=\omega_r/2\pi$ and $f_\theta=\omega_\theta/2\pi$ correspond to the radial and vertical epicyclic frequencies, respectively.
Substituting \ref{S3-7} in the constraint condition $g_{\mu\nu}u^{\mu}u^{\nu}=-1$, we get,
\begin{align}
-g_{rr}(u^{0}2\pi f_{r}\delta r)^2-g_{\theta\theta}(u^{0} 2\pi f_{\theta})^2+E^{2}\Big[U\bigg(r_c,\frac{\pi}{2}\bigg)+\frac{1}{2}\frac{\partial^2 U}{\partial r^2}\bigg(r_c,\frac{\pi}{2}\bigg)\delta r^2+\frac{1}{2}\frac{\partial^2 U}{\partial\theta^2}\bigg(r_c,\frac{\pi}{2}\bigg)\delta\theta^2\Big]=0 
\label{S3-8}
\end{align}
We assume the oscillations in the radial and vertical directions to be uncoupled which allows us to equate the coefficients of $\delta r^2$ and $\delta\theta^2$ to zero, which yields,
\begin{align}
f_{r}^2=\frac{c^6}{G^2M^2}\Bigg[\frac{(g_{tt}+g_{t\phi}\Omega)^2}{2(2\pi)^{2}g_{rr}}\bigg(\frac{\partial^2{U}}{\partial r^2}\bigg)_{r_c,\frac{\pi}{2}}\Bigg]
\label{S3-9}
\end{align}
\begin{align}
f_{\theta}^2=\frac{c^6}{G^2M^2}\Bigg[\frac{(g_{tt}+g_{t\phi}\Omega)^2}{2(2\pi)^{2}g_{\theta\theta}}\bigg(\frac{\partial^2{U}}{\partial\theta^2}\bigg)_{r_c,\frac{\pi}{2}}\Bigg]
\label{S3-10}
\end{align}
From \ref{S3-6}, \ref{S3-9} and \ref{S3-10}, we note that the three fundamental frequencies depend on the metric parameters $r_2$ and $a$, the circular radius $r$ at which the frequencies are excited and the black hole mass $M$.
The quantities in the square bracket in \ref{S3-9} and \ref{S3-10} are dimensionless as $r_2$, $a$ and $r$ are all scaled with the black hole mass. To bring back the dimensions, a factor of $\frac{c^6}{G^2M^2}$ is multiplied in the RHS of \ref{S3-9} and \ref{S3-10}, which shows that the fundamental frequencies of oscillation associated with the motion of test particles in a curved background scale inversely with the mass of the black hole. 
In \ref{nur} and \ref{nutheta}, we respectively consider radial variation of $f_{r}$ and $f_{\theta}$ for $r_2=0$, $r_2=0.6$ and $r_2= 1.2$ for different choices of spin, $a=0$ (red), $a=0.8$ for $r_2=0$ and $a=\frac{1}{2}(1-\frac{r_2}{2})$ for non-zero $r_2$ (orange), and $a\approx a_{max}=1-\frac{r_2}{2}$ (blue). Once again, the figures are plotted from the marginally stable circular orbit $r_{ms}$ to $10 R_g$.
The upper panel is plotted for a $M=10 M_\odot$ BH, while the lower panel corresponds to a $M=10^6 M_\odot$ BH.
\ref{nur} elucidates that for a given dilaton charge and spin, the radial epicyclic frequency increases up to a maximum as we move closer to the black hole and then drops to zero at the $r_{ms}$ where $\frac{\partial^2 U (r,\frac{\pi}{2})}{\partial r^2}=0$. Further, we note that for a given $r_2$, $f_r$ increases with an increase in spin, and for a given spin, $f_r$ increases with an increase in $r_2$. The amount of increase in $f_r$ with $r_2$ is more for a lower spin than a higher one. 
From \ref{nutheta}, we note that the vertical epicyclic frequency $f_\theta$ increases monotonically as we move closer to the BH. Further, for a given $r_2$, $f_\theta$ decreases with an increase in spin, while for a given spin, $f_\theta$ increases with an increase in $r_2$.

We next discuss the various theoretical models existing in the literature which are proposed to explain the high frequency QPOs (HFQPOs) in black holes and neutron star sources. Before we discuss the physics associated with each of these models, we enlist them in \ref{Tab1}. The HFQPOs are observed  in commensurable pairs and interestingly the ratio between the so-called twin-peak HFQPOs is observed to be 3:2. From \ref{Tab1} we note that $f_1$ and $f_2$ are the model dependent HFQPOs while $f_3$ corresponds to the model dependent low frequency QPO (LFQPO). Interestingly, the model dependent QPO frequencies are linear combinations of the three fundamental frequencies $f_\phi$, $f_r$ and $f_\theta$ described above. The QPO models can be broadly classified into two categories, namely, the kinematic and the resonant models, where it is assumed that the model dependent HFQPOs are generated at the same circular radius $r$, which we denote by $r_{em}$ \cite{2016A&A...586A.130S,Yagi:2016jml,Kotrlova:2020pqy}. Apart from these, some diskoseismic models \cite{1980PASJ...32..377K,Perez:1996ti,Silbergleit:2000ck} also exist in the literature which assume that the oscillatory modes giving rise to the twin-peak QPOs are generated at different radii of the accretion disk. These models often fail to reproduce the twin-peak HFQPOs in the 3:2 ratio, as revealed by magnetohydrodynamic simulations \cite{Tsang:2008fz,Fu:2008iw,Fu:2010tf} and hence we do not discuss those models in this paper.
Note that, although these models are mainly proposed to explain the HFQPOs, the Relativistic Precession Model (RPM) does propound a mechanism to explain both the HFQPOs and the LFQPO simultaneously. 

We also enlist the black hole sources which exhibit the HFQPOs in their power spectrum in \ref{Tab2}. We have used  4 stellar-mass BHs and only 1 SMBH in our analysis because these are the BHs that exhibit twin-peak HFQPOs in their power spectrum. In this regard, we would like to mention that many BHs exhibit QPOs in their power spectrum, but only a handful of these BHs exhibit twin-peak HFQPOs in their power spectrum, which are listed in \ref{Tab2}. Since the present work focuses on deciphering the signatures of strong gravity, we consider only those sources where HFQPOs are observed \cite{Torok:2004xs,Abramowicz:2011xu,Aschenbach:2004kj,Kotrlova:2017wyq,Torok:2011qy}.
Note that, for stellar mass black holes these frequencies are in the range of hundreds of Hz while for supermassive black holes like Sgr A*, the HFQPOs are $\sim \rm mHz$ \cite{Torok:2004xs,Aschenbach:2004kj}. This change in the order of magnitude of frequencies can be attributed to the order of magnitude difference in the masses of these objects, (see, e.g. \ref{S3-9}, \ref{S3-10}).
The observed HFQPOs are denoted by $\nu_1$ and $\nu_2$ with $\Delta \nu_1$ and $\Delta \nu_2$ the associated errors, while $\nu_3$ corresponds to the LFQPO which incidentally is only observed in the source GRO J1655-40 along with the twin-peak HFQPOs. Interestingly, one can note from \ref{Tab2} that $\nu_1$ and $\nu_2$ are indeed in the 3:2 ratio.

\begin{table}[htp!]
    \centering
\begin{tabular}[width=\textwidth]{|c|c|c|c|}
\hline
     Models & $f_1$ & $f_2$ & $f_3$  \\
\hline
\hline
Relativistic Precession Model (kinematic)\cite{Stella:1998mq,stella1997lense,stella1999correlations}&$f_\phi$&$f_\phi-f_r$&$f_\phi-f_\theta$ \\
\hline
Tidal Disruption Model (kinematic)\cite{Cadez:2008iv,Kostic:2009hp,Germana:2009ce}&$f_\phi+f_r$&$f_\phi$&- \\
\hline
Parametric Resonance Model (resonance)\cite{Kluzniak:2002bb,Abramowicz:2003xy,Rebusco:2004ba}&$f_\theta$&$f_r$&-\\
\hline
Forced Resonance Model 1 (resonance)\cite{Kluzniak:2002bb} &$f_\theta$&$f_{\theta}-f_r$&-\\
\hline
Forced Resonance Model 2 (resonance)\cite{Kluzniak:2002bb}&$f_\theta+f_r$&$f_\theta$&- \\
\hline
Keplerian Resonance Model 1 (resonance)\cite{Nowak:1996hg} &$f_\phi$&$ f_r$&-\\
\hline
Keplerian Resonance Model 2 (resonance)\cite{Nowak:1996hg} &$f_\phi$&$2f_r$&-\\
\hline
Keplerian Resonance Model 3 (resonance)\cite{Nowak:1996hg} &$3f_r$&$f_\phi$&-\\
\hline
Warped Disk Oscillation Model (resonance)\cite{2001PASJ...53....1K,2004PASJ...56..559K,2004PASJ...56..905K,2005PASJ...57..699K,2008PASJ...60..111K} &$2f_\phi-f_r$&$2(f_\phi-f_r)$&-\\
\hline
Non-axisymmetric Disk Oscillation Model 1 (resonance) \cite{2004ApJ...617L..45B,2005AN....326..849B,2005ragt.meet...39B}&$f_\theta$&$f_\phi-f_r$&-\\
\hline
Non-axisymmetric Disk Oscillation Model 2 (resonance)\cite{Torok:2010rk,Torok:2011qy,Kotrlova:2020pqy} &$2f_\phi-f_\theta$&$f_\phi-f_r$&-\\
\hline
\end{tabular}
    \caption{In the above table we enlist the different theoretical models existing in previous literature to explain the HFQPOs in black hole sources.}
    \label{Tab1}
\end{table}

\begin{table}[h]
\begin{center}
\begin{tabular}{|c|c|c|c|c|}
\hline
$\rm Source $ & $\rm Mass$ & $ \nu_{1} \pm  \Delta \nu_{1}$ & $ \nu_{2} \pm {\Delta} \nu_{2}$ & $ \nu_3 \pm \Delta \nu_3$\\
& $\rm (\rm M_{\rm BH ~} in ~M_\odot)$ & $(\rm in~ Hz)$ & $\rm (\rm in ~Hz)$ & $\rm (\rm in~ Hz)$\\
\hline 
$\rm GRO ~J1655-40$ & $\rm 5.4\pm 0.3$ \cite{Beer:2001cg} & $\rm 441  \rm \pm 2 $ \cite{Motta:2013wga} & $\rm 298 \rm \pm 4 $ \cite{Motta:2013wga} & $\rm 17.3 \pm \rm 0.1 $ \cite{Motta:2013wga}\\ 
\hline
$\rm XTE ~J1550-564$ & $\rm 9.1\pm 0.61$ \cite{Orosz:2011ki} & $\rm 276 \rm \pm 3 $ & $\rm 184  \pm 5 $ & $ -$\\
\hline
$\rm GRS ~1915+105$ & $\rm 12.4^{+2.0}_{-1.8}$ \cite{Reid:2014ywa} & $\rm 168  \pm 3 $ & $\rm 113  \pm 5 $ & $\rm - $\\
\hline
$\rm H ~1743+322$ & $\rm 8.0-14.07$ \cite{Pei:2016kka,Bhattacharjee:2019vyy,Petri:2008jc} & $\rm 242 \pm 3 $ & $\rm 166  \pm 5 $ & $\rm - $\\
\hline
$\rm Sgr~A^*$ & $\rm (3.5-4.9)$ & $\rm (1.445 \pm 0.16)$ & $\rm (0.886 \pm 0.04)$ & $ - $\\
 & $\rm ~\times 10^6$ \cite{Ghez:2008ms,Gillessen:2008qv} & $\rm ~\times 10^{-3} $ \cite{Torok:2004xs,Stuchlik:2008fy} & $\rm ~\times 10^{-3} $ \cite{Torok:2004xs,Stuchlik:2008fy} & $ - $\\
\hline
\end{tabular}
\caption{The above table enlists the black hole sources where high frequency QPOs (HFQPOs) are observed.}
\label{Tab2}
\end{center}

\end{table}

\section{Constraining the dilaton charge of black holes with the HFQPO data}
\label{S5}
In this section, we compare the theoretical HFQPO frequencies from each of the eleven models with the observed HFQPO data of black holes listed in \ref{Tab2}.
We carry out a $\chi^2$ analysis, which in turn enables us to obtain the most favored dilaton charge from the HFQPO data. The $\chi^2$ function is defined as:
\begin{align}
\mathbf{\chi ^2_{i} (r_2)= \frac{\lbrace \nu_{\textrm{1}{,i}}-f_1(r_2,a_{\rm min},M_{\rm min},r_{\rm em,\rm min}) \rbrace ^2}{\sigma_{\nu_{\rm 1},i}^2}  
+  \frac{\lbrace \nu_{\textrm{2}{,i}}-f_2(r_2,a_{\rm min},M_{\rm min},r_{\rm em,\rm min}) \rbrace ^2}{\sigma_{\nu_{\rm 2}, i}^2}~,}
\label{S5-1}
\end{align}
which obtains the difference between the theoretical and the observed HFQPO frequencies. In \ref{S5-1}, $\nu_{1,i}$ and $\nu_{2,i}$ represent the observed upper and lower HFPQOs, respectively, for the $i$-th source listed in \ref{Tab2}. The uncertainties $\sigma_{\nu_{1},i}$ and $\sigma_{\nu_{2},i}$ correspond to the observational errors in $\nu_{1,i}$ and $\nu_{2,i}$, also reported in \ref{Tab2}. Theoretical frequencies $f_1$ and $f_2$ for each model are considered from \ref{Tab1}. These are linear functions of $f_r$, $f_\theta$ and $f_\phi$, which in turn depend on the dilaton charge $r_2$, the spin $a$, the emission radius $r_{em}$ and the mass of the black hole $M$. In order to obtain the observationally favored $r_2$, we minimize the $\chi^2$ in two stages \cite{1976ApJ...210..642A}. First, we choose a model, then we fix a source and a $r_2$ and vary the mass, spin, and $r_{em}$ to get the minimum $\chi^2_i$ (\ref{S5-1}), for that $r_2$. We repeat this procedure for all the allowed dilaton parameters $r_2$ (i.e., $0\leq r_2 \leq 2$, which is obtained from \ref{S2-15}). This gives us the variation of $\chi^2_i$ with $r_2$ for the  chosen source with respect to the selected model and enables us to obtain the $\chi^2_{i,min}$, which is the minimum of $\chi^2_i$ for that source, with respect to the given model. 
Below, we outline our method more elaborately:
\begin{enumerate}
\item First of all, we need to choose a QPO model from \ref{Tab1} which gives us the dependence of $f_1$ and $f_2$ on $r_2$, $a$, $M$ and $r_{em}$.
\item Then we select a source from \ref{Tab2}. 
\item We now fix a value of $r_2$ in the allowed range: $0\leq r_2 \leq 2$, such that the black hole is cloaked by an event horizon (see \ref{S2}, \ref{fig1}). For the chosen $r_2$, the spin can vary from $-a_{max}\leq a \leq a_{max}$, where $a_{max}=1-r_2/2$ (from \ref{S2-15}).
\item For the chosen $r_2$, the spin can be varied between $-a_{max}\leq a \leq a_{max}$, the mass between $(M-\Delta M) \leq M \leq (M + \Delta M)$ where $\Delta M$ is the error in the mass (obtained from \ref{Tab2}) and the $r_{em}$ between $r_{ms}(r_2,a)\leq r_{\rm em} \leq r_{ms}(r_2,a) + 20 R_{g}$, where $r_{em}$ is the radius at which the QPOs are generated, $R_g=GM/c^2$ is the gravitational radius and $r_{ms}$ is the radius of the marginally stable circular orbit obtained by solving for the roots of $\frac{\partial^2 U (r,\frac{\pi}{2})}{\partial r^2}=0$ (discussed in \ref{S3}).
\item For each combination of $M$, $a$, and $r_{em}$ at the given $r_2$, the $\chi^2_i$ is calculated for the chosen source. The values of $M$, $a$, and $r_{em}$ where $\chi^2_i$ minimizes correspond to the most favored mass, spin, and emission radius for the chosen $r_2$. These are denoted by $M_{min}$, $r_{em,min}$ and $a_{min}$ in \ref{S5-1}. 
\item We repeat steps 3-5 for all the values of $r_2$ in the allowed range. This gives us a variation of $\chi^2_i$ with $r_2$ for the chosen source, with respect to the selected model. The magnitude of $r_2$ where $\chi^2_i$ minimizes gives us the most favored $r_2$ for the $i^{th}$ source with respect to that model. This dilaton parameter is denoted by $r_{2, min}$ and the corresponding value of $\chi^2_i$ by $\chi^2_{i, min}$.\\
\item Steps 3-6 are repeated for the remaining sources, with the same model. 
\item The $M_{\rm min}$ and $a_{\rm min}$ corresponding to $r_{2,\rm min}$ are reported in \ref{Tab3} to \ref{Tab7}, for all the five sources and with respect to all the eleven models.
\item Now, the above procedure is done for the remaining ten models, which gives us $r_{2,\rm min}$ for each source and with respect to each model.
\item For models like RPM, which can also explain the LFQPO in GRO J1655-40, the $\chi^2$ function is given by,
\begin{align}
\label{S5-2}
\mathbf{\chi ^2_{j} (r_2)}&= \mathbf{\frac{\lbrace \nu_{\textrm{1}{,j}}-f_1(r_2,a_{\rm min},M_{\rm min},r_{\rm min}) \rbrace ^2}{\sigma_{\nu_{\rm 1}, j}^2}  
+  \frac{\lbrace \nu_{\textrm{2}{,j}}-f_2(r_2,a_{\rm min},M_{\rm min},r_{\rm min}) \rbrace ^2}{\sigma_{\nu_{\rm 2}, j}^2} }
\nonumber 
\\
&\hskip 4 cm \mathbf{+\frac{\lbrace \nu_{3,{\rm GRO}}-f_3(r_2,a_{\rm min},M_{\rm min},r_{\rm min}) \rbrace ^2}{\sigma_{\nu_{3},j}^2}~.}
\end{align}
\item In our work, we wish to obtain the observationally favored magnitude of $r_2$ from the observed QPO data, but the model-dependent QPO frequencies depend on $r_2$, $a$, $M$, and $r_{em}$. Thus, we divide the parameters into two classes: (a) the ``interesting parameters" (which here is $r_2$) and (b) the ``uninteresting parameters" (which are $a$, ${M}$, and $r_{\rm em}$). The ``uninteresting parameters" are derived from $\chi^2$ minimization for various choices of the ``interesting parameters" \cite{1976ApJ...210..642A}. With one interesting parameter, the confidence intervals (i.e., $\Delta \chi^{2}$ from $\chi^{2}_{\rm min}$) corresponding to 68\%, 90\% and 99\% confidence levels are equal to 1, 2.71, and 6.63 \cite{1976ApJ...210..642A}.

\end{enumerate}

In what follows, we will compare each of these models with the HFQPO data of the five black holes to obtain the observationally favored dilaton charge. In \ref{fig5}-\ref{fig15} we consider each model and plot the variation of $\rm{ln} \chi^{2}$ or $\chi^2$  with the dilaton charge $r_2$ for all the five black holes (in the left panel, i.e., \ref{fig5a}-\ref{fig15a}) and for a subset of the five BH sources where the values of $\chi^2$ are high, and the variation is also substantial, such that $r_2$ can be constrained within the 1-$\sigma$, 2-$\sigma$, and 3-$\sigma$ confidence intervals (in the right panel, i.e., \ref{fig5b}-\ref{fig15b}). We discuss this in greater detail in the next two subsections.

\subsection{Kinematic Models}
The kinematic models aim to explain the QPOs in terms of local motion of plasma blobs in the accretion disk orbiting the black hole near the marginally stable circular orbit $r_{ms}$.
The most well studied kinematic models are the Relativistic Precession Model (RPM) and the Tidal Disruption Model (TDM). We discuss the implications of each of these models from the QPO data below:
\subsubsection{Relativistic Precession Model (RPM)}This model \cite{Stella:1998mq,stella1997lense,stella1999correlations} was originally proposed to explain the twin-peak KHz QPOs observed in neutron star sources. According to RPM, the upper HFQPO $f_1$ is attributed to the Keplerian motion of matter inhomogeneities orbiting the BH, while the lower HFQPO $f_2$ is associated with the periastron precession frequency $f_{per}=f_\phi - f_r$ of the accreting blob. The motion of the orbiting blob is assumed to be slightly eccentric and since the three fundamental frequencies are different at a given $r$, the blob does not return to the same position after a complete revolution. In other words, there is a radial precession and also a precession of the orbital plane. The frequency of radial precession is denoted by $f_{per}$ (the periastron precession frequency) while that of the orbital precession is denoted by $f_{nod}=f_\phi - f_\theta$ (the nodal precession frequency). The association of the lower HFQPO $f_2$ with the periastron precession frequency was motivated from the observation that as the twin-peak HFQPOs in neutron star sources gradually drift towards higher frequencies, the difference between these frequencies decrease.
From \ref{nur} one may note that $f_r$ attains a maxima and goes to zero at the $r_{ms}$, while the behavior of $f_\phi$ in \ref{fig2} is monotonically increasing with decreasing $r$. Thus, as the accreting blob moves closer to the central object $\Delta \nu =f_1- f_2=f_\phi -f_{per}$ decreases, in accordance with the aforesaid observations. 
RPM also explains the LFQPO $\nu_3$ observed in the BH sources in terms of the nodal precession frequency, i.e $f_3=f_{nod}$. 
This model gave excellent agreement with the data of several LMXRBs \cite{Stella:1998mq}. RPM was first applied to black holes in \cite{stella1997lense} while Motta et al. \cite{Motta:2013wga} obtained the spin and mass of GRO J1655-40 (where two HFQPOs and a LFQPO was simultaneously detected) using RPM. In \cite{Motta:2013wga} the black hole was assumed to be described by the Kerr metric and the mass of GRO J1655-40 obtained by fitting the data with RPM was in agreement with previous estimates based on optical observations. The spin however was very different from earlier results and we will discuss the implications of this finding in the next section.

The goal of this work is to constrain the dilaton parameter $r_2$ and the spin $a$ from HFQPO based observations and we assume the black holes in \ref{Tab2} to be described by the Kerr-Sen spacetime. We do not determine the mass of these black holes from our analysis, but assume their previously determined masses based on optical/NIR photometry, also reported in \ref{Tab2}. 

\begin{figure}[htp]
  \begin{subfigure}{0.5\textwidth}
    \centering
    \includegraphics[width=\linewidth]{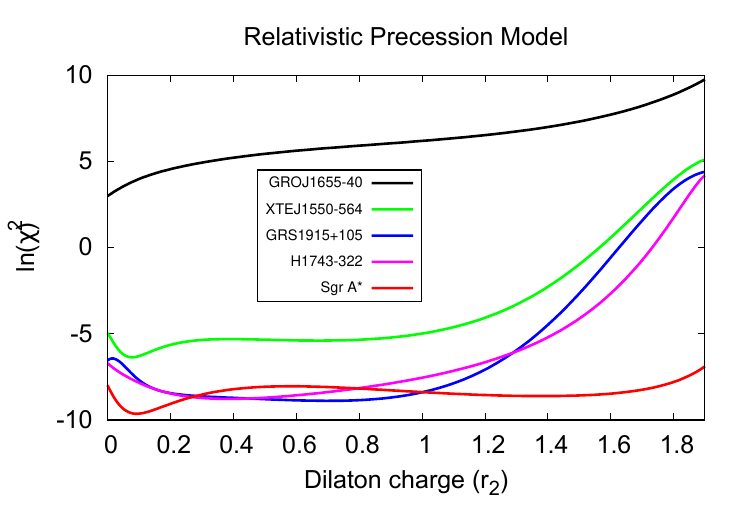}
    \caption{}
    \label{fig5a}
  \end{subfigure}
  \begin{subfigure}{0.5\textwidth}
    \centering
    \includegraphics[width=\linewidth]{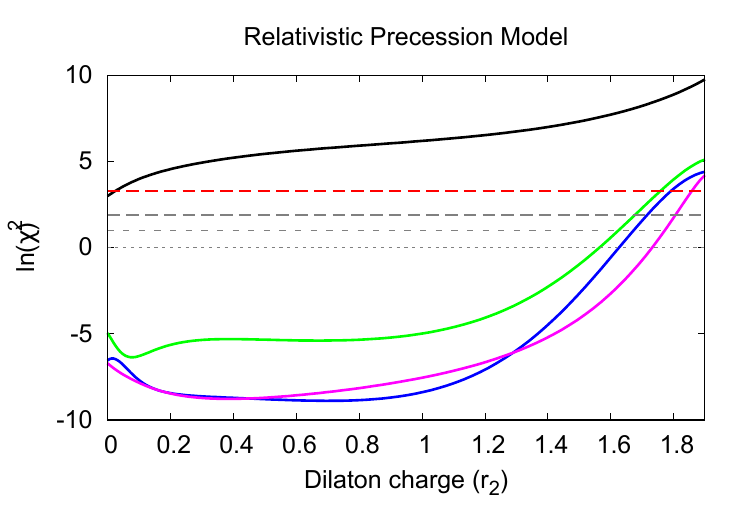}
    \caption{}
    \label{fig5b}
  \end{subfigure}
  \caption{The above figure (a) demonstrates the variation of $\chi^2$ with the dilaton charge $r_2$ for the five black hole sources in \ref{Tab2} assuming RPM. Figure (b) plots the the variation of $\chi^2$ with $r_2$ (assuming the same model) but for a subset of those five BHs  where the $\chi^2$ values are large and the variation with $r_2$ is also substantial such that the confidence lines can be drawn. The grey dotted line corresponds to the 1-$\sigma$ contour, the grey short-dashed line corresponds to the 2-$\sigma$ contour and the grey long-dashed line is associated with the 3-$\sigma$ contour for XTE J1550-564, GRS 1915+105 and H1743-322. The red long-dashed line is the 3-$\sigma$ contour corresponding to the source GRO J1655-40. }
  \label{fig5}
\end{figure}

In \ref{fig5a}, we plot the variation of $\rm{ln} \chi^{2}$ with the dilaton charge $r_2$ for all the five black holes. We consider $\rm{ln} \chi^{2}$ so that we can show the variation of $\chi^2$ of all the black holes clearly in the same frame, although they  differ substantially from each other. We note from \ref{fig5a} that $\chi^2$ hardly varies with $r_2$ for Sgr A* (red curve). This data therefore, does not allow us to establish strong constraints on $r_2$. For XTE J1550-564 (green curve in \ref{fig5}), GRS 1915+105 (blue curve in \ref{fig5}) and H1743-322 (magenta curve in \ref{fig5}), the $\chi^2$ minimizes at $r_2\sim 0.1$, $r_2\sim 0.6$ and $r_2\sim 0.3$ respectively. However, the magnitude of $\chi^2_{min}\sim 0$. Therefore, $\Delta \chi^2$ values from $\chi^2_{min}$ corresponding to 1-$\sigma$, 2-$\sigma$, and 3-$\sigma$ contours are 1, 2.71 and 6.63, respectively. These are plotted with the grey lines in \ref{fig5b}, such that the grey dotted line corresponds to the 1-$\sigma$ contour, the grey short-dashed line corresponds to the 2-$\sigma$ contour and the grey long-dashed line is associated with the 3-$\sigma$ contour. We note from \ref{fig5b} that the HFQPO data of XTE J1550-564, GRS 1915+105 and H1743-322 respectively rule out $r_2>1.6$, $r_2>1.65$ and $r_2>1.7$ outside 1-$\sigma$. If the 3-$\sigma$ interval is considered, then XTE J1550-564, GRS 1915+105 and H1743-322 data rule $r_2>1.7, 1.75, ~\rm{and}~1.8$, respectively. Although the data for these three black holes exhibit a preference towards a non-zero dilaton charge, the constraints are not very strong, due to the slow variation of $\chi^2$. All three aforesaid BHs however, rule out near extremal values of $r_2$. 

The strongest constraint on the dilaton charge is established by the QPO data of GRO J1655-40 (black curve in \ref{fig5}). This black hole exhibits a very strong variation of $\chi^2$ with $r_2$ and also high $\chi^2$ values. From \ref{fig5b}, one may note that the minimum of $\chi^2$ occurs at $r_2\sim 0$ and the 3-$\sigma$ contour (denoted by the red-dashed line) rules out $r_2>0.01$. Thus, this black hole strongly favors $r_2\sim0$, if RPM is assumed to explain the QPO data.

\subsubsection{Tidal Disruption Model} 
The Tidal Disruption Model (TDM) aims to explain the twin-peak HFQPOs in terms of tidal disruption of spherical blobs orbiting the black hole \cite{Cadez:2008iv,Kostic:2009hp,Germana:2009ce}. Through extensive numerical simulations it was shown \cite{Cadez:2008iv,Kostic:2009hp,Germana:2009ce} that as these blobs approach the event horizon, they get tidally stretched which in turn results in an overall increase in luminosity as observed in the light curves of black holes.
\begin{figure}[t!]
  \begin{subfigure}{0.5\textwidth}
    \centering
    \includegraphics[width=\linewidth]{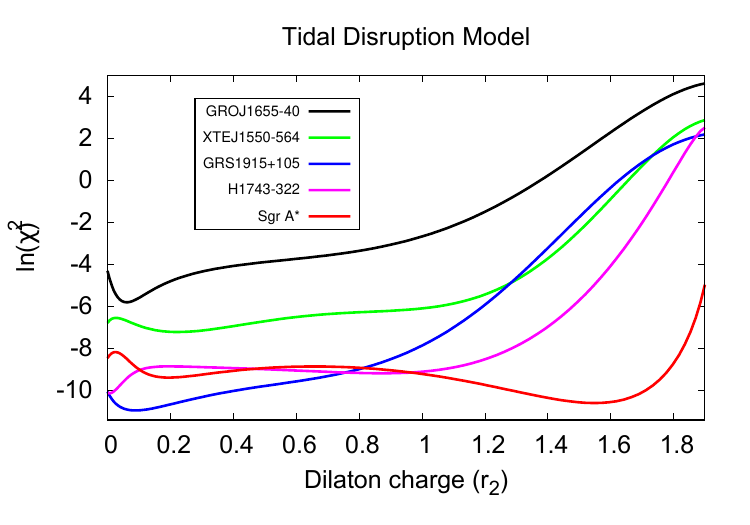}
    \caption{}
    \label{fig6a}
  \end{subfigure}
  \begin{subfigure}{0.5\textwidth}
    \centering
    \includegraphics[width=\linewidth]{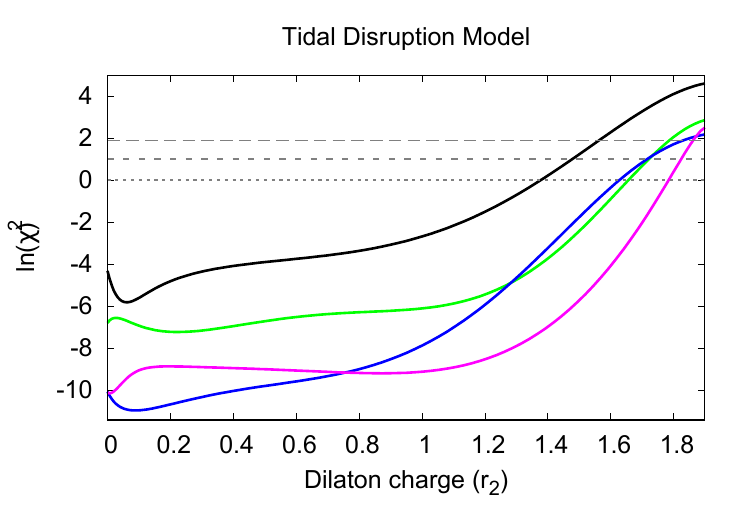}
    \caption{}
    \label{fig6b}
  \end{subfigure}
  \caption{The above figure (a) demonstrates the variation of $\chi^2$ with the dilaton charge $r_2$ for the five black hole sources in \ref{Tab2} assuming TDM. Figure (b) plots the the variation of $\chi^2$ with $r_2$ (assuming the same model) but for a subset of those five BHs  where the $\chi^2$ values are large and the variation with $r_2$ is also substantial such that the confidence lines can be drawn. The grey dotted line corresponds to the 1-$\sigma$ contour, the grey short-dashed line corresponds to the 2-$\sigma$ contour and the grey long-dashed line is associated with the 3-$\sigma$ contour. The confidence contour lines are the same for all the four BHs as $\chi^2_{min}\sim 0$ for all of them. }
  \label{fig6}
\end{figure}
This can be attributed to the overall increase in area of the blob due to tidal interactions. They could also simulate the twin-peak HFQPOs observed in the power spectrum. Their analysis revealed that the lower peak $f_2$ corresponds to $f_\phi$ while the upper HFQPO $f_1$ corresponds to $f_\phi+f_r$.

Once again in \ref{fig6a} we plot the variation of $\rm{ln} \chi^{2}$ with the dilaton charge $r_2$ for all the five black holes assuming now the TDM. Similar to the results of the RPM, here also the variation of $\chi^2$ with $r_2$ for Sgr A* is minimal. Therefore, Sgr A* does not establish strong constrains on $r_2$ if TDM is used to explain its HFQPO data. The remaining four black holes moderately constrain the theory as GRO J1655-40 (black curve) rules out $r_2>1.4$, XTE J1550-564 (green curve) and GRS 1915+105 (blue curve) rule out $r_2>1.65$ and H1743-322 (magenta curve) rules out $r_2>1.8$ outside 1-$\sigma$ interval (grey dotted line in \ref{fig6b}). When the 3-$\sigma$ interval (grey long-dashed line in \ref{fig6b}) is considered, GRO J1655-40 data rules out $r_2>1.6$ while XTE J1550-564 data rules out $r_2>1.8$. The remaining two stellar mass BH sources (GRS 1915+105 and H1743-322) allow $0\lesssim r_2\lesssim 1.85$ within 3-$\sigma$. The above discussion elucidates that if the TDM is considered, then the source GRO J1655-40 establishes strongest constrain on $r_2$ as it rules out the maximum allowed parameter space. Note that the $\Delta\chi^2$ intervals are virtually the same for all the four stellar mass black holes as $\chi^2_{min}\sim 0$ for all of them although the $\chi^2$ minimizes at slightly different values of the dilaton parameter $r_2$. In fact, for GRO J1655-40, XTE J1550-564 and GRS 1915+105 the $\chi^2$ minimizes for a non-zero dilaton charge, $r_2\sim 0.1$ (GRO J1655-40), $r_2\sim 0.2$ (XTE J1550-564) and $r_2\sim 0.1$ (GRS 1915+105) although $r_2\sim 0$ is allowed within 1-$\sigma$ interval.

\subsection{Resonant Models}
Resonant models assume that the HFQPOs occur as a result of resonance between various
oscillation modes in the accretion disk \cite{2001A&A...374L..19A,Abramowicz:2003xy,2005A&A...436....1T,Torok:2011qy,Kluzniak:2002bb,2001AcPPB..32.3605K}. Earlier we discussed the epicyclic frequencies assuming matter to move in nearly circular geodesics along the equatorial plane. Slight perturbations from the circular orbit $\delta r=r-r_c$ (where $r_c$ is the radius of a circular orbit) and the equatorial plane $\delta \theta=\theta -\pi/2$ are of the form $e^{i\omega_j\theta}$ where $j=r,\theta$ (see \ref{S3-7}) which implies that $\delta r$ and $\delta \theta$ obey 
\begin{eqnarray}
\delta\ddot{r}+\omega_{r}^{2}\delta r=0\nonumber\\
\delta\ddot{\theta}+\omega_{\theta}^{2}\delta\theta=0
\label{S4-1}
\end{eqnarray}
corresponding to equations of simple harmonic motion. \ref{S4-1} describes two simple harmonic motions which are uncoupled and also has no forcing terms, which is often the scenario for thin disks or accretion tori  \cite{2001A&A...374L..19A,Kluzniak:2002bb}. When one takes into account the effects of pressure and dissipation in the accretion flow, then one needs to include the forcing terms in \ref{S4-1},
\begin{eqnarray}
\delta\ddot{r}+\omega_{r}^{2}\delta r=\omega_{r}^{2}F_r(\delta r,\delta\theta,\delta\dot r,\delta\dot\theta)\nonumber\\
\delta\ddot\theta+\omega_{\theta}^{2}\delta\theta=\omega_{\theta}^{2}F_{\theta}(\delta r,\delta\theta,\delta\dot r,\delta\dot\theta)
\label{S4-2}
\end{eqnarray}
where the forcing terms $F_r$ and $F_\theta$ can be some non-linear functions of their arguments. The exact mathematical forms of $F_r$ and $F_\theta$ are not very well-known since the dissipation and pressure effects in the accretion flow are not quite generic. However, in order to describe different physical situations one assumes different mathematical ansatz for $F_r$ and $F_\theta$ \cite{Abramowicz:2003xy, Horak:2004hm}. The resonant models can better explain the commensurability of the QPO frequencies and based on different forcing terms describing different accretion scenarios, different resonant models are proposed in the literature.

\subsubsection{Parametric Resonance Model}
The Parametric Resonance Model assumes that the radial epicyclic mode excites the vertical epicyclic mode which is based on the fact that random fluctuations in thin disks are expected to have $\delta r \gg \delta \theta$ \cite{Kluzniak:2002bb,2001A&A...374L..19A,Abramowicz:2003xy,2005A&A...436....1T,Rebusco:2004ba,torok2005orbital}. Under this assumption, the perturbation $\delta r$ satisfies the equation for simple harmonic motion but the perturbation equation for $\delta \theta$ has a forcing term $\propto \delta r\delta\theta$, i.e., 
\begin{eqnarray}
\delta\ddot r+\omega_r^2\delta r=0\nonumber\\
\delta\ddot\theta+\omega_{\theta}^{2}\delta\theta=-\omega_{\theta}^{2}\delta r\delta\theta
\label{S4-3}
\end{eqnarray}
We get such type of excitations only when $\frac{f_r}{f_{\theta}}=\frac{2}{n}$ where n is a positive integer.
This resonance becomes strongest when the value of n is the smallest. Now, from \ref{nur} and \ref{nutheta} one may note that $f_{\theta}>f_r$ for the Kerr-Sen black hole. Therefore, the minimum value of n has to be 3 which in turn explains the 3:2 ratio observed in twin-peak HFPQOs. 
Thus, according to this model $f_1=f_{\theta}$ and $f_2=f_r$ and this model can explain the commensurability of the QPO frequencies which has been further confirmed in numerical simulations \cite{Abramowicz:2003xy,Rebusco:2004ba,Horak:2004hm}.

\begin{figure}[htp]
  \begin{subfigure}{0.5\textwidth}
    \centering
    \includegraphics[width=\linewidth]{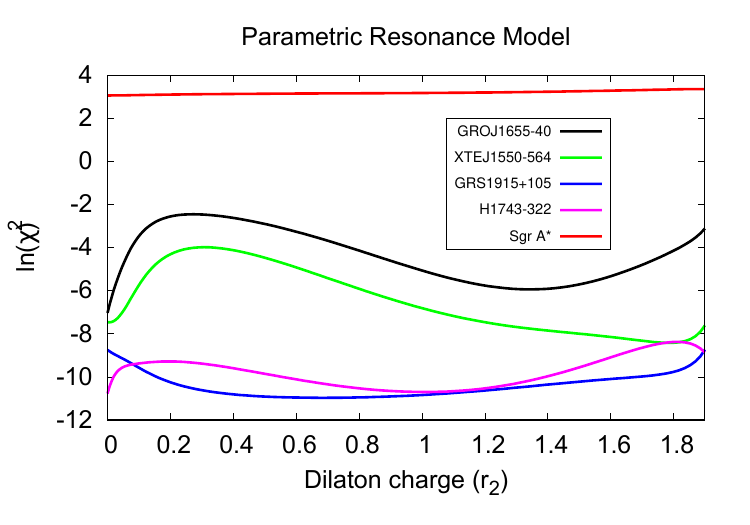}
    \caption{}
    \label{fig7a}
  \end{subfigure}
  \begin{subfigure}{0.5\textwidth}
    \centering
    \includegraphics[width=\linewidth]{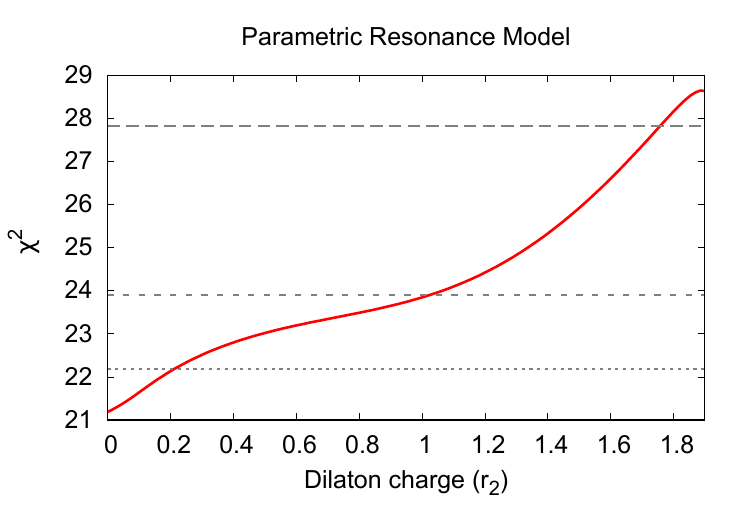}
    \caption{}
    \label{fig7b}
  \end{subfigure}
  \caption{The above figure (a) demonstrates the variation of $\chi^2$ with the dilaton charge $r_2$ for the five black hole sources in \ref{Tab2} assuming PRM. Figure (b) plots the the variation of $\chi^2$ with $r_2$ (assuming the same model) for Sgr A*  where the $\chi^2$ values are large and the variation with $r_2$ is also substantial such that the confidence lines can be drawn. The grey dotted line corresponds to the 1-$\sigma$ contour, the grey short-dashed line corresponds to the 2-$\sigma$ contour and the grey long-dashed line is associated with the 3-$\sigma$ contour. }
  \label{fig7}
\end{figure}
We compare this model with the observed data of five black holes and study the variation of $\chi^2$ with $r_2$. We note from \ref{fig7a} that for the four stellar mass black holes, the variation of $\chi^2$ with $r_2$ is very less and no strong constrains on $r_2$ can be established from these data. Sgr A* however, exhibits reasonable variation of $\chi^2$ with $r_2$ which is separately plotted in \ref{fig7b}. From \ref{fig7b} we note that the $\chi^2$ minimizes at $r_2\sim 0$ and the allowed range of dilaton charge within 1-$\sigma$ (grey dotted line) is $0\lesssim r_2 \lesssim 0.2$, within 2-$\sigma$ (grey short dashed line) is $0\lesssim r_2 \lesssim 1$ and within 3-$\sigma$ (grey long dashed line) is $0\lesssim r_2 \lesssim 1.75$. Sgr A* data rules out $r_2>1.75$ outside 3-$\sigma$. Thus, if PRM is used to explain the HFQPO data of these black holes, strong constrains on $r_2$ can be provided only from the Sgr A* data which shows a preference towards $r_2\sim 0$. 

\subsubsection{Forced Resonance model}
When the accretion flow can be described by a nearly Keplerian accretion disk, the parametric resonance seems to be a natural explanation of the HFQPOs \cite{2001A&A...374L..19A,Kluzniak:2002bb,2001AcPPB..32.3605K}. However, when forces due to pressure, viscosity and/or magnetic stresses are present in the accretion flow, then one may not neglect the non-linear couplings between $\delta r$ and $\delta\theta$, in addition to the parametric resonance \cite{2005A&A...436....1T}. With the help of numerical simulations Lee et al.\cite{2004ApJ...603L..93L} showed that pressure forces can often give rise to resonant forcing  between vertical oscillations due to radial oscillations \cite{2001A&A...374L..19A,2004ApJ...603L..93L}, in agreement with the previous findings by Abramowicz \& Kluzniak \cite{2001A&A...374L..19A}. In such a scenario, the equations governing the perturbations $\delta r$ and $\delta \theta$ given in \ref{S4-3} gets modified to,
\begin{eqnarray}
\delta\ddot r+\omega_r^2\delta r=0\nonumber\\
\delta\ddot\theta+\omega_{\theta}^{2}\delta\theta=-\omega_{\theta}^{2}\delta r\delta\theta + \mathcal{F_\theta}(\delta\theta) 
\label{S4-4}
\end{eqnarray}
In \ref{S4-4} $\mathcal{F_\theta}$ corresponds to non-linear terms in $\delta\theta$, while 
$\delta r=\mathcal{A} cos(\omega_r t)$ still holds. The above equations admit solutions of the form,
\begin{eqnarray}
\frac{f_{\theta}}{f_r}=\frac{m}{n}
\end{eqnarray}
where m and n are natural numbers. When $\frac{m}{n}=\frac{3}{2}$ we get parametric resonance. But, this model not only allows resonance between $f_r$ and $f_\theta$ but also between combinations of frequencies like $f_{\theta}+f_r$ and $f_{\theta}-f_r$. Two widely used models of forced resonances correspond to $\frac{f_{\theta}}{f_r}=3$ {(Forced Resonance Model 1 or FRM1)} and $\frac{f_{\theta}}{f_r}=2$ { (Forced Resonance Model 2 or FRM2).} It can be shown that for 3:1 Forced Resonance Model the upper HFQPO corresponds to $f_{1}=f_\theta$ while the lower HFQPO corresponds to $f_{2}= f_\theta-f_r$ while for 2:1 Forced Resonance Model the upper HFQPO is associated with 
$f_1=f_{\theta}+f_r$ and the lower HFQPO corresponds to $f_2=f_{\theta}$.
Thus, in case of both FRM1 and FRM2 we get our desired ratio 3:2 which is observed in the twin peak HFQPOs.

\begin{figure}[t!]
  \begin{subfigure}{0.5\textwidth}
    \centering
    \includegraphics[width=\linewidth]{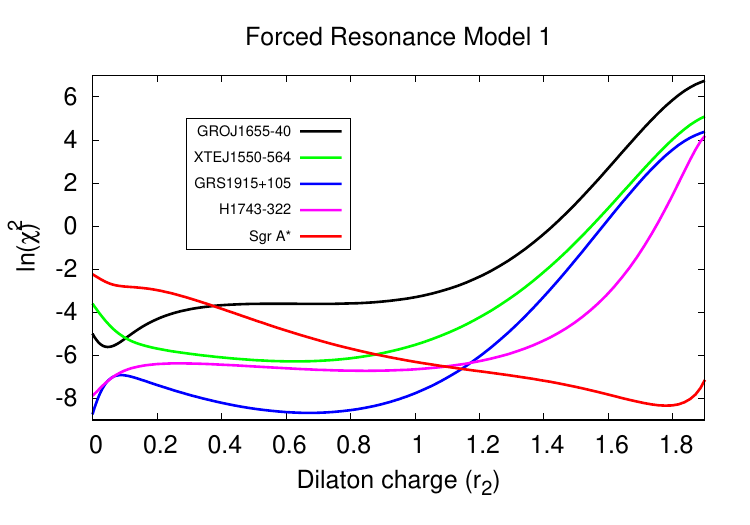}
    \caption{}
    \label{fig8a}
  \end{subfigure}
  \begin{subfigure}{0.5\textwidth}
    \centering
    \includegraphics[width=\linewidth]{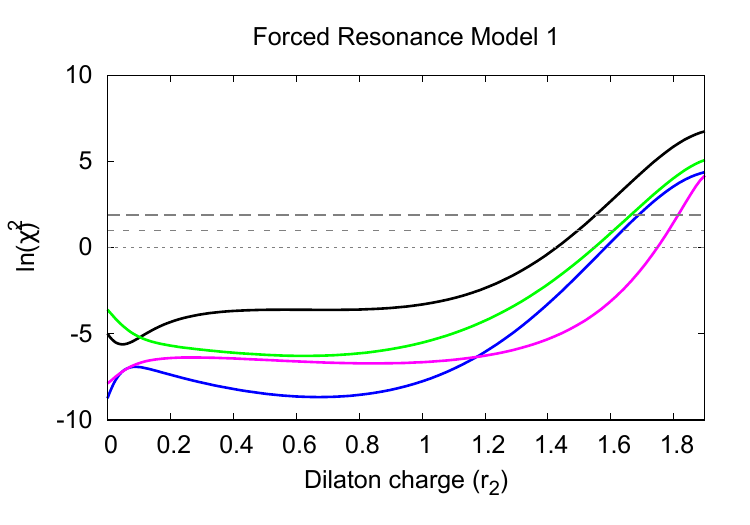}
    \caption{}
    \label{fig8b}
  \end{subfigure}
  \caption{The above figure (a) demonstrates the variation of $\chi^2$ with the dilaton charge $r_2$ for the five black hole sources in \ref{Tab2} assuming FRM1. Figure (b) plots the the variation of $\chi^2$ with $r_2$ (assuming the same model) but for a subset of those five BHs  where the $\chi^2$ values are large and the variation with $r_2$ is also substantial such that the confidence lines can be drawn. The grey dotted line corresponds to the 1-$\sigma$ contour, the grey short-dashed line corresponds to the 2-$\sigma$ contour and the grey long-dashed line is associated with the 3-$\sigma$ contour. The confidence contour lines are the same for all the four BHs as $\chi^2_{min}\sim 0$ for all of them. }
  \label{fig8}
\end{figure}

In \ref{fig8a} the variation of ln$\chi^2$ with $r_2$ is plotted for all the five BHs assuming FRM1. The figure reveals that for Sgr A* the $\chi^2$ minimizes at $r_2\sim 1.8$. This however, cannot be considered as a strong signature of EMDA gravity as $\chi^2_{min}\sim 0$ and the $\chi^2$ for this source does not vary substantially such that the $\Delta \chi^2$ intervals corresponding to 68\%, 90\% and 99\% confidence lines can rule out certain range of the allowed parameter space of $r_2$. In other words, the Kerr scenario is allowed within 1-$\sigma$.
For the remaining four black holes the variation of $\chi^2$ is reasonable such that GRO J1655-40, XTE J1550-564, GRS 1915+105 and H 1743-322 respectively rule out $r_2>1.3$, $r_2>1.6$,  $r_2>1.6$ and  $r_2>1.7$ outside 1-$\sigma$ (\ref{fig8b}). When the 3-$\sigma$ interval is considered GRO J1655-40, XTE J1550-564, GRS 1915+105 and H 1743-322 rule out $r_2>1.5$, $r_2>1.7$,  $r_2>1.7$ and  $r_2>1.8$ respectively (\ref{fig8b}). Out of these four black holes, the $\chi^2$ minimizes at a non-zero dilaton charge, $r_2\sim 0.05$ and $r_2\sim 0.6$ for GRO J1655-40 and XTE J1550-564 respectively, while for GRS 1915+105 and H1743-322 the $\chi^2$ minimizes at $r_2\sim 0$ (\ref{fig8b}). However, for GRO J1655-40 and XTE J1550-564, $r_2\sim 0$ is allowed within 68\% confidence interval.

\begin{figure}[t!]
  \begin{subfigure}{0.5\textwidth}
    \centering
    \includegraphics[width=\linewidth]{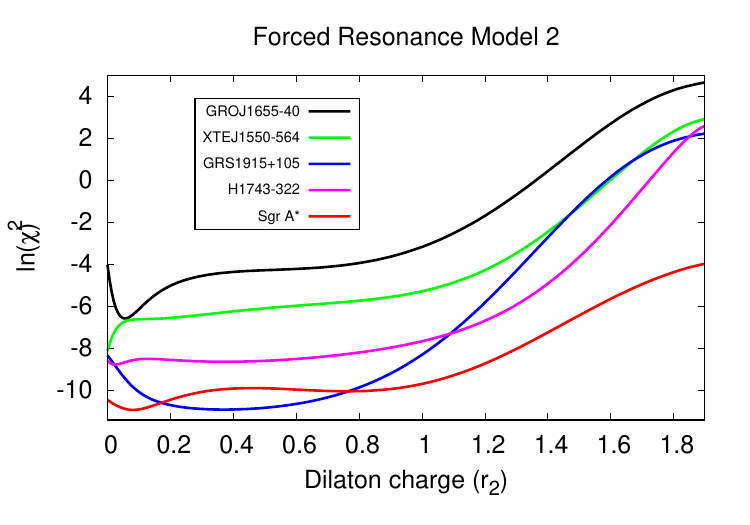}
    \caption{}
    \label{fig9a}
  \end{subfigure}
  \begin{subfigure}{0.5\textwidth}
    \centering
    \includegraphics[width=\linewidth]{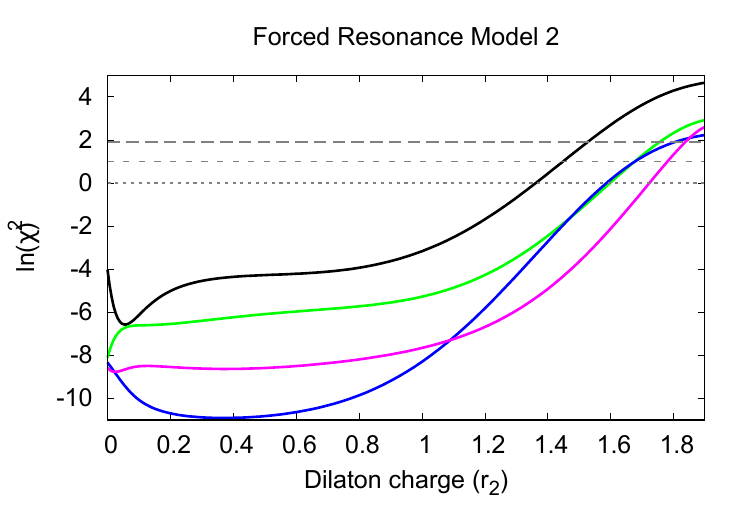}
    \caption{}
    \label{fig9b}
  \end{subfigure}
  \caption{The above figure (a) demonstrates the variation of $\chi^2$ with the dilaton charge $r_2$ for the five black hole sources in \ref{Tab2} assuming FRM2. Figure (b) plots the the variation of $\chi^2$ with $r_2$ (assuming the same model) but for a subset of those five BHs  where the $\chi^2$ values are large and the variation with $r_2$ is also substantial such that the confidence lines can be drawn. The grey dotted line corresponds to the 1-$\sigma$ contour, the grey short-dashed line corresponds to the 2-$\sigma$ contour and the grey long-dashed line is associated with the 3-$\sigma$ contour. The confidence contour lines are the same for all the four BHs as $\chi^2_{min}\sim 0$ for all of them.}
  \label{fig9}
\end{figure}

In \ref{fig9a} we plot the variation of ln$\chi^2$ with $r_2$ for all the five BHs assuming FRM2. Like most of the previous models, the HFQPO data of Sgr A* does not impose strong observational constrains on $r_2$ as the variation of $\chi^2$ with $r_2$ is minimal. GRO J1655-40 on the other hand shows strongest variation of $\chi^2$ with $r_2$ and rules out $r_2>1.4$ outside 1-$\sigma$ and $r_2>1.55$ outside 3-$\sigma$ (\ref{fig9b}).

 The $\chi^2$ minimizes for $r_2\sim 0.05$, although the Kerr scenario is allowed within 1-$\sigma$. The remaining three BHs impose weaker constrains on $r_2$, e.g., for XTE J1550-564 and GRS 1915+105 the allowed range of $r_2$ is $0\lesssim r_2\lesssim 1.6$ while for H1743-322 it is 
$0\lesssim r_2\lesssim 1.7$ if the 1-$\sigma$ interval is considered. When the 3-$\sigma$ interval is taken into account, the allowed range increases to $0\lesssim r_2\lesssim 1.75$ for XTE J1550-564 and $0\lesssim r_2\lesssim 1.8$ for GRS 1915+105 and H1743-322. Out of these three black holes GRS 1915+105 shows a minimum $\chi^2$ for $r_2\sim 0.4$ although GR is allowed within 1-$\sigma$ interval.

\subsubsection{Keplerian Resonance Model}

The resonant models considered so far assume coupling between the vertical and radial epicyclic frequencies. However, there can also be non-linear resonances between the radial epicyclic modes and the orbital angular motion, which forms the basis of the Keplerian Resonance Model \cite{2005A&A...436....1T,2001PASJ...53L..37K,2001A&A...374L..19A, Nowak:1996hg}. Such couplings arise due to trapping of non-axisymmetric g-mode oscillations induced by a corotation resonance \cite{2001PASJ...53L..37K} which may be prevalant in the inner region of the relativistic thin accretion disks. However, numerical simulations \cite{Li:2002yi,2003PASJ...55..257K} reveal that g-mode oscillations rather gets damped due to corotation resonance, instead of being excited which weakens the prospect of Keplerian Resonance Models in explaining the HFQPOs. 
Keplerian Resonance may also occur when a pair of spatially separated vortices with opposite vorticities oscillating with radial epicyclic frequencies, couple with the orbital angular frequencies \cite{2005A&A...436....1T,2010tbha.book.....A} which are different at the two different locations. There may be three variants of Keplerian Resonance: (a) when $f_1=f_\phi$ and $f_2=f_r$ (which is denoted by Keplerian Resonance Model 1 or KRM1), (b) when $f_1=f_\phi$ and $f_2=2f_r$ (Keplerian Resonance Model 2 or KRM2) and (iii) $f_1=3f_r$  and $f_2=f_\phi$ (Keplerian Resonance Model 3 or KRM3).

\begin{figure}[t!]
  \begin{subfigure}{0.5\textwidth}
    \centering
    \includegraphics[width=\linewidth]{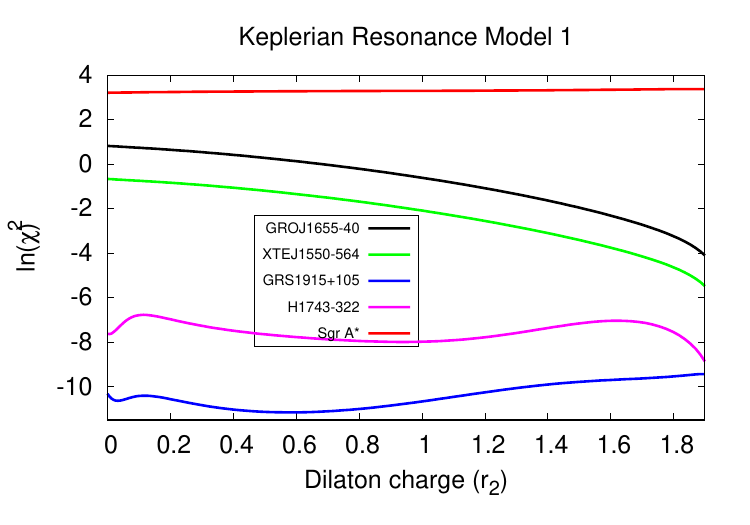}
    \caption{}
    \label{fig10a}
  \end{subfigure}
  \begin{subfigure}{0.5\textwidth}
    \centering
    \includegraphics[width=\linewidth]{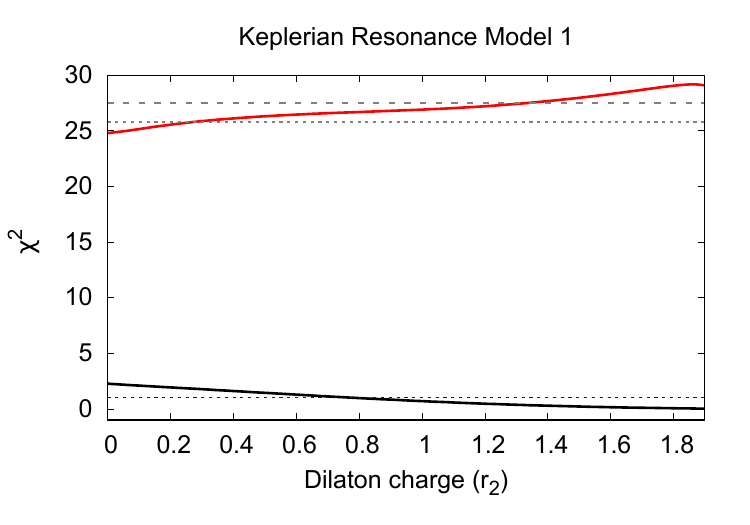}
    \caption{}
    \label{fig10b}
  \end{subfigure}
  \caption{The above figure (a) demonstrates the variation of $\chi^2$ with the dilaton charge $r_2$ for the five black hole sources in \ref{Tab2} assuming KRM1. Figure (b) plots the the variation of $\chi^2$ with $r_2$ (assuming the same model) but for a subset of those five BHs  where the $\chi^2$ values are large and the variation with $r_2$ is also substantial such that the confidence lines can be drawn. The grey dotted line corresponds to the 1-$\sigma$ contour while the grey short-dashed line corresponds to the 2-$\sigma$ contour for Sgr A*. The blue dotted line on the other hand corresponds to the 1-$\sigma$ confidence contour for GRO J1655-40.}
  \label{fig10}
\end{figure}

\ref{fig10a} demonstrates the variation of $\chi^2$ with the dilaton charge $r_2$ for the five BH sources assuming KRM1. $\chi^2$ is extremely small for all values of $r_2$ for H1743-322 and GRS 1915+105, hence these sources fail to impose strong constrains on $r_2$. The source XTE J1550-564 interestingly exhibits a monotonically decreasing $\chi^2$ with increasing $r_2$, such that $\chi^2_{min}$ occurs at $r_2\sim 1.9$. This indicates that the HFQPO data of this source shows a preference towards the Kerr-Sen scenario compared to \gr. 
However, the $\chi^2_{min}\sim 0$ and hence the 68\%, 90\% and 99\% contour lines do not intersect the $\chi^2$ plot for this source, indicating that although XTE J1550-564 shows a preference towards the EMDA gravity, \gr\ is allowed within 1-$\sigma$. GRO J1655-40 shows a trend very similar to XTE J1550-564, but now the $\chi^2$ values are higher. As a consequence, the 1-$\sigma$ line corresponding to this source (blue dotted line in \ref{fig10b}) intersects the $\chi^2$ curve, thereby ruling out $0\lesssim r_2 \lesssim 0.7$. Thus assuming KRM1, the HFQPO data of GRO J1655-40 favors the EMDA scenario (with $\chi^2_{min}$ occuring at $r_2\sim 1.9$) and ruling out GR outside 1-$\sigma$ interval but including $r_2=0$ within 3-$\sigma$. Although this is very interesting, one may note that
numerical simulations \cite{Li:2002yi,2003PASJ...55..257K} do not provide a strong evidence of  
Keplerian Resonance from g-mode oscillations although Keplerian Resonance from a pair of spatially separated vortices may be present.
Finally, the HFQPO data of Sgr A* shows a preference towards the Kerr scenario, with $\chi^2$ minimizing at $r_2\sim 0$ (\ref{fig10}). The values of $\chi^2$ are large and the variation of $\chi^2$ is also substantial such that the allowed range of dilaton charge within 1-$\sigma$ is $0\lesssim r_2 \lesssim 0.3$ and this source rules out $r_2\gtrsim 1.4$ outside 2-$\sigma$ (see \ref{fig10b}). It may be worthwhile to mention that the Kerr metric is also a black hole solution in several alternative gravity scenarios \cite{Sen:1992ua,Campbell:1992hc,Psaltis:2007cw}, including EMDA gravity, and hence observational evidence of the Kerr solution does not necessarily imply that the underlying gravity theory is GR.

\begin{figure}[t!]
  \begin{subfigure}{0.5\textwidth}
    \centering
    \includegraphics[width=\linewidth]{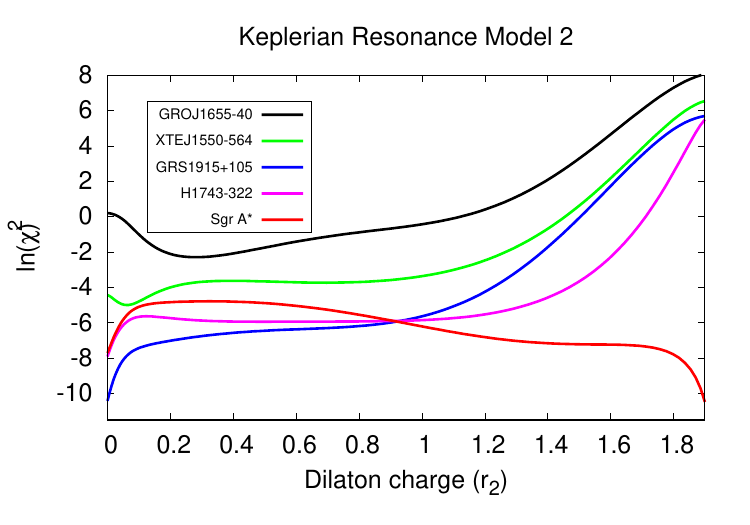}
    \caption{}
    \label{fig11a}
  \end{subfigure}
  \begin{subfigure}{0.5\textwidth}
    \centering
    \includegraphics[width=\linewidth]{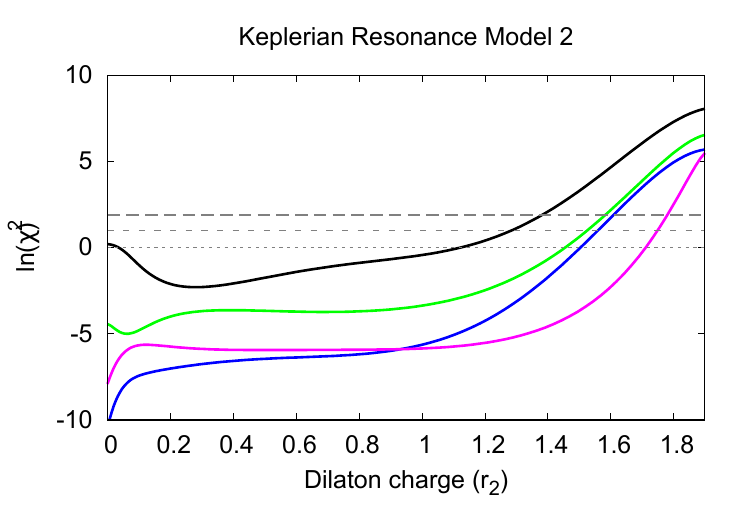}
    \caption{}
    \label{fig11b}
  \end{subfigure}
  \caption{The above figure (a) demonstrates the variation of $\chi^2$ with the dilaton charge $r_2$ for the five black hole sources in \ref{Tab2} assuming KRM2. Figure (b) plots the the variation of $\chi^2$ with $r_2$ (assuming the same model) but for a subset of those five BHs  where the $\chi^2$ values are large and the variation with $r_2$ is also substantial such that the confidence lines can be drawn. The grey dotted line corresponds to the 1-$\sigma$ contour, the grey short-dashed line corresponds to the 2-$\sigma$ contour and the grey long-dashed line is associated with the 3-$\sigma$ contour. The confidence contour lines are the same for all the four BHs.}
  \label{fig11}
\end{figure}

We next discuss constrains on $r_2$ assuming KRM2. From \ref{fig11a} we note that for Sgr A* (red curve) the $\chi^2$ values are very small and the variation of $\chi^2$ with $r_2$ is also very less. Thus, although the $\chi^2$ minimizes at $r_2\sim 1.9$, GR is within 1-$\sigma$ and this source does not impose strong bounds on $r_2$. The strongest bound on $r_2$ is imposed by GRO J1655-40 where the $\chi^2$ minimizes for $r_2\sim 0.2$ (\ref{fig11b}). Interestingly, the Kerr scenario ($r_2\sim 0$) is just outside 1-$\sigma$ line but within 2-$\sigma$ and this source rules out $r_2>1.1$ and $r_2>1.4$ outside 1-$\sigma$ (grey dotted line) and 3-$\sigma$ (grey long-dashed line) respectively (\ref{fig11b}).
The remaining three BH sources impose moderate constrains on $r_2$. For GRS 1915+105 and H1743-322 the $\chi^2$ minimizes at $r_2\sim 0$. While GRS 1915+105 rules out $r_2>1.45$ outside 1-$\sigma$ and $r_2>1.6$ outside 3-$\sigma$, H1743-322 rules out $r_2>1.65$ outside 1-$\sigma$ and $r_2>1.8$ outside 3-$\sigma$ (\ref{fig11b}). For XTE J1550-564, the $\chi^2$ minimizes at $r_2\sim 0.05$ but $r_2\sim0$ falls within 1-$\sigma$ and this source rules out $r_2>1.45$ outside 1-$\sigma$ and $r_2>1.6$ outside 3-$\sigma$ (\ref{fig11b}). All the stellar-mass BH sources rule out near extremal dilaton charges.

\begin{figure}[H]
  \begin{subfigure}{0.5\textwidth}
    \centering
    \includegraphics[width=\linewidth]{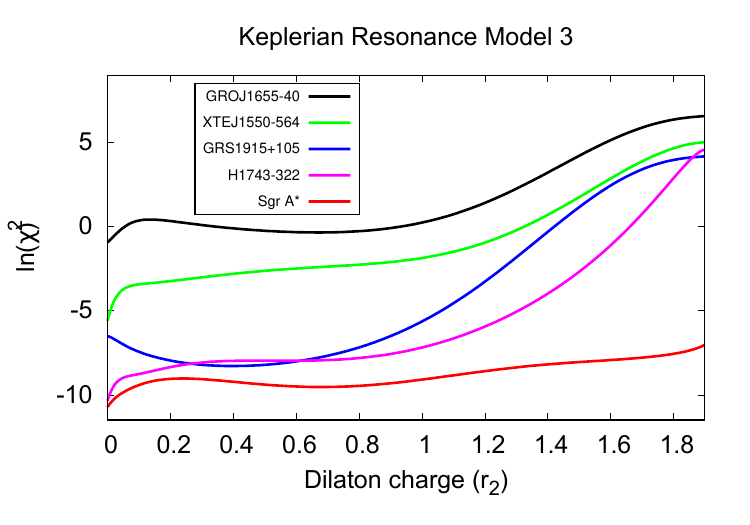}
    \caption{}
    \label{fig12a}
  \end{subfigure}
  \begin{subfigure}{0.5\textwidth}
    \centering
    \includegraphics[width=\linewidth]{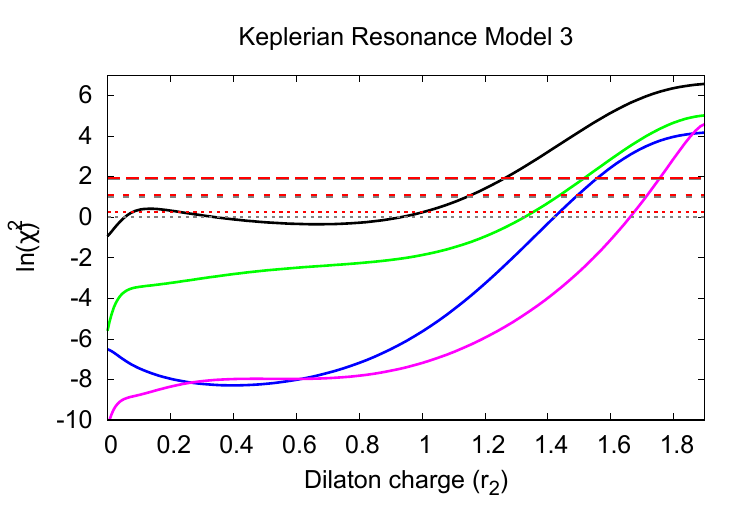}
    \caption{}
    \label{fig12b}
  \end{subfigure}
  \caption{The above figure (a) demonstrates the variation of $\chi^2$ with the dilaton charge $r_2$ for the five black hole sources in \ref{Tab2} assuming KRM3. Figure (b) plots the the variation of $\chi^2$ with $r_2$ (assuming the same model) but for a subset of those five BHs  where the $\chi^2$ values are large and the variation with $r_2$ is also substantial such that the confidence lines can be drawn. The red dotted line corresponds to the 1-$\sigma$ contour, the red short-dashed line corresponds to the 2-$\sigma$ contour and the red long-dashed line is associated with the 3-$\sigma$ contour for GRO J1655-40. The grey lines on the other hand correspond to the confidence intervals for the remaining three stellar-mass BHs.}
  \label{fig12}
\end{figure}

Finally, we discuss the viability of the KRM3 model in explaining the twin-peak HFQPOs in black holes. \ref{fig12} illustrates the variation of $\rm{ln}\chi^2$ with $r_2$ for the five BH sources. As before, the values of $\chi^2$ as well as its variation with $r_2$ is very less for Sgr A*, which indicates that this source cannot impose strong bounds on $r_2$. GRO J1655-40 once again imposes the strongest constrain on $r_2$ with its $\chi^2$ minimizing at $r_2\sim 0$. This source rules out $0.1\lesssim r_2 \lesssim 0.25$ and $r_2 \gtrsim 1$ outside 1-$\sigma$ (red dotted line in \ref{fig12b}) and $r_2\gtrsim 1.25$ outside 3-$\sigma$ (red long-dashed line in \ref{fig12b}). XTE J1550-564 and H1743-322 exhibits a $\chi^2$ minima at $r_2\sim 0$ while GRS 1915+105 exhibits a $\chi^2$ minima at $r_2\sim 0.4$. Since the magnitude of $\chi^2_{min}$ for all these three black holes are really small, their 1-$\sigma$ (grey dotted line in \ref{fig12b}), 2-$\sigma$ (grey short-dashed line in \ref{fig12b})and 3-$\sigma$ contour lines (grey long-dashed line in \ref{fig12b}) overlap and correspond to 1, 2.71 and 6.63, respectively. \ref{fig12b} reveal that HFQPO data of XTE J1550-564, GRS 1915+105 and H1743-322 rule out $r_2\gtrsim 1.35$, $r_2\gtrsim 1.4$ and $r_2\gtrsim 1.65$ outside 1-$\sigma$ and $r_2\gtrsim 1.5$, $r_2\gtrsim 1.6$ and $r_2\gtrsim 1.8$ outside 3-$\sigma$. All stellar-mass BH sources rule out near extremal dilaton charges from the HFQPO data.

\subsubsection{Non-axisymmetric Disk-Oscillation Models}
The Non-axisymmetric Disk-Oscillation Models (NADO) consider non-geodesic effects in the accretion flow due to pressure forces \cite{Torok:2015tpu,Sramkova:2015bha,Kotrlova:2020pqy}
by modelling the accretion disk in terms of pressure-supported tori filled with a perfect fluid. In these accretion flows there can be resonance between the vertical epicyclic frequency $f_\theta$ and the $m=-1$ non-axisymmetric radial epicyclic frequency which corresponds to the periastron precession frequency $f_{per}$. This model (NADO1), also termed as the Vertical Precession Resonance Model \cite{2004ApJ...617L..45B,2005AN....326..849B,2005ragt.meet...39B} was used to determine the spin of GRO J1655-40 assuming the background spacetime to be given by the Kerr metric and $f_1=f_\theta$ and $f_2=f_{per}$. Although it is difficult to explain the coupling between an axisymmetric and a non-axisymmetric mode \cite{Horak:2008zg,Torok:2011qy}, this model could reproduce the spin of GRO J1655-40 estimated using the Continuum Fitting method. Note that the spin of GRO J1655-40 determined using the RPM is not consistent with the Continuum Fitting method \cite{Motta:2013wga}.

\begin{figure}[htp]
  \begin{subfigure}{0.5\textwidth}
    \centering
    \includegraphics[width=\linewidth]{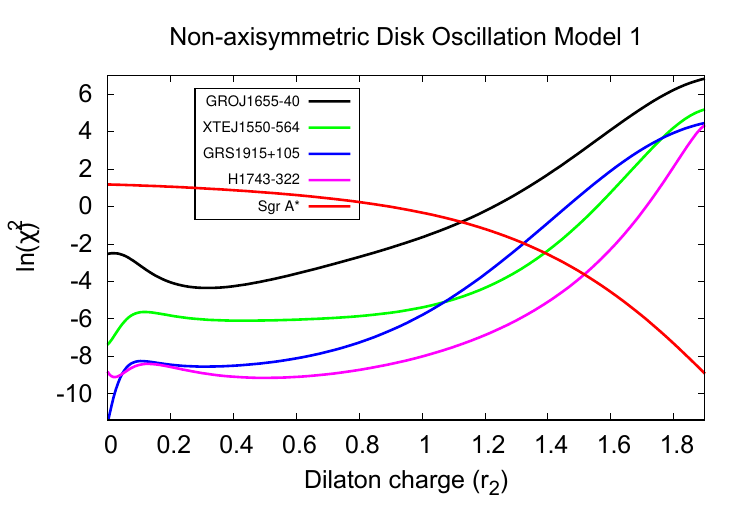}
    \caption{}
    \label{fig13a}
  \end{subfigure}
  \begin{subfigure}{0.5\textwidth}
    \centering
    \includegraphics[width=\linewidth]{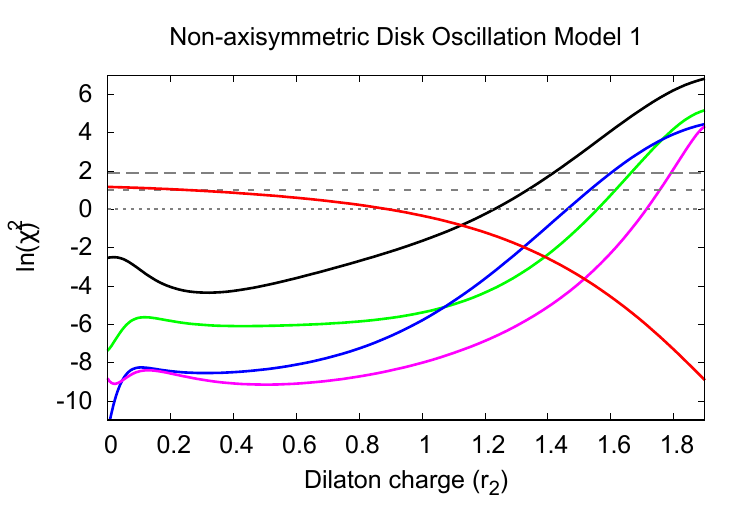}
    \caption{}
    \label{fig13b}
  \end{subfigure}
  \caption{The above figure (a) demonstrates the variation of $\chi^2$ with the dilaton charge $r_2$ for the five black hole sources in \ref{Tab2} assuming NADO1. Figure (b) plots the the variation of $\chi^2$ with $r_2$ (assuming the same model) along with the confidence lines. The grey dotted line corresponds to the 1-$\sigma$ contour, the grey short-dashed line corresponds to the 2-$\sigma$ contour and the grey long-dashed line is associated with the 3-$\sigma$ contour. The confidence contour lines are the same for all the BHs.}
  \label{fig13}
\end{figure}

\ref{fig13} shows the variation of $\chi^2$ with $r_2$ for the five BH sources assuming NADO1. For GRO J1655-40, the $\chi^2$ minimizes at $r_2\sim 0.3$ and its HFQPO data rules out $r_2\gtrsim 1.2$ outside 1-$\sigma$ (grey dotted line in \ref{fig13b}) and $r_2\gtrsim 1.4$ outside 3-$\sigma$ (grey long-dashed line in \ref{fig13b}). For  XTE J1550-564 and GRS 1915+105 the $\chi^2$ minimizes at $r_2\sim 0$ while for H1743-322 minimum $\chi^2$ occurs at $r_2\sim 0.03$. HFQPO data of XTE J1550-564, GRS 1915+105 and H1743-322 respectively rule out $r_2\gtrsim 1.5$, $r_2\gtrsim 1.4$ and $r_2\gtrsim 1.7$ outside 1-$\sigma$ (grey dotted line) and $r_2\gtrsim 1.7$, $r_2\gtrsim 1.6$ and $r_2\gtrsim 1.8$ outside 3-$\sigma$ (grey long-dashed line). Thus, all the 4 stellar mass BHs rule out near extremal dilaton charge. In this regard Sgr A* is exceptional, as its $\chi^2$ minimizes at $r_2\sim 1.9$ and it rules out $r_2\sim 0$ outside 2-$\sigma$. When the 3-$\sigma$ interval is considered $r_2\sim 0$ is allowed. 

\begin{figure}[htp]
  \begin{subfigure}{0.5\textwidth}
    \centering
    \includegraphics[width=\linewidth]{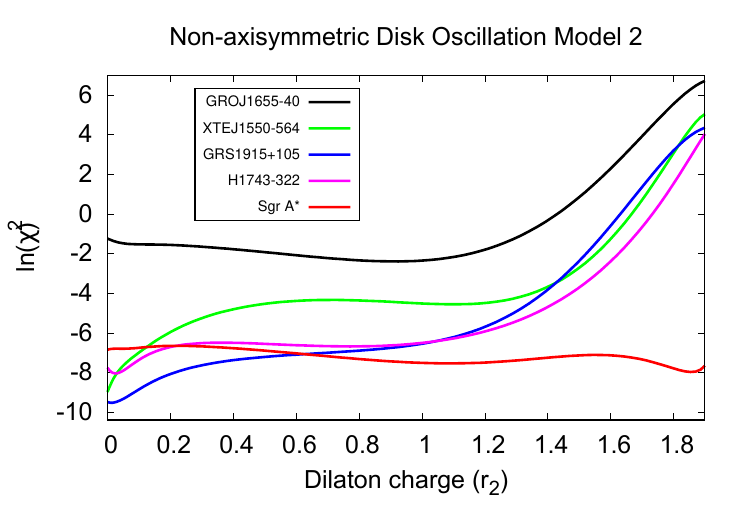}
    \caption{}
    \label{fig14a}
  \end{subfigure}
  \begin{subfigure}{0.5\textwidth}
    \centering
    \includegraphics[width=\linewidth]{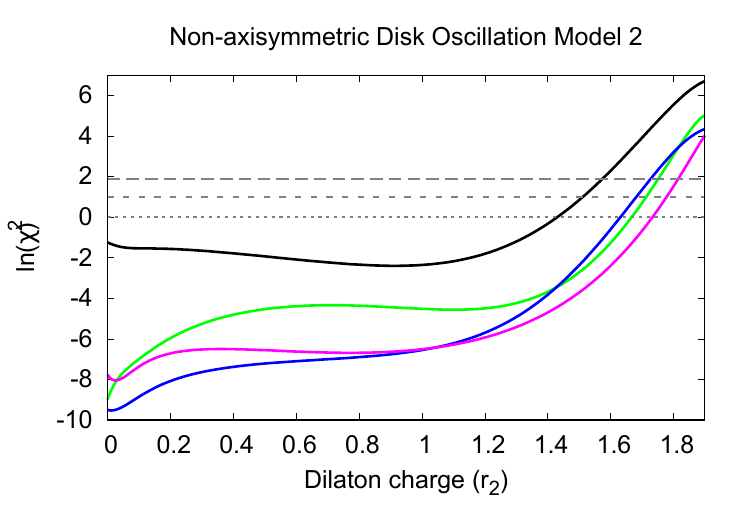}
    \caption{}
    \label{fig14b}
  \end{subfigure}
  \caption{The above figure (a) demonstrates the variation of $\chi^2$ with the dilaton charge $r_2$ for the five black hole sources in \ref{Tab2} assuming NADO2. Figure (b) plots the the variation of $\chi^2$ with $r_2$ (assuming the same model) but for a subset of those five BHs  where the $\chi^2$ values are large and the variation with $r_2$ is also substantial such that the confidence lines can be drawn. The grey dotted line corresponds to the 1-$\sigma$ contour, the grey short-dashed line corresponds to the 2-$\sigma$ contour and the grey long-dashed line is associated with the 3-$\sigma$ contour. The confidence contour lines are the same for all the four BHs.}
  \label{fig14}
\end{figure}

One may also consider resonance between two non-axi-symmetric modes, namely, the $m=-1$ non-axisymmetric radial epicyclic frequency with the $m=-2$ non-axisymmetric vertical epicyclic frequency such that $f_1=2f_\phi-f_\theta$ and $f_2=f_{per}$ (NADO2) \cite{Torok:2010rk,Torok:2011qy,Kotrlova:2020pqy}. Although the non-axi-symmetric modes may couple and give rise to the HFQPOs, the exact physics leading to such couplings remains ill-understood. \ref{fig14} demonstrates the variation of $\chi^2$ with $r_2$ for the five BH sources assuming NADO2. We note from \ref{fig14a} that the magnitude of $\chi^2$ and the variation of $\chi^2$ are both very less for Sgr A*, as a result of which it does not impose strong bounds on $r_2$. The remaining four black holes do constrain the model and hence they are separately    plotted in \ref{fig14b}. GRO J1655-40 does the best in this regard as it rules out the largest parameter space of $r_2$, the allowed range being $0\lesssim r_2 \lesssim 1.4$ within 1-$\sigma$ and $0\lesssim r_2 \lesssim 1.6$ within 3-$\sigma$. Thus, $r_2\gtrsim 1.6$ is ruled out outside 3-$\sigma$. Also for this source $\chi^2$ minimizes at $r_2\sim 1$ although $r_2\sim 0$ is allowed within 1-$\sigma$. For XTE J1550-564 and GRS 1915+105 the $\chi^2$ minimizes at $r_2\sim 0$ while for H1743-322 the $\chi^2_{min}\sim 0.02$. These sources rule out $r_2\gtrsim 1.65$ (XTE J1550-564), $r_2\gtrsim 1.6$ (GRS 1915+105) and $r_2\gtrsim 1.7$ (H1743-322) outside 1-$\sigma$. When the 3-$\sigma$ interval is considered, XTE J1550-564, GRS 1915+105 and H1743-322 respectively rule out $r_2\gtrsim 1.8$, $r_2\gtrsim 1.75$ and $r_2\gtrsim 1.83$. Once again, all the four stellar-mass BH sources universally rule out near extremal $r_2$ when NADO2 is used to explain their HFQPO data.

\subsubsection{Warped Disc Oscillation Model}
The fact that the radial epicyclic frequency attains a maximum and then declines to zero at the ISCO can give rise to certain non-linear resonances between the various disk oscillation modes when the relativistic disk is deformed by a warp \cite{2001PASJ...53....1K,2004PASJ...56..559K,2004PASJ...56..905K,2005PASJ...57..699K,2008PASJ...60..111K}. The horizontal resonances can excite the p-mode and the g-mode oscillations while the vertical resonances can induce only the g-mode oscillations \cite{2004PASJ...56..559K}.
This forms the basis of the Warped Disc Oscillation Model which assumes the upper high frequency QPO to be $f_{1}=2f_\phi -f_r$ while the lower HFQPO is taken to be $f_{2}=2(f_\phi -f_r)$.

\begin{figure}[t!]
  \begin{subfigure}{0.5\textwidth}
    \centering
    \includegraphics[width=\linewidth]{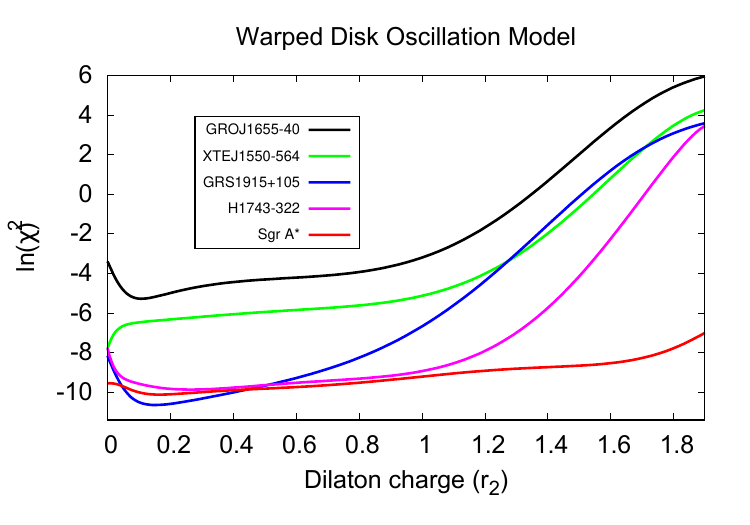}
    \caption{}
    \label{fig15a}
  \end{subfigure}
  \begin{subfigure}{0.5\textwidth}
    \centering
    \includegraphics[width=\linewidth]{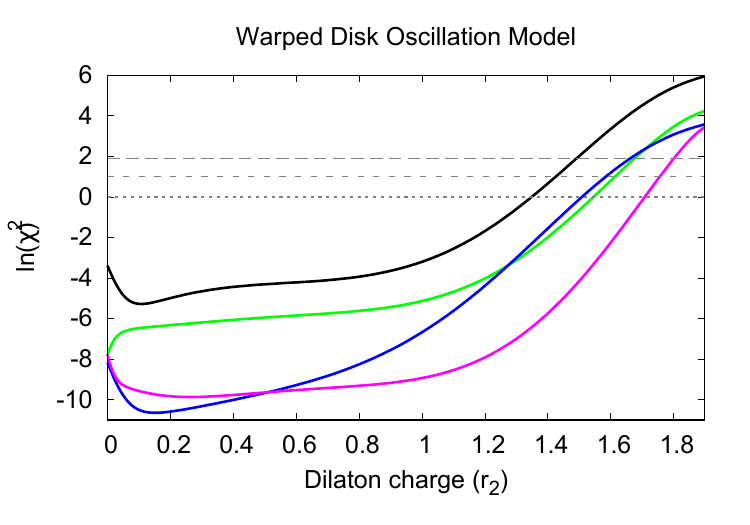}
    \caption{}
    \label{fig15b}
  \end{subfigure}
  \caption{The above figure (a) demonstrates the variation of $\chi^2$ with the dilaton charge $r_2$ for the five black hole sources in \ref{Tab2} assuming WDOM. Figure (b) plots the the variation of $\chi^2$ with $r_2$ (assuming the same model) but for a subset of those five BHs  where the $\chi^2$ values are large and the variation with $r_2$ is also substantial such that the confidence lines can be drawn. The grey dotted line corresponds to the 1-$\sigma$ contour, the grey short-dashed line corresponds to the 2-$\sigma$ contour and the grey long-dashed line is associated with the 3-$\sigma$ contour. The confidence contour lines are the same for all the four BHs.}
  \label{fig15}
\end{figure}

Assuming WDOM we plot in \ref{fig15} the variation of $\chi^2$ with $r_2$ for all the five BH sources. Like NADO2, Sgr A* data fails to constrain the model for reasons discussed earlier while GRO J1655-40 imposes the strongest constrain. The confidence intervals are plotted with grey lines in \ref{fig15b}. For GRO J1655-40 the $\chi^2$ minimizes at $r_2\sim 0.1$ and the data rules out $r_2\gtrsim 1.3$ outside 1-$\sigma$ and $r_2\gtrsim 1.5$ outside 3-$\sigma$. However, $r_2\sim 0$ is allowed within 1-$\sigma$. For XTE J1550-564, $\chi^2_{min}$ occurs at $r_2\sim 0$, and the data rules out $r_2\gtrsim 1.55$ and $r_2\gtrsim 1.7$ outside 1-$\sigma$ and 3-$\sigma$ respectively. When the data of GRS 1915+105 is considered, $\chi^2$ minimizes at $r_2\sim 0.1$ although $r_2\sim 0$ is allowed within 1-$\sigma$. The data rules out $r_2\gtrsim 1.5$ and $r_2\gtrsim 1.7$ outside 1-$\sigma$ and 3-$\sigma$ respectively.
For H1743-322, the $\chi^2$ minimizes at $r_2\sim 0.2$ while $r_2\sim 0$ is allowed within 68\% confidence interval. When the 68\% and 99\% confidence intervals are considered, the data rules out $r_2\gtrsim 1.7$ and $r_2\gtrsim 1.8$ respectively.

We summarize the main results of our analysis in \ref{Tab3} to \ref{Tab7} where we enlist the constraints on the dilaton charge, spin and mass of the five BHs from their HFQPO data (LFQPO data is taken only for GRO J1655-40 with the RPM model), considering the  different QPO models. In the next section we estimate the observationally favored model parameters using the Markov chain Monte Carlo (MCMC) simulations. The results obtained from MCMC simulations are also summarized in the aforesaid tables for comparison purpose.
For each source we report the constraints on $r_2$, $a$ and $M$ only for those models which could rule out a certain part of the parameter space of $r_2$ using the HFQPO data (\ref{fig5b}-\ref{fig15b}).

\section{Constraining the model parameters using the MCMC method}
\label{S6-1}
In the previous section, we used the grid-search method to derive the best-fit model parameters. In this method, we have varied the model parameters (the dilaton charge $r_2$, the spin $a$, the emission radius $r_{em}$, and the mass of the black hole $M$) across their entire allowed range, and for every combination of values of these four parameters, we have calculated the $\chi^2$, which is a function of these four parameters. 
While using the grid-search method, the chances of getting stuck at a local minima can be substantially reduced if the model parameters are varied in very small intervals, i.e., fine gridding is used. Such a fine gridding ensures evaluating the $\chi^2$ for a very large number of combinations of values of the four model parameters, and it is clear that the greater the number of combinations, the higher is the probability of locating the global minima of $\chi^2$. 
To ensure that our gridding is sufficiently fine, in this section, we verify our results with the Bayesian approach, which employs Markov chain Monte Carlo (MCMC) simulations. For each source, we have compared the observed HFQPO frequencies with all eleven model-dependent HFQPO frequencies and obtained constraints on the model parameters. We have estimated the best-fit model parameters along with their confidence intervals, for the different BH sources considered in our work, taking into account all the eleven HFQPO models.  

The Bayesian posterior distribution for the model parameters is given by,
\begin{align}
\mathcal{P}(\theta|D)=\frac{\mathcal{P}(D|\theta)\mathcal{P}(\theta)}{\mathcal{P}(D)}
\label{3}
\end{align}
where, $\mathcal{P}(\theta)$ represents the prior distribution for the model parameters $\theta\equiv \lbrace r_2, a, M, r_{em}\rbrace$, and $\mathcal{P}(D|\theta)$ is associated with the likelihood function and $\mathcal{P}(D)=1$ since the data is known \cite{Verde:2009tu}. We assume flat priors for $r_2$, $a$, $r_{em}$ and $M$ (for GRS 1915 + 105, H 1743 + 322 and Sgr A*) in the allowed range (discussed in \ref{S5}) and Gaussian priors for the masses of GRO J1655-40 and XTE J1550-564 (see \ref{Tab2}), based on previous estimates. 
The likelihood function considers the contribution from both the upper and lower HFQPO frequencies and also the LFQPO if exhibited by the source and addressed by the model (e.g., GRO J1655-40 exhibits twin-peak HFQPOs and an LFQPO, all of which can be explained by the RPM model). This is given by,
\begin{align}
\log \mathcal{L}=\log \mathcal{L}_{U_1} + \log \mathcal{L}_{U_2} + \log \mathcal{L}_{L}
\label{4}
\end{align}
where, 
\begin{align}
\log \mathcal{L}_{U_1} =-\frac{1}{2} \frac{\lbrace \nu_{\textrm{1}{,i}}-f_1(r_2,a,M,r_{\rm em}) \rbrace ^2}{\sigma_{\nu_{\rm 1},i}^2}
\label{5}
\end{align}
\begin{align}
\log \mathcal{L}_{U_2} =-\frac{1}{2} \frac{\lbrace \nu_{\textrm{2}{,i}}-f_2(r_2,a,M,r_{\rm em}) \rbrace ^2}{\sigma_{\nu_{\rm 2},i}^2}
\label{6}
\end{align}
\begin{align}
\log \mathcal{L}_{L} =-\frac{1}{2} \frac{\lbrace \nu_{\textrm{3}{,i}}-f_3(r_2,a,M,r_{\rm em}) \rbrace ^2}{\sigma_{\nu_{\rm 3},i}^2}
\label{6-1}
\end{align}
Note that except for RPM with source GRO J1655-40 $\log \mathcal{L}_{L}=0$. For each source, we take the HFQPO observational data from previous literature, which are also listed in \ref{Tab2}. We consider $64$ chains, and in each chain, we draw $10,000$ samples for each of the four model parameters based on the prior information, which ensures a thorough investigation of the multidimensional parameter space. 

While performing the MCMC simulations, we have noted that the likelihood function is very sensitive to the variations in $M$ and $r_{em}$. Thus, although determining $r_{em}$ and $M$ is not the main goal of this work, marginalizing the likelihood function with respect to $r_{em}$ and $M$ would average out the effects of these two parameters, thereby lowering the maxima of the marginalized likelihood function. Further, this would lead to loss of information regarding $M_{min}$ and $r_{em,min}$. This does not seem to be desirable as we lose the opportunity to determine the BH mass from the HFQPO data and end up in a region of the parameter space where the likelihood function does not attain the global maxima. 
Also, HFQPOs are expected to be incited near the BH and indeed, the $r_{em,min}$ we are reporting in the corner plots confirm this. If we had marginalized over $r_{em}$, then all emission radius between $r_{ms}\leq r_{em} \leq r_{ms}+20 R_g$ seem to be on the same footing. Thus, keeping $r_{em}$ and $M$ free, but setting their priors suitably seems to be more appropriate in the present situation. 

A similar argument is applicable for the spin parameter. 
Although the likelihood function is often not very sensitive to spin (one of the exceptions being GRO J1655-40 when tested with RPM), marginalization over spin will average out its effect, and we will not be able to tell from the HFQPO data whether the source is slowly/moderately/rapidly rotating. Since, the effect of a BH's spin gains prominence in its vicinity, spin plays a crucial role in strong gravity physics. Hence, marginalization over the spin will lead to misinterpretation of the data and will also prevent us from providing a spin estimate, even though MCMC gives us the provision to do so. 
Thus, maximizing the likelihood function without marginalization over any of the parameters seems more reasonable in our case. A similar analysis has been performed in \cite{Hazarika:2025axz,Wang:2021gtd} where marginalization has not been used.
In what follows, we present our MCMC results source-wise.

\begin{figure}[H]
\vspace{-1.2cm}
{\bf \underline{GRO J1655-40}}
    \centering

    \begin{subfigure}[b]{0.42\textwidth}
        \includegraphics[width=\linewidth]{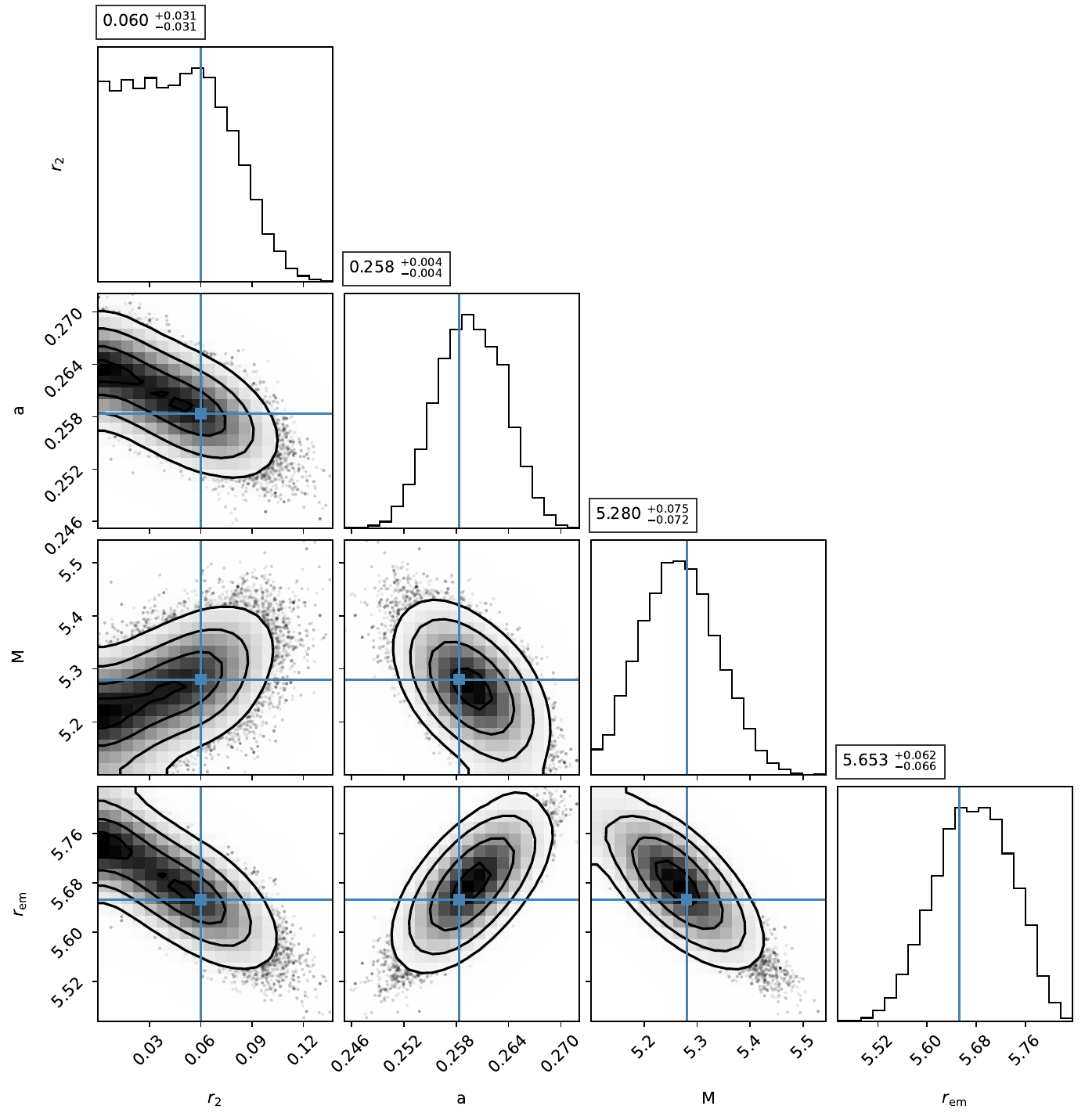}
        \caption*{(a) Relativistic Precession Model}
    \end{subfigure}
    \hspace{0cm}
    \begin{subfigure}[b]{0.42\textwidth}
        \includegraphics[width=\linewidth]{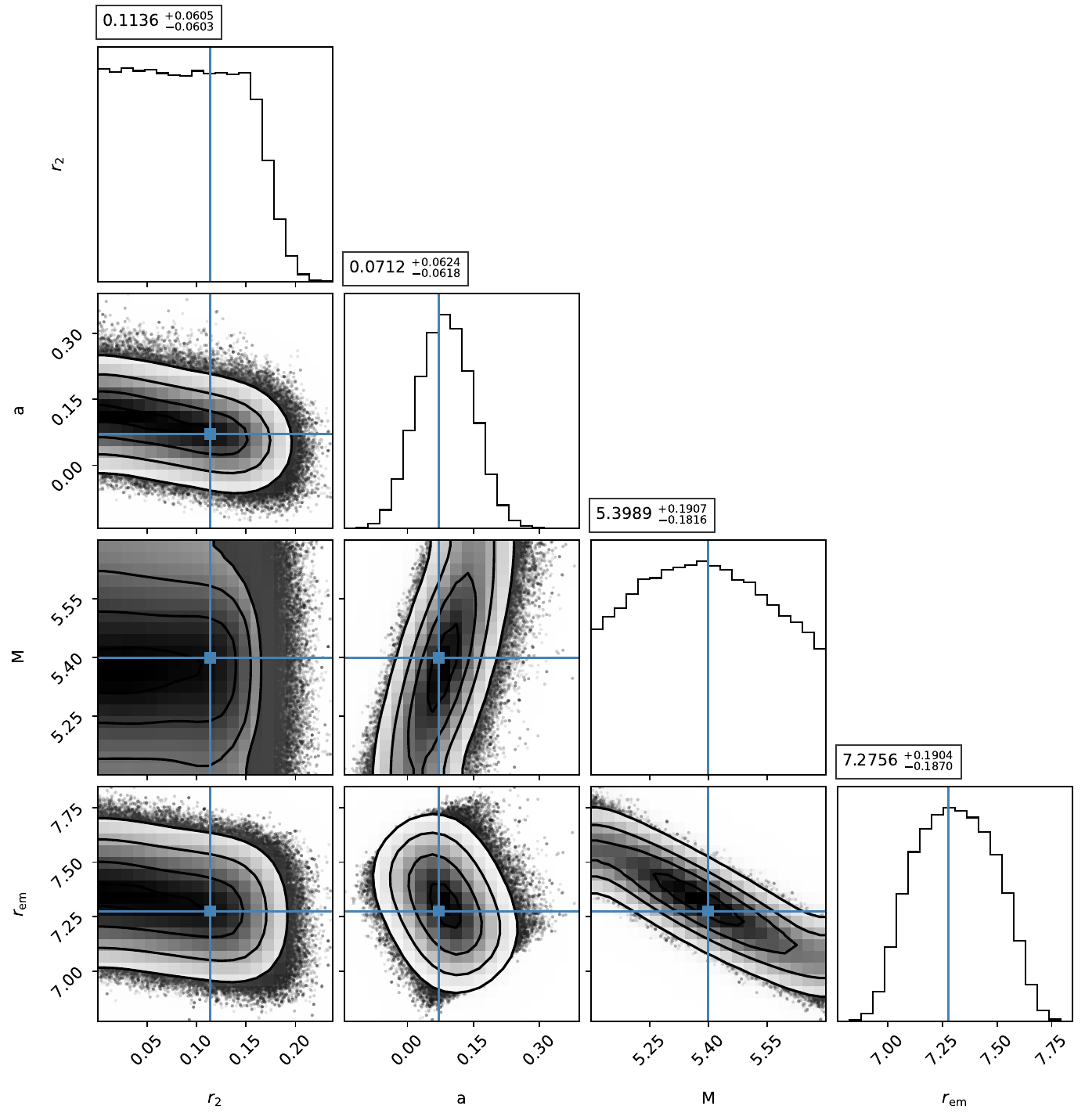}
        \caption*{(b) Tidal Disruption Model}
    \end{subfigure}
    \hspace{0cm }
    \begin{subfigure}[b]{0.42\textwidth}
        \includegraphics[width=\linewidth]{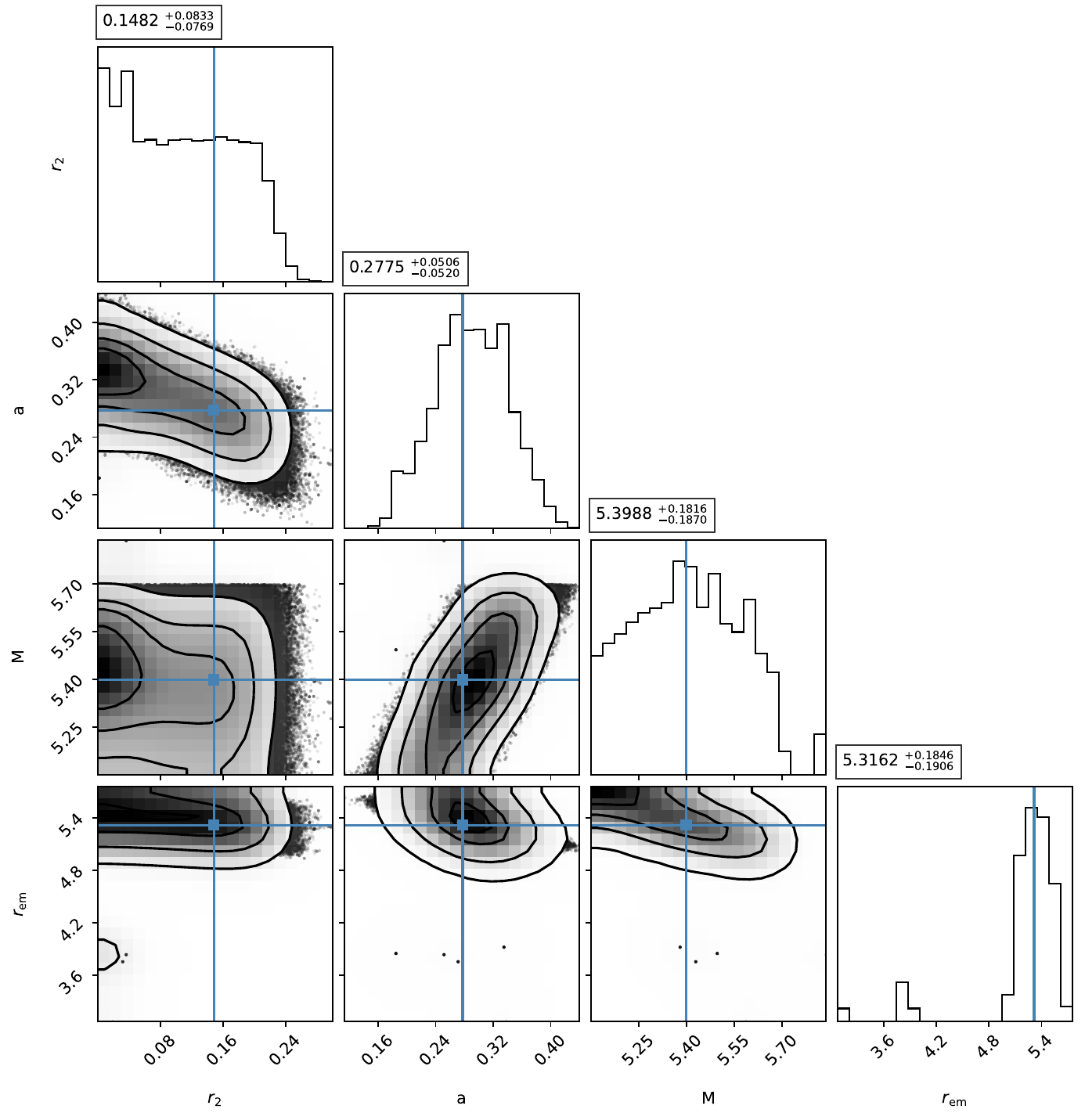}
        \caption*{(c) Forced Resonance Model 1} 
    \end{subfigure}

    \vspace{0.5cm} 

    \begin{subfigure}[b]{0.42\textwidth}
        \hspace{0cm} 
        \includegraphics[width=\linewidth]{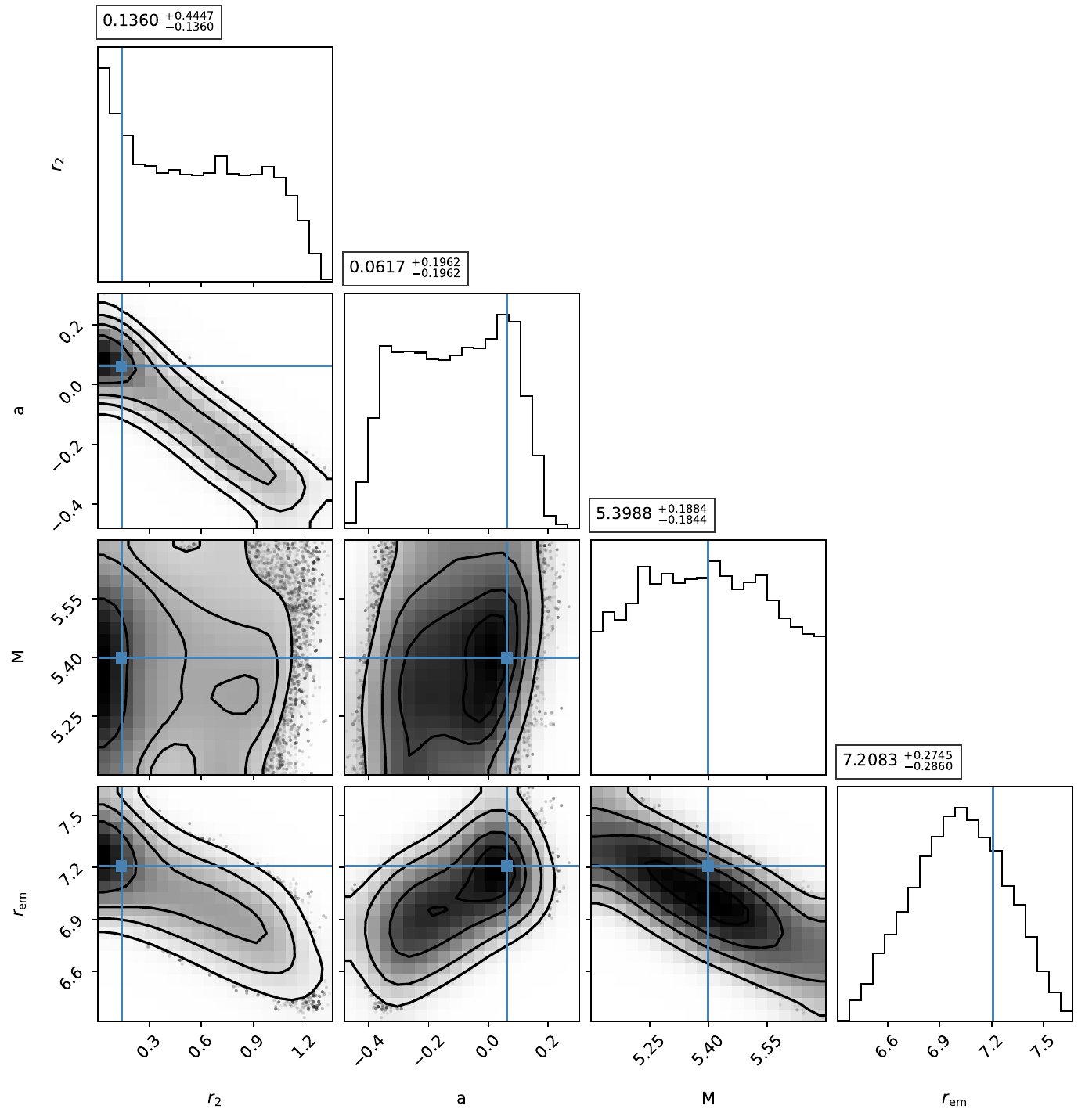}
        \caption*{(d) Forced Resonance Model 2}
    \end{subfigure}
    \hspace{2cm}
    \begin{subfigure}[b]{0.42\textwidth}
        \includegraphics[width=\linewidth]{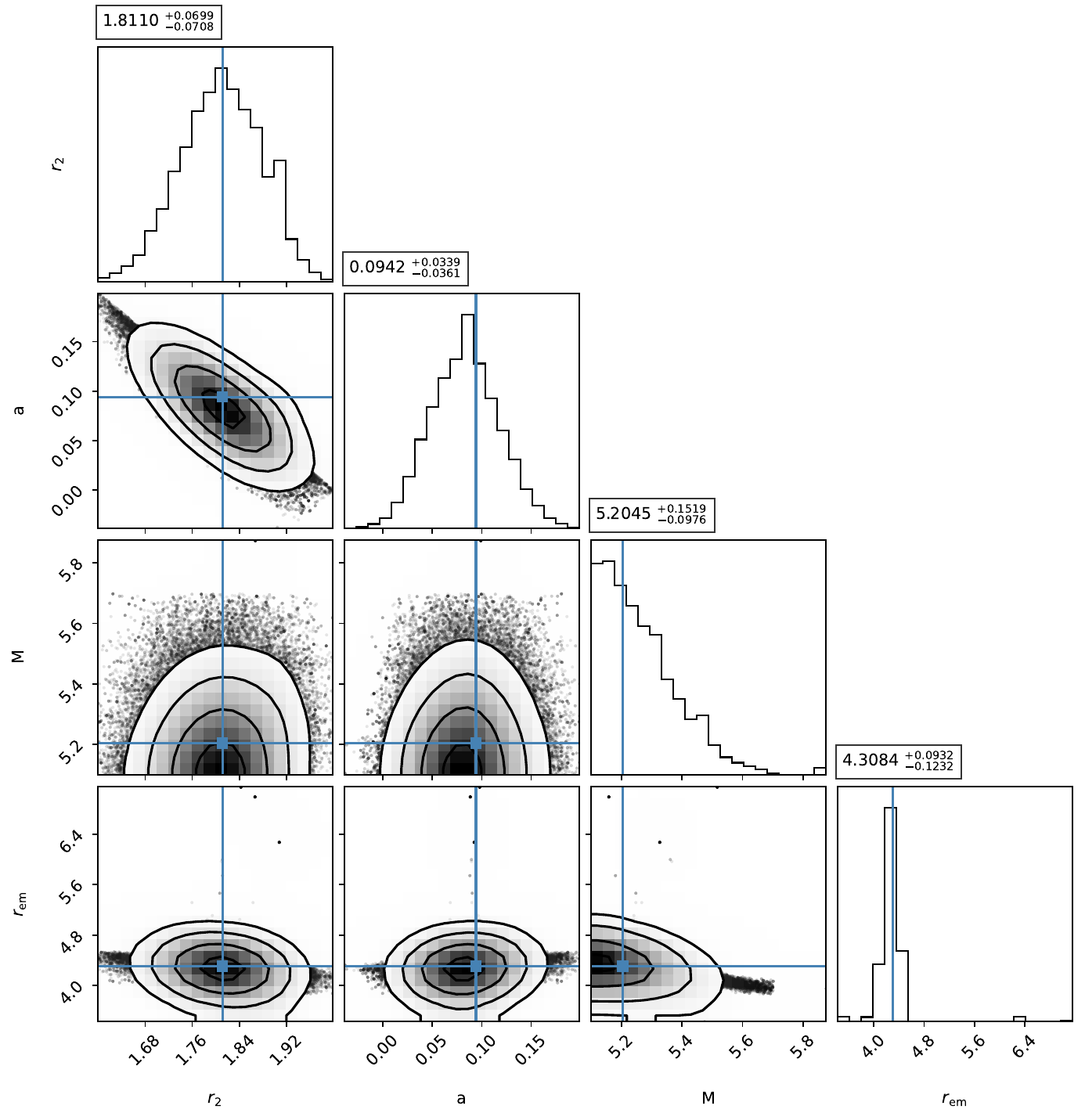}
        \caption*{(e) Keplerian Resonance Model 1}
    \end{subfigure}

\caption{Constraints on the model parameters using the QPO data of GRO J1655-40 considering (a) the Relativistic Precession Model, (b) the Tidal Disruption Model, (c) the Forced Resonance Model 1, (d) the Forced Resonance Model 2, and (e) the Keplerian Resonance Model 1.}
\label{Fig1}
\end{figure}

\begin{figure}[H]
\vspace{-1.2cm}
{\bf \underline{GRO J1655-40}}
    \centering

    \begin{subfigure}[b]{0.42\textwidth}
        \includegraphics[width=\linewidth]{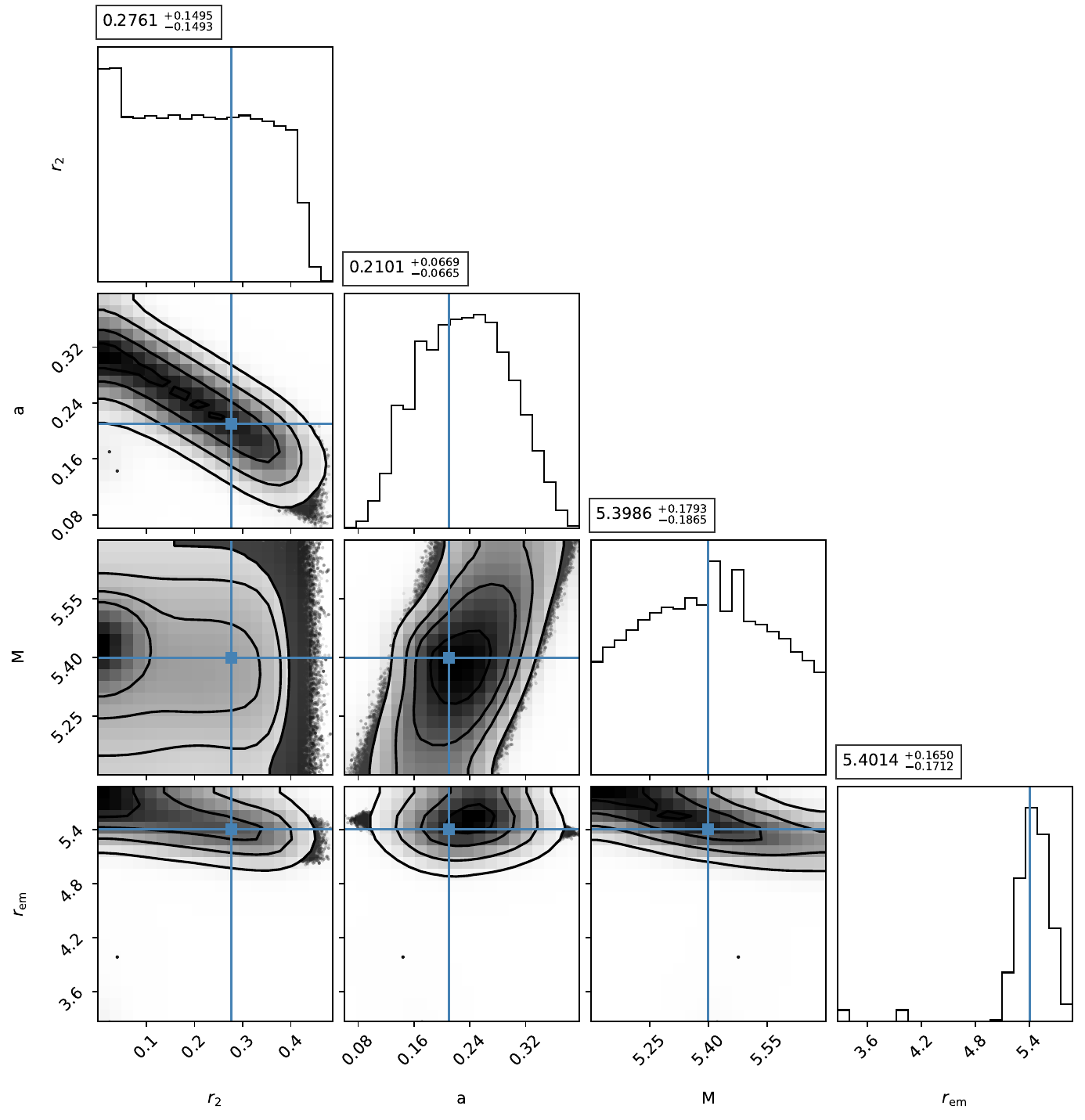}
        \caption*{(a) Keplerian Resonance Model 2}
    \end{subfigure}
    \hspace{0cm}
    \begin{subfigure}[b]{0.42\textwidth}
        \includegraphics[width=\linewidth]{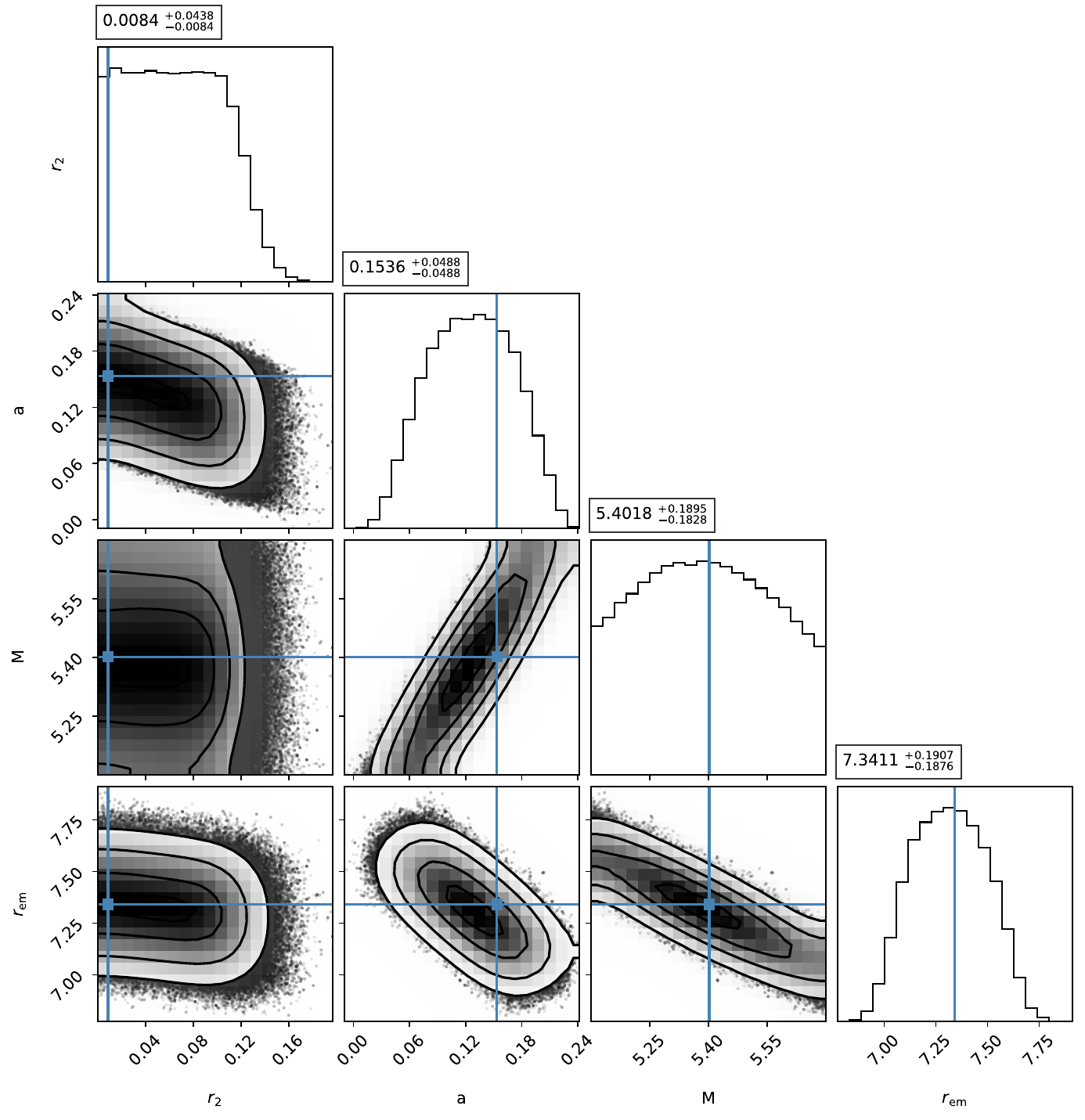}
        \caption*{(b) Keplerian Resonance Model 3}
    \end{subfigure}
    \hspace{0cm }
    \begin{subfigure}[b]{0.42\textwidth}
        \includegraphics[width=\linewidth]{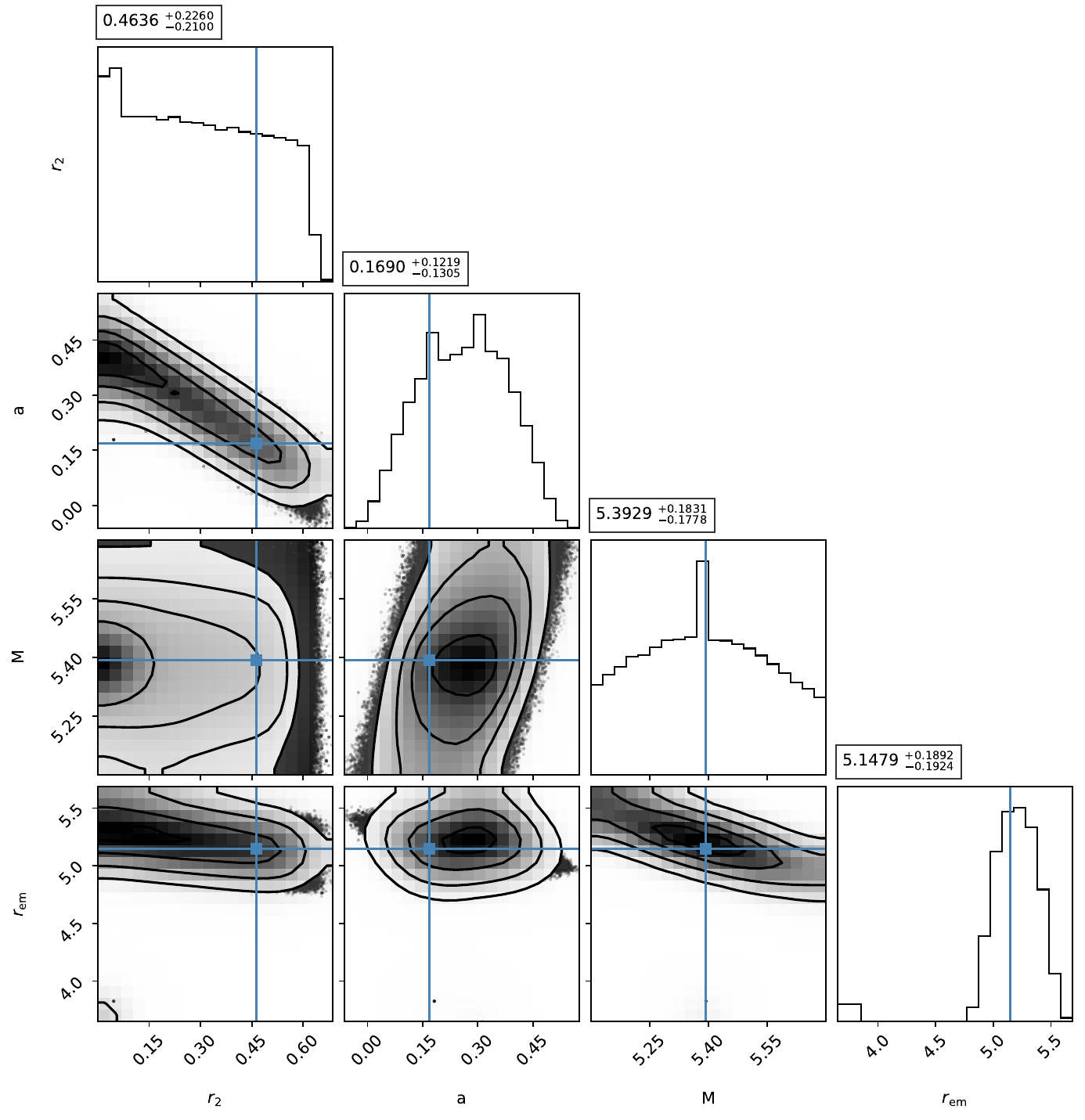}
        \caption*{\hspace{-0.36cm}(c) Non-axisymmetric Disk Oscillation Model 1} 
    \end{subfigure}

    \vspace{0.3cm} 

    \begin{subfigure}[b]{0.42\textwidth}
        \hspace{0cm} 
        \includegraphics[width=\linewidth]{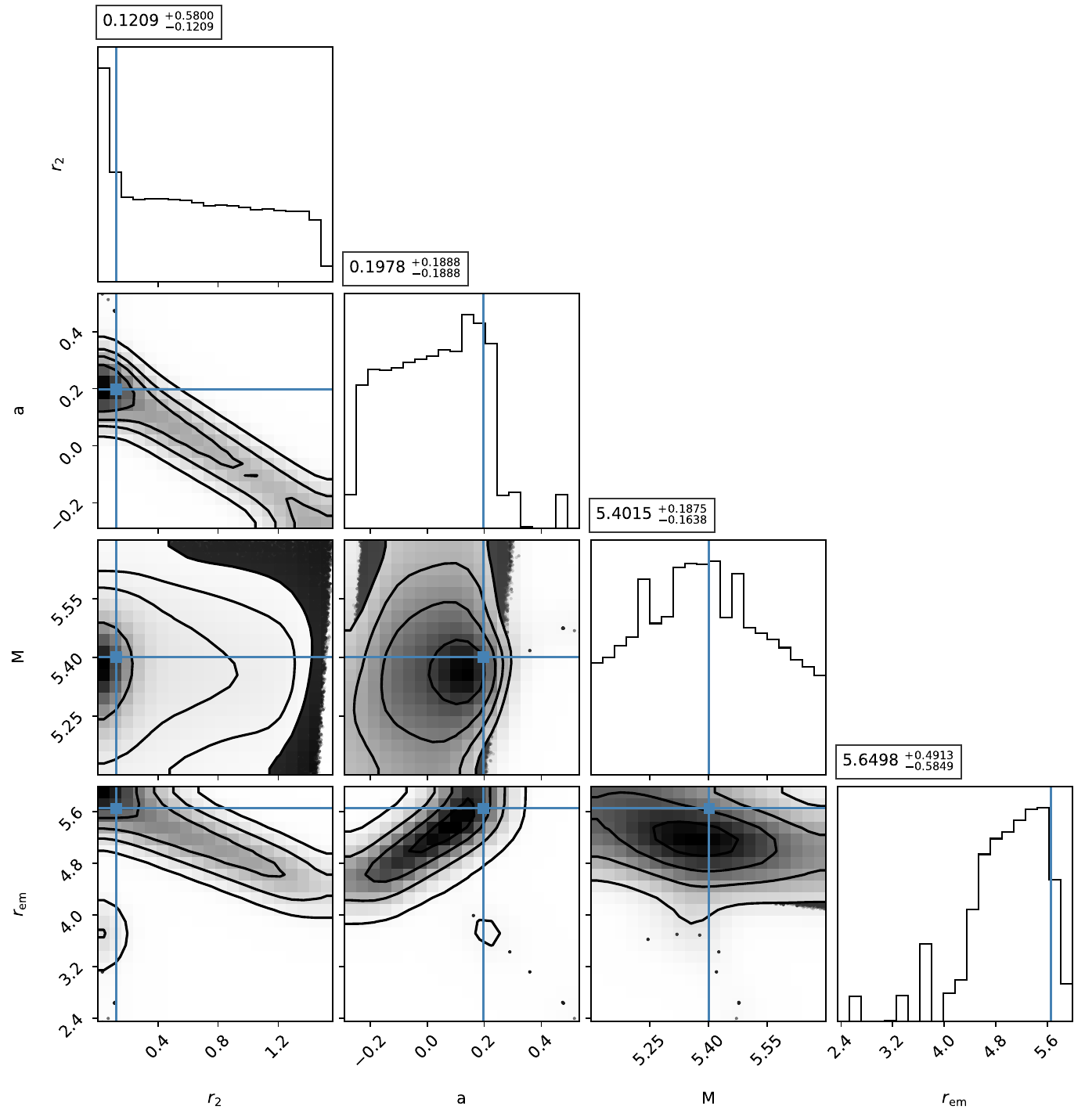}
        \caption*{(d) Non-axisymmetric Disk Oscillation Model 2}
    \end{subfigure}
    \hspace{2cm}
    \begin{subfigure}[b]{0.42\textwidth}
        \includegraphics[width=\linewidth]{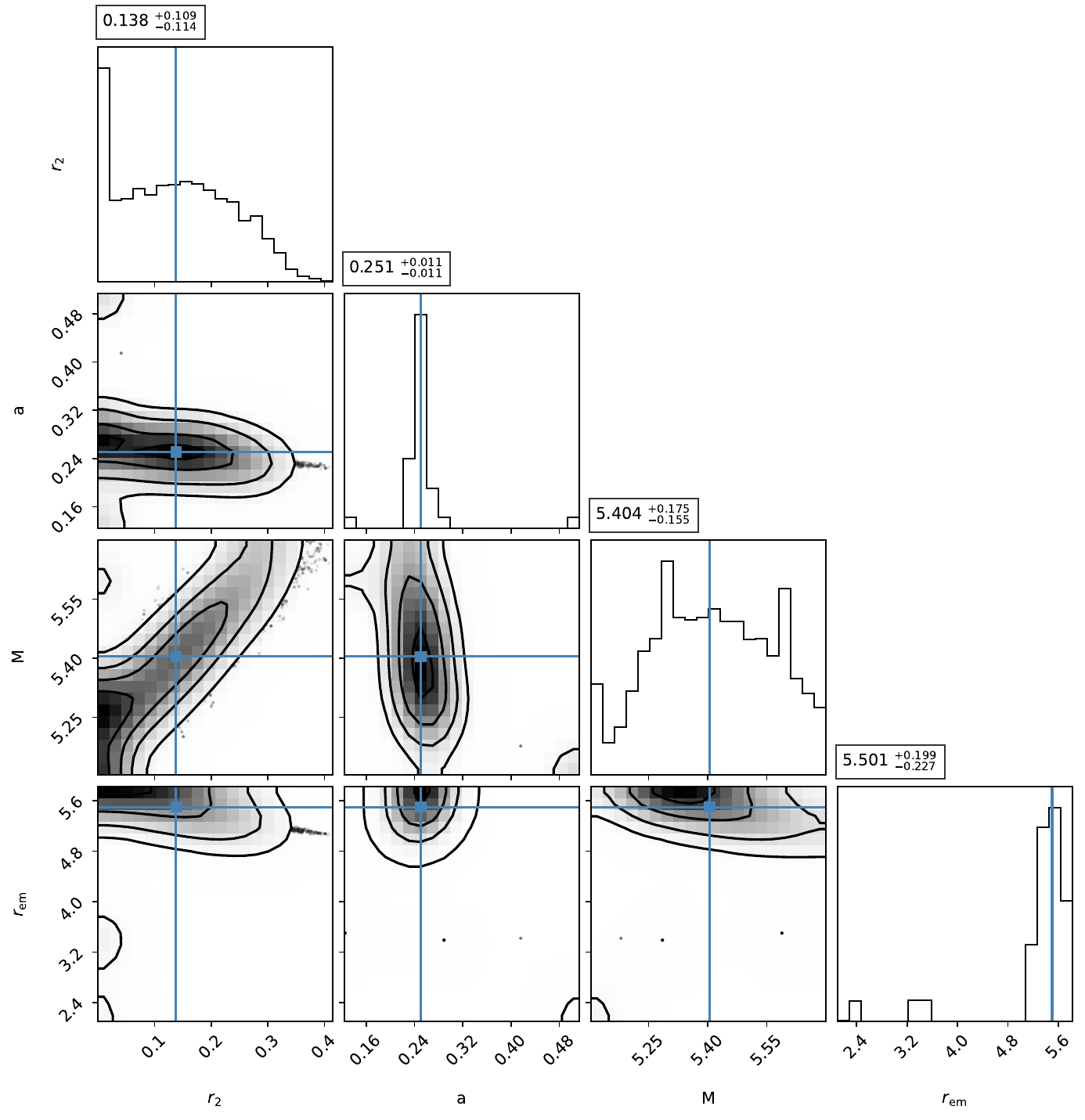}
        \caption*{(e) Warped Disc Oscillation Model}
    \end{subfigure}

\caption{Constraints on the model parameters using the QPO data of GRO J1655-40 considering (a) the Keplerian Resonance Model 2, (b) the Keplerian Resonance Model 3, (c) the Non-axisymmetric Disk Oscillation Model 1, (d) the Non-axisymmetric Disk Oscillation Model 2, and (e) the Warped Disc Oscillation Model.}
\label{Fig2}
\end{figure}

\begin{figure}[htp]
\vspace*{-3.6cm }
{\bf \underline{XTE J1550-564}}
\centering

\begin{minipage}[b]{0.49\textwidth}
\centering
\includegraphics[width=\textwidth]{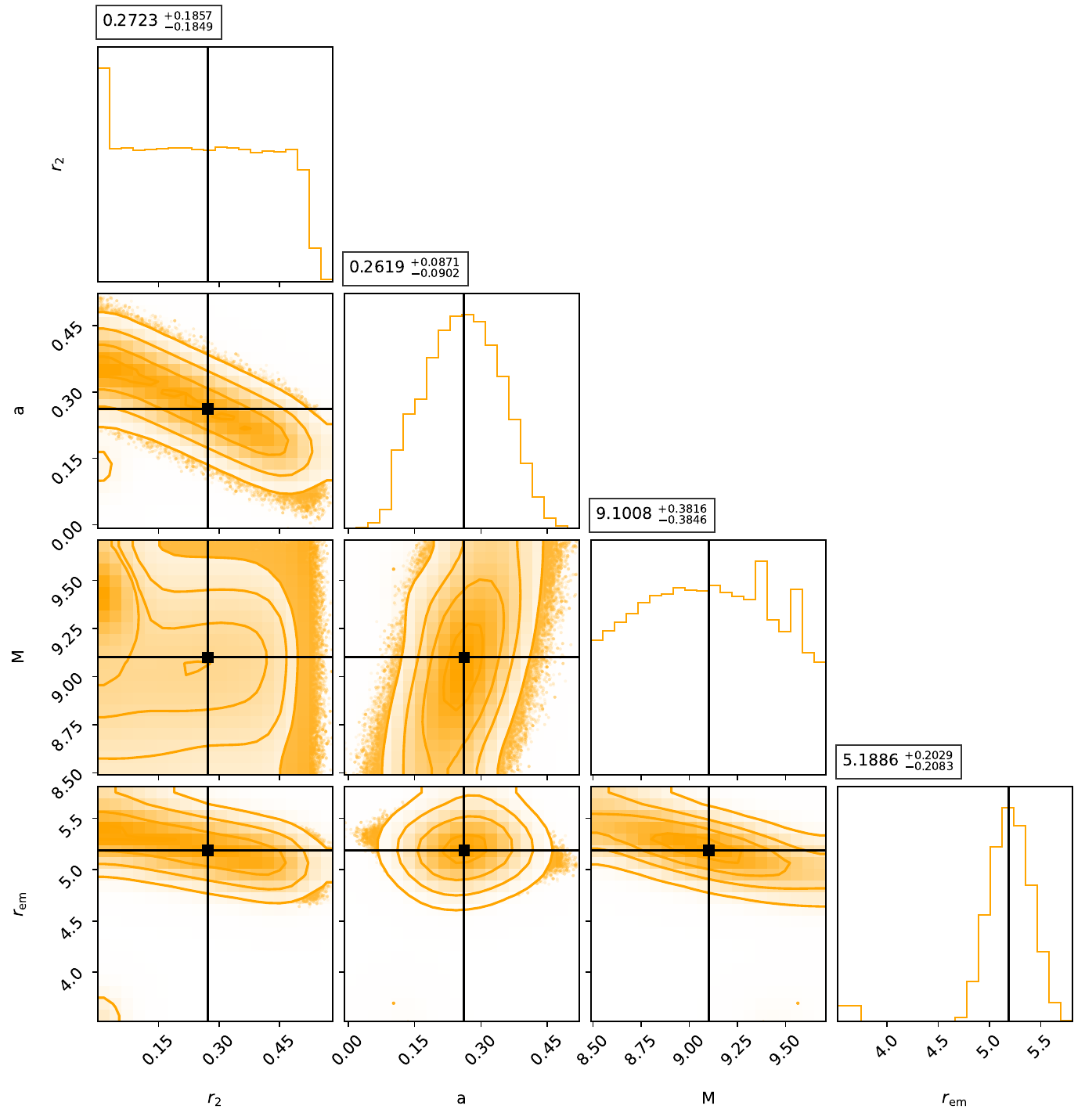}
\caption*{(a) Relativistic Precession Model}
\end{minipage}
\hfill
\begin{minipage}[b]{0.49\textwidth}
\centering
\includegraphics[width=\textwidth]{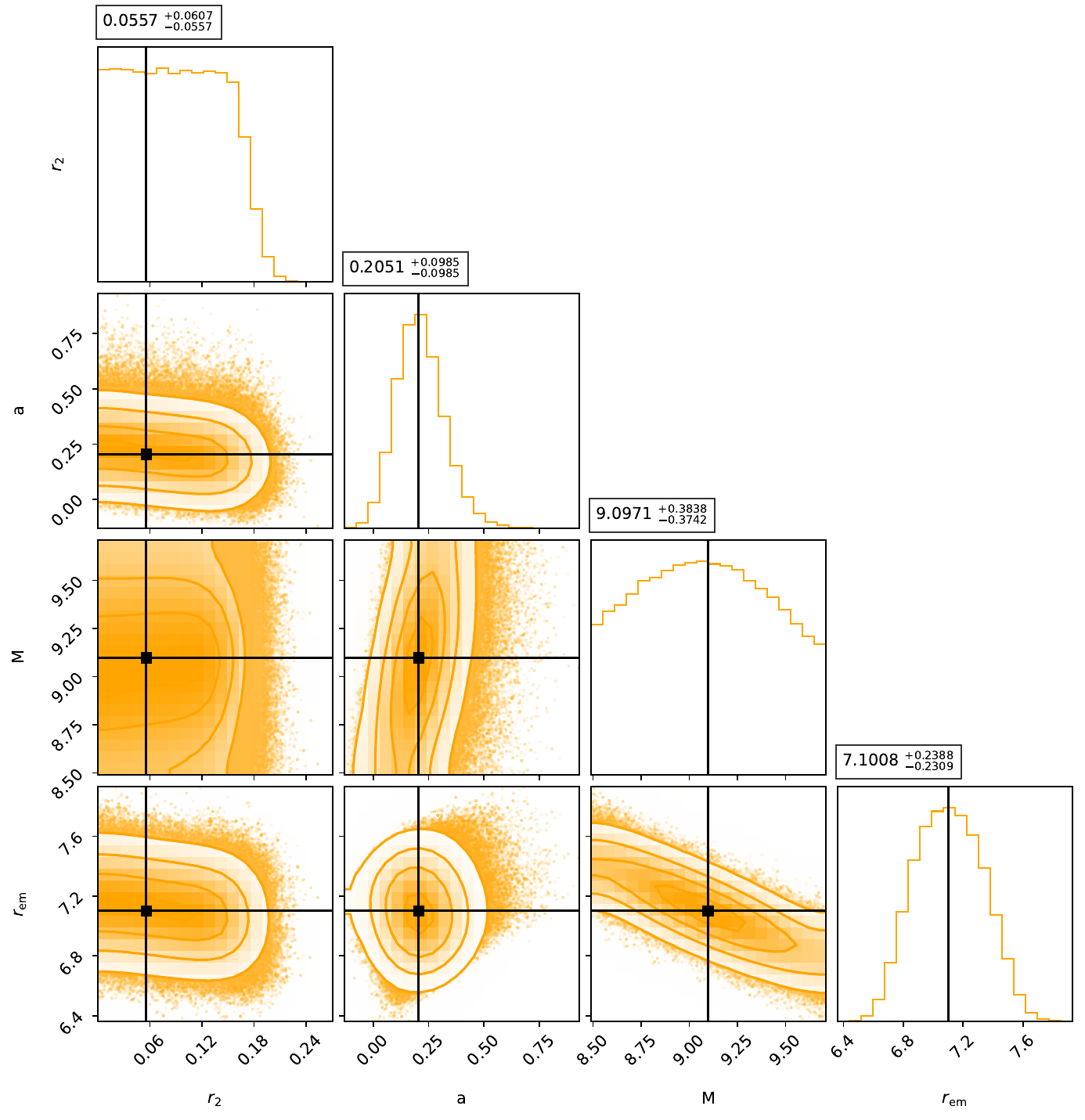}
\caption*{(b) Tidal Disruption Model}
\end{minipage}

\vspace{0.3cm}

\begin{minipage}[b]{0.49\textwidth}
\centering
\includegraphics[width=\textwidth]{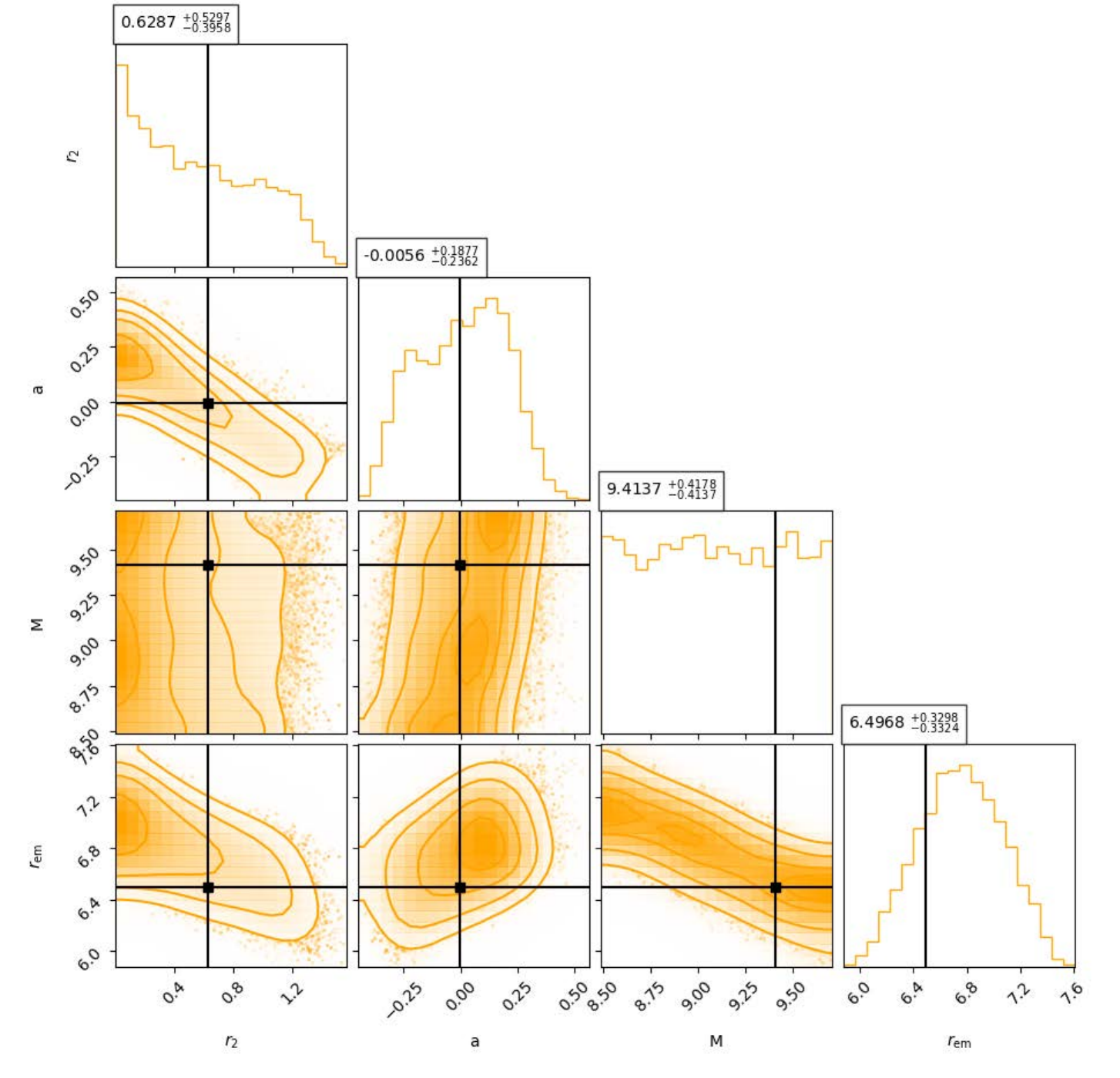}
\caption*{(c) Forced Resonance Model 1}
\end{minipage}
\hfill
\begin{minipage}[b]{0.49\textwidth}
\centering
\includegraphics[width=\textwidth]{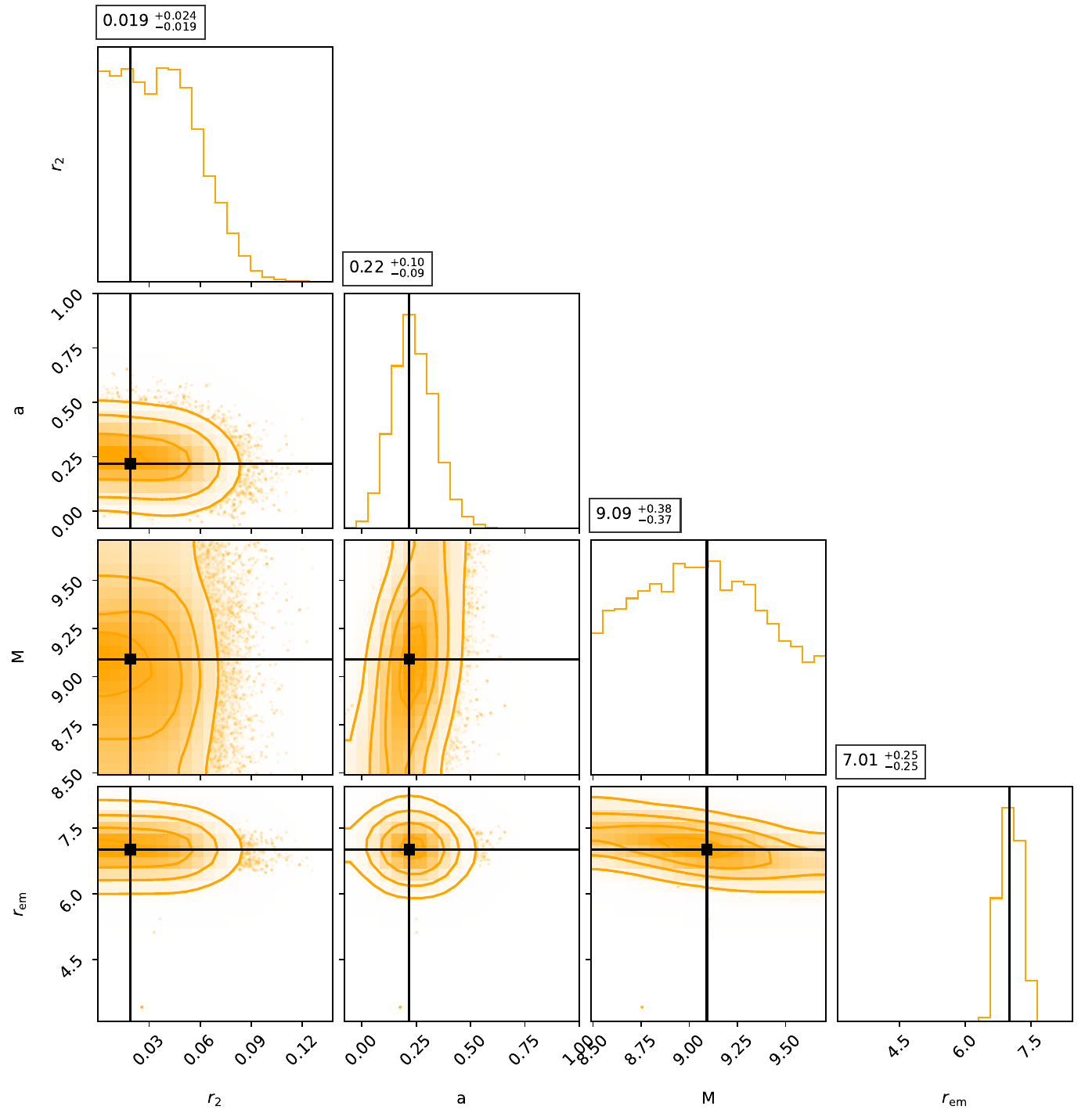}
\caption*{(d) Forced Resonance Model 2}
\end{minipage} 

\caption{Constraints on the model parameters using the HFQPO data of XTE J1550-564 considering (a) the Relativistic Precession Model, (b) the Tidal Disruption Model, (c) the Forced Resonance Model 1 and (d) the Forced Resonance Model 2.}
\label{Fig3}
\end{figure}

\begin{figure}[H]
\vspace*{-1.2cm }
{\bf \underline{XTE J1550-564}}
    \centering

    \begin{subfigure}[b]{0.42\textwidth}
        \includegraphics[width=\linewidth]{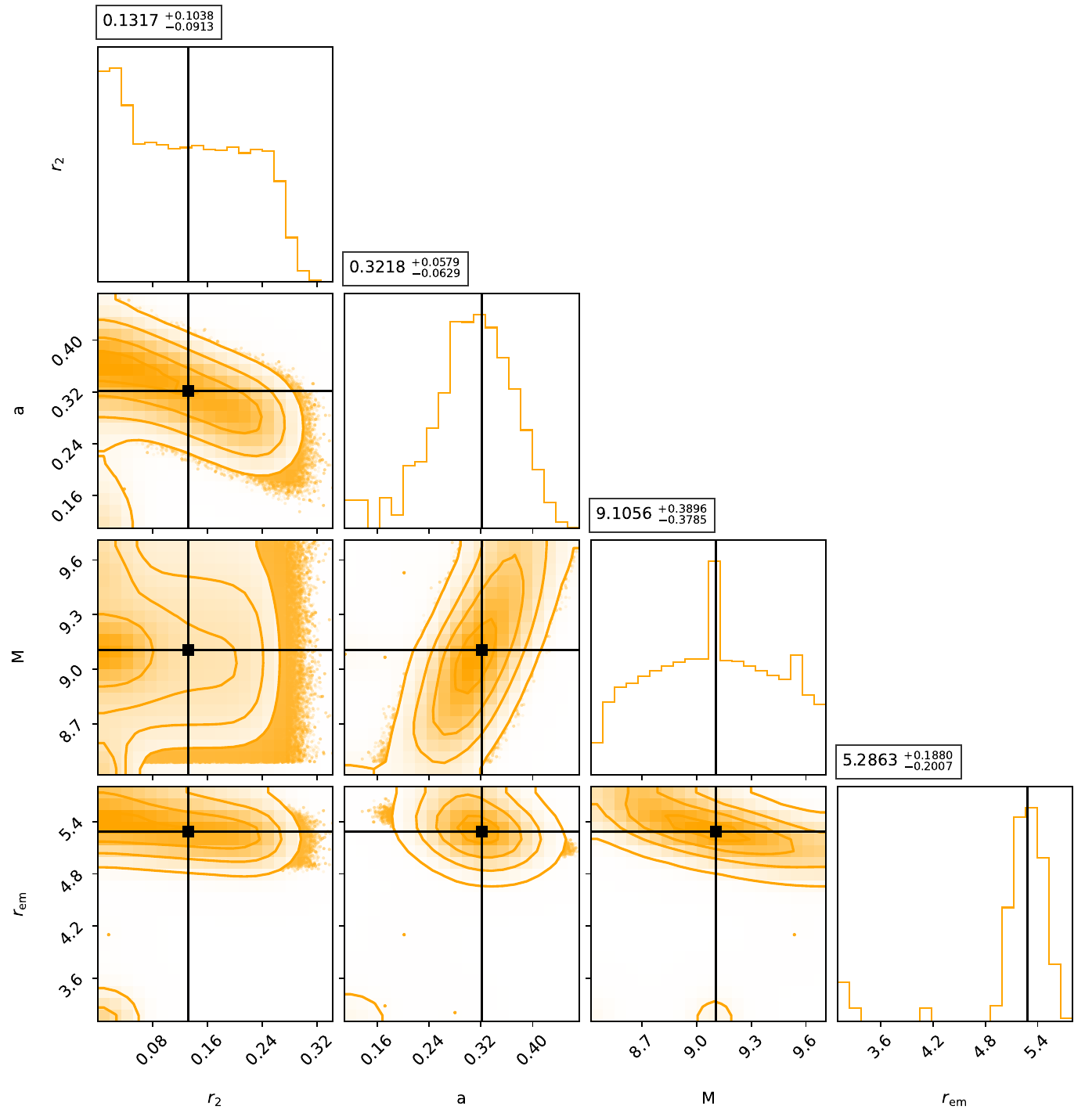}
        \caption{Keplerian Resonance Model 2}
    \end{subfigure}
    \hspace{0cm}
    \begin{subfigure}[b]{0.42\textwidth}
        \includegraphics[width=\linewidth]{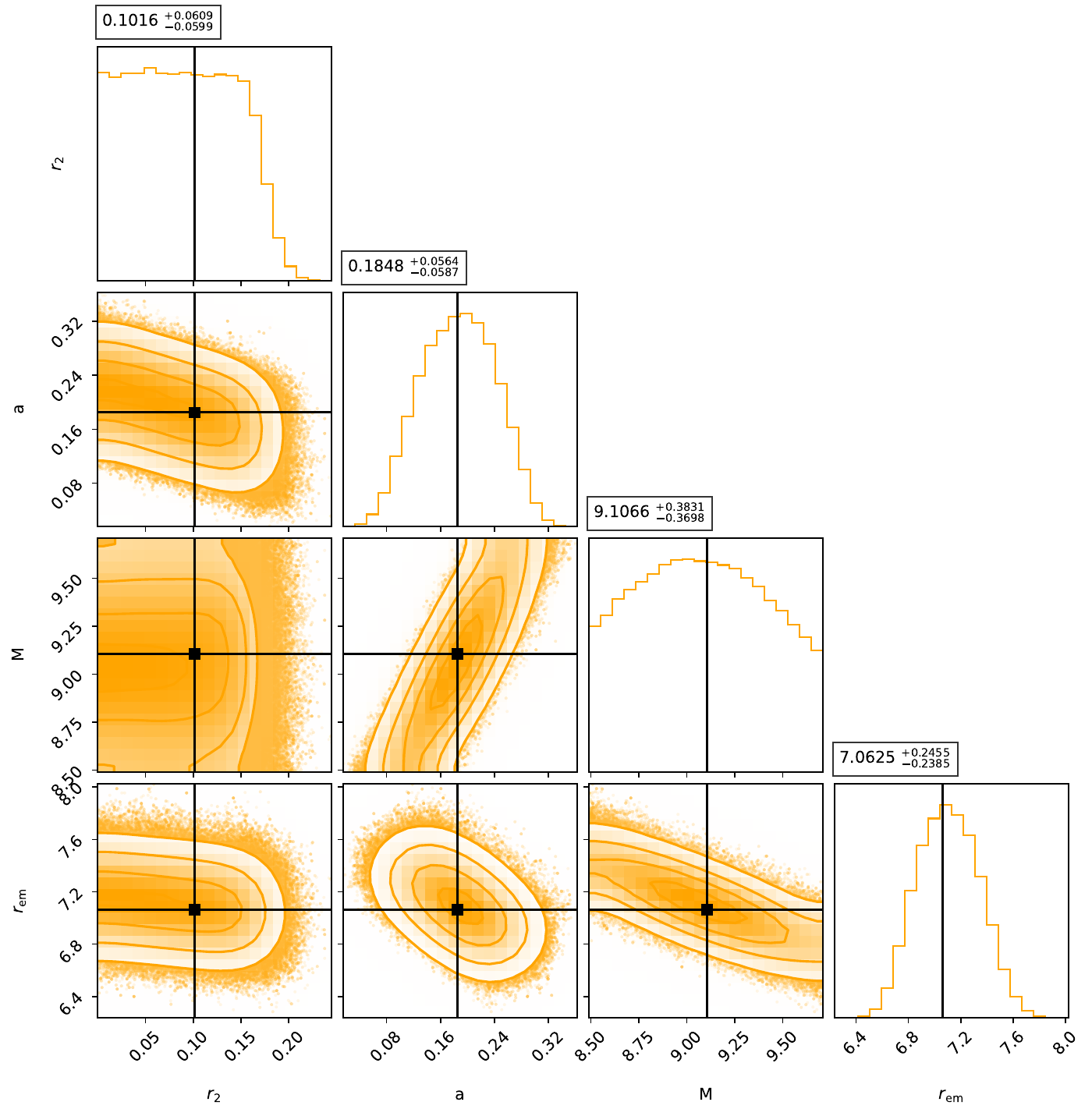}
        \caption{Keplerian Resonance Model 3}
    \end{subfigure}
    \hspace{0cm }
    \begin{subfigure}[b]{0.42\textwidth}
        \includegraphics[width=\linewidth]{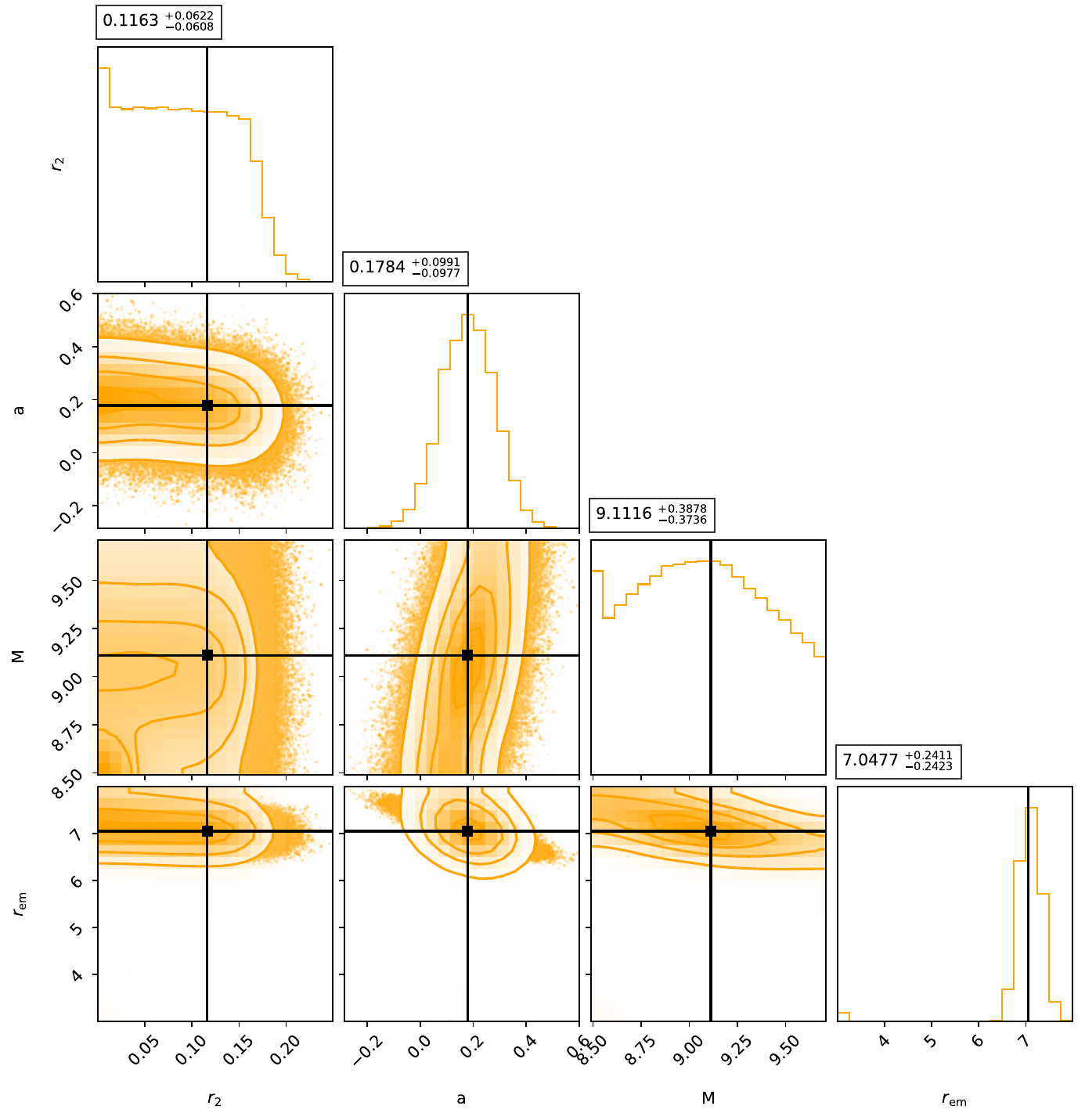}
        \caption{Warped Disc Oscillation Model} 
    \end{subfigure}

    \vspace{0.36cm} 

    \begin{subfigure}[b]{0.42\textwidth}
        \hspace{0cm} 
        \includegraphics[width=\linewidth]{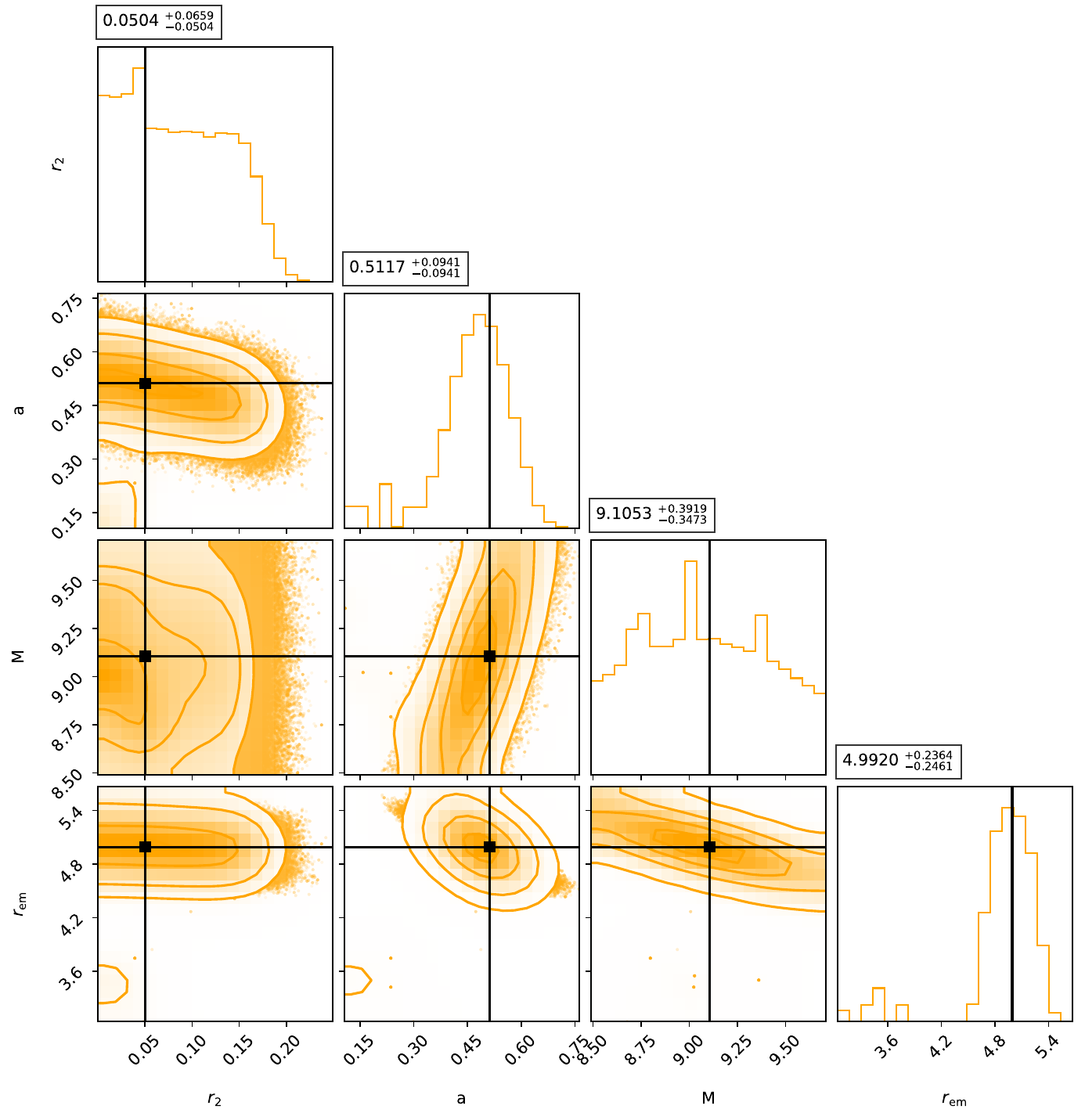}
        \caption{Non-axisymmetric Disk Oscillation Model 1}
    \end{subfigure}
    \hspace{2cm}
    \begin{subfigure}[b]{0.42\textwidth}
        \includegraphics[width=\linewidth]{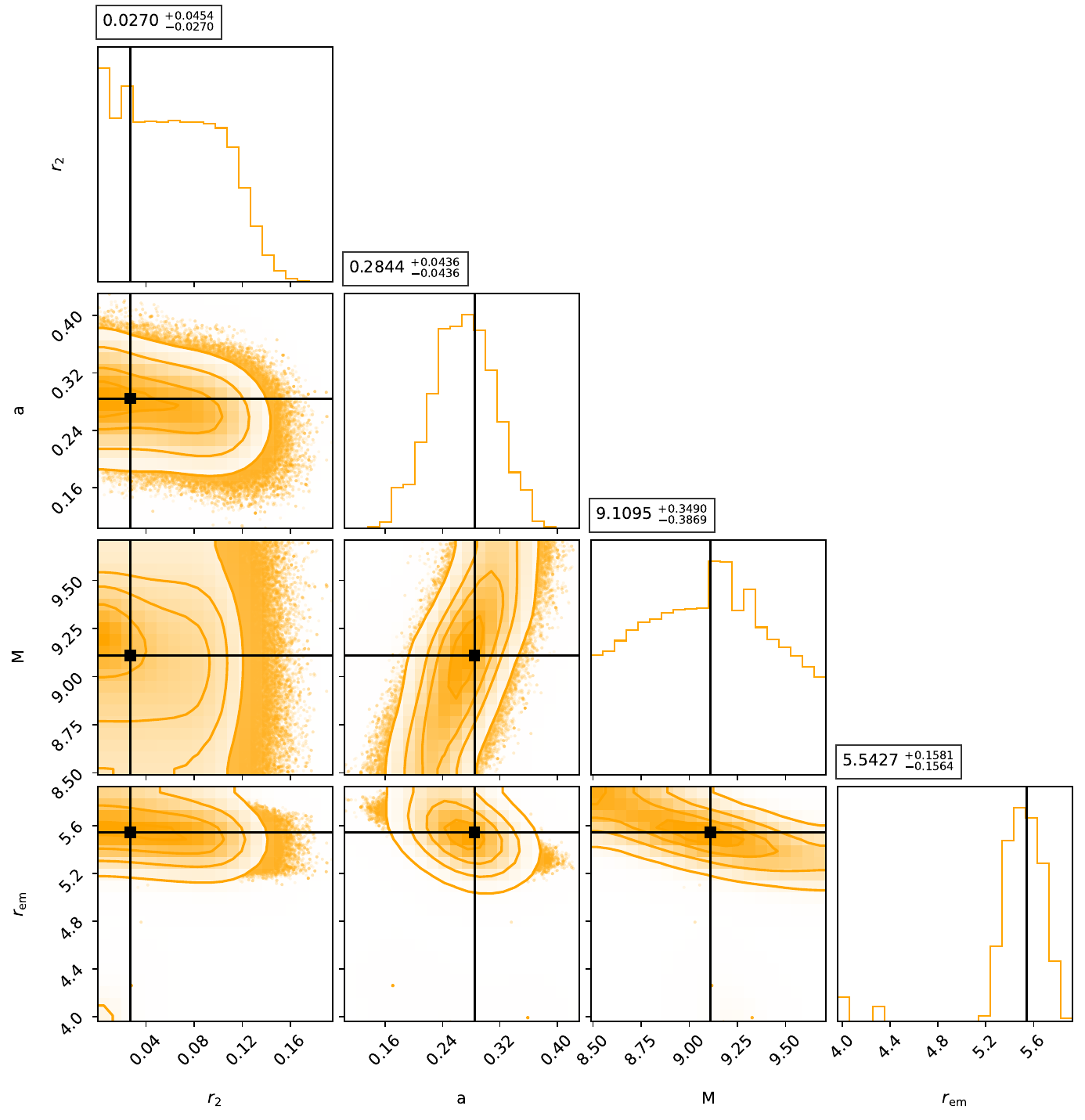}
        \caption{Non-axisymmetric Disk Oscillation Model 2}
    \end{subfigure}

\caption{Constraints on the model parameters using the HFQPO data of XTE J1550-564 considering (a) the Keplerian Resonance Model 2, (b) the Keplerian Resonance Model 3, (c) the Warped Disc Oscillation Model, (d) the Non-axisymmetric Disk Oscillation Model 1 and (e) the Non-axisymmetric Disk Oscillation Model 2.}
\label{Fig4}
\end{figure}

\begin{figure}[htpb]
{\bf \underline{GRS 1915+105}}
\centering

\begin{minipage}[b]{0.49\textwidth}
\centering
\includegraphics[width=\textwidth]{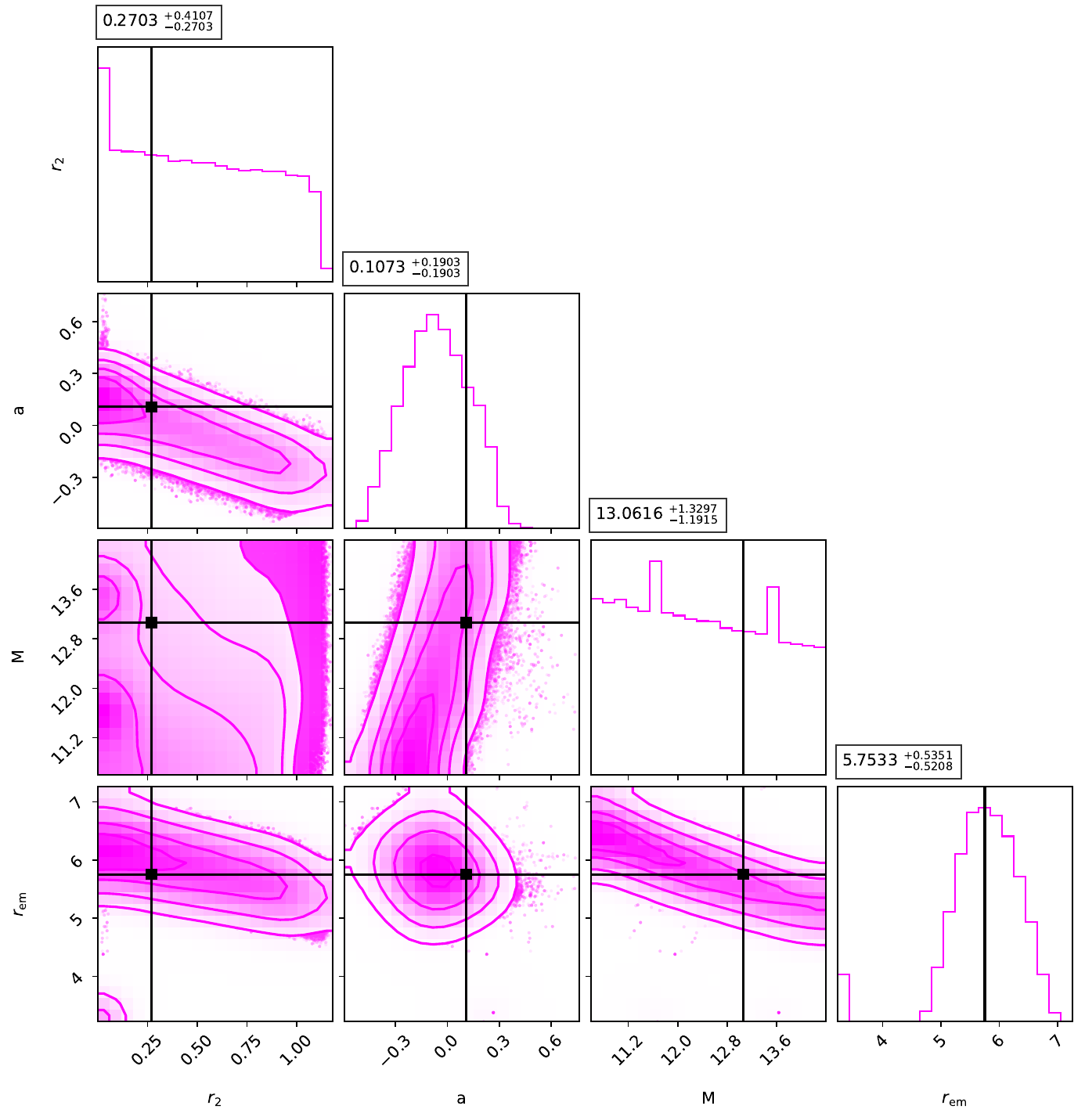}
\caption*{(a) Relativistic Precession Model}
\end{minipage}
\hfill
\begin{minipage}[b]{0.49\textwidth}
\centering
\includegraphics[width=\textwidth]{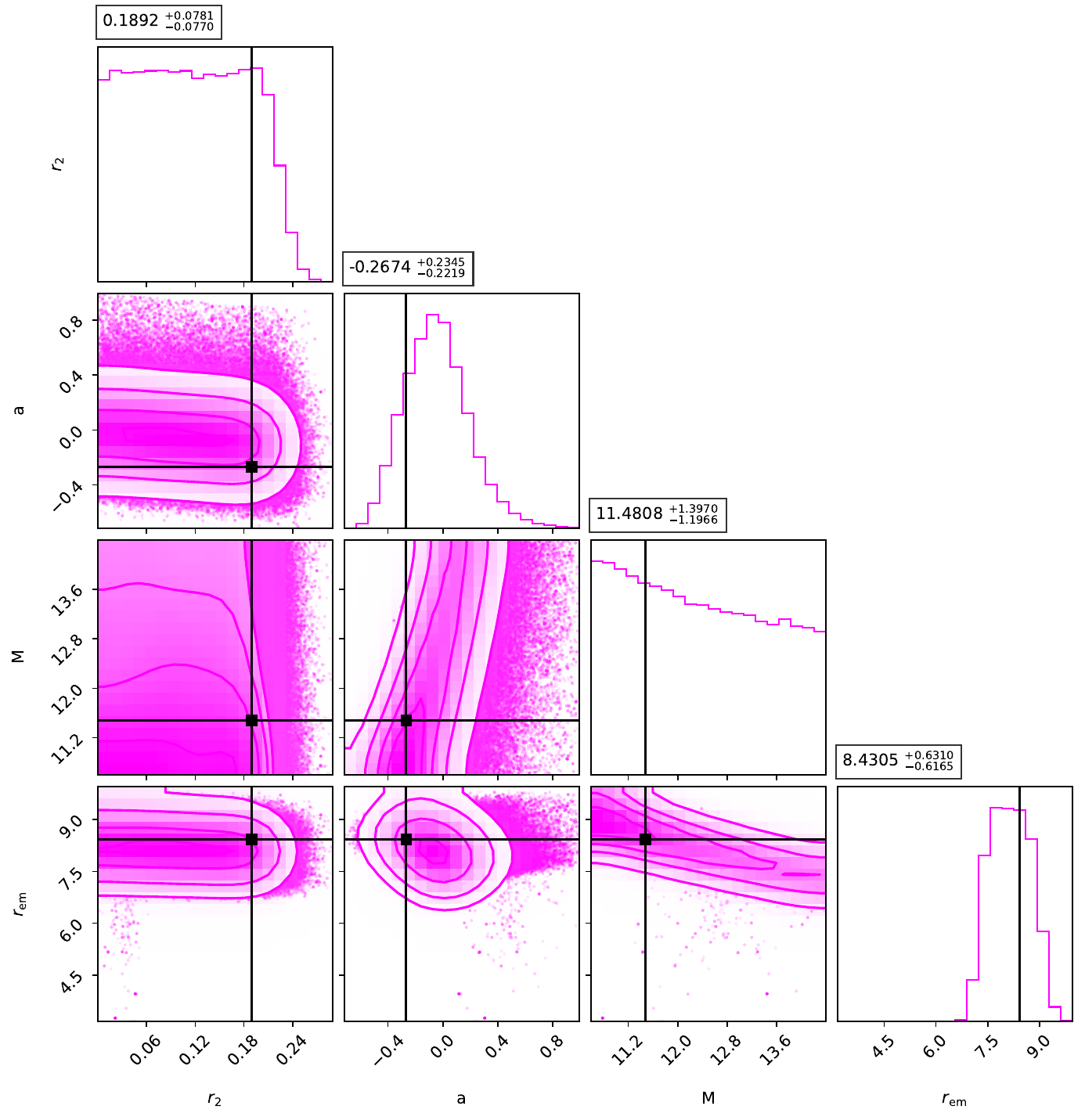}
\caption*{(b) Tidal Disruption Model}
\end{minipage}

\vspace{0.3cm}

\begin{minipage}[b]{0.49\textwidth}
\centering
\includegraphics[width=\textwidth]{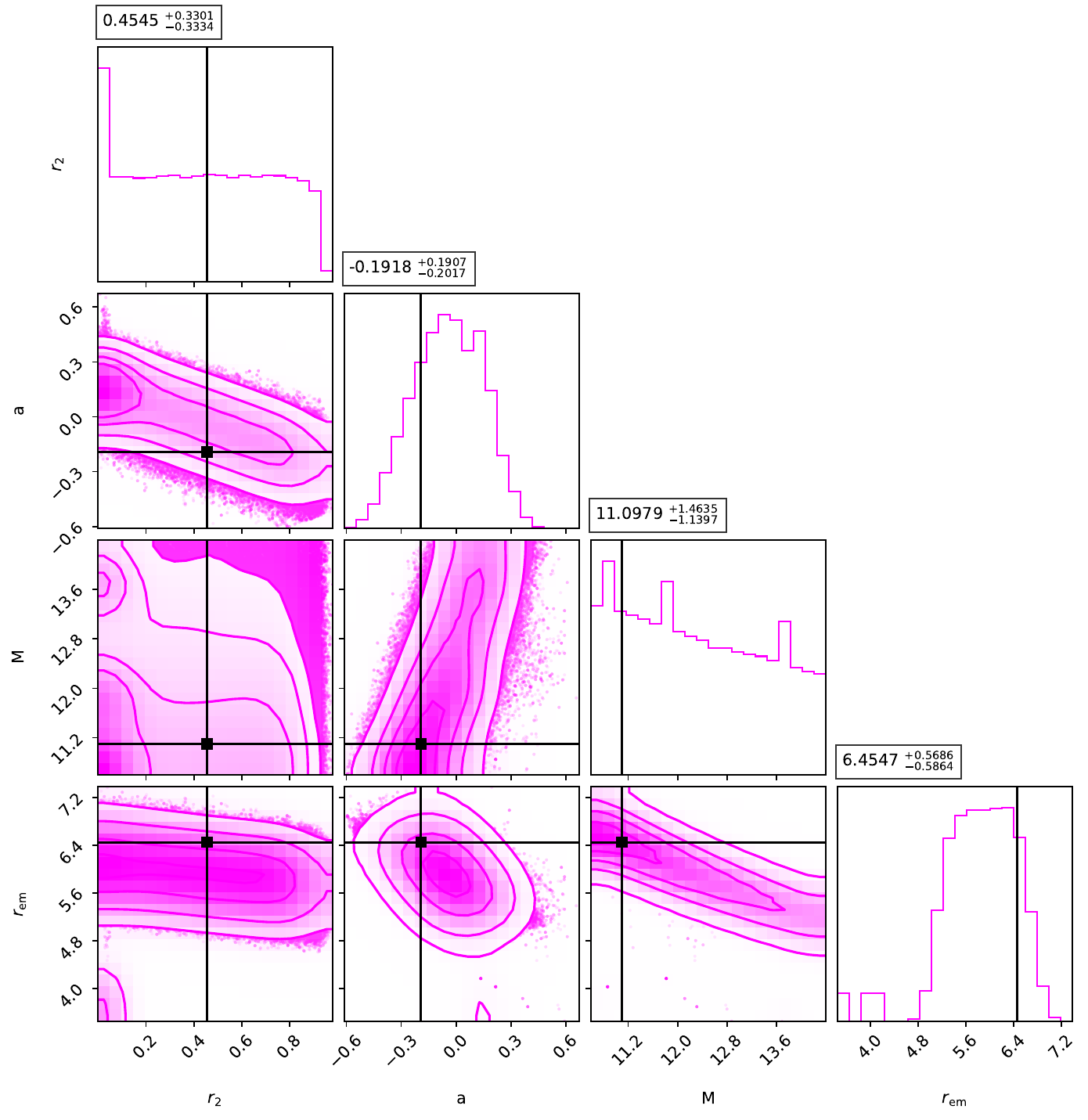}
\caption*{(c) Forced Resonance Model 1}
\end{minipage}
\hfill
\begin{minipage}[b]{0.49\textwidth}
\centering
\includegraphics[width=\textwidth]{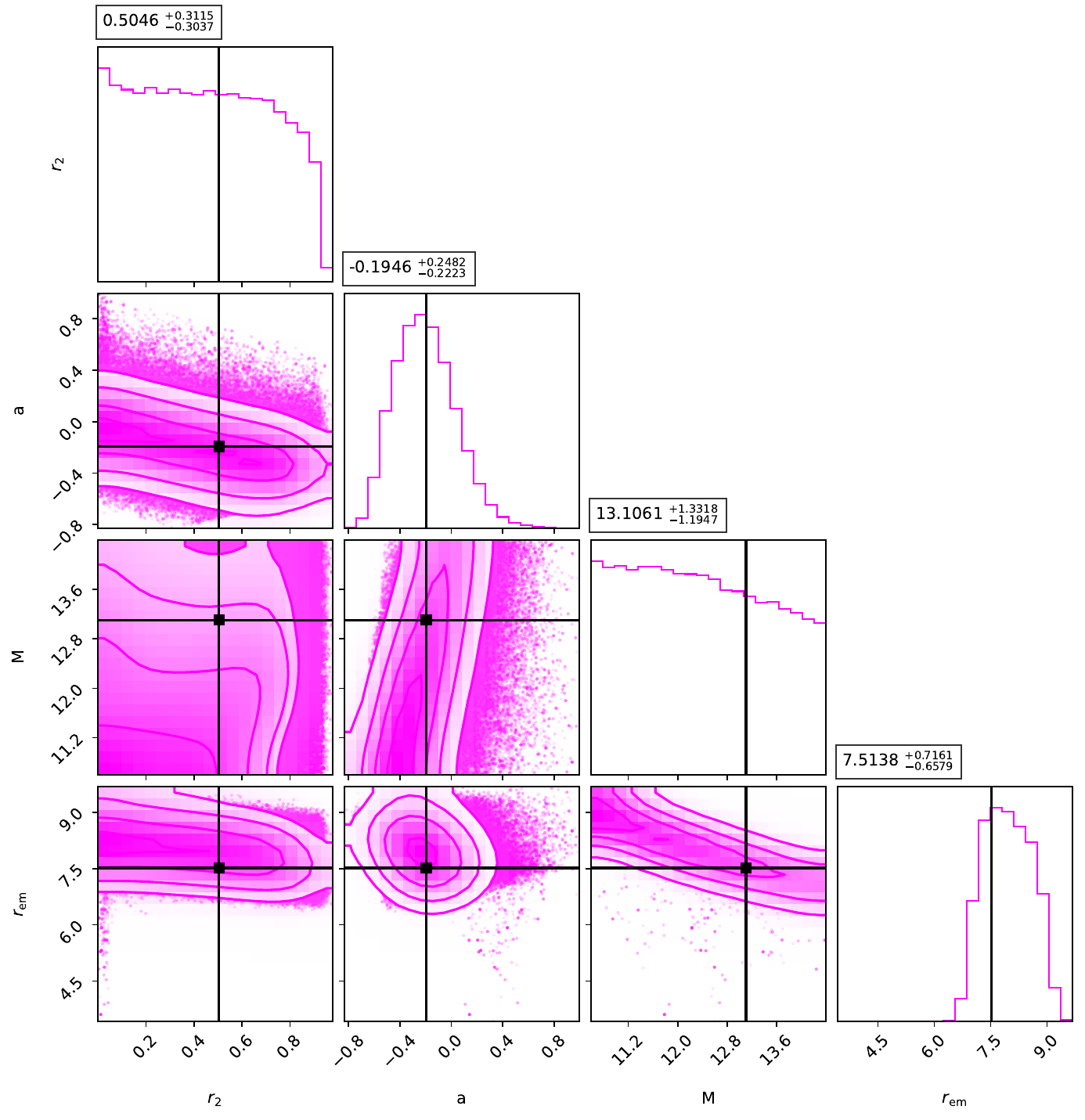}
\caption*{(d) Forced Resonance Model 2}
\end{minipage} 

\caption{Constraints on the model parameters using the HFQPO data of GRS 1915+105 considering (a) the Relativistic Precession Model, (b) the Tidal Disruption Model, (c) the Forced Resonance Model 1 and (d) the Forced Resonance Model 2.}
\label{Fig5}
\end{figure}

\begin{figure}[H]
\vspace*{-1.2cm }
{\bf \underline{GRS 1915+105}}
    \centering

    \begin{subfigure}[b]{0.42\textwidth}
        \includegraphics[width=\linewidth]{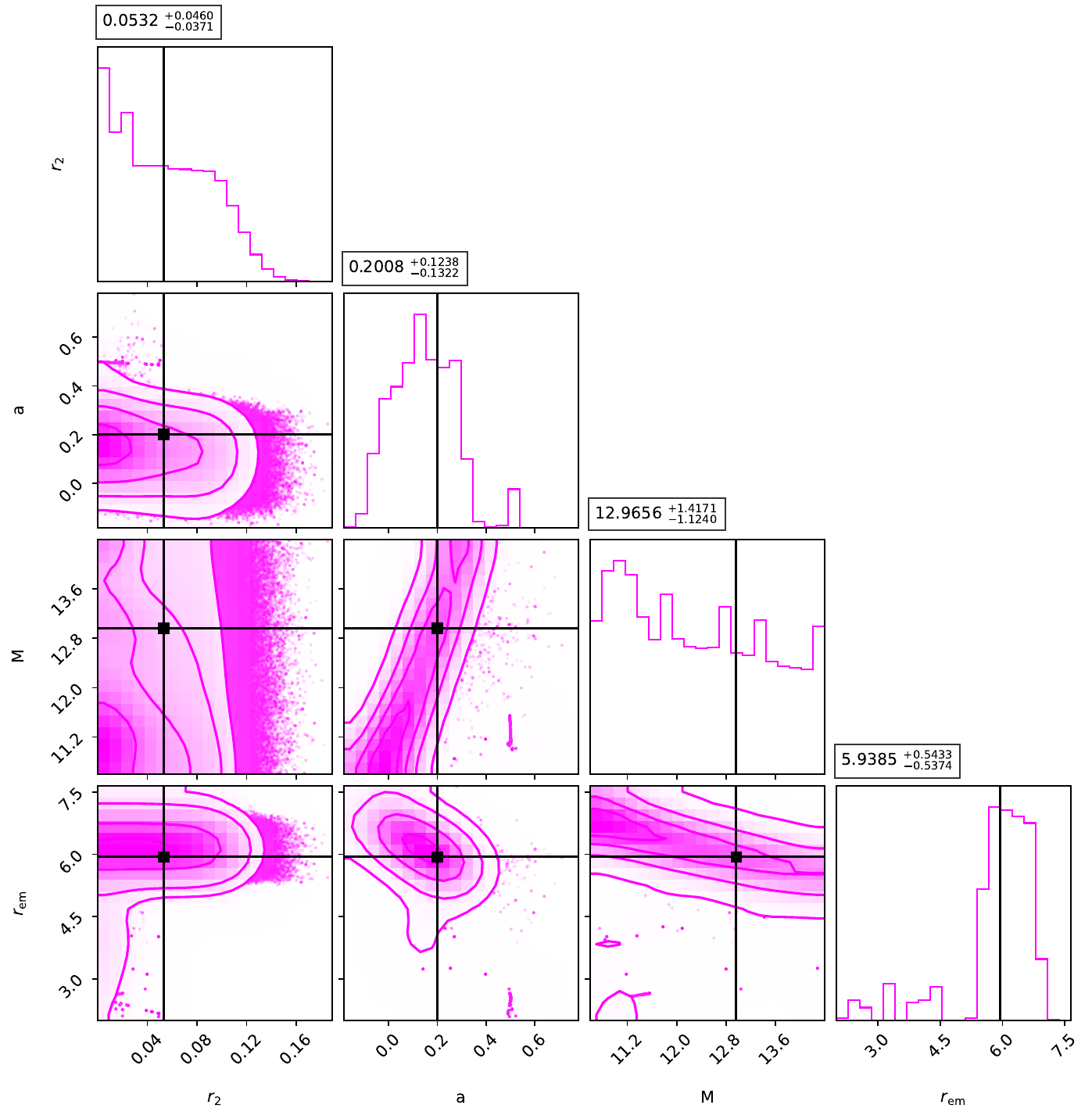}
        \caption{Keplerian Resonance Model 2}
    \end{subfigure}
    \hspace{0cm}
    \begin{subfigure}[b]{0.42\textwidth}
        \includegraphics[width=\linewidth]{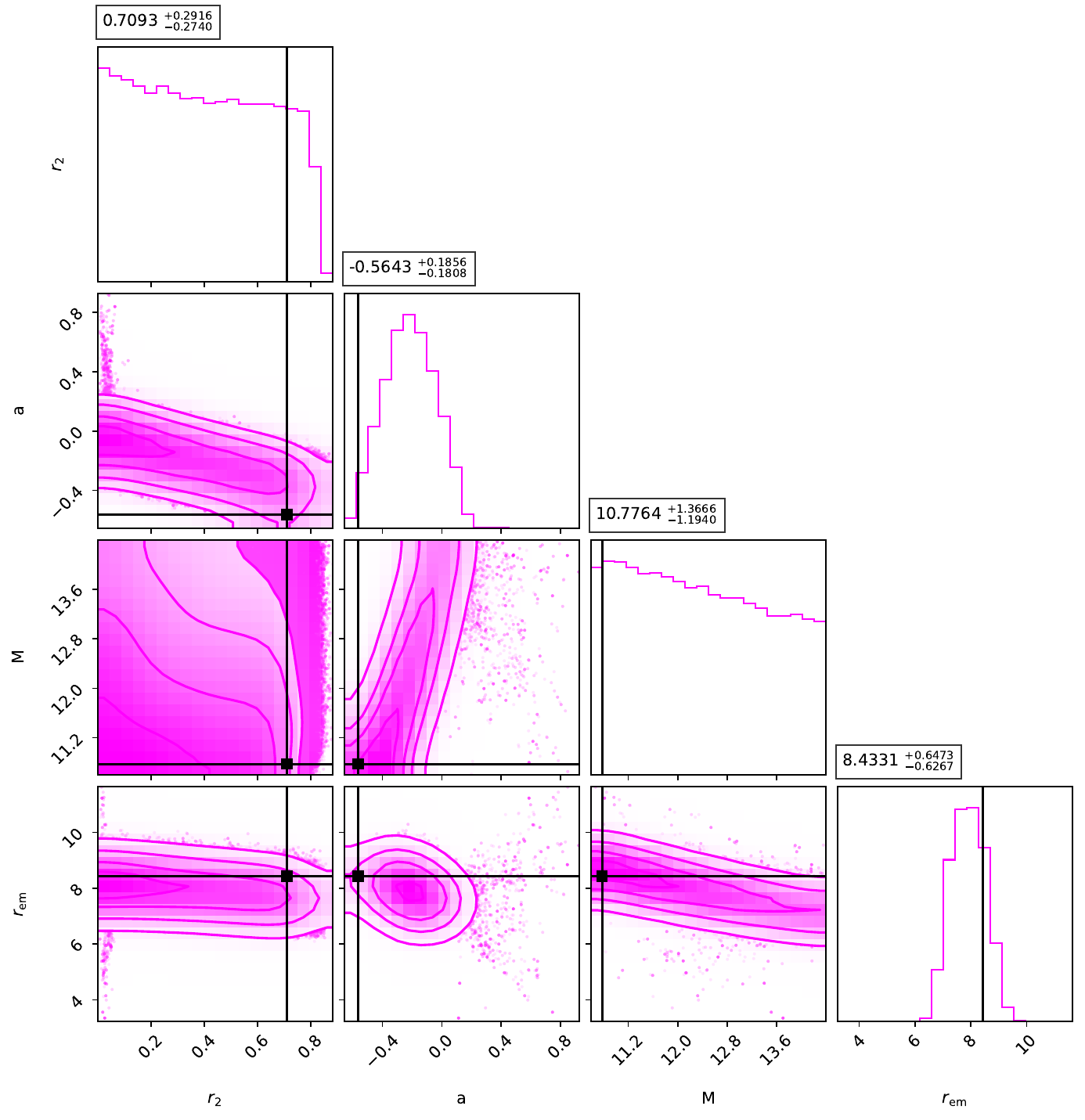}
        \caption{Keplerian Resonance Model 3}
    \end{subfigure}
    \hspace{0cm }
    \begin{subfigure}[b]{0.42\textwidth}
        \includegraphics[width=\linewidth]{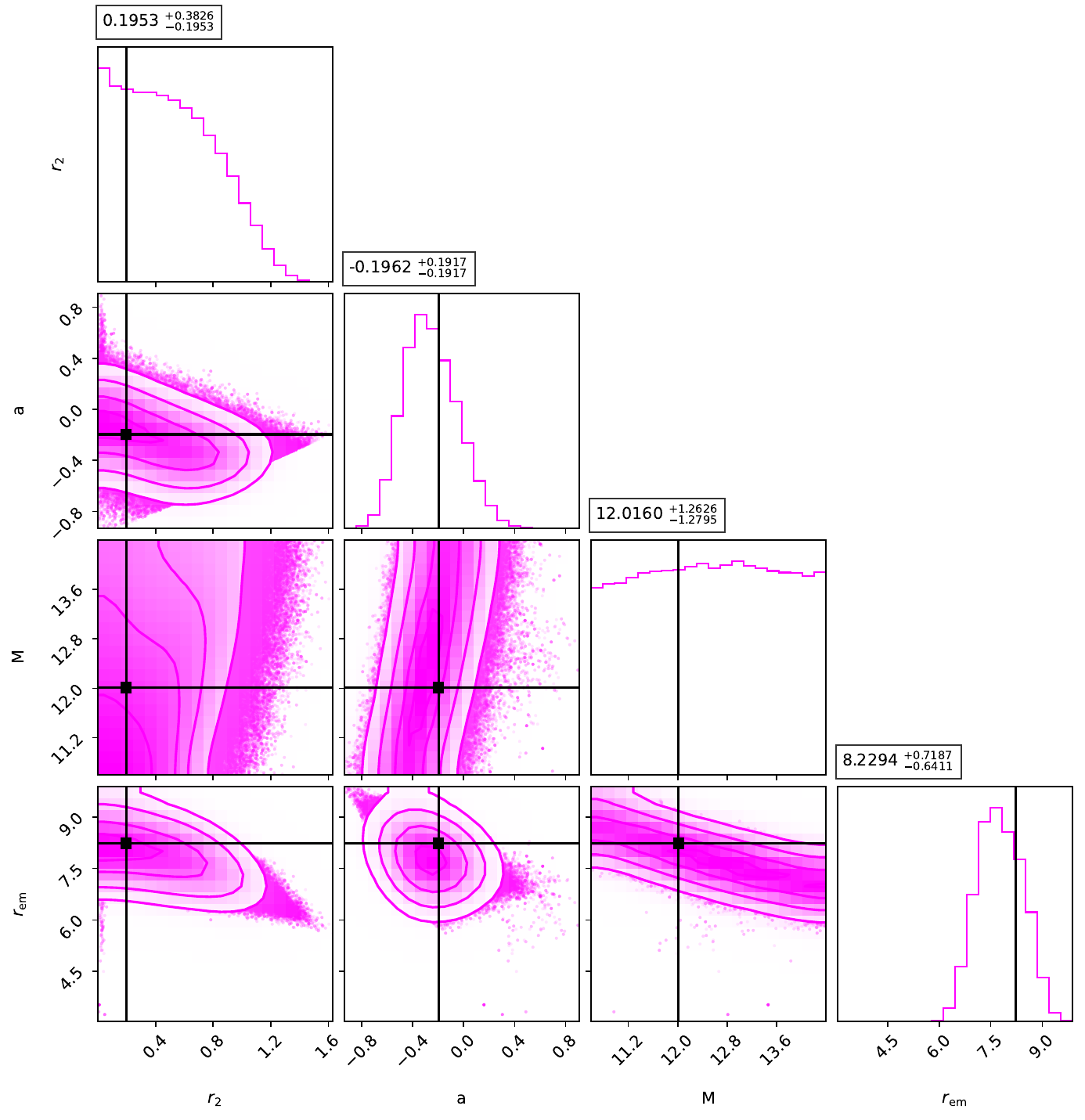}
        \caption{Warped Disc Oscillation Model} 
    \end{subfigure}

    \vspace{0.36cm} 

    \begin{subfigure}[b]{0.42\textwidth}
        \hspace{0cm} 
        \includegraphics[width=\linewidth]{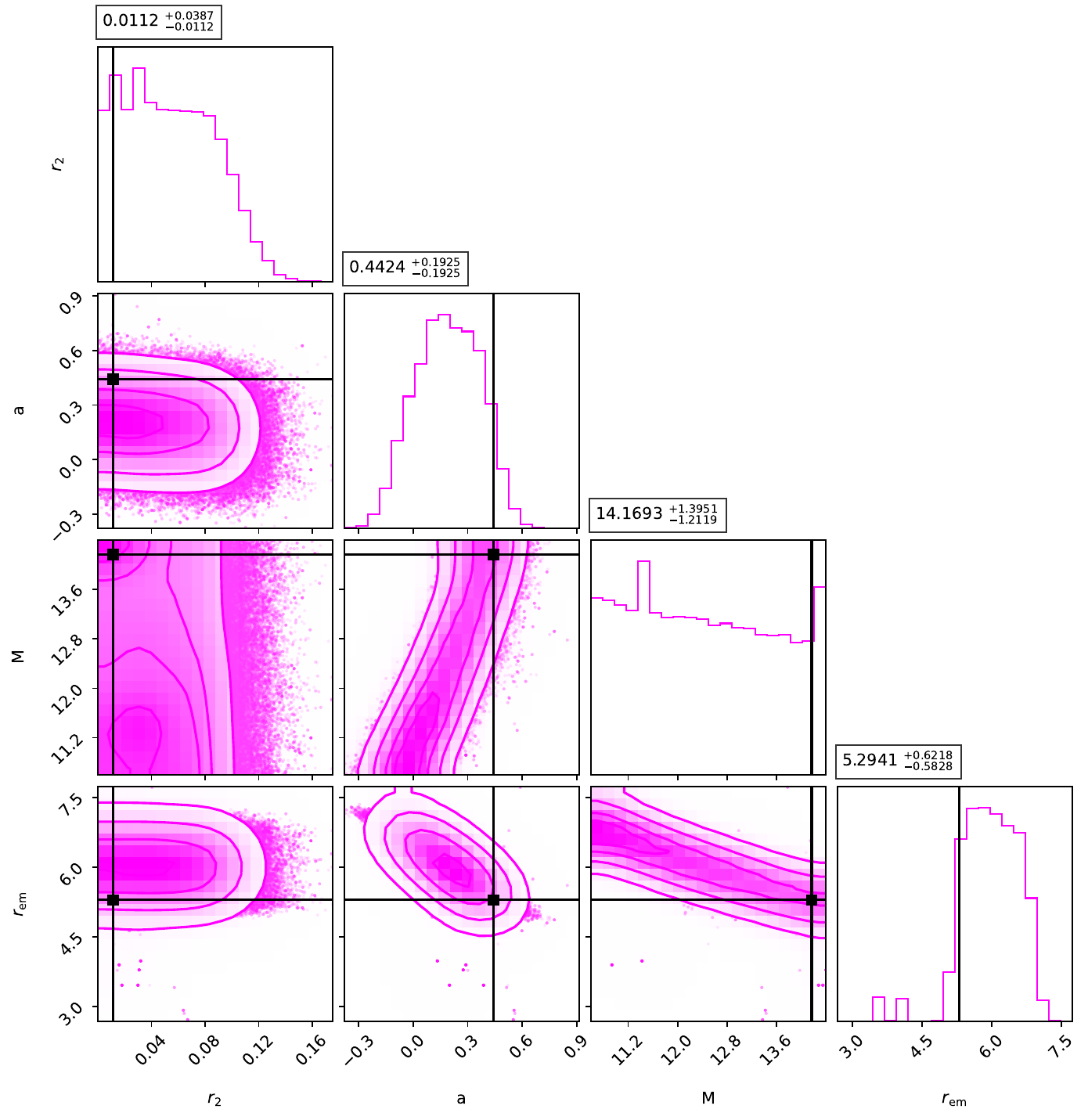}
        \caption{Non-axisymmetric Disk Oscillation Model 1}
    \end{subfigure}
    \hspace{2cm}
    \begin{subfigure}[b]{0.42\textwidth}
        \includegraphics[width=\linewidth]{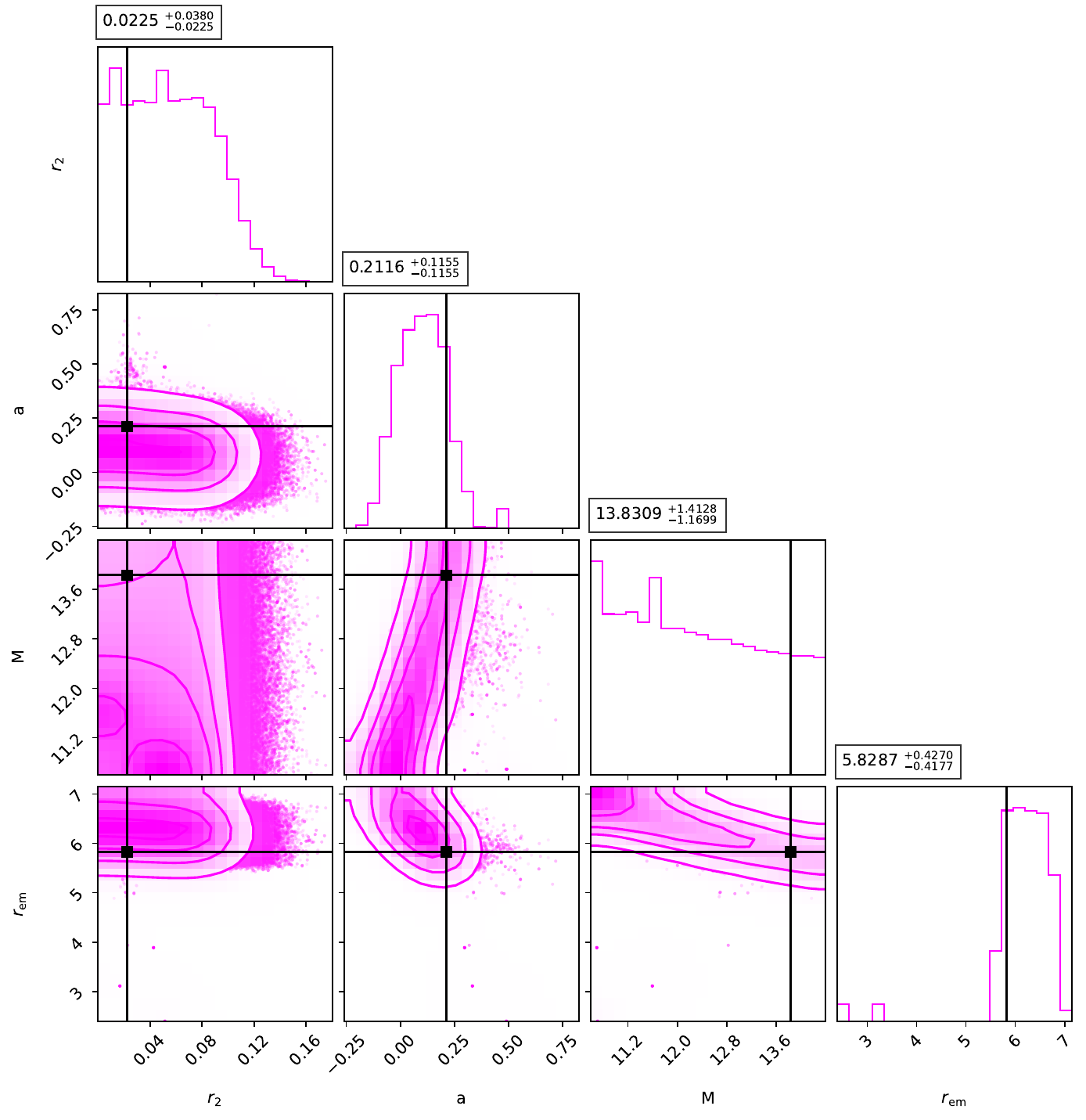}
        \caption{Non-axisymmetric Disk Oscillation Model 2}
    \end{subfigure}

\caption{Constraints on the model parameters using the HFQPO data of GRS 1915+105 considering (a) the Keplerian Resonance Model 2, (b) the Keplerian Resonance Model 3, (c) the Warped Disc Oscillation Model, (d) the Non-axisymmetric Disk Oscillation Model 1 and (e) the Non-axisymmetric Disk Oscillation Model 2.}
\label{Fig6}
\end{figure}

\begin{figure}[htpb]
\vspace*{-1.2cm }
{\bf \underline{H1743-322}}
\centering

\begin{minipage}[b]{0.49\textwidth}
\centering
\includegraphics[width=\textwidth]{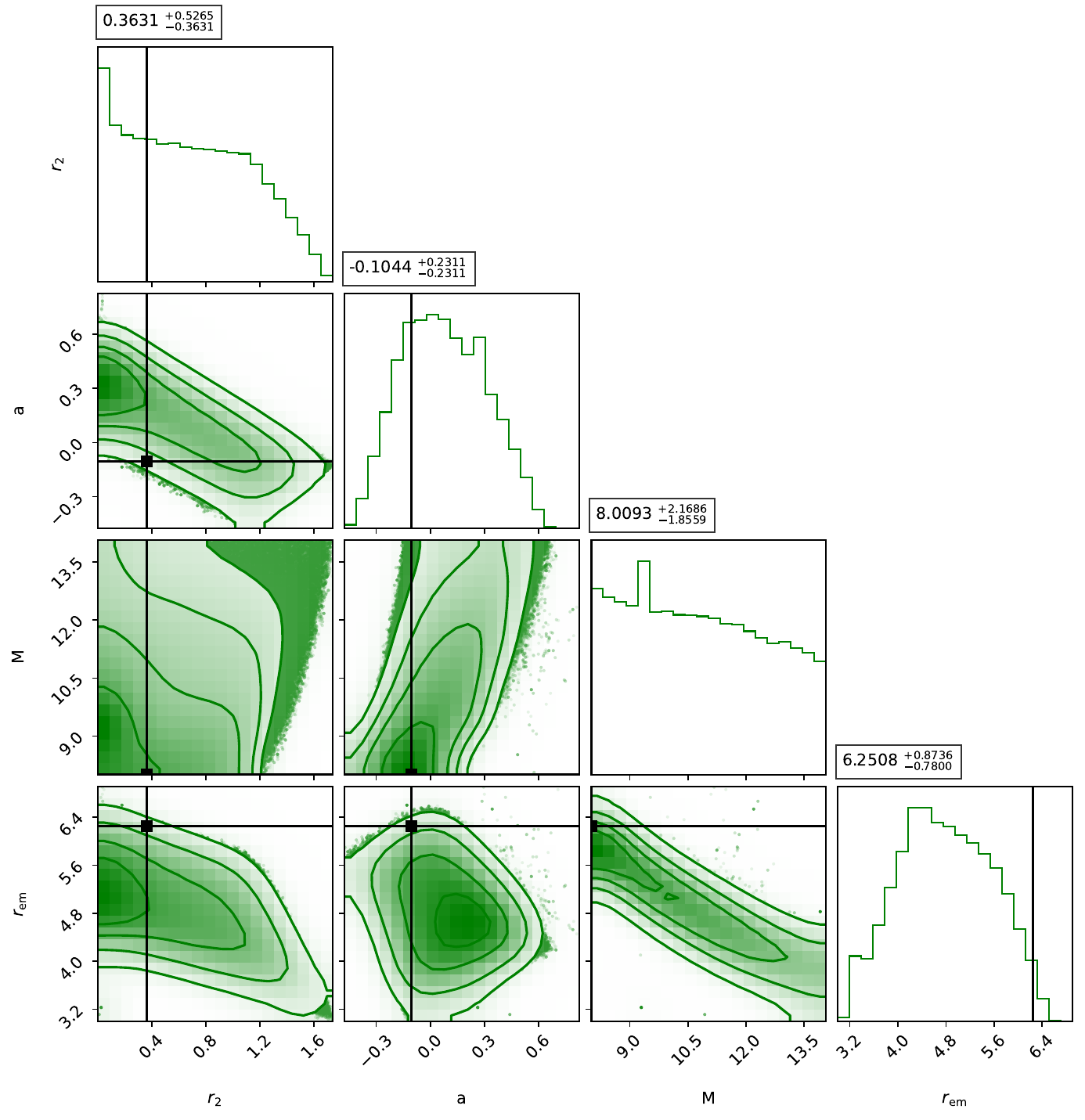}
\caption*{(a) Relativistic Precession Model}
\end{minipage}
\hfill
\begin{minipage}[b]{0.49\textwidth}
\centering
\includegraphics[width=\textwidth]{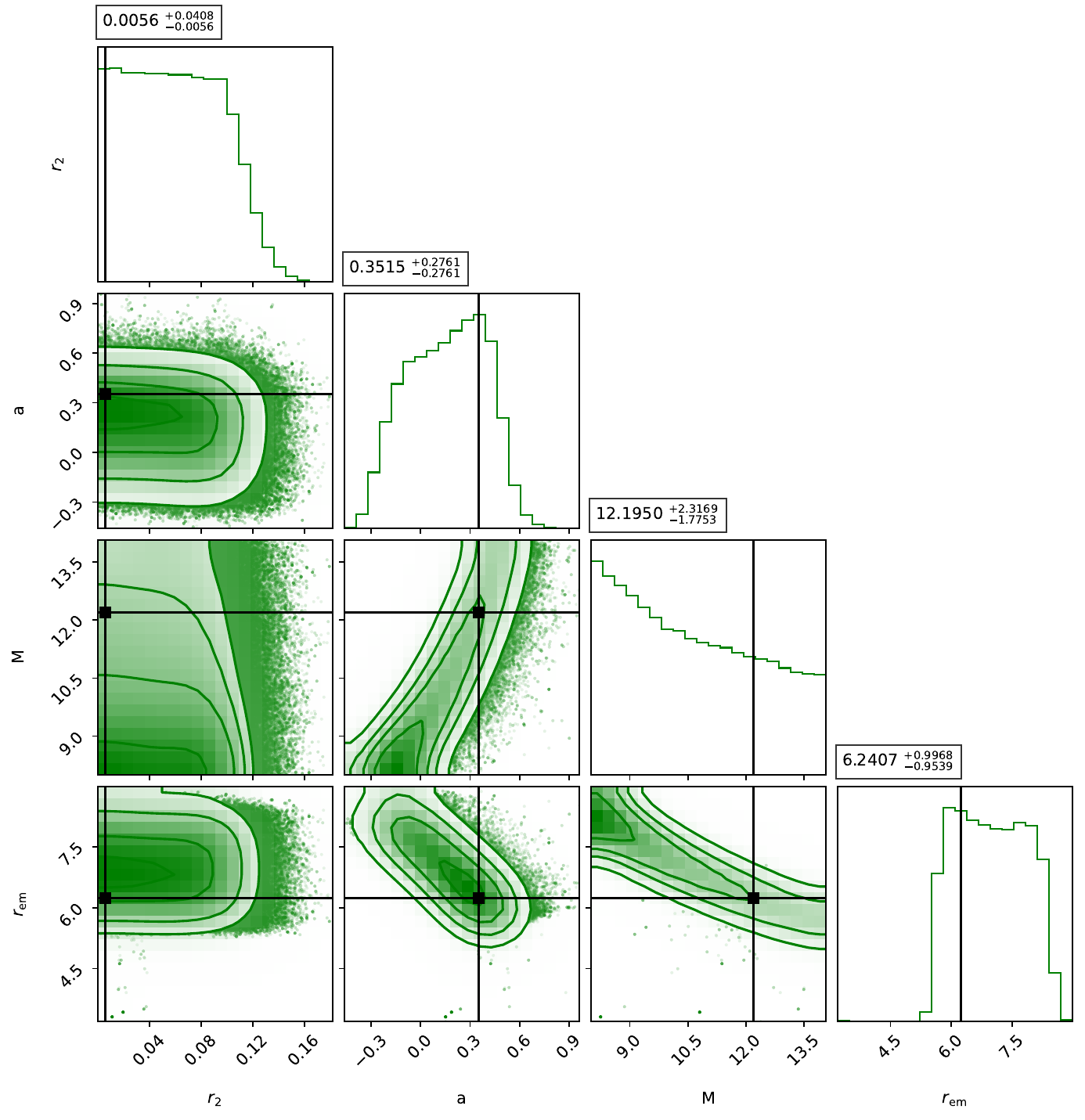}
\caption*{(b) Tidal Disruption Model}
\end{minipage}

\vspace{0.3cm}

\begin{minipage}[b]{0.49\textwidth}
\centering
\includegraphics[width=\textwidth]{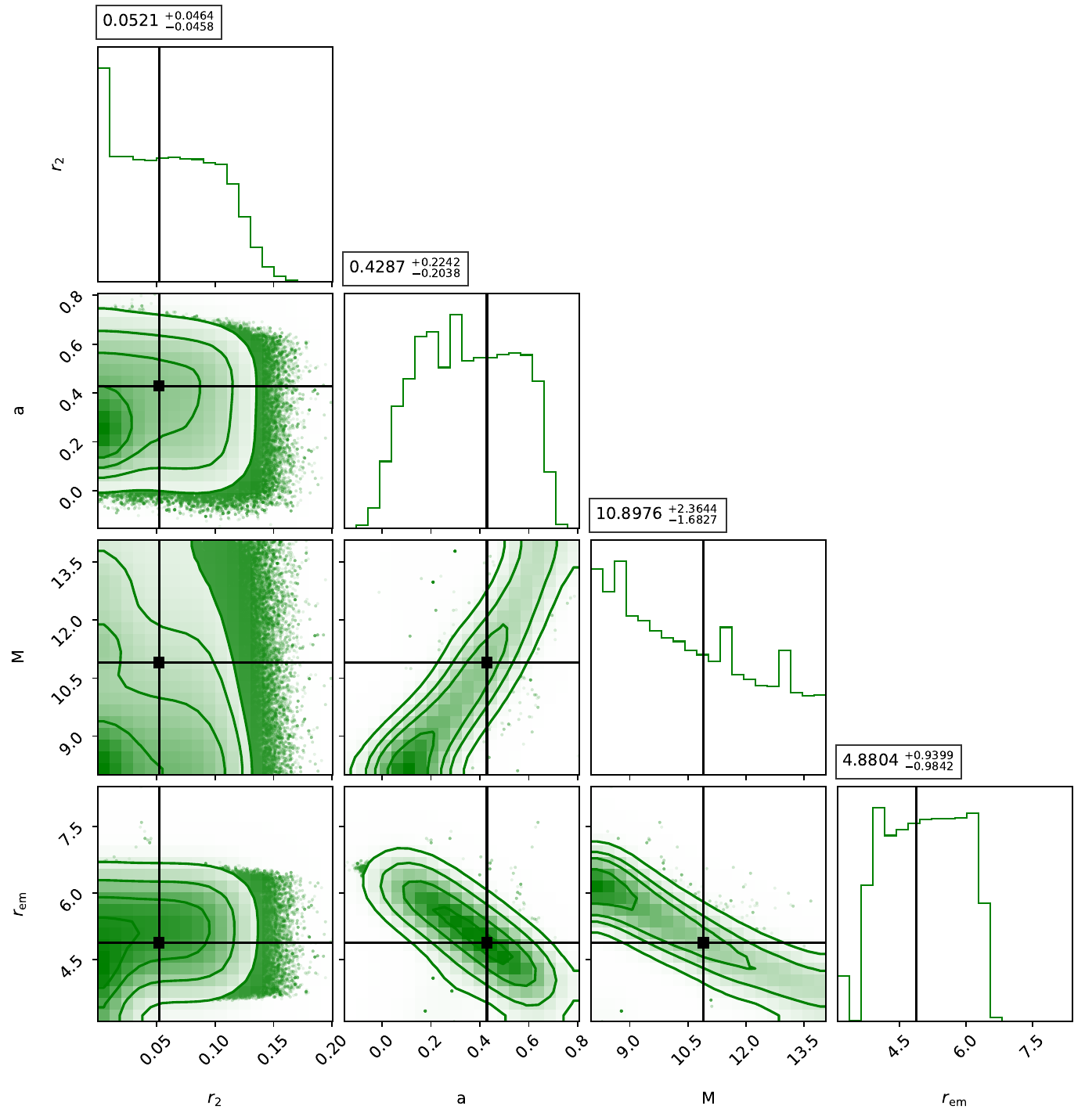}
\caption*{(c) Forced Resonance Model 1}
\end{minipage}
\hfill
\begin{minipage}[b]{0.49\textwidth}
\centering
\includegraphics[width=\textwidth]{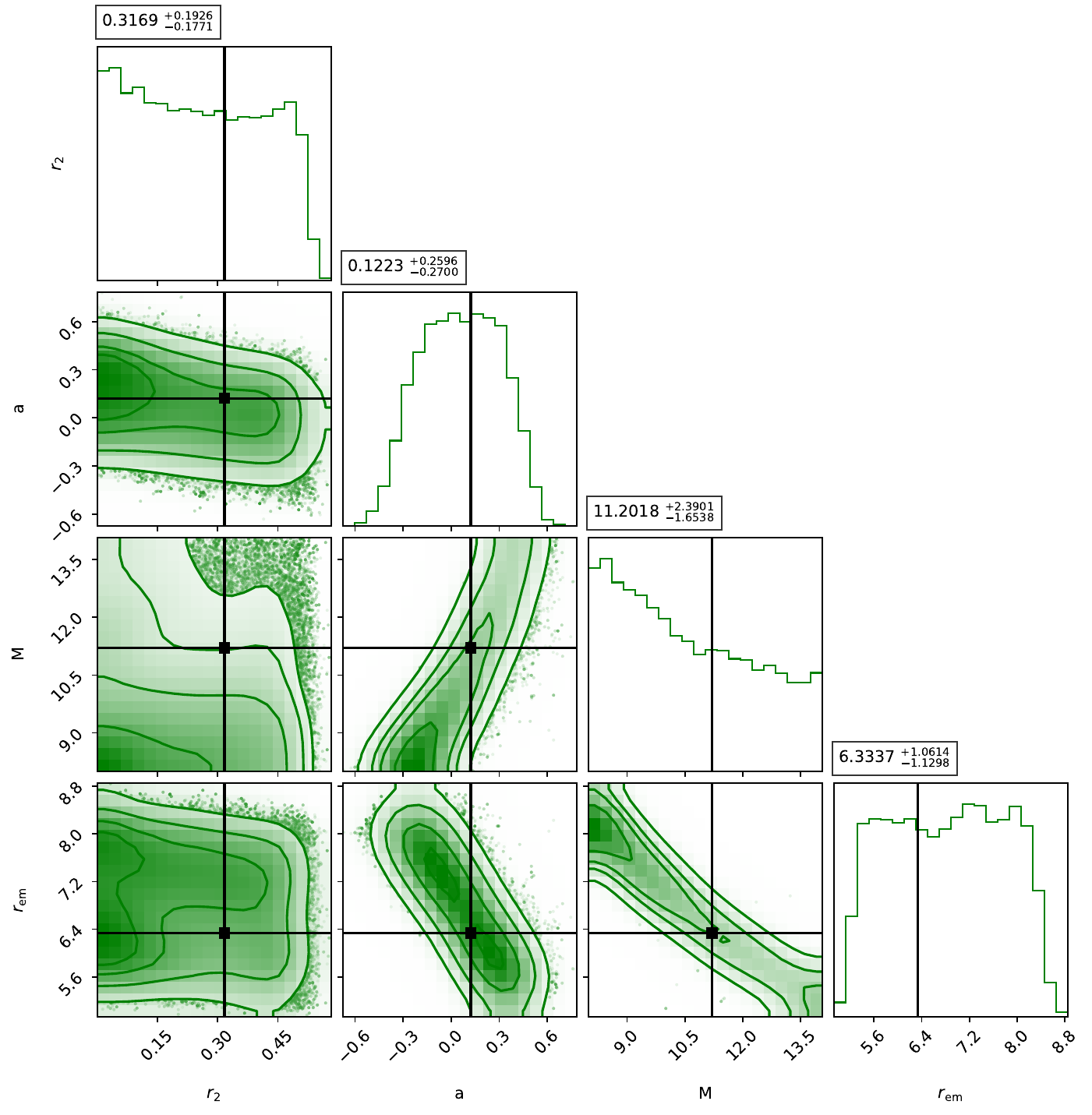}
\caption*{(d) Forced Resonance Model 2}
\end{minipage} 

\caption{Constraints on the model parameters using the HFQPO data of H1743-322 considering (a) the Relativistic Precession Model, (b) the Tidal Disruption Model, (c) the Forced Resonance Model 1 and (d) the Forced Resonance Model 2.}
\label{Fig7}
\end{figure}

\begin{figure}[H]
\vspace*{-1.2cm }
{\bf \underline{H1743-322}}
    \centering

    \begin{subfigure}[b]{0.42\textwidth}
        \includegraphics[width=\linewidth]{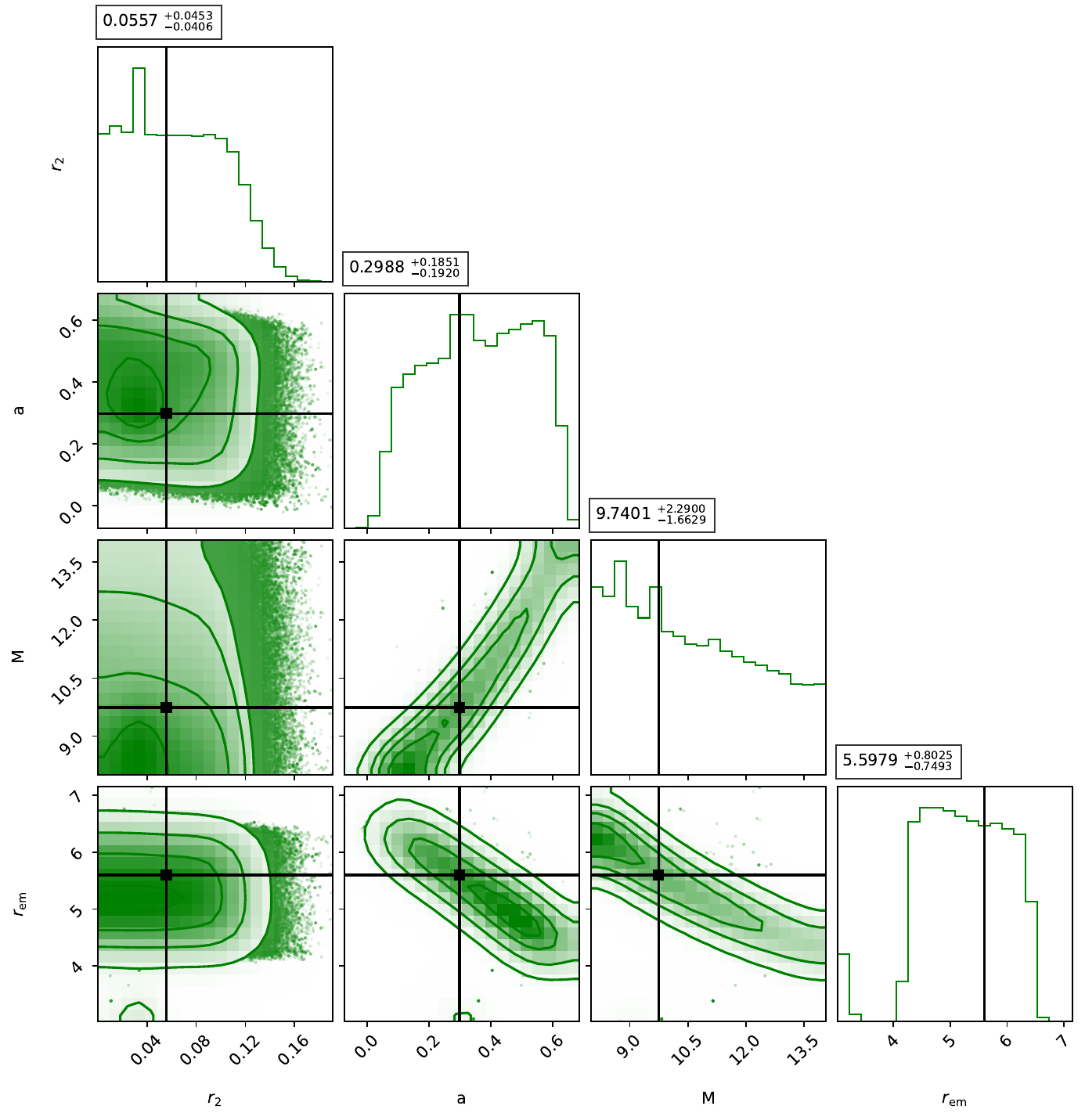}
        \caption{Keplerian Resonance Model 2}
    \end{subfigure}
    \hspace{0cm}
    \begin{subfigure}[b]{0.42\textwidth}
        \includegraphics[width=\linewidth]{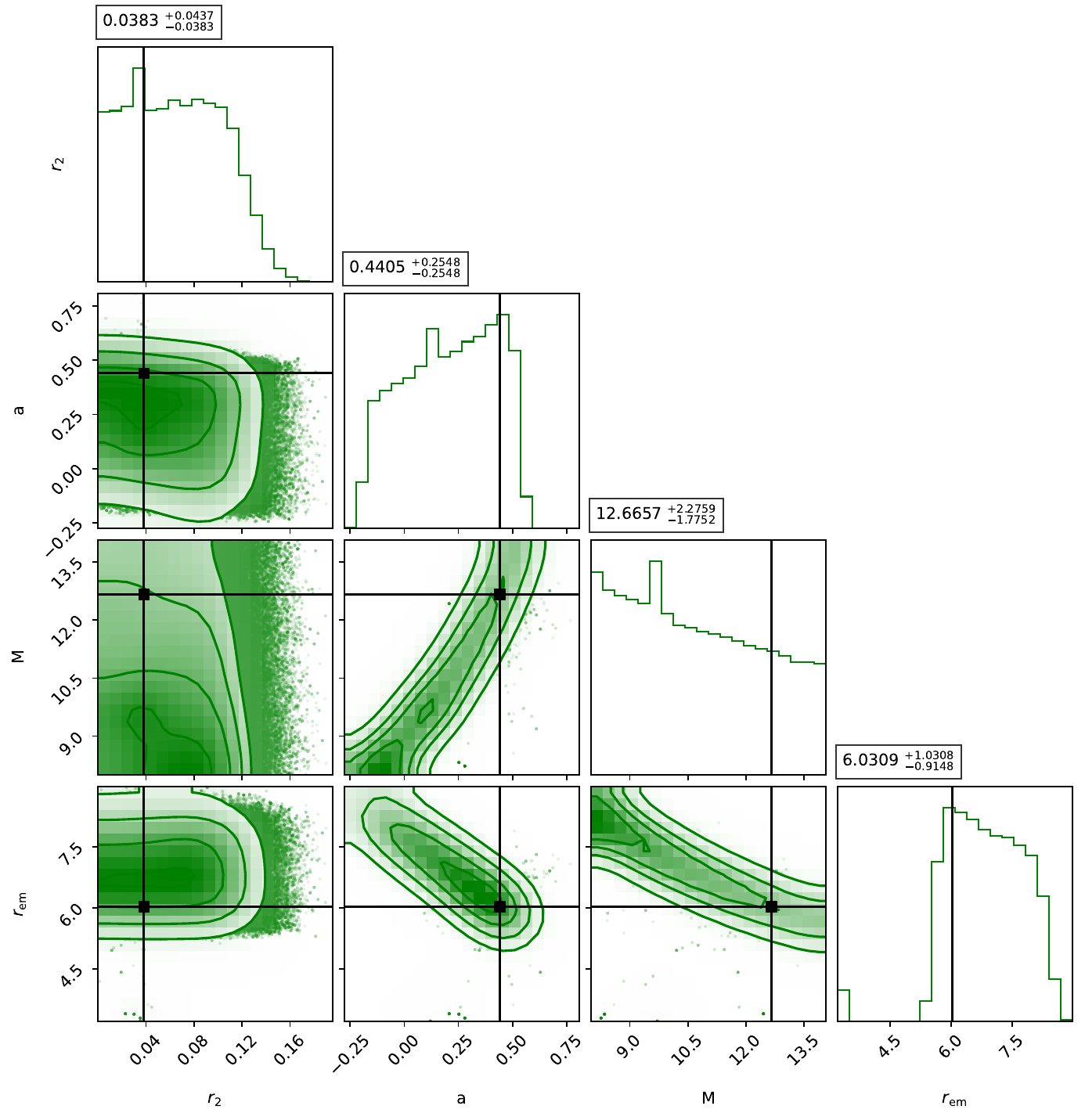}
        \caption{Keplerian Resonance Model 3}
    \end{subfigure}
    \hspace{0cm }
    \begin{subfigure}[b]{0.42\textwidth}
        \includegraphics[width=\linewidth]{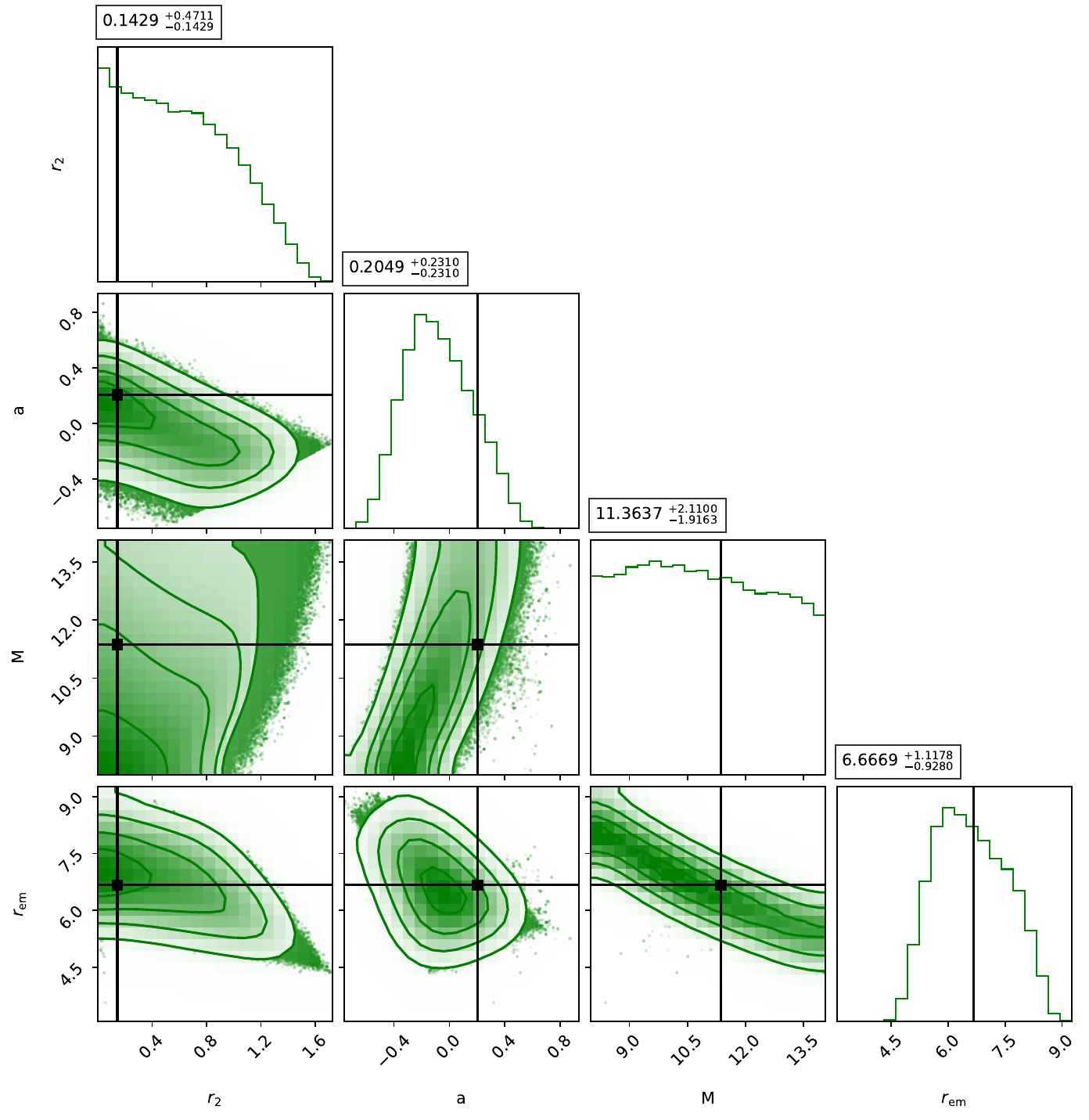}
        \caption{Warped Disc Oscillation Model} 
    \end{subfigure}

    \vspace{0.36cm} 

    \begin{subfigure}[b]{0.42\textwidth}
        \hspace{0cm} 
        \includegraphics[width=\linewidth]{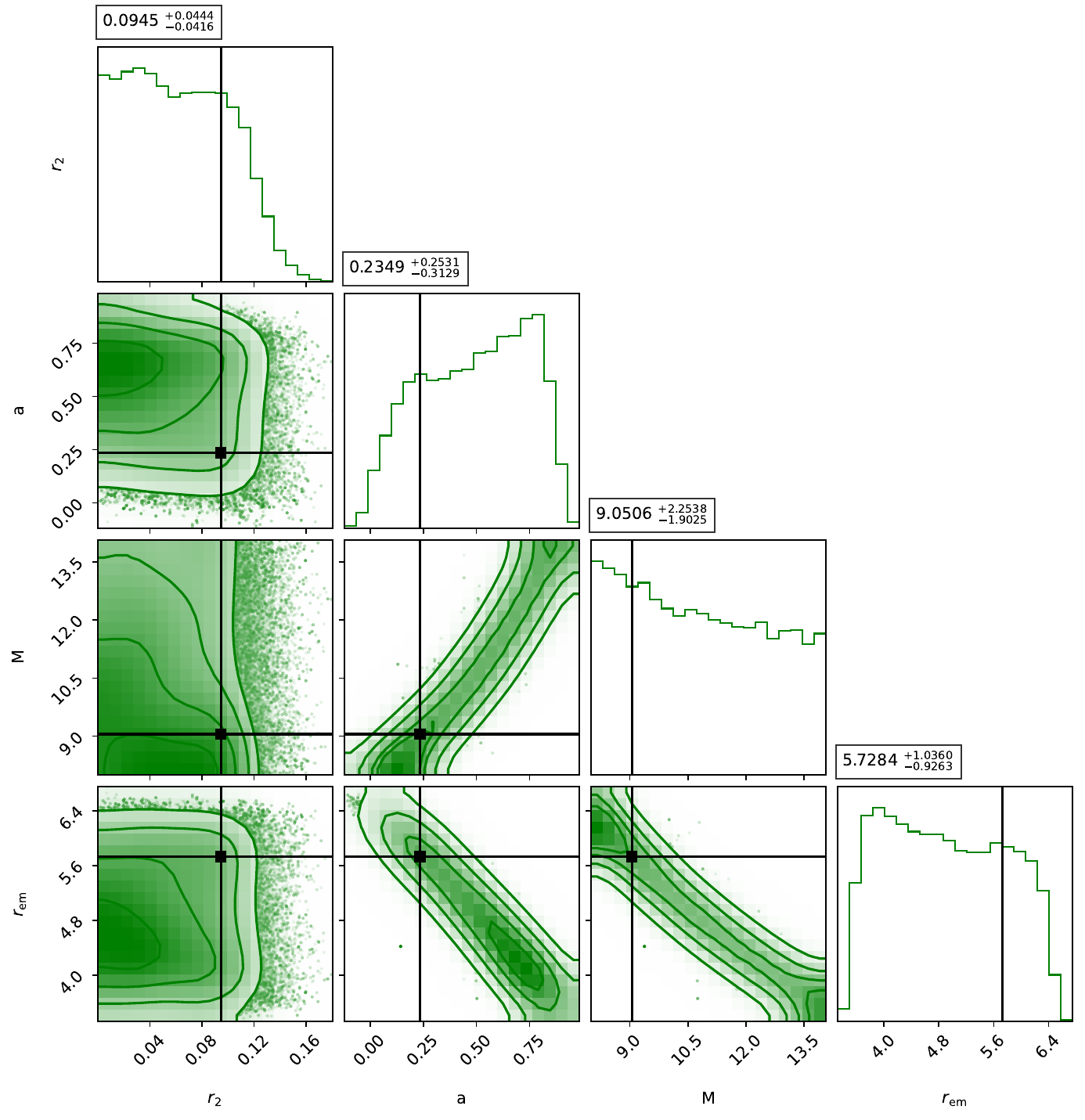}
        \caption{Non-axisymmetric Disk Oscillation Model 1}
    \end{subfigure}
    \hspace{2cm}
    \begin{subfigure}[b]{0.42\textwidth}
        \includegraphics[width=\linewidth]{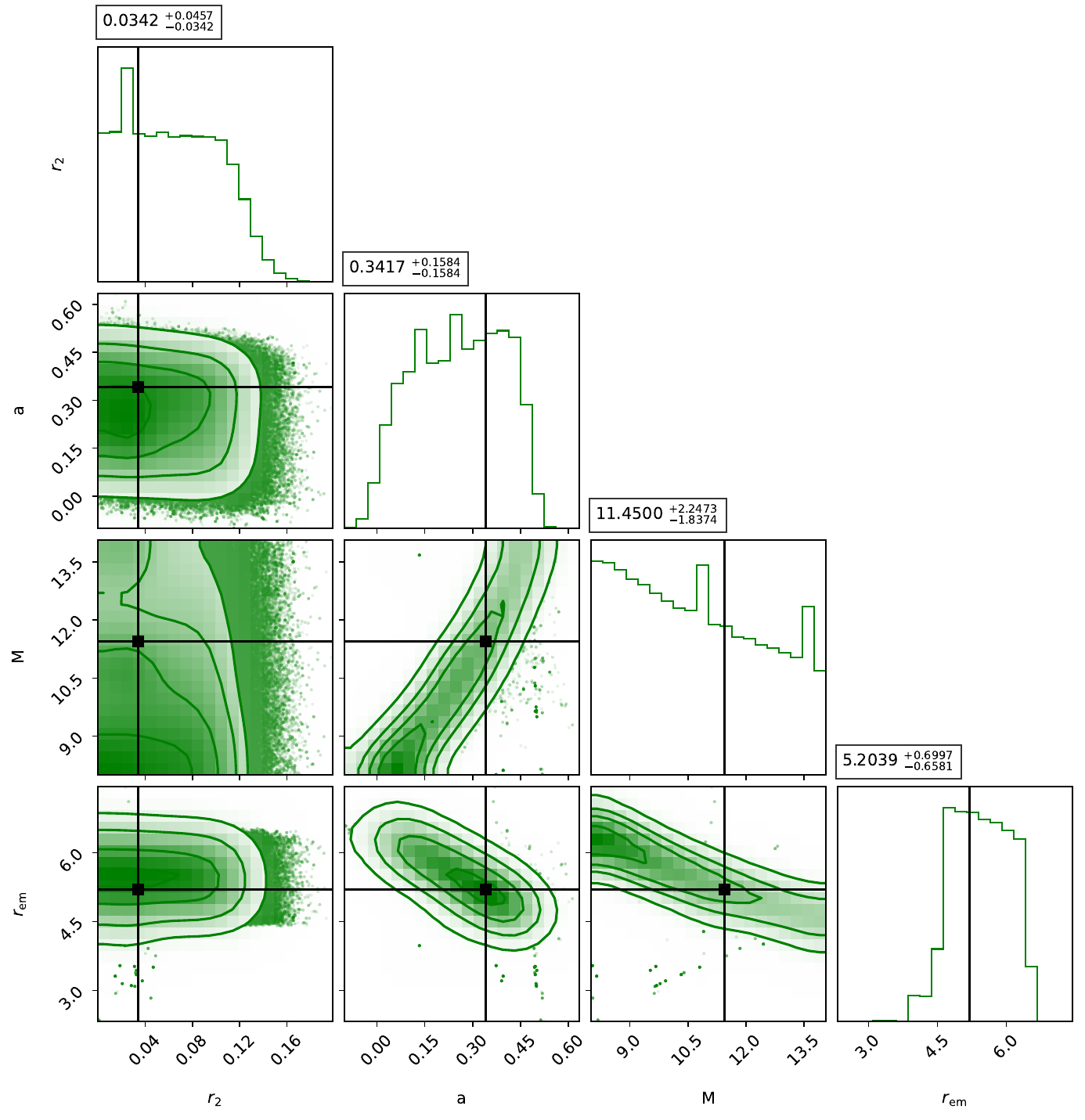}
        \caption{Non-axisymmetric Disk Oscillation Model 2}
    \end{subfigure}

\caption{Constraints on the model parameters using the HFQPO data of H1743-322 considering (a) the Keplerian Resonance Model 2, (b) the Keplerian Resonance Model 3, (c) the Warped Disc Oscillation Model, (d) the Non-axisymmetric Disk Oscillation Model 1 and (e) the Non-axisymmetric Disk Oscillation Model 2}
\label{Fig8}
\end{figure}

\begin{figure}[t!]
\vspace*{-0.5cm }
{\bf \underline{Sgr A*}}
    \centering

    \begin{subfigure}[b]{0.42\textwidth}
        \includegraphics[width=\linewidth]{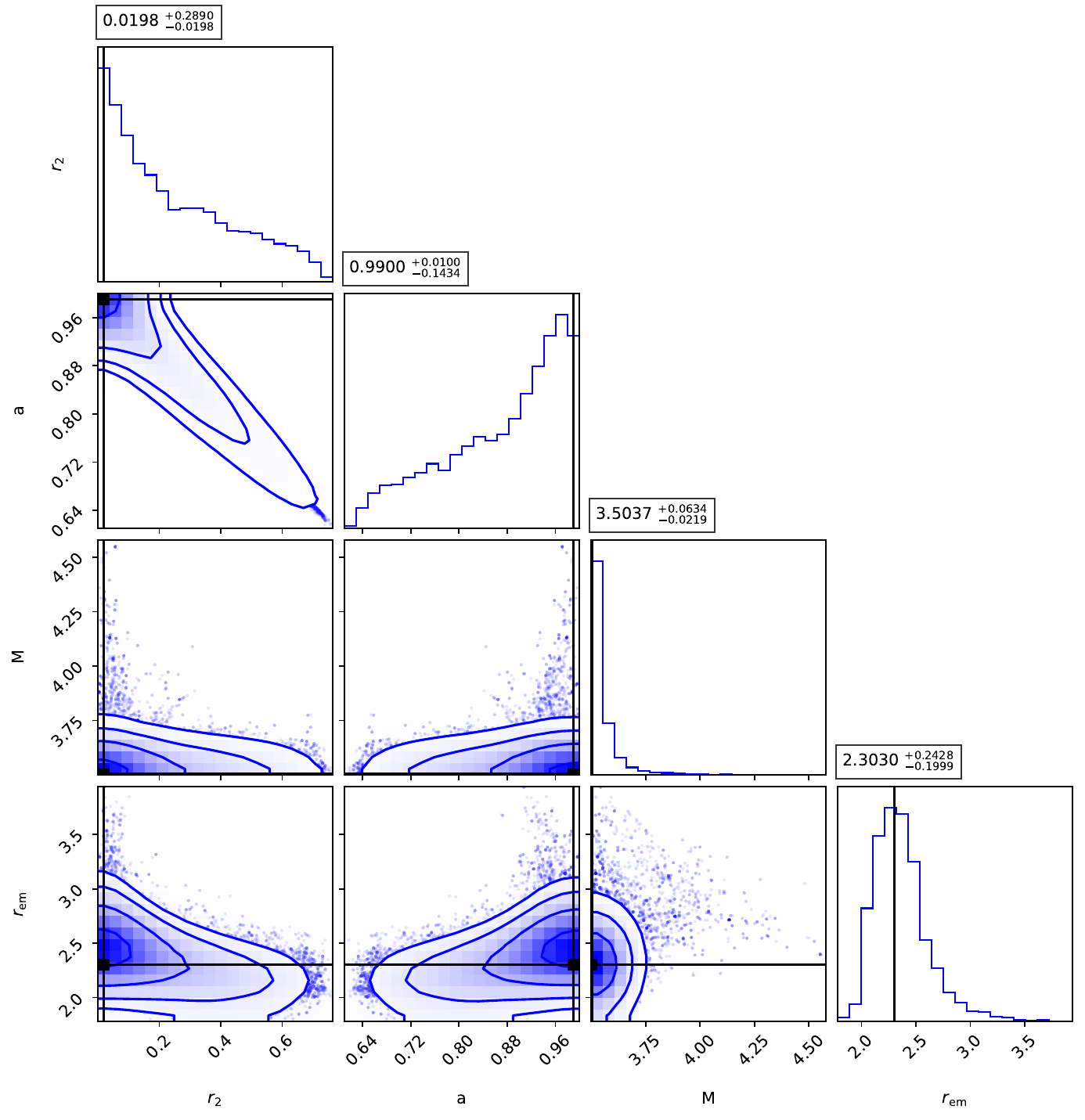}
        \caption{Parametric Resonance Model}
    \end{subfigure}
    \hspace{0cm}
    \begin{subfigure}[b]{0.42\textwidth}
        \includegraphics[width=\linewidth]{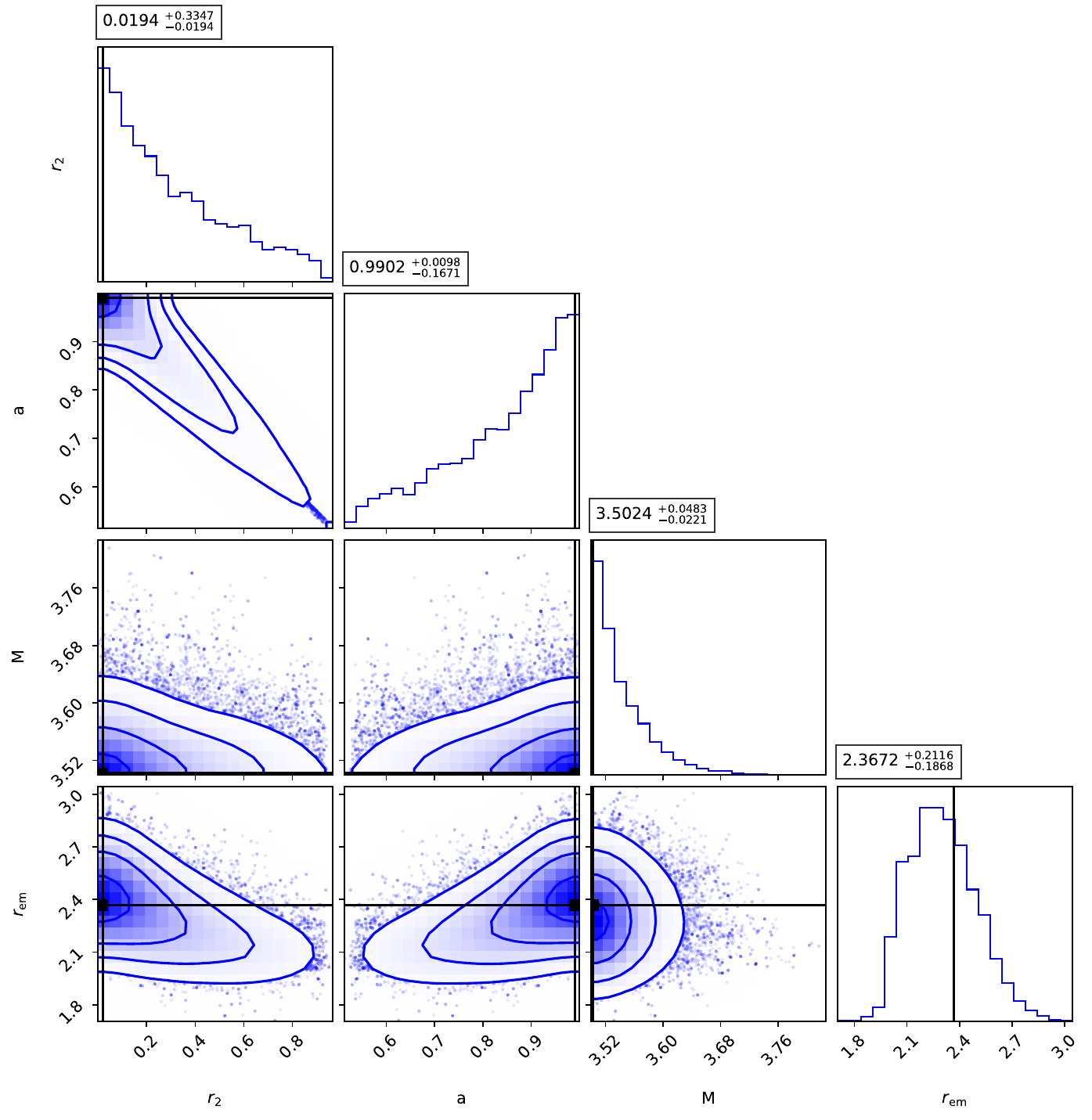}
        \caption{Keplerian Resonance Model 1}
    \end{subfigure}
    \hspace{0cm }
    \begin{subfigure}[b]{0.42\textwidth}
        \includegraphics[width=\linewidth]{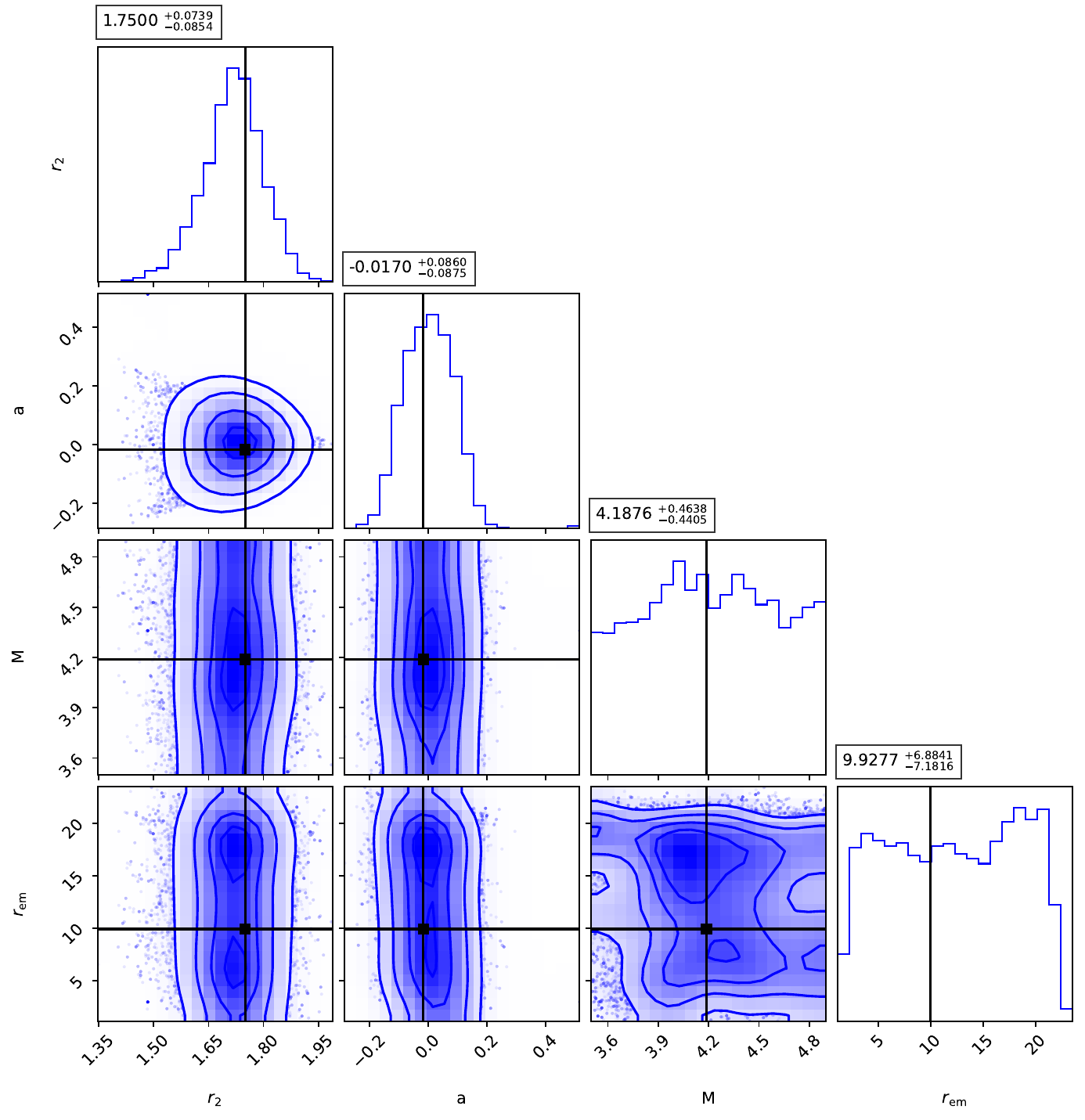}
        \caption{\hspace{-0.295cm} Non-axisymmetric Disk Oscillation Model 1} 
    \end{subfigure}

    \caption{Constraints on the model parameters using the HFQPO data of Sgr A* considering (a) the Parametric Resonance Model, (b) the Keplerian Resonance Model 1 and (c) the Non-axisymmetric Disk Oscillation Model 1.}
\label{Fig9}
\end{figure}

In \ref{Fig1}-\ref{Fig9}, we present the corner plots obtained from our MCMC simulations with the shaded regions representing the  1-$\sigma$, 2-$\sigma$, and 3-$\sigma$ confidence intervals for the posterior distributions of the model parameters. We present the corner plots for all the five BH sources in \ref{Fig1}-\ref{Fig9}.  
For a given source, we report the MCMC results only for those models where we could provide 1-$\sigma$ bounds on $r_2$ based on our earlier grid-search method (\ref{fig5b}-\ref{fig15b}). We have verified that for the remaining models, the metric parameters cannot be well constrained, in agreement with our earlier grid-search analysis.
Once the corner plots from all models are presented for a given source a table summarizing the best-fit model parameters is reported from both the MCMC and the grid-search method. From the figures and the tables, one may note that the results based on the two methods are consistent. For sources where we could give tight constrains on $r_2$ using the grid-search method (e.g., GRO J1655-40 with models like RPM, TDM, NADO1, and Sgr A* with models like PRM, KRM1, NADO1), the agreement is even better. For sources where tight constraints on $r_2$ could not be established by the grid-search method, the best-fit parameters when compared with MCMC exhibit moderate deviation. This is attributed to the insensitivity of the $\chi^2$ or the likelihood function on the metric parameters ($r_2$ and $a$) for some sources with respect to certain models, also reflected in the $\chi^2$ versus $r_2$ variation plots in the previous section. Interestingly, the MCMC results are pleasantly in agreement with the grid-search approach.

%


\begin{table}[htpb]
\centering
\setlength{\tabcolsep}{5pt}               
\renewcommand{\arraystretch}{1.6}         
\footnotesize
\begin{adjustbox}{max width=\textwidth}
\begin{tabular}{|l|l|l|l|l|l|l|l|l|}
\hline
\multicolumn{9}{|c|}{\textbf{GRO J1655-40}} \\ \hline
\multirow{2}{*}{\textbf{Previous constrains}}    & \multicolumn{2}{c|}{\textbf{ $\mathbf{r_2}$}}  & \multicolumn{3}{c|}{\textbf{Spin}} & \multicolumn{3}{c|}{\textbf{Mass $(M_{\odot})$} }\\  \cline{2-9} 
& \multicolumn{2}{c|}{$0\lesssim r_2 \lesssim 2$ \cite{Banerjee:2020ubc}} & {$\rm a\sim 0.65-0.75 $ \cite{Shafee_2005}}  & {$\rm a\sim 0.94-0.98$ \cite{Miller:2009cw}} & {$ \rm a=0.29\pm 0.003  $ \cite{Motta:2013wga}}   &  \multicolumn{3}{|c|}{\textbf{$\rm 5.4\pm 0.3$} \cite{Beer:2001cg} } \\

   \hline \hline                 
\multirow{2}{*}{\textbf{Model}} & \multicolumn{5}{c|}{\textbf{Grid Search}} & \multicolumn{3}{c|}{\textbf{MCMC}} \\ \cline{2-9} 
& \textbf{ $\mathbf{r_2}$} & \textbf{1-$\sigma$} & \textbf{3-$\sigma$} & \textbf{Spin} & \textbf{Mass $(M_{\odot})$} & \textbf{$\mathbf{r_2}$} & \textbf{Spin} & \textbf{Mass $(M_{\odot})$} \\ \hline
RPM   & 0    & $0 \lesssim r_2 \lesssim 0.005$ & $0\lesssim r_2\lesssim 0.01$  & 0.3 $\rm ({r_2}_{,\rm min}\sim 0)$ & 5.1 $\rm ({r_2}_{,\rm min}\sim 0)$ & $0.060^{+0.031}_{-0.031}$ & $0.258^{+0.004}_{-0.004}$ & $5.280^{+0.075}_{-0.072}$ \\ \hline
TDM   & 0.1  & $0 \lesssim r_2 \lesssim 1.4$   & $0\lesssim r_2\lesssim 1.6$ & $ \rm 0.1 ~ ({r_2}_{,\rm min}\sim 0.1)$  & $\rm 5.6~ ({r_2}_{,\rm min}\sim 0.1)$  & $0.1136^{+0.0605}_{-0.0603}$ & $0.0712^{+0.0624}_{-0.0618}$ & $5.3989^{+0.1907}_{-0.1816}$ \\ 
&  &  &  & $ \rm 0.1 ~ (r_2\sim 0)$   & $\rm 5.4 ~(r_2\sim 0)$ &  & & \\
\hline

FRM1  & 0.05 & $0 \lesssim r_2 \lesssim 1.3$   & $0\lesssim r_2\lesssim 1.5$  &  $ \rm 0.3 ({r_2}_{,\rm min}\sim 0.05)$  & $\rm 5.3 ({r_2}_{,\rm min}\sim 0.05)$  & $0.1482^{+0.0833}_{-0.0769}$ & $0.2775^{+0.0506}_{-0.0520}$ & $5.3988^{+0.1816}_{-0.1870}$ \\ 
&   &   &   &  $ \rm 0.3 (r_2\sim 0.0)$   & $\rm 5.2 (r_2\sim 0)$ & & & \\

\hline
FRM2  & 0.05 & $0 \lesssim r_2 \lesssim 1.4$   & $0\lesssim r_2\lesssim 1.55$  &  $ \rm 0.1 ({r_2}_{,\rm min}\sim 0.05)$  & $\rm 5.4 ({r_2}_{,\rm min}\sim 0.05)$ & $0.1360^{+0.4447}_{-0.1360}$ & $0.0617^{+0.1962}_{-0.1962}$ & $5.3988^{+0.1884}_{-0.1844}$ \\ 
&   &  &  &   $ \rm 0.1 (r_2\sim 0)$ &  $\rm 5.3 (r_2\sim 0)$  &   &   &  \\

\hline
KRM1  & 1.9  & $0.7 \lesssim r_2 \lesssim 1.9$  &  U & $ \rm 0.04 ({r_2}_{,\rm min}\sim 1.9)$ &   $\rm 5.1 ({r_2}_{,\rm min}\sim 1.9)$  & $1.8110^{+0.0699}_{-0.0708}$ & $0.0942^{+0.0339}_{-0.0361}$ & $5.2045^{+0.1519}_{-0.0976}$ \\ 
&   &   &   &   $\rm  0.99 (r_2\sim 0)$  &  $\rm 5.1 (r_2\sim 0)$  &  &  &  \\

\hline
KRM2  & 0.2  & $0.03 \lesssim r_2 \lesssim 1.1$& $0\lesssim r_2\lesssim 1.4$  & $ \rm 0.3 ({r_2}_{,\rm min}\sim 0.2)$   &  $\rm 5.7 ({r_2}_{,\rm min}\sim 0.2)$  &  $0.2761^{+0.1495}_{-0.1493}$ & $0.2101^{+0.0669}_{-0.0665}$ & $5.3986^{+0.1793}_{-0.1865}$ \\ 
&   &   &   &   $ \rm 0.3 (r_2\sim 0)$  &  $\rm 5.3 (r_2\sim 0)$  &  &  &  \\

\hline
KRM3  & 0    & $0 \lesssim r_2 \lesssim 0.1$ \& $0.25 \lesssim r_2 \lesssim 1$   & $0\lesssim r_2\lesssim 1.25$  &  0.1 $\rm ({r_2}_{,\rm min}\sim 0)$ & 5.2 $\rm ({r_2}_{,\rm min}\sim 0)$ & $0.0084^{+0.0438}_{-0.0084}$ & $0.1536^{+0.0488}_{-0.0488}$ & $5.4018^{+0.1895}_{-0.1828}$ \\ 
\hline

NADO1 & 0.3  & $0 \lesssim r_2 \lesssim 1.2$   & $0\lesssim r_2\lesssim 1.4$  &  $\rm  0.2 ({r_2}_{,\rm min}\sim 0.3)$  &  $\rm 5.4 ({r_2}_{,\rm min}\sim 0.3)$  & $0.4636^{+0.2260}_{-0.2100}$ & $0.1690^{+0.1219}_{-0.1305}$ & $5.3929^{+0.1831}_{-0.1778}$ \\ 
&   &  &  &  $\rm  0.4 (r_2\sim 0)$  & $\rm 5.3 (r_2\sim 0)$  &  &  &  \\

\hline
NADO2 & 1    & $0 \lesssim r_2 \lesssim 1.4$   & $0\lesssim r_2\lesssim 1.6$  & $\rm  -0.1 ({r_2}_{,\rm min}\sim 1)$ & $\rm 5.3 ({r_2}_{,\rm min}\sim 1)$  & $0.1209^{+0.5800}_{-0.1209}$ & $0.1978^{+0.1888}_{-0.1888}$ & $5.4015^{+0.1875}_{-0.1638}$ \\ 
&    &   &   &   $\rm  0.2 (r_2\sim 0)$ &  $\rm 5.1 (r_2\sim 0)$  &  &  &  \\

\hline
WDOM  & 0.1  & $0 \lesssim r_2 \lesssim 1.3$   &  $0\lesssim r_2\lesssim 1.5$ &  $\rm  0.1 ({r_2}_{,\rm min}\sim 0.1)$  & $\rm 5.5 ({r_2}_{,\rm min}\sim 0.1)$  & $0.138^{+0.109}_{-0.114}$ & $0.251^{+0.011}_{-0.011}$ & $5.404^{+0.175}_{-0.155}$ \\ 
&  &  &  &  $\rm 0.1 (r_2\sim 0)$  &  $\rm 5.3 (r_2\sim 0)$  &  &  &  \\

\hline
\end{tabular}
\end{adjustbox}
\caption{Comparison of best-fit model parameters for GRO J1655-40 derived using the grid-search and the MCMC methods.}
\label{Tab3}
\end{table}
In \ref{Tab3} we enlist the constraints on $r_2$, $a$ and $M$ for the source GRO J1655-40 assuming the various QPO models. The source exhibits strong jets and based on its jet power, independent constraints have been imposed on $r_2$ and $a$ \cite{Banerjee:2020ubc}, also mentioned in \ref{Tab3}. We will eventually compare the constraints of $r_2$ and $a$ for GRO J1655-40, from both the observations related to HFQPO and jet power. From \ref{Tab3}, we note that the RPM imposes the strongest constrain on $r_2$, it strongly favors $r_2\sim 0-0.1$ and rules out large values of $r_2$. In that event, the observationally favored spin is $\sim 0.3$ (from grid-search) and $\sim 0.26$ (from MCMC), in agreement with \cite{Motta:2013wga} (which again is based on the QPO data) but it is inconsistent with the other independent spin estimates based on the Continuum-Fitting method ($\rm a\sim 0.65-0.75 $\cite{Shafee_2005}) or the Fe-line method ($\rm a\sim 0.94-0.98$ \cite{Miller:2009cw}). Even the jet power of this source predicts a spin $a\gtrsim 0.7$ (for $r_2\sim 0$)\cite{Banerjee:2020ubc}, in agreement with \cite{Shafee_2005,Miller:2009cw}, although it does not constrain $r_2$ significantly. All the three previous estimates of spin \cite{Shafee_2005,Miller:2009cw,Motta:2013wga} assumed the background geometry to be governed by the Kerr metric. Our analysis based on RPM reveals that even if we allow a deviation from the Kerr scenario (here the dilaton charge), the HFQPO observations indicate a preference towards the Kerr scenario. But the spin derived by fitting the RPM with the HFQPO data is not in agreement with previous estimates. This plausibly indicates that RPM may not be the most suitable model to describe the HFQPO data of 
GRO J1655-40. The other models listed in \ref{Tab3} rule out a certain region of the parameter space of $r_2$, but none of them yield a spin consistent with previous estimates based on independent observations \cite{Shafee_2005,Miller:2009cw}, except KRM1. Interestingly, KRM1 rules out $r_2\sim 0$ outside 1-$\sigma$ (using both MCMC and grid-search) and exhibits a $\chi^2_{min}$ at $r_2\sim 1.9$ (using grid-search) and at $r_2\sim 1.81$ (using MCMC) indicating a strong preference towards the Kerr-Sen scenario. In case, one assumes the Kerr metric to explain its twin-peak HFQPOs, the spin turns out to be $\sim 0.99$, in agreement with \cite{Miller:2009cw} (see \ref{Tab3}). In case we assume $r_{2,\min}\sim 1.9$ to explain the data, the spin is again near-maximal $a\sim 0.04$ (note that for $r_2\sim 1.9$, $a_{\rm max}\sim 0.05$). However, Keplerian resonance due to g-mode oscillations was shown to get damped due to corotation resonance \cite{Li:2002yi,2003PASJ...55..257K}, although  Keplerian resonance due to coupling of the orbital angular frequencies of a pair of spatially separated vortices with opposite vorticities oscillating with radial epicyclic frequencies may still be operational \cite{2005A&A...436....1T,2010tbha.book.....A}.

We further note from \ref{Tab3} that even if we keep $r_2\sim 0$, none of the models can explain the spin of GRO J1655-40, in agreement with at least one of the previous estimates, except KRM1. This may indicate that apart from KRM1 all the models in \ref{Tab3} are probably not suitable for GRO J1655-40. Note that, PRM was not included in \ref{Tab3} because it does not rule out any allowed range of $r_2$ (see \ref{fig7a}). But, if we assume PRM and $r_2\sim 0$, the best estimate of spin turns out to be $a_{\rm min}\sim 0.92$ which is close to the spin measured by the Fe-line method \cite{Miller:2009cw}. But if PRM is used to explain the HFQPO data, non-zero values of $r_2$ are equally preferred. Also, KRM1 shows a $\chi^2_{min}$ at $r_2\sim 1.8-1.9$. Our analysis seems to reveal that possibly KRM1 or PRM are suitable QPO models to explain the HFQPO data of GRO J1655-40. Also, the fact that the previous two spin estimates \cite{Shafee_2005,Miller:2009cw} are inconsistent might indicate that the black hole may have some additional hair (e.g. dilaton charge) which when incorporated might give consistent values of that hair parameter and the spin by both the Continuum-Fitting and the Fe-line method. This however needs to be confirmed with the availability of more precise X-ray spectral and timing data of the source.

\begin{table}[htbp]
\centering
\setlength{\tabcolsep}{5pt}               
\renewcommand{\arraystretch}{1.6}         
\footnotesize
\begin{adjustbox}{max width=\textwidth}
\begin{tabular}{|l|l|l|l|l|l|l|l|l|}
\hline
\multicolumn{9}{|c|}{\textbf{XTE J1550-564}} \\ \hline
\multirow{2}{*}{\textbf{Previous constrains}}    & \multicolumn{2}{c|}{\textbf{ $\mathbf{r_2}$}}  & \multicolumn{3}{c|}{\textbf{Spin}} & \multicolumn{3}{c|}{\textbf{Mass $(M_{\odot})$} }\\  \cline{2-9} 
& \multicolumn{2}{c|}{$0\lesssim r_2 \lesssim 2$ \cite{Banerjee:2020ubc}} & \multicolumn{2}{c|}{$ \rm -0.11<a<0.71$ \cite{Steiner:2010bt}}  & {$ \rm a= 0.55^{+0.15}_{-0.22}$\cite{Steiner:2010bt}}    &  \multicolumn{3}{c|}{\textbf{$ \rm 9.1\pm 0.61$} \cite{Orosz:2011ki}} \\
\hline\hline
\multirow{2}{*}{\textbf{Model}} 
& \multicolumn{5}{c|}{\textbf{Grid Search}} 
& \multicolumn{3}{c|}{\textbf{MCMC}} \\ \cline{2-9}
& \textbf{$\mathbf{r_2}$} 
& \textbf{1-$\sigma$} 
&\textbf{3-$\sigma$} 
& \textbf{Spin} 
& \textbf{Mass $(M_{\odot})$} 
& \textbf{$\mathbf{r_2}$} 
& \textbf{Spin} 
& \textbf{Mass $(M_{\odot})$} \\ \hline

RPM    & 0.1  & $0 \lesssim r_2 \lesssim 1.6$  & $0\lesssim r_2\lesssim 1.7$ & $ \rm 0.4 ({r_2}_{,\rm min}\sim 0.1)$   & $\rm 9.71 ({r_2}_{,\rm min}\sim 0.1)$  & $0.2723^{+0.1857}_{-0.1849}$ & $0.2619^{+0.0871}_{-0.0902}$ & $9.1008^{+0.3816}_{-0.3846}$ \\ 
&  &  &  &  $ 0.4 (r_2\sim 0)$ &  $\rm 9.31 (r_2\sim 0)$ &  &   &  \\

\hline
TDM    & 0.2  & $0 \lesssim r_2 \lesssim 1.65$ & $0\lesssim r_2\lesssim 1.8$  &  $\rm 0.1 ({r_2}_{,\rm min}\sim 0.2)$ & $\rm 9.01 ({r_2}_{,\rm min}\sim 0.2)$  & $0.0557^{+0.0607}_{-0.0557}$ & $0.2051^{+0.0985}_{-0.0985}$ & $9.0971^{+0.3838}_{-0.3742}$ \\ 
&  &  &  &  $ 0.2 (r_2\sim 0)$ &  $\rm 9.31 (r_2\sim 0)$  &  &  & \\

\hline
FRM1   & 0.5  & $0 \lesssim r_2 \lesssim 1.6$  & $0\lesssim r_2\lesssim 1.7$ &  $\rm 0.1 ({r_2}_{,\rm min}\sim 0.5)$  & $ \rm 8.81 ({r_2}_{,\rm min}\sim 0.5)$ & $0.6287^{+0.5297}_{-0.3958}$ & $-0.0056^{+0.1877}_{-0.2362}$ & $9.4137^{+0.4178}_{-0.4137}$ \\ 
&  &  &  &  $ 0.4 (r_2\sim 0)$ &  $\rm 8.91 (r_2\sim 0)$ &  &  & \\

\hline
FRM2   & 0    & $0 \lesssim r_2 \lesssim 1.6$  & $\lesssim r_2\lesssim 1.75$ & 0.2 $\rm ({r_2}_{,\rm min}\sim 0)$ & 8.91 $\rm ({r_2}_{,\rm min}\sim 0)$ & $0.019^{+0.024}_{-0.019}$    & $0.22^{+0.10}_{-0.09}$        & $9.09^{+0.38}_{-0.37}$       \\ \hline

KRM2   & 0.05 & $0 \lesssim r_2 \lesssim 1.45$ & $0\lesssim r_2\lesssim 1.6$ &  $\rm 0.3 ({r_2}_{,\rm min}\sim 0.05)$   & $\rm 8.61 ({r_2}_{,\rm min}\sim 0.05)$   &  $0.1317^{+0.1038}_{-0.0913}$ & $0.3218^{+0.0579}_{-0.0629}$ & $9.1056^{+0.3896}_{-0.3785}$ \\ 
&  &   &   &  $ 0.4 (r_2\sim 0)$  &  $\rm 9.31 (r_2\sim 0)$ &  &  &  \\

\hline
KRM3   & 0    & $0 \lesssim r_2 \lesssim 1.35$ & $0\lesssim r_2\lesssim 1.5$ & 0.2 $\rm ({r_2}_{,\rm min}\sim 0)$ & 8.91  $\rm ({r_2}_{,\rm min}\sim 0)$ & $0.1016^{+0.0609}_{-0.0599}$ & $0.1848^{+0.0564}_{-0.0587}$ & $9.1066^{+0.3831}_{-0.3698}$ \\ \hline

NADO1  & 0    & $0 \lesssim r_2 \lesssim 1.5$  &  $0\lesssim r_2\lesssim 1.7$ & 0.5 $\rm ({r_2}_{,\rm min}\sim 0)$  & 8.91  $\rm ({r_2}_{,\rm min}\sim 0)$  & $0.0504^{+0.0659}_{-0.0504}$ & $0.5117^{+0.0941}_{-0.0941}$ & $9.1053^{+0.3919}_{-0.3473}$ \\ \hline
NADO2  & 0    & $0 \lesssim r_2 \lesssim 1.65$ &   $0\lesssim r_2\lesssim 1.8$ & 0.3 $\rm ({r_2}_{,\rm min}\sim 0)$ & 9.11 $\rm ({r_2}_{,\rm min}\sim 0)$  & $0.0270^{+0.0454}_{-0.0270}$ & $0.2844^{+0.0436}_{-0.0436}$ & $9.1095^{+0.3490}_{-0.3869}$ \\ \hline
WDOM   & 0    & $0 \lesssim r_2 \lesssim 1.55$ &  $0\lesssim r_2\lesssim 1.7$ & 0.2 $\rm ({r_2}_{,\rm min}\sim 0)$ & 8.91 $\rm ({r_2}_{,\rm min}\sim 0)$  & $0.1163^{+0.0622}_{-0.0608}$ & $0.1784^{+0.0991}_{-0.0977}$ & $9.1116^{+0.3878}_{-0.3736}$ \\ \hline

\end{tabular}
\end{adjustbox}
\caption{Comparison of best-fit model parameters for XTE J1550-564 derived using the grid-search and the MCMC methods.}
\label{Tab4}
\end{table}

In \ref{Tab4} we present the constraints on $r_2$, $a$ and $M$ for the source XTE J1550-564 assuming the various QPO models. From the table it is evident that most models mildly constrain the dilaton charge and all of them rule out near extremal values of $r_2$ outside 3-$\sigma$. KRM1 and PRM on the other hand allow all values of $r_2$ within 1-$\sigma$ and hence are not listed in \ref{Tab4}. In order to compare the spin estimates of this source with previous results \cite{Steiner:2010bt,Banerjee:2020ubc}, we have to consider the $r_2\sim 0$ case, even for those models where $\chi^2$ does not minimize at $r_2\sim 0$, as previous estimates were made assuming the Kerr geometry.
Based on our grid-search analysis, the spin of XTE J1550-564 obtained from the models NADO1, RPM, FRM1 and KRM2 exhibit the best agreement with previous estimates \cite{Steiner:2010bt,Banerjee:2020ubc} which are $\sim 0.34$ from the Continuum-Fitting \cite{Steiner:2010bt}, $ \rm a= 0.55^{+0.15}_{-0.22}$ from the Fe-line \cite{Steiner:2010bt} and $0.3\lesssim a \lesssim 0.6$\cite{Banerjee:2020ubc} from the jet power. 
The spins predicted from the remaining five models listed in \ref{Tab4} also fall within the error bars mentioned in \cite{Steiner:2010bt}. PRM and KRM1 do not constrain $r_2$ and predict $a\sim 0.95$ and $a\sim 0.999$ respectively, when the Kerr scenario is considered, which are not consistent with \cite{Steiner:2010bt,Banerjee:2020ubc}. Hence, these two models may not be the right description of the HFQPO data of XTE J1550-564. The remaining nine models seem to fit the HFQPO data reasonably well but none of them constrain $r_2$ very strongly. Hence, it is difficult to conclusively establish or rule out any of the nine models in \ref{Tab4}. 

Using MCMC, we report that the models TDM, FRM2, NADO1 and NADO2 prefer $r_2\sim 0$ or a BH with a very small dilation charge. The spin predictions from these models when compared with grid-search are consistent, but when compared with previous estimates the agreement with NADO1 is the best. For the remaining five models MCMC exhibits a preference towards a positive $r_2$, more or less in agreement with the grid-search method.
One may note that the HFQPO observations of different BH sources may be governed by different models, as the accretion environment and the related physics may not be identical for all the sources.  
Hence, there may not be a single QPO model which may be applicable to all sources. 
Also, one may note that the spin of the source predicted based on the Continuum-Fitting and the Fe-line method are again not fully consistent, although they agree within the error bar which however is very large for the Continuum-Fitting method. This might be an indication towards some beyond GR phenomenon at play in the strong gravity regime near black holes (as the data does not strongly rule out non-zero $r_2$, if one considers the present case) or possibly the methods of determining the BH spins by the two aforesaid methods may require further investigation. 


\begin{table}[htbp]
\centering
\setlength{\tabcolsep}{5pt}               
\renewcommand{\arraystretch}{1.6}         
\footnotesize
\begin{adjustbox}{max width=\textwidth}
\begin{tabular}{|l|l|l|l|l|l|l|l|l|}
\hline
\multicolumn{9}{|c|}{\textbf{GRS 1915+105}} \\ \hline
\multirow{2}{*}{\textbf{Previous constrains}}    & \multicolumn{2}{c|}{\textbf{ $\mathbf{r_2}$}}  & \multicolumn{4}{c|}{\textbf{Spin}} & \multicolumn{2}{c|}{\textbf{Mass $(M_{\odot})$} }\\  \cline{2-9} 
& \multicolumn{2}{c|}{$0\lesssim r_2 \lesssim 2$ \cite{Banerjee:2020ubc}} & {$ a\sim \rm 0.98 $ \cite{McClintock:2006xd}} & {$ a\sim \rm 0.7 $ \cite{2006MNRAS.373.1004M}} & {$ \rm a\sim 0.6-0.98 $ \cite{Blum:2009ez}} & {$ \rm a\sim 0.4-0.98$ \cite{Mills:2021dxs}}   &  \multicolumn{2}{c|}{\textbf{$ \rm 12.4^{+2.0}_{-1.8} $} \cite{Reid:2014ywa}} \\
\hline\hline
\multirow{2}{*}{\textbf{Model}} 
& \multicolumn{5}{c|}{\textbf{Grid Search}} 
& \multicolumn{3}{c|}{\textbf{MCMC}} \\ \cline{2-9}
& \textbf{Best $r_2$} 
& \textbf{1-$\sigma$} 
& \textbf{3-$\sigma$}
& \textbf{Spin} 
& \textbf{Mass $(M_{\odot})$} 
& \textbf{$r_2$} 
& \textbf{Spin} 
& \textbf{Mass $(M_{\odot})$} \\ \hline

RPM    & 0.6  & $0 \lesssim r_2 \lesssim 1.65$ & $0\lesssim r_2\lesssim 1.75$ & $\rm -0.1 ({r_2}_{,\rm min}\sim 0.6)$ & $\rm 12.3 ({r_2}_{,\rm min}\sim 0.6)$ & $0.2703^{+0.4107}_{-0.2703}$ & $0.1073^{+0.1903}_{-0.1903}$ & $13.0616^{+1.3297}_{-1.1915}$ \\ 
&   &  &  &  $\rm 0 (r_2\sim 0)$ & $\rm 11 (r_2\sim 0)$  &  &  & \\
\hline

TDM    & 0.1  & $0 \lesssim r_2 \lesssim 1.65$ &  $0\lesssim r_2\lesssim 1.85$  &  $\rm  0 ({r_2}_{,\rm min}\sim 0.1)$    &  $\rm 13.3 ({r_2}_{,\rm min}\sim 0.1)$  & $0.1892^{+0.0781}_{-0.0770}$ & $-0.2674^{+0.2345}_{-0.2219}$ & $11.4808^{+1.3970}_{-1.1966}$ \\ 
&  &  &  &  $\rm 0.1 (r_2\sim 0)$ &  $\rm 14.1 (r_2\sim 0)$  &  &  & \\

\hline
FRM1   & 0.6  & $0 \lesssim r_2 \lesssim 1.6$  & $0\lesssim r_2\lesssim 1.7$   & $\rm  -0.1 ({r_2}_{,\rm min}\sim 0.6)$ &  $\rm 12.4 ({r_2}_{,\rm min}\sim 0.6)$  & $0.4545^{+0.3301}_{-0.3334}$ & $-0.1918^{+0.1907}_{-0.2017}$ & $11.0979^{+1.4635}_{-1.1397}$ \\ 
&   &   &   &   $\rm 0 (r_2\sim 0)$ & $\rm 11.0 (r_2\sim 0)$  &  &  &  \\

\hline
FRM2   & 0.3  & $0 \lesssim r_2 \lesssim 1.6$  &  $0\lesssim r_2\lesssim 1.8$  & $\rm  -0.3 ({r_2}_{,\rm min}\sim 0.3)$  & $\rm 11.9 ({r_2}_{,\rm min}\sim 0.3)$ & $0.5046^{+0.3115}_{-0.3037}$ & $-0.1946^{+0.2482}_{-0.2223}$ & $13.1061^{+1.3318}_{-1.1947}$ \\ 
&   &   &   &  $\rm -0.3 (r_2\sim 0)$  &  $\rm 10.8 (r_2\sim 0)$  &  &  &  \\

\hline
KRM2   & 0    & $0 \lesssim r_2 \lesssim 1.45$ &  $0\lesssim r_2\lesssim 1.6$ & 0.2 $\rm  ({r_2}_{,\rm min}\sim 0)$  & 12.7 $\rm  ({r_2}_{,\rm min}\sim 0)$  & $0.0532^{+0.0460}_{-0.0371}$ & $0.2008^{+0.1238}_{-0.1322}$ & $12.9656^{+1.4171}_{-1.1240}$ \\ 
\hline

KRM3   & 0.4  & $0 \lesssim r_2 \lesssim 1.4$  &   $0\lesssim r_2\lesssim 1.6$    &  $\rm -0.4 ({r_2}_{,\rm min}\sim 0.4)$ & $\rm 11.0 ({r_2}_{,\rm min}\sim 0.4)$  & $0.7093^{+0.2916}_{-0.2740}$ & $-0.5643^{+0.1856}_{-0.1808}$ & $10.7764^{+1.3666}_{-1.1940}$ \\ 
&   &   &   &  $\rm 0 (r_2\sim 0)$  &  $\rm 12.7 (r_2\sim 0)$ &  &   &  \\

\hline
NADO1  & 0    & $0 \lesssim r_2 \lesssim 1.4$  &  $0\lesssim r_2\lesssim 1.6$   & 0.1 $\rm  ({r_2}_{,\rm min}\sim 0)$ & 11.6 $\rm  ({r_2}_{,\rm min}\sim 0)$ & $0.0112^{+0.0387}_{-0.0112}$ & $0.4424^{+0.1925}_{-0.1925}$ & $14.1693^{+1.3951}_{-1.2119}$ \\ 
\hline
NADO2  & 0    & $0 \lesssim r_2 \lesssim 1.6$  &  $0\lesssim r_2\lesssim 1.75$    & 0.1 $\rm  ({r_2}_{,\rm min}\sim 0)$ & 12.1 $\rm  ({r_2}_{,\rm min}\sim 0)$  & $0.0225^{+0.0380}_{-0.0225}$ & $0.2116^{+0.1155}_{-0.1155}$ & $13.8309^{+1.4128}_{-1.1699}$ \\ 
\hline
WDOM   & 0.1  & $0 \lesssim r_2 \lesssim 1.5$  &  $0\lesssim r_2\lesssim 1.7$    &  $\rm -0.2({r_2}_{,\rm min}\sim 0.1)$   & $\rm 11.7 ({r_2}_{,\rm min}\sim 0.1)$  & $0.1953^{+0.3826}_{-0.1953}$ & $-0.1962^{+0.1917}_{-0.1917}$ & $12.0160^{+1.2626}_{-1.2795}$ \\ 
&    &   &   &  $\rm -0.2 (r_2\sim 0)$ &  $\rm 11.4 (r_2\sim 0)$ &  &  &  \\

\hline

\end{tabular}
\end{adjustbox}
\caption{Comparison of best-fit model parameters for GRS 1915+105 derived using the grid-search and the MCMC methods.}
\label{Tab5}
\end{table}

In \ref{Tab5} we present the estimates of dilaton charge, mass and spin for GRS 1915+105 from different QPO models and also compare the same with previous results. The table reveals that the nine models which impose some constrain on $r_2$ rule out only the near extremal values of $r_2$ outside 3-$\sigma$ (from grid-search). The remaining two models (not mentioned in \ref{Tab5}), namely, PRM and KRM1, allow all values of the dilaton charge, $0\lesssim r_2\lesssim 2$ within 1-$\sigma$. Once again the table reveals a lot of discrepancy in the earlier spin estimates, e.g. the Fe-line method gives $ \rm a\sim 0.6-0.98 $ \cite{Blum:2009ez} while the Continuum-Fitting method yields intermediate $ a\sim \rm 0.7 $ \cite{2006MNRAS.373.1004M} as well as near maximal spin $ a\sim \rm 0.98 $ \cite{McClintock:2006xd}. More recently, when the revised mass and inclination of GRS 1915+105 is considered, its spin turns out to be in the range $ \rm a\sim 0.4-0.98$ \cite{Mills:2021dxs}. From the observed jet power, the spin of this source turns out to be $\rm a\sim 0.6-0.9$ \cite{Banerjee:2020ubc}. 
Using grid-search we note that all the models in \ref{Tab5} estimate zero to very low and sometimes negative spins for this source (when $r_2\sim 0$ is assumed), in contradiction with previous estimates. This indicates that the nine models listed in \ref{Tab5} may not be the correct description of the HFQPO data for this source. Using MCMC however, NADO1 shows a preference towards $r_2\sim 0$ and $a\sim 0.44$, within the range predicted by \cite{Mills:2021dxs}.

If we consider PRM and KRM1 (not listed in \ref{Tab5}) to explain the HFQPO data (the background geometry being Kerr, i.e. $r_2\sim 0$), the spin turns out to be $a\sim 0.8$ and $a\sim 0.97$ respectively. The spin predicted by these two models are in agreement with previous independent measurements \cite{Banerjee:2020ubc,Blum:2009ez,McClintock:2006xd,2006MNRAS.373.1004M,Mills:2021dxs} but these models impose no constrain on $r_2$ (\ref{fig7a} and \ref{fig10a}). Note that even from the observed jet power of this source, the dilaton charge cannot be constrained \cite{Banerjee:2020ubc}. Thus the HFQPO data possibly indicates NADO1/parametric resonance/ or Keplerian resonance at play in the accretion disk of this source, but this is subject to further investigation as the spin estimates of this source exhibit a lot of disparity.
Also, with the current precision the present HFQPO data equally favors the Kerr and the Kerr-Sen scenario.

\vspace{-1cm}
\vspace{-1cm}
\begin{table}[htbp]
\centering
\setlength{\tabcolsep}{5pt}               
\renewcommand{\arraystretch}{1.4}         
\footnotesize
\begin{adjustbox}{max width=\textwidth}
\begin{tabular}{|l|l|l|l|l|l|l|l|l|}
\hline
\multicolumn{9}{|c|}{\textbf{H1743-322}} \\ \hline

\multirow{2}{*}{\textbf{Previous constrains}}    & \multicolumn{2}{c|}{\textbf{ $\mathbf{r_2}$}}  & \multicolumn{4}{c|}{\textbf{Spin}} & \multicolumn{2}{c|}{\textbf{Mass $(M_{\odot})$} }\\  \cline{2-9} 
& \multicolumn{2}{c|}{$0\lesssim r_2 \lesssim 2$ \cite{Banerjee:2020ubc}} & \multicolumn{4}{c|} {$ a=\rm 0.2 \pm 0.3 $ \cite{Steiner:2011kd} }  &  \multicolumn{2}{c|}{\textbf{$\rm 8.0-14.07$} \cite{Pei:2016kka,Bhattacharjee:2019vyy,Petri:2008jc}} \\
\hline\hline

\multirow{2}{*}{\textbf{Model}} 
& \multicolumn{5}{c|}{\textbf{Grid Search}} 
& \multicolumn{3}{c|}{\textbf{MCMC}} \\ \cline{2-9}
& \textbf{$r_2$} 
& \textbf{1-$\sigma$} 
& \textbf{3-$\sigma$} 
& \textbf{Spin} 
& \textbf{Mass $(M_{\odot})$} 
& \textbf{$r_2$} 
& \textbf{Spin} 
& \textbf{Mass $(M_{\odot})$} \\ \hline

RPM    & 0.3  & $0 \lesssim r_2 \lesssim 1.7$  &  $0\lesssim r_2\lesssim 1.8$   &  $ 0.1 ({r_2}_{,\rm min}\sim 0.3)$   & $\rm 9.57 ({r_2}_{,\rm min}\sim 0.3)$  & $0.3631^{+0.5265}_{-0.3631}$ & $-0.1044^{+0.2311}_{-0.2311}$ & $8.0093^{+2.1686}_{-1.8559}$ \\ 
&   &    &   &  $ 0.5 (r_2\sim 0)$  &  $\rm 12.07 (r_2\sim 0)$ &  &  & \\
\hline

TDM    & 0    & $0 \lesssim r_2 \lesssim 1.8$  &  $0\lesssim r_2\lesssim 1.85$   &  -0.1 $({r_2}_{,\rm min}\sim 0)$  & 8.57 $({r_2}_{,\rm min}\sim 0)$ & $0.0056^{+0.0408}_{-0.0056}$ & $0.3515^{+0.2761}_{-0.2761}$  & $12.1950^{+2.3169}_{-1.7753}$ \\ \hline

FRM1   & 0    & $0 \lesssim r_2 \lesssim 1.7$  &  $0\lesssim r_2\lesssim 1.8$     &   0.7$({r_2}_{,\rm min}\sim 0)$   & 14.07 $({r_2}_{,\rm min}\sim 0)$ & $0.0521^{+0.0464}_{-0.0458}$ & $0.4287^{+0.2242}_{-0.2038}$  & $10.8976^{+2.3644}_{-1.6827}$ \\ \hline

FRM2   & 0    & $0 \lesssim r_2 \lesssim 1.7$  &  $0\lesssim r_2\lesssim 1.8$   &   0.3 $({r_2}_{,\rm min}\sim 0)$  & 11.57 $({r_2}_{,\rm min}\sim 0)$ & $0.3169^{+0.1926}_{-0.1771}$ & $0.1223^{+0.2596}_{-0.2700}$  & $11.2018^{+2.3901}_{-1.6538}$ \\ \hline

KRM2   & 0    & $0 \lesssim r_2 \lesssim 1.65$ &  $0\lesssim r_2\lesssim 1.8$   &   0.4 $({r_2}_{,\rm min}\sim 0)$ & 10.47  $({r_2}_{,\rm min}\sim 0)$ & $0.0557^{+0.0453}_{-0.0406}$ & $0.2988^{+0.1851}_{-0.1920}$  & $9.7401^{+2.2900}_{-1.6629}$ \\ \hline

KRM3   & 0    & $0 \lesssim r_2 \lesssim 1.65$ &  $0\lesssim r_2\lesssim 1.8$    &  0.2 $({r_2}_{,\rm min}\sim 0)$  & 10.17 $({r_2}_{,\rm min}\sim 0)$ & $0.0383^{+0.0437}_{-0.0383}$ & $0.4405^{+0.2548}_{-0.2548}$ & $12.6657^{+2.2759}_{-1.7752}$ \\ \hline

NADO1  & 0.03 & $0 \lesssim r_2 \lesssim 1.7$  &  $0\lesssim r_2\lesssim 1.8$    &  $ 0.8 ({r_2}_{,\rm min}\sim 0.03)$ & $\rm 13.27 ({r_2}_{,\rm min}\sim 0.03)$ & $0.0945^{+0.0444}_{-0.0416}$ & $0.2349^{+0.2531}_{-0.3129}$  & $9.0506^{+2.2538}_{-1.9025}$ \\ 
&   &   &   &  $ 0.1 (r_2\sim 0)$ &  $\rm 8.17 (r_2\sim 0)$ &  &  &   \\

\hline
NADO2  & 0.02 & $0 \lesssim r_2 \lesssim 1.7$  & $0\lesssim r_2\lesssim 1.83$    &  $ 0.1 ({r_2}_{,\rm min}\sim 0.02)$  &  $\rm 8.67 ({r_2}_{,\rm min}\sim 0.02)$  & $0.0342^{+0.0457}_{-0.0342}$ & $0.3417^{+0.1584}_{-0.1584}$  & $11.4500^{+2.2473}_{-1.8374}$ \\ 
&  &   &   &  $ 0.2 (r_2\sim 0)$ &  $\rm 9.47 (r_2\sim 0)$ &  &  &  \\

\hline
WDOM   & 0.2  & $0 \lesssim r_2 \lesssim 1.7$  &   $0\lesssim r_2\lesssim 1.8$  &  $ 0.1 ({r_2}_{,\rm min}\sim 0.2)$   & $\rm 10.67 ({r_2}_{,\rm min}\sim 0.2)$  & $0.1429^{+0.4711}_{-0.1429}$ & $0.2049^{+0.2310}_{-0.2310}$  & $11.3637^{+2.1100}_{-1.9163}$ \\ 
&  &   &   &  $ 0.4 (r_2\sim 0)$ &  $\rm 12.77 (r_2\sim 0)$ &  &  &  \\
\hline

\end{tabular}
\end{adjustbox}
\caption{Comparison of best-fit model parameters for H1743-322 derived using the grid-search and the MCMC methods.}
\label{Tab6}
\end{table}
\ref{Tab6} presents the constrains on the dilaton charge, mass and spin from the various QPO models for the source H1743-322 considering its HFQPO data. Once again, the table enlists only those models which rule out a certain region of the allowed parameter space of $r_2$ and it is clear that all the nine models in \ref{Tab6} rule out near extremal values of $r_2$ outside 3-$\sigma$ (from grid-search). From the observed jet power also $r_2$ could not be constrained \cite{Banerjee:2020ubc}. The spin of this source has been estimated based on the Continuum-Fitting method which turns out to be $ a=\rm 0.2 \pm 0.3 $ (with 68\% confidence) and $a<0.92$ with 99.7\% confidence \cite{Steiner:2011kd}. From the observed jet power the spin is constrained in the range $0.25\lesssim a \lesssim 0.5$ \cite{Banerjee:2020ubc}, more or less consistent with \cite{Steiner:2011kd}.
Based on the grid-search method, we note that the HFQPO models which exhibit best agreement with Steiner et al. \cite{Steiner:2011kd} are KRM3 and NADO2, although the spins predicted by RPM, TDM, FRM2, KRM2, NADO1 and WDOM fall within the 1-$\sigma$ error bar of \cite{Steiner:2011kd} (\ref{Tab6}). The spin predicted by FRM1 is slightly higher ($\sim 0.7$) but is within the 3-$\sigma$ interval predicted by \cite{Steiner:2011kd}. 

When MCMC method is used, most models in \ref{Tab6} report a nearly vanishing $r_2$ except RPM, FRM2 and WDOM. If we consider those models which favor $r_2\sim 0$, then the spin estimate is within the 1-$\sigma$ error bar as reported previously in \cite{Steiner:2011kd}. For this source, the parameter estimates using grid-search and MCMC method exhibit some deviation but they agree within the error-bars (\ref{Tab6}).

The models PRM and KRM1 (not reported in \ref{Tab6}) fail to constrain $r_2$ (\ref{fig7a} and \ref{fig10a}) and predict $a\sim 0.95$ and $a\sim 0.99$ respectively when $r_2\sim 0$ is considered, which are clearly in contradiction with previous estimates \cite{Steiner:2011kd,Banerjee:2020ubc}. This might indicate that these two models may not be the right description of the HFQPO data of H1743-322, although from the present data it is difficult to conclusively point out the most favored HFQPO model for this source. Most of the models presented in \ref{Tab6} predict the observationally favored dilaton charge to be $r_2\sim 0$ or nearly zero, except RPM, WDOM and FRM2 . All the nine models however include $r_2\sim 0$ within 1-$\sigma$ and from the present data one cannot conclusively rule out the Kerr-Sen scenario but the data seems to indicate a preference towards the Kerr-scenario.

\vspace{-1cm}

\begin{table}[t!]
\centering
\setlength{\tabcolsep}{5pt}               
\renewcommand{\arraystretch}{1.6}         
\footnotesize
\begin{adjustbox}{max width=\textwidth}
\begin{tabular}{|l|l|l|l|l|l|l|l|l|l|}
\hline
\multicolumn{10}{|c|}{\textbf{Sgr A*}} \\ \hline

\multirow{2}{*}{\textbf{Previous constrains}}    & \multicolumn{1}{c|}{\textbf{ $\mathbf{r_2}$}}  & \multicolumn{7}{c|}{\textbf{Spin}} & \multicolumn{1}{c|}{\textbf{Mass $\rm 10^6 ~(M_{\odot})$} }\\  \cline{2-10} 
& {$0\lesssim r_2 \lesssim 0.5$ \cite{Sahoo:2023czj}} &  {$ \rm a\sim 0.92$ \cite{Moscibrodzka:2009gw}} & {$ \rm a\sim 0.5 $ \cite{Shcherbakov:2010ki}} &  {$\rm a \lesssim 0.1 $ \cite{Fragione:2020khu}}                                                                           &  {$ \rm a= 0.9\pm 0.06$ \cite{Daly:2023axh}}   &  {$ \rm a> 0.4$ \cite{Meyer:2006fd}}  & {$ \rm a\sim 0.52$ \cite{Genzel:2003as}}  & {$ \rm a\sim 0.22$ \cite{Belanger:2006gm}} &  {$ \rm (3.5-4.9) $ \cite{Ghez:2008ms,Gillessen:2008qv}}\\
\hline\hline

\multirow{2}{*}{\textbf{Model}} 
& \multicolumn{5}{c|}{\textbf{Grid Search}} 
& \multicolumn{4}{c|}{\textbf{MCMC}} \\ \cline{2-10}
& \textbf{ $r_2$} 
& \textbf{1-$\sigma$}
 & \textbf{3-$\sigma$}
& \textbf{ Spin} 
& \textbf{Mass $( 10^6 M_{\odot})$} 
& \multicolumn{1}{c|} {\textbf{$r_2$}}
& \multicolumn{1}{c|} {\textbf{Spin}} 
& \multicolumn{2}{c|} {\textbf{Mass $( 10^6 M_{\odot})$}} 
\\ 
\hline

PRM    & 0   & $0 \lesssim r_2 \lesssim 0.2$ &  $0\lesssim r_2\lesssim 1.75$  &  0.99 $\rm ({r_2}_{,\rm min}\sim 0)$ & $3.5 $ $\rm ({r_2}_{,\rm min}\sim 0)$ & \multicolumn{1}{c|} {$0.0198^{+0.2890}_{-0.0198}$} & \multicolumn{1}{c|} {$0.9900^{+0.0100}_{-0.1434}$} & \multicolumn{2}{c|}{$3.5037^{+0.0634}_{-0.0219}$}  \\ \hline
KRM1   & 0   & $0 \lesssim r_2 \lesssim 0.3$ &  U   &  0.99 $\rm ({r_2}_{,\rm min}\sim 0)$ & $3.5 $  $\rm ({r_2}_{,\rm min}\sim 0)$ & \multicolumn{1}{c|}{$0.0194^{+0.3347}_{-0.0194}$} & \multicolumn{1}{c|}{$0.9902^{+0.0098}_{-0.1671}$} & \multicolumn{2}{c|}{$3.5024^{+0.0483}_{-0.0221}$}  \\ \hline
NADO1  & 1.9 & $0.9 \lesssim r_2 \lesssim 2$ &  U   &  $ \rm 0.04 ({r_2}_{,\rm min}\sim 1.9)$ &   $\rm  4 ({r_2}_{,\rm min}\sim 1.9)$    & \multicolumn{1}{c|} {$1.7500^{+0.0739}_{-0.0854}$} &  \multicolumn{1}{c|}{$-0.0170^{+0.0860}_{-0.0875}$} & \multicolumn{2}{c|}{$4.1876^{+0.4638}_{-0.4405}$} \\ 
 &  &  &  &  $ 0.99 (r_2\sim 0)$  &  $\rm 3.5  (r_2\sim 0)$ & \multicolumn{1}{c|} {} &  \multicolumn{1}{c|}{} &  \multicolumn{2}{c|} {}
\\
\hline

\end{tabular}
\end{adjustbox}
\caption{Comparison of best-fit model parameters for Sgr A* derived using the grid-search and the MCMC methods.}
\label{Tab7}
\end{table}
\newpage
In \ref{Tab7} we enlist the QPO models which rule out a certain range of the allowed parameter space of $r_2$ from the HFQPO data of Sgr A*. The parameter estimates based on the grid-search and the MCMC methods are almost consistent.
Considering the grid-search method we note that,
PRM establishes the strongest constrain on $r_2$ thereby ruling out $r_2\gtrsim 1.75$ outside 3-$\sigma$. According to this model, the most favored dilaton charge is $r_2\sim 0$ and it rules out $r_2\gtrsim 0.2$ outside 1-$\sigma$. KRM1 also favors $r_2\sim 0$ and rules out $r_2\gtrsim 0.3$ outside 1-$\sigma$, although allows all values of $r_2$ when the 3-$\sigma$ interval is considered. Interestingly, if NADO1 is used to explain the HFQPO data of Sgr A*, the observationally favored dilaton charge seems to be near extremal, i.e. $r_2\sim 1.9$ and it rules out GR within 1-$\sigma$ (in agreement with MCMC), but includes the Kerr scenario if the 3-$\sigma$ interval is considered. 
The EMDA scenario has been tested with the shadow observations of this source provided by the EHT collaboration \cite{EventHorizonTelescope:2022xnr,EventHorizonTelescope:2022xqj}, which provides a constrain $0\lesssim r_2\lesssim 0.5$ for the dilaton charge and $|a|\lesssim 0.95$ considering its different mass and distance estimates \cite{Sahoo:2023czj}.

PRM and KRM1 predict a near maximal spin of Sgr A*, i.e., $a\sim 0.99$ when $r_2\sim 0$ is considered, which is allowed only in \cite{Meyer:2006fd}. \ref{Tab7} reveals that previous spin estimates of this source exhibit a huge disparity, while some advocate near zero spin \cite{Fragione:2020khu}, some mention that it has intermediate spin \cite{Genzel:2003as,Shcherbakov:2010ki} while others predict high to near maximal spin \cite{Daly:2023axh,Moscibrodzka:2009gw}. Apart from the models reported in \ref{Tab7}, the remaining eight models cannot distinguish the Kerr scenario and the Kerr-Sen scenario from the HFQPO data. From these eight models if $r_2\sim 0$ is considered, the observationally favored spin seem to be $a\sim 0.92$ (for RPM), $a\sim 0.8$ (for TDM), $a\sim 0.99$ (for FRM1), $a\sim 0.99$ (for FRM2), $a\sim 0.92$ (for KRM2), $a\sim 0.99$ (for KRM3), $a\sim 0.8$ (for NADO2) and  $a\sim 0.99$ (for WDOM). The spins predicted from RPM and KRM2 are in agreement with \cite{Daly:2023axh,Sahoo:2023czj,Moscibrodzka:2009gw} which might indicate that these two models may be the best representation of the data, but they equally favour the GR and the EMDA scenario. The fact that the spin estimates of this source are so diverse might signal some beyond GR physics at play in the near horizon regime of this source, which needs to be subjected to further investigation.





\section{Concluding Remarks}
\label{S7}
In this work, we explore observations related to high-frequency quasi-periodic oscillations (HFQPOs) in black holes (BHs) in understanding the nature of strong gravity. We consider the black holes to be governed by the Einstein-Maxwell-dilaton-axion (EMDA) gravity, which manifests a departure from the standard general relativistic scenario through the presence of a dilaton charge. The EMDA scenario appears in the low-energy effective action of the heterotic string theory and finds interesting applications in the early-time and late-time cosmologies. Therefore, it is worthwhile to explore the astrophysical signatures of such a theory from the available observations. 
Rotating black hole solutions in EMDA gravity correspond to the Kerr-Sen solution, whose electric charge and rotation are imparted by the dilaton and axion fields, respectively. Previously, we studied the implications of the EMDA gravity from observations related to black hole shadow \cite{Sahoo:2023czj}, jet power \cite{Banerjee:2020ubc}, and the continuum spectrum \cite{Banerjee:2020qmi}, which motivates us to investigate the role of this theory in explaining yet another astrophysical observation, the high-frequency quasi-periodic oscillations in black holes. 

In order to accomplish our goal, we compare the available HFQPO data related to BHs with the various kinematic and resonant QPO models presented in the literature. It is believed that local or collective motion of accreting plasma near the innermost stable circular orbit gives rise to HFQPOs. Since these are governed primarily by the spacetime curvature and not significantly by the accretion processes, HFQPOs seem to be a cleaner probe to the background spacetime compared to the continuum-spectrum or the Fe-line. Like HFQPOs, the images of black holes are also governed chiefly by the background metric and interestingly, in our black hole sample we have Sgr A*, whose image is available and also exhibits HFQPOs in its power spectrum. We shall discuss the constraints on the dilaton charge and spin from both these observations. 

In our present work, our sample size is restricted to five black hole sources (GRO J1655-40, XTE J1550+564, GRS 1915+105, H 1743-322, and Sgr A*), based on the available observations. When the HFQPO data of GRO J1655-40 is compared with the eleven models considered here, the Parametric Resonance Model (PRM) or the Keplerian Resonance Model 1 (KRM1) seem to be the best explanation of the data as the spin of the black hole predicted by these two models (assuming Kerr geometry) are in agreement with the spin estimated previously by the Fe-line method \cite{Miller:2009cw} and also from the jet power \cite{Banerjee:2020ubc}, but differs from the Continuum-Fitting method \cite{Shafee_2005}. But PRM establishes no constraint on the dilaton charge, i.e., using PRM, the HFQPO data of GRO J1655-40 cannot distinguish between the Kerr and the Kerr-Sen scenario. KRM1 on the other hand, exhibits a preference towards the Kerr-Sen scenario preferring a high dilaton charge and ruling out GR outside 1-$\sigma$. The other nine models rule out near extremal dilaton charge for this source outside 3-$\sigma$ but fail to predict the spin in agreement with at least one of the previous estimates \cite{Shafee_2005,Miller:2009cw,Banerjee:2020ubc}. Interestingly, the Relativistic Precession Model (RPM), which strongly favors \gr, predicts a black hole spin $a\sim 0.3$ in contradiction with all the previous estimates \cite{Shafee_2005,Miller:2009cw,Banerjee:2020ubc}. This plausibly indicates that RPM may not be a suitable QPO model for GRO J1655-40 \cite{Bambi:2013fea} and the models which seem to be more suitable (PRM/KRM1) cannot rule out the EMDA scenario from the present data. The present results and the fact that the spin of this source estimated by the Continuum-Fitting and the Fe-line method are quite disparate, might indicate signatures of physics beyond GR at play in the strong gravity regime of this source, which needs to be subjected to further investigation.

When the source XTE J1550-564 is considered, the QPO models which best explain the HFQPO data are NADO1, RPM, FRM1, and KRM2 (using the grid-search method when $r_2\sim 0$ is considered) as they exhibit the best agreement with previous spin estimates \cite{Steiner:2010bt,Banerjee:2020ubc}, which are $\sim 0.34$ from the Continuum-Fitting \cite{Steiner:2010bt}, $ \rm a= 0.55^{+0.15}_{-0.22}$ from the Fe-line \cite{Steiner:2010bt} and $0.3\lesssim a \lesssim 0.6$\cite{Banerjee:2020ubc} from the jet power. The MCMC method predicts a non-zero dilaton charge for this source using RPM, FRM1 and KRM2. When MCMC is implemented on this source assuming NADO1, $r_2\sim 0$ is favored (in agreement with grid-search) and the predicted spin is also in agreement with previous estimates \cite{Steiner:2010bt,Banerjee:2020ubc}. Hence, NADO1 or RPM/FRM1/KRM2 seem to be the best description of the HFQPO data for this source.\\
The spins predicted by the models TDM, FRM2, KRM3, NADO2 and WDOM also fall within the error bars reported using the Continuum-Fitting method \cite{Steiner:2010bt}, which however, is pretty large $-0.11\lesssim a \lesssim 0.71$. PRM and KRM1 do not seem to be suitable for this source as they predict near-maximal spins. The nine models that predict the black hole spin in agreement with earlier estimates cannot constrain the dilaton charge very strongly and hence fail to clearly distinguish the EMDA scenario from \gr, although they consistently rule out extremal and near extremal values of dilaton charge for this source.

For the source GRS 1915+105, the models PRM and/or KRM1 seem to be the best explanation of the HFQPO data as they predict spins $a\sim 0.8$ and $a\sim 0.97$ respectively, in agreement with previous estimates which are $a\sim 0.6-0.98$ (from the Fe-line method) \cite{Blum:2009ez}, $a\sim 0.7$ \cite{2006MNRAS.373.1004M}, $a\sim 0.98$ \cite{McClintock:2006xd}, and more recently $a\sim 0.4-0.98$ \cite{Mills:2021dxs} (from the Continuum-Fitting method) and $a\sim 0.6-0.9$ from the observed jet power \cite{Banerjee:2020ubc}. These models, however fail to constrain the dilaton charge and hence cannot rule out either the Kerr or the Kerr-Sen scenario. The remaining nine models predict zero to very low or even retrograde spins for this source, in contradiction with previous estimates, and hence do not seem to be the correct description of the data. 

For the source H1743-322, the earlier estimates of spin, $a=0.2\pm0.3$ (1-$\sigma$) and $a<0.92$ (3-$\sigma$) from the Continuum-Fitting method \cite{Steiner:2011kd} and $0.25\lesssim a \lesssim 0.5$ from the observed jet power \cite{Banerjee:2020ubc}, reveal that this source is moderately spinning. For this source, the models PRM/KRM1 predict near extremal spins and hence seem to be unsuitable. The best agreement with \cite{Steiner:2011kd} is exhibited by KRM3 and NADO2, although the spins predicted by RPM, TDM, FRM2, KRM2, NADO1 and WDOM also fall within the 1-$\sigma$ error bar. Using both the grid-search and the MCMC methods we note that the aforesaid models strongly rule out near extremal dilaton charge for this source, although equally favour the Kerr and the Kerr-Sen scenario with mild dilaton charges. The spin estimates of this source by independent observations \cite{Steiner:2011kd,Banerjee:2020ubc} are more or less consistent, but for a better understanding of the best QPO model and the underlying nature of strong gravity, more precise data is required.

Finally for Sgr A*, the spin estimates exhibit a lot of discrepancy, $ \rm a\sim 0.5 $ \cite{Shcherbakov:2010ki}, $ \rm a\sim 0.92$ \cite{Moscibrodzka:2009gw}, $\rm a \lesssim 0.1 $ \cite{Fragione:2020khu}, $ \rm a\sim 0.22$ \cite{Belanger:2006gm}, $ \rm a\sim 0.52$\cite{Genzel:2003as}, $ \rm a> 0.4$ \cite{Meyer:2006fd}, $a=0.9\pm 0.06$ \cite{Daly:2023axh} and $|a|<0.95$ \cite{Sahoo:2023czj}. All the previous estimates rule out near extremal spin for this source except Meyer et al. \cite{Meyer:2006fd}. The latest results reveal that the spin can be as high as $a\sim 0.96$ \cite{Daly:2023axh,Sahoo:2023czj}, which implies that this source is probably not maximally spinning. In that event, PRM and KRM1 do not seem to be suitable QPO models for this source, as these predict $a\sim 0.99$. NADO1 on the other hand, indicates a preference towards high dilaton charges ($r_2\sim 1.65-1.9$) and rules out GR outside 1-$\sigma$ using both MCMC and grid-search method. This model also indicates that Sgr A* is an extremal BH.\\
The spins predicted from RPM and KRM2 are in agreement with \cite{Daly:2023axh,Sahoo:2023czj,Moscibrodzka:2009gw}, which might indicate that these two models may be the best representation of the data, but they equally favour the GR and the EMDA scenario. When the EMDA scenario was tested with observations related to the shadow of Sgr A* \cite{EventHorizonTelescope:2022xnr,EventHorizonTelescope:2022xqj}, the dilaton charge was constrained between $0\lesssim r_2\lesssim 0.5$ while the spin $|a|\lesssim 0.95$ was obtained considering its different mass and distance estimates \cite{Sahoo:2023czj}.
The fact that the spin estimates of this source are so diverse might again signal some beyond GR physics at play in the near horizon regime of this source.

Based on spin estimates from the different HFQPO models and previous estimates by other independent methods, the above discussion elucidates that:
\begin{itemize}
\item GRO J1655-40 data seems to rule out RPM and favors PRM and/or KRM1.
\item PRM and KRM1 are not suitable for XTE J1550-564, rather NADO1, RPM, FRM1 and KRM2 provide consistent spin estimates. 
\item PRM and/or KRM1 seem to be the best description of the data for GRS 1915+105.
\item PRM and KRM1 are not suitable for H1743-322 rather, KRM3 and/or NADO2 seem to be favored.
\item PRM, KRM1, and NADO1 do not seem to be suitable for Sgr A*, rather RPM and/or KRM2 are favored.
\end{itemize}
From the above discussion, one may note that it is difficult to explain the HFQPO data of all the available sources with a single QPO model. This is probably the reason why there is no agreement on the correct choice of the QPO model among researchers, despite the fact that QPOs have been observed for decades. Although there is a belief that a single QPO model, yet to be determined, should explain the HFQPO data of all black holes, others suggest that the correct choice of the QPO model may be source-dependent.
Further, for all the above sources, the models that provide consistent spin estimates with at least one of the previous measurements fail to strongly constrain the dilaton charge.
They either allow all values of $r_2$ or at most rule out only near extremal values of $r_2$ outside 3-$\sigma$. This implies that from the present HFQPO data, one cannot rule out the EMDA scenario from GR. This, when combined with the fact that for most black hole sources the spin measurements by different methods often yield very diverse estimates, might indicate signatures of additional hairs in black holes which need to be explored further. This has also been mentioned in \cite{Bambi:2013fea,Bambi:2012pa}.
The present analysis is limited due to poor statistics (as we have only a handful of black hole sources that exhibit QPOs in their power spectrum) and lack of precise data. The precision is expected to improve by an order of magnitude as the ESA (European Space Agency) X-ray mission LOFT (Large Observatory for X-ray Timing) becomes operational, when stronger constraints on the spin and additional black hole hairs can be established based on QPO-based observations.

\section*{Acknowledgements}
Research of I.B. is funded by the Start-Up
Research Grant from SERB, DST, Government of India
(Reg. No. SRG/2021/000418).\\

\bibliography{Re,QPO,IB,new-ref,SgrA,Gravity_1_full,Gravity_2_full,Gravity_3_partial,Brane,Black_Hole_Shadow,KN-ED,EMDA-Jet,nohair}

\providecommand{\href}[2]{#2}\begingroup\raggedright\begin{thebibliography}{100}

\bibitem{1972Natur.235..271B}
C.~T. {Bolton}, ``{Identification of Cygnus X-1 with HDE 226868},''
  \href{http://dx.doi.org/10.1038/235271b0}{{\em Nature} {\bfseries 235}
  no.~5336, (Feb., 1972) 271--273}.

\bibitem{Webster:1972bsw}
B.~L. Webster and P.~Murdin, ``{Cygnus X-1-a Spectroscopic Binary with a Heavy
  Companion ?},'' \href{http://dx.doi.org/10.1038/235037a0}{{\em Nature}
  {\bfseries 235} (1972) 37--38}.

\bibitem{Abbott:2017vtc}
{\bfseries LIGO Scientific, VIRGO} Collaboration, B.~P. Abbott {\em et~al.},
  ``{GW170104: Observation of a 50-Solar-Mass Binary Black Hole Coalescence at
  Redshift 0.2},'' \href{http://dx.doi.org/10.1103/PhysRevLett.118.221101,
  10.1103/PhysRevLett.121.129901}{{\em Phys. Rev. Lett.} {\bfseries 118}
  no.~22, (2017) 221101}, \href{http://arxiv.org/abs/1706.01812}{{\ttfamily
  arXiv:1706.01812 [gr-qc]}}.
[Erratum: Phys. Rev. Lett.121,no.12,129901(2018)].

\bibitem{Abbott:2016nmj}
{\bfseries LIGO Scientific, Virgo} Collaboration, B.~P. Abbott {\em et~al.},
  ``{GW151226: Observation of Gravitational Waves from a 22-Solar-Mass Binary
  Black Hole Coalescence},''
  \href{http://dx.doi.org/10.1103/PhysRevLett.116.241103}{{\em Phys. Rev.
  Lett.} {\bfseries 116} no.~24, (2016) 241103},
\href{http://arxiv.org/abs/1606.04855}{{\ttfamily arXiv:1606.04855 [gr-qc]}}.

\bibitem{Abbott:2016blz}
{\bfseries Virgo, LIGO Scientific} Collaboration, B.~P. Abbott {\em et~al.},
  ``{Observation of Gravitational Waves from a Binary Black Hole Merger},''
  \href{http://dx.doi.org/10.1103/PhysRevLett.116.061102}{{\em Phys. Rev.
  Lett.} {\bfseries 116} no.~6, (2016) 061102},
\href{http://arxiv.org/abs/1602.03837}{{\ttfamily arXiv:1602.03837 [gr-qc]}}.

\bibitem{Akiyama:2019cqa}
{\bfseries Event Horizon Telescope} Collaboration, K.~Akiyama {\em et~al.},
  ``{First M87 Event Horizon Telescope Results. I. The Shadow of the
  Supermassive Black Hole},''
  \href{http://dx.doi.org/10.3847/2041-8213/ab0ec7}{{\em Astrophys. J.}
  {\bfseries 875} no.~1, (2019) L1},
\href{http://arxiv.org/abs/1906.11238}{{\ttfamily arXiv:1906.11238
  [astro-ph.GA]}}.

\bibitem{EventHorizonTelescope:2022xnr}
{\bfseries Event Horizon Telescope} Collaboration, K.~Akiyama {\em et~al.},
  ``{First Sagittarius A* Event Horizon Telescope Results. I. The Shadow of the
  Supermassive Black Hole in the Center of the Milky Way},''
  \href{http://dx.doi.org/10.3847/2041-8213/ac6674}{{\em Astrophys. J. Lett.}
  {\bfseries 930} no.~2, (2022) L12}.

\bibitem{EventHorizonTelescope:2022xqj}
{\bfseries Event Horizon Telescope} Collaboration, K.~Akiyama {\em et~al.},
  ``{First Sagittarius A* Event Horizon Telescope Results. VI. Testing the
  Black Hole Metric},'' \href{http://dx.doi.org/10.3847/2041-8213/ac6756}{{\em
  Astrophys. J. Lett.} {\bfseries 930} no.~2, (2022) L17}.

\bibitem{Bekenstein:1971hc}
J.~D. Bekenstein, ``{Nonexistence of baryon number for static black holes},''
  \href{http://dx.doi.org/10.1103/PhysRevD.5.1239}{{\em Phys. Rev. D}
  {\bfseries 5} (1972) 1239--1246}.

\bibitem{Israel:1967wq}
W.~Israel, ``{Event horizons in static vacuum space-times},''
  \href{http://dx.doi.org/10.1103/PhysRev.164.1776}{{\em Phys. Rev.} {\bfseries
  164} (1967) 1776--1779}.

\bibitem{Carter:1971zc}
B.~Carter, ``{Axisymmetric Black Hole Has Only Two Degrees of Freedom},''
  \href{http://dx.doi.org/10.1103/PhysRevLett.26.331}{{\em Phys. Rev. Lett.}
  {\bfseries 26} (1971) 331--333}.

\bibitem{Robinson:1975bv}
D.~C. Robinson, ``{Uniqueness of the Kerr black hole},''
  \href{http://dx.doi.org/10.1103/PhysRevLett.34.905}{{\em Phys. Rev. Lett.}
  {\bfseries 34} (1975) 905--906}.

\bibitem{Mazur:1982db}
P.~O. Mazur, ``{PROOF OF UNIQUENESS OF THE KERR-NEWMAN BLACK HOLE SOLUTION},''
  \href{http://dx.doi.org/10.1088/0305-4470/15/10/021}{{\em J. Phys. A}
  {\bfseries 15} (1982) 3173--3180}.

\bibitem{Bizon:1990sr}
P.~Bizon, ``{Colored black holes},''
  \href{http://dx.doi.org/10.1103/PhysRevLett.64.2844}{{\em Phys. Rev. Lett.}
  {\bfseries 64} (1990) 2844--2847}.

\bibitem{Garfinkle:1990qj}
D.~Garfinkle, G.~T. Horowitz, and A.~Strominger, ``{Charged black holes in
  string theory},'' \href{http://dx.doi.org/10.1103/PhysRevD.43.3140}{{\em
  Phys. Rev. D} {\bfseries 43} (1991) 3140}. [Erratum: Phys.Rev.D 45, 3888
  (1992)].

\bibitem{Greene:1992fw}
B.~R. Greene, S.~D. Mathur, and C.~M. O'Neill, ``{Eluding the no hair
  conjecture: Black holes in spontaneously broken gauge theories},''
  \href{http://dx.doi.org/10.1103/PhysRevD.47.2242}{{\em Phys. Rev. D}
  {\bfseries 47} (1993) 2242--2259},
  \href{http://arxiv.org/abs/hep-th/9211007}{{\ttfamily arXiv:hep-th/9211007}}.

\bibitem{Lavrelashvili:1992ia}
G.~V. Lavrelashvili and D.~Maison, ``{Regular and black hole solutions of
  Einstein Yang-Mills Dilaton theory},''
  \href{http://dx.doi.org/10.1016/0550-3213(93)90441-Q}{{\em Nucl. Phys. B}
  {\bfseries 410} (1993) 407--422}.

\bibitem{Torii:1993vm}
T.~Torii and K.-i. Maeda, ``{Black holes with nonAbelian hair and their
  thermodynamical properties},''
  \href{http://dx.doi.org/10.1103/PhysRevD.48.1643}{{\em Phys. Rev. D}
  {\bfseries 48} (1993) 1643--1651}.

\bibitem{Herdeiro:2014goa}
C.~A.~R. Herdeiro and E.~Radu, ``{Kerr black holes with scalar hair},''
  \href{http://dx.doi.org/10.1103/PhysRevLett.112.221101}{{\em Phys. Rev.
  Lett.} {\bfseries 112} (2014) 221101},
  \href{http://arxiv.org/abs/1403.2757}{{\ttfamily arXiv:1403.2757 [gr-qc]}}.

\bibitem{Berti:2013gfa}
E.~Berti, V.~Cardoso, L.~Gualtieri, M.~Horbatsch, and U.~Sperhake, ``{Numerical
  simulations of single and binary black holes in scalar-tensor theories:
  circumventing the no-hair theorem},''
  \href{http://dx.doi.org/10.1103/PhysRevD.87.124020}{{\em Phys. Rev. D}
  {\bfseries 87} no.~12, (2013) 124020},
  \href{http://arxiv.org/abs/1304.2836}{{\ttfamily arXiv:1304.2836 [gr-qc]}}.

\bibitem{Penrose:1964wq}
R.~Penrose, ``{Gravitational collapse and space-time singularities},''
\href{http://dx.doi.org/10.1103/PhysRevLett.14.57}{{\em Phys. Rev. Lett.}
  {\bfseries 14} (1965) 57--59}.

\bibitem{Hawking:1976ra}
S.~W. Hawking, ``{Breakdown of Predictability in Gravitational Collapse},''
\href{http://dx.doi.org/10.1103/PhysRevD.14.2460}{{\em Phys. Rev.} {\bfseries
  D14} (1976) 2460--2473}.

\bibitem{Christodoulou:1991yfa}
D.~Christodoulou, ``{The formation of black holes and singularities in
  spherically symmetric gravitational collapse},''
\href{http://dx.doi.org/10.1002/cpa.3160440305}{{\em Commun. Pure Appl. Math.}
  {\bfseries 44} no.~3, (1991) 339--373}.

\bibitem{Bekenstein:1984tv}
J.~Bekenstein and M.~Milgrom, ``{Does the missing mass problem signal the
  breakdown of Newtonian gravity?},''
  \href{http://dx.doi.org/10.1086/162570}{{\em Astrophys. J.} {\bfseries 286}
  (1984) 7--14}. \url {https://10.1086/162570}.

\bibitem{Perlmutter:1998np}
{\bfseries Supernova Cosmology Project} Collaboration, S.~Perlmutter {\em
  et~al.}, ``{Measurements of Omega and Lambda from 42 high redshift
  supernovae},'' \href{http://dx.doi.org/10.1086/307221}{{\em Astrophys. J.}
  {\bfseries 517} (1999) 565--586},
\href{http://arxiv.org/abs/astro-ph/9812133}{{\ttfamily arXiv:astro-ph/9812133
  [astro-ph]}}.

\bibitem{Riess:1998cb}
{\bfseries Supernova Search Team} Collaboration, A.~G. Riess {\em et~al.},
  ``{Observational evidence from supernovae for an accelerating universe and a
  cosmological constant},'' \href{http://dx.doi.org/10.1086/300499}{{\em
  Astron. J.} {\bfseries 116} (1998) 1009--1038},
\href{http://arxiv.org/abs/astro-ph/9805201}{{\ttfamily arXiv:astro-ph/9805201
  [astro-ph]}}.

\bibitem{Shiromizu:1999wj}
T.~Shiromizu, K.-i. Maeda, and M.~Sasaki, ``{The Einstein equation on the
  3-brane world},'' \href{http://dx.doi.org/10.1103/PhysRevD.62.024012}{{\em
  Phys.Rev.} {\bfseries D62} (2000) 024012},
\href{http://arxiv.org/abs/gr-qc/9910076}{{\ttfamily arXiv:gr-qc/9910076
  [gr-qc]}}.

\bibitem{Dadhich:2000am}
N.~Dadhich, R.~Maartens, P.~Papadopoulos, and V.~Rezania, ``{Black holes on the
  brane},'' \href{http://dx.doi.org/10.1016/S0370-2693(00)00798-X}{{\em
  Phys.Lett.} {\bfseries B487} (2000) 1--6},
\href{http://arxiv.org/abs/hep-th/0003061}{{\ttfamily arXiv:hep-th/0003061
  [hep-th]}}.

\bibitem{Harko:2004ui}
T.~Harko and M.~Mak, ``{Vacuum solutions of the gravitational field equations
  in the brane world model},''
  \href{http://dx.doi.org/10.1103/PhysRevD.69.064020}{{\em Phys.Rev.}
  {\bfseries D69} (2004) 064020},
\href{http://arxiv.org/abs/gr-qc/0401049}{{\ttfamily arXiv:gr-qc/0401049
  [gr-qc]}}.

\bibitem{Carames:2012gr}
T.~R.~P. Carames, M.~E.~X. Guimaraes, and J.~M. Hoff~da Silva, ``{Effective
  gravitational equations for $f(R)$ braneworld models},''
  \href{http://dx.doi.org/10.1103/PhysRevD.87.106011}{{\em Phys. Rev.}
  {\bfseries D87} no.~10, (2013) 106011},
\href{http://arxiv.org/abs/1205.4980}{{\ttfamily arXiv:1205.4980 [gr-qc]}}.

\bibitem{Chakraborty:2015bja}
S.~Chakraborty and S.~SenGupta, ``{Effective gravitational field equations on
  $m$-brane embedded in n-dimensional bulk of Einstein and $f(\mathcal {R})$
  gravity},'' \href{http://dx.doi.org/10.1140/epjc/s10052-015-3768-z}{{\em Eur.
  Phys. J. C} {\bfseries 75} no.~11, (2015) 538},
  \href{http://arxiv.org/abs/1504.07519}{{\ttfamily arXiv:1504.07519 [gr-qc]}}.
  \url { https://doi.org/10.1140/epjc/s10052-015-3768-z}.

\bibitem{Nojiri:2006gh}
S.~Nojiri and S.~D. Odintsov, ``{Modified f(R) gravity consistent with
  realistic cosmology: From matter dominated epoch to dark energy universe},''
  \href{http://dx.doi.org/10.1103/PhysRevD.74.086005}{{\em Phys. Rev.}
  {\bfseries D74} (2006) 086005},
\href{http://arxiv.org/abs/hep-th/0608008}{{\ttfamily arXiv:hep-th/0608008
  [hep-th]}}.

\bibitem{Lanczos:1932zz}
C.~Lanczos, ``{Electricity as a natural property of Riemannian geometry},''
  \href{http://dx.doi.org/10.1103/RevModPhys.39.716}{{\em Rev. Mod. Phys.}
  {\bfseries 39} (1932) 716--736}. \url {
  https://doi.org/10.1103/RevModPhys.39.716}.

\bibitem{Lovelock:1971yv}
D.~Lovelock, ``{The Einstein tensor and its generalizations},''
  \href{http://dx.doi.org/10.1063/1.1665613}{{\em J. Math. Phys.} {\bfseries
  12} (1971) 498--501}. \url { https://doi.org/10.1063/1.1665613}.

\bibitem{Padmanabhan:2013xyr}
T.~Padmanabhan and D.~Kothawala, ``{Lanczos-Lovelock models of gravity},''
  \href{http://dx.doi.org/10.1016/j.physrep.2013.05.007}{{\em Phys.Rept.}
  {\bfseries 531} (2013) 115--171},
\href{http://arxiv.org/abs/1302.2151}{{\ttfamily arXiv:1302.2151 [gr-qc]}}.

\bibitem{Horndeski:1974wa}
G.~W. Horndeski, ``{Second-order scalar-tensor field equations in a
  four-dimensional space},'' \href{http://dx.doi.org/10.1007/BF01807638}{{\em
  Int. J. Theor. Phys.} {\bfseries 10} (1974) 363--384}. \url
  {https://doi.org/10.1007/BF0180763}.

\bibitem{Sotiriou:2013qea}
T.~P. Sotiriou and S.-Y. Zhou, ``{Black hole hair in generalized scalar-tensor
  gravity},'' \href{http://dx.doi.org/10.1103/PhysRevLett.112.251102}{{\em
  Phys. Rev. Lett.} {\bfseries 112} (2014) 251102},
  \href{http://arxiv.org/abs/1312.3622}{{\ttfamily arXiv:1312.3622 [gr-qc]}}.
  \url {https://doi.org/10.1103/PhysRevLett.112.251102}.

\bibitem{Charmousis:2015txa}
C.~Charmousis and M.~Tsoukalas, ``{Lovelock Galileons and black holes},''
  \href{http://dx.doi.org/10.1103/PhysRevD.92.104050}{{\em Phys. Rev.}
  {\bfseries D92} no.~10, (2015) 104050},
\href{http://arxiv.org/abs/1506.05014}{{\ttfamily arXiv:1506.05014 [gr-qc]}}.

\bibitem{Sen:1992ua}
A.~Sen, ``{Rotating charged black hole solution in heterotic string theory},''
  \href{http://dx.doi.org/10.1103/PhysRevLett.69.1006}{{\em Phys. Rev. Lett.}
  {\bfseries 69} (1992) 1006--1009},
  \href{http://arxiv.org/abs/hep-th/9204046}{{\ttfamily arXiv:hep-th/9204046}}.
  \url { https://doi.org/10.1103/PhysRevLett.69.1006}.

\bibitem{Rogatko:2002qe}
M.~Rogatko, ``{Positivity of energy in Einstein-Maxwell axion dilaton
  gravity},'' \href{http://dx.doi.org/10.1088/0264-9381/19/20/303}{{\em Class.\
  Quant.\ Grav.} {\bfseries 19} (2002) 5063--5072},
  \href{http://arxiv.org/abs/hep-th/0209126}{{\ttfamily arXiv:hep-th/0209126}}.

\bibitem{Sonner:2006yn}
J.~Sonner and P.~K. Townsend, ``{Recurrent acceleration in dilaton-axion
  cosmology},'' \href{http://dx.doi.org/10.1103/PhysRevD.74.103508}{{\em Phys.
  Rev. D} {\bfseries 74} (2006) 103508},
  \href{http://arxiv.org/abs/hep-th/0608068}{{\ttfamily arXiv:hep-th/0608068}}.

\bibitem{Catena:2007jf}
R.~Catena and J.~Moller, ``{Axion-dilaton cosmology and dark energy},''
  \href{http://dx.doi.org/10.1088/1475-7516/2008/03/012}{{\em JCAP} {\bfseries
  03} (2008) 012}, \href{http://arxiv.org/abs/0709.1931}{{\ttfamily
  arXiv:0709.1931 [hep-ph]}}.

\bibitem{Gibbons:1987ps}
G.~Gibbons and K.-i. Maeda, ``{Black Holes and Membranes in Higher Dimensional
  Theories with Dilaton Fields},''
  \href{http://dx.doi.org/10.1016/0550-3213(88)90006-5}{{\em Nucl. Phys. B}
  {\bfseries 298} (1988) 741--775}.

\bibitem{Horowitz:1991cd}
G.~T. Horowitz and A.~Strominger, ``{Black strings and P-branes},''
  \href{http://dx.doi.org/10.1016/0550-3213(91)90440-9}{{\em Nucl. Phys. B}
  {\bfseries 360} (1991) 197--209}.

\bibitem{Kallosh:1993yg}
R.~Kallosh and T.~Ortin, ``{Charge quantization of axion - dilaton black
  holes},'' \href{http://dx.doi.org/10.1103/PhysRevD.48.742}{{\em Phys. Rev. D}
  {\bfseries 48} (1993) 742--747},
  \href{http://arxiv.org/abs/hep-th/9302109}{{\ttfamily arXiv:hep-th/9302109}}.

\bibitem{Thorne:1986iy}
K.~S. Thorne, R.~Price, and D.~Macdonald, eds., {\em {BLACK HOLES: THE MEMBRANE
  PARADIGM}}.
\newblock 1986.

\bibitem{Campbell:1992hc}
B.~A. Campbell, N.~Kaloper, R.~Madden, and K.~A. Olive, ``{Physical properties
  of four-dimensional superstring gravity black hole solutions},''
  \href{http://dx.doi.org/10.1016/0550-3213(93)90620-5}{{\em Nucl. Phys. B}
  {\bfseries 399} (1993) 137--168},
  \href{http://arxiv.org/abs/hep-th/9301129}{{\ttfamily arXiv:hep-th/9301129}}.
  \url {https://doi.org/10.1016/0550-3213(93)90620-5}.

\bibitem{Psaltis:2007cw}
D.~Psaltis, D.~Perrodin, K.~R. Dienes, and I.~Mocioiu, ``{Kerr Black Holes are
  Not Unique to General Relativity},''
  \href{http://dx.doi.org/10.1103/PhysRevLett.100.091101}{{\em Phys. Rev.
  Lett.} {\bfseries 100} (2008) 091101},
  \href{http://arxiv.org/abs/0710.4564}{{\ttfamily arXiv:0710.4564
  [astro-ph]}}.

\bibitem{Gyulchev:2006zg}
G.~N. Gyulchev and S.~S. Yazadjiev, ``{Kerr-Sen dilaton-axion black hole
  lensing in the strong deflection limit},''
  \href{http://dx.doi.org/10.1103/PhysRevD.75.023006}{{\em Phys. Rev. D}
  {\bfseries 75} (2007) 023006},
  \href{http://arxiv.org/abs/gr-qc/0611110}{{\ttfamily arXiv:gr-qc/0611110}}.

\bibitem{An:2017hby}
J.~An, J.~Peng, Y.~Liu, and X.-H. Feng, ``{Kerr-Sen Black Hole as Accelerator
  for Spinning Particles},''
  \href{http://dx.doi.org/10.1103/PhysRevD.97.024003}{{\em Phys. Rev. D}
  {\bfseries 97} no.~2, (2018) 024003},
  \href{http://arxiv.org/abs/1710.08630}{{\ttfamily arXiv:1710.08630 [gr-qc]}}.

\bibitem{Younsi:2016azx}
Z.~Younsi, A.~Zhidenko, L.~Rezzolla, R.~Konoplya, and Y.~Mizuno, ``{New method
  for shadow calculations: Application to parametrized axisymmetric black
  holes},'' \href{http://dx.doi.org/10.1103/PhysRevD.94.084025}{{\em Phys. Rev.
  D} {\bfseries 94} no.~8, (2016) 084025},
  \href{http://arxiv.org/abs/1607.05767}{{\ttfamily arXiv:1607.05767 [gr-qc]}}.

\bibitem{Hioki:2008zw}
K.~Hioki and U.~Miyamoto, ``{Hidden symmetries, null geodesics, and photon
  capture in the Sen black hole},''
  \href{http://dx.doi.org/10.1103/PhysRevD.78.044007}{{\em Phys. Rev. D}
  {\bfseries 78} (2008) 044007},
  \href{http://arxiv.org/abs/0805.3146}{{\ttfamily arXiv:0805.3146 [gr-qc]}}.

\bibitem{Mizuno:2018lxz}
Y.~Mizuno, Z.~Younsi, C.~M. Fromm, O.~Porth, M.~De~Laurentis, H.~Olivares,
  H.~Falcke, M.~Kramer, and L.~Rezzolla, ``{The Current Ability to Test
  Theories of Gravity with Black Hole Shadows},''
  \href{http://dx.doi.org/10.1038/s41550-018-0449-5}{{\em Nat. Astron.}
  {\bfseries 2} no.~7, (2018) 585--590},
\href{http://arxiv.org/abs/1804.05812}{{\ttfamily arXiv:1804.05812
  [astro-ph.GA]}}.

\bibitem{Narang:2020bgo}
A.~Narang, S.~Mohanty, and A.~Kumar, ``{Test of Kerr-Sen metric with black hole
  observations},'' \href{http://arxiv.org/abs/2002.12786}{{\ttfamily
  arXiv:2002.12786 [gr-qc]}}.

\bibitem{Banerjee:2020qmi}
I.~Banerjee, B.~Mandal, and S.~SenGupta, ``{Implications of
  Einstein\textendash{}Maxwell dilaton\textendash{}axion gravity from the black
  hole continuum spectrum},''
  \href{http://dx.doi.org/10.1093/mnras/staa3232}{{\em Mon. Not. Roy. Astron.
  Soc.} {\bfseries 500} no.~1, (2020) 481--492},
  \href{http://arxiv.org/abs/2007.13980}{{\ttfamily arXiv:2007.13980 [gr-qc]}}.

\bibitem{Banerjee:2020ubc}
I.~Banerjee, B.~Mandal, and S.~SenGupta, ``{Signatures of Einstein-Maxwell
  dilaton-axion gravity from the observed jet power and the radiative
  efficiency},'' \href{http://dx.doi.org/10.1103/PhysRevD.103.044046}{{\em
  Phys. Rev. D} {\bfseries 103} no.~4, (2021) 044046},
  \href{http://arxiv.org/abs/2007.03947}{{\ttfamily arXiv:2007.03947 [gr-qc]}}.

\bibitem{Tripathi:2021rwb}
A.~Tripathi, B.~Zhou, A.~B. Abdikamalov, D.~Ayzenberg, and C.~Bambi,
  ``{Constraints on Einstein-Maxwell dilaton-axion gravity from X-ray
  reflection spectroscopy},''
  \href{http://dx.doi.org/10.1088/1475-7516/2021/07/002}{{\em JCAP} {\bfseries
  07} (2021) 002}, \href{http://arxiv.org/abs/2103.07593}{{\ttfamily
  arXiv:2103.07593 [astro-ph.HE]}}.

\bibitem{Sahoo:2023czj}
S.~K. Sahoo, N.~Yadav, and I.~Banerjee, ``{Imprints of
  Einstein-Maxwell-dilaton-axion gravity in the observed shadows of Sgr A* and
  M87*},'' \href{http://dx.doi.org/10.1103/PhysRevD.109.044008}{{\em Phys. Rev.
  D} {\bfseries 109} no.~4, (2024) 044008},
  \href{http://arxiv.org/abs/2305.14870}{{\ttfamily arXiv:2305.14870 [gr-qc]}}.

\bibitem{2006csxs.book.....L}
W.~H.~G. {Lewin} and M.~{van der Klis}, {\em {Compact Stellar X-ray Sources}},
  vol.~39.
\newblock 2006.

\bibitem{vanderKlis:2000ca}
M.~van~der Klis, ``{Millisecond oscillations in x-ray binaries},''
  \href{http://dx.doi.org/10.1146/annurev.astro.38.1.717}{{\em Ann. Rev.
  Astron. Astrophys.} {\bfseries 38} (2000) 717--760},
  \href{http://arxiv.org/abs/astro-ph/0001167}{{\ttfamily
  arXiv:astro-ph/0001167}}.

\bibitem{Torok:2004xs}
G.~Torok, ``{A Possible 3:2 orbital epicyclic resonance in QPOs frequencies of
  Sgr A*},'' \href{http://dx.doi.org/10.1051/0004-6361:20042558}{{\em Astron.
  Astrophys.} {\bfseries 440} (2005) 1},
  \href{http://arxiv.org/abs/astro-ph/0412500}{{\ttfamily
  arXiv:astro-ph/0412500}}. \url {https://doi.org/ 10.1051/0004-6361:20042558}.

\bibitem{Maselli:2014fca}
A.~Maselli, L.~Gualtieri, P.~Pani, L.~Stella, and V.~Ferrari, ``{Testing
  Gravity with Quasi Periodic Oscillations from accreting Black Holes: the Case
  of the Einstein-Dilaton-Gauss-Bonnet Theory},''
  \href{http://dx.doi.org/10.1088/0004-637X/801/2/115}{{\em Astrophys. J.}
  {\bfseries 801} no.~2, (2015) 115},
  \href{http://arxiv.org/abs/1412.3473}{{\ttfamily arXiv:1412.3473
  [astro-ph.HE]}}.

\bibitem{Abramowicz:2011xu}
M.~A. Abramowicz and P.~C. Fragile, ``{Foundations of Black Hole Accretion Disk
  Theory},'' \href{http://dx.doi.org/10.12942/lrr-2013-1}{{\em Living Rev.
  Rel.} {\bfseries 16} (2013) 1},
  \href{http://arxiv.org/abs/1104.5499}{{\ttfamily arXiv:1104.5499
  [astro-ph.HE]}}.

\bibitem{Aschenbach:2004kj}
B.~Aschenbach, ``{Measuring mass and angular momentum of black holes with
  high-frequency quasiperiodic oscillations},''
  \href{http://dx.doi.org/10.1051/0004-6361:20041412}{{\em Astron. Astrophys.}
  {\bfseries 425} (2004) 1075--1082},
  \href{http://arxiv.org/abs/astro-ph/0406545}{{\ttfamily
  arXiv:astro-ph/0406545}}. \url {https://doi.org/10.1051/0004-6361:20041412}.

\bibitem{Kotrlova:2017wyq}
A.~Kotrlov{\'a}, E.~{\v{S}}r{\'a}mkov{\'a}, G.~T{\"o}r{\"o}k, Z.~Stuchl{\'\i}k,
  and K.~Goluchov{\'a}, ``{Super-spinning compact objects and models of
  high-frequency quasi-periodic oscillations observed in Galactic microquasars.
  II. Forced resonances},''
  \href{http://dx.doi.org/10.1051/0004-6361/201730585}{{\em Astron. Astrophys.}
  {\bfseries 607} (2017) A69},
  \href{http://arxiv.org/abs/1708.04300}{{\ttfamily arXiv:1708.04300
  [astro-ph.HE]}}.

\bibitem{Torok:2011qy}
G.~Torok, A.~Kotrlova, E.~Sramkova, and Z.~Stuchlik, ``{Confronting the models
  of 3:2 quasiperiodic oscillations with the rapid spin of the microquasar GRS
  1915+105},'' \href{http://dx.doi.org/10.1051/0004-6361/201015549}{{\em
  Astron. Astrophys.} {\bfseries 531} (2011) A59},
  \href{http://arxiv.org/abs/1103.2438}{{\ttfamily arXiv:1103.2438
  [astro-ph.HE]}}.

\bibitem{1971SvA....15..377S}
V.~F. {Shvartsman}, ``{Halos around ``Black Holes''.},'' {\em sovast}
  {\bfseries 15} (Dec., 1971) 377.

\bibitem{1973SvA....16..941S}
R.~A. {Syunyaev}, ``{Variability of X Rays from Black Holes with Accretion
  Disks.},'' {\em sovast} {\bfseries 16} (June, 1973) 941.

\bibitem{PhysRevLett8217}
L.~Stella and M.~Vietri, ``khz quasiperiodic oscillations in low-mass x-ray
  binaries as probes of general relativity in the strong-field regime,''
  \href{http://dx.doi.org/10.1103/PhysRevLett.82.17}{{\em Phys. Rev. Lett.}
  {\bfseries 82} (Jan, 1999) 17--20}.
  \url{https://link.aps.org/doi/10.1103/PhysRevLett.82.17}.

\bibitem{Stella:1997tc}
L.~Stella and M.~Vietri, ``{Lense-Thirring precession and QPOS in low mass
  x-ray binaries},'' \href{http://dx.doi.org/10.1086/311075}{{\em Astrophys. J.
  Lett.} {\bfseries 492} (1998) L59},
  \href{http://arxiv.org/abs/astro-ph/9709085}{{\ttfamily
  arXiv:astro-ph/9709085}}.

\bibitem{Stella_1999}
L.~Stella, M.~Vietri, and S.~M. Morsink, ``Correlations in the quasi-periodic
  oscillation frequencies of low-mass x-ray binaries and the relativistic
  precession model,'' \href{http://dx.doi.org/10.1086/312291}{{\em The
  Astrophysical Journal} {\bfseries 524} no.~1, (Oct, 1999) L63--L66}.
  \url{https://doi.org/10.1086/312291}.

\bibitem{Cadez:2008iv}
A.~Cadez, M.~Calvani, and U.~Kostic, ``{On the tidal evolution of the orbits of
  low-mass satellites around black holes},''
  \href{http://dx.doi.org/10.1051/0004-6361:200809483}{{\em Astron. Astrophys.}
  {\bfseries 487} (2008) 527--532},
  \href{http://arxiv.org/abs/0809.1783}{{\ttfamily arXiv:0809.1783
  [astro-ph]}}. \url {https://doi.org/10.1051/0004-6361:200809483}.

\bibitem{Kostic:2009hp}
U.~Kostic, A.~Cadez, M.~Calvani, and A.~Gomboc, ``{Tidal effects on small
  bodies by massive black holes},''
  \href{http://dx.doi.org/10.1051/0004-6361/200811059}{{\em Astron. Astrophys.}
  {\bfseries 496} (2009) 307}, \href{http://arxiv.org/abs/0901.3447}{{\ttfamily
  arXiv:0901.3447 [astro-ph.HE]}}. \url {https://10.1051/0004-6361/200811059}.

\bibitem{Germana:2009ce}
C.~Germana, U.~Kostic, A.~Cadez, and M.~Calvani, ``{Tidal disruption of small
  satellites orbiting black holes},''
  \href{http://dx.doi.org/10.1063/1.3149456}{{\em AIP Conf. Proc.} {\bfseries
  1126} no.~1, (2009) 367--369},
  \href{http://arxiv.org/abs/0902.2134}{{\ttfamily arXiv:0902.2134
  [astro-ph.HE]}}. \url {https://10.1063/1.3149456}.

\bibitem{Kluzniak:2002bb}
W.~Kluzniak and M.~A. Abramowicz, ``{Parametric epicyclic resonance in black
  hole disks: qpos in micro-quasars},''
  \href{http://arxiv.org/abs/astro-ph/0203314}{{\ttfamily
  arXiv:astro-ph/0203314}}. \url {arXiv:astro-ph/0203314}.

\bibitem{Abramowicz:2003xy}
M.~A. Abramowicz, V.~Karas, W.~Kluzniak, W.~H. Lee, and P.~Rebusco,
  ``{Nonlinear resonance in nearly geodesic motion in low mass x-ray
  binaries},'' \href{http://dx.doi.org/10.1093/pasj/55.2.467}{{\em Publ.
  Astron. Soc. Jap.} {\bfseries 55} (2003) 466--467},
  \href{http://arxiv.org/abs/astro-ph/0302183}{{\ttfamily
  arXiv:astro-ph/0302183}}. \url {https://10.1093/pasj/55.2.467}.

\bibitem{Rebusco:2004ba}
P.~Rebusco, ``{Twin peaks kHz QPOs: Mathematics of the 3:2 orbital
  resonance},'' \href{http://dx.doi.org/10.1093/pasj/56.3.553}{{\em Publ.
  Astron. Soc. Jap.} {\bfseries 56} (2004) 553},
  \href{http://arxiv.org/abs/astro-ph/0403341}{{\ttfamily
  arXiv:astro-ph/0403341}}. \url {https://10.1093/pasj/56.3.553}.

\bibitem{Nowak:1996hg}
M.~A. Nowak, R.~V. Wagoner, M.~C. Begelman, and D.~E. Lehr, ``{The 67 hz
  feature in the black hole candidate grs 1915+105 as a possible
  ``diskoseismic'' mode},'' \href{http://dx.doi.org/10.1086/310534}{{\em
  Astrophys. J. Lett.} {\bfseries 477} (1997) L91},
  \href{http://arxiv.org/abs/astro-ph/9612142}{{\ttfamily
  arXiv:astro-ph/9612142}}.

\bibitem{Torok:2010rk}
G.~Torok, P.~Bakala, E.~Sramkova, Z.~Stuchlik, and M.~Urbanec, ``{On
  mass-constraints implied by the relativistic precession model of twin-peak
  quasi-periodic oscillations in Circinus X-1},''
  \href{http://dx.doi.org/10.1088/0004-637X/714/1/748}{{\em Astrophys. J.}
  {\bfseries 714} (2010) 748--757},
  \href{http://arxiv.org/abs/1008.0088}{{\ttfamily arXiv:1008.0088
  [astro-ph.HE]}}. \url {https://doi.org/10.1088/0004-637X/714/1/748}.

\bibitem{Kotrlova:2020pqy}
A.~Kotrlov\'a, E.~\v{S}r\'amkov\'a, G.~T\"or\"ok, K.~Goluchov\'a, J.~Hor\'ak,
  O.~Straub, D.~Lancov\'a, Z.~Stuchl\'\i{}k, and M.~A. Abramowicz, ``{Models of
  high-frequency quasi-periodic oscillations and black hole spin estimates in
  Galactic microquasars},''
  \href{http://dx.doi.org/10.1051/0004-6361/201937097}{{\em Astron. Astrophys.}
  {\bfseries 643} (2020) A31},
  \href{http://arxiv.org/abs/2008.12963}{{\ttfamily arXiv:2008.12963
  [astro-ph.HE]}}.

\bibitem{1980PASJ...32..377K}
S.~{Kato} and J.~{Fukue}, ``{Trapped Radial Oscillations of Gaseous Disks
  around a Black Hole},'' {\em Publ. Astron. Soc. Jap.} {\bfseries 32} (1980)
  377.

\bibitem{Perez:1996ti}
C.~A. Perez, A.~S. Silbergleit, R.~V. Wagoner, and D.~E. Lehr, ``{Relativistic
  diskoseismology. 1. Analytical results for 'gravity modes'},''
  \href{http://dx.doi.org/10.1086/303658}{{\em Astrophys. J.} {\bfseries 476}
  (1997) 589--604}, \href{http://arxiv.org/abs/astro-ph/9601146}{{\ttfamily
  arXiv:astro-ph/9601146}}. \url {https://10.1086/303658}.

\bibitem{Silbergleit:2000ck}
A.~S. Silbergleit, R.~V. Wagoner, and M.~Ortega-Rodriguez, ``{Relativistic
  diskoseismology. 2. Analytical results for C modes},''
  \href{http://dx.doi.org/10.1086/318659}{{\em Astrophys. J.} {\bfseries 548}
  (2001) 335--347}, \href{http://arxiv.org/abs/astro-ph/0004114}{{\ttfamily
  arXiv:astro-ph/0004114}}. \url {https://10.1086/318659}.

\bibitem{Dexter:2013sxa}
J.~Dexter and O.~Blaes, ``{A model of the steep power law spectra and
  high-frequency quasi-periodic oscillations in luminous black hole X-ray
  binaries},'' \href{http://dx.doi.org/10.1093/mnras/stu121}{{\em Mon. Not.
  Roy. Astron. Soc.} {\bfseries 438} no.~4, (2014) 3352--3357},
  \href{http://arxiv.org/abs/1312.0941}{{\ttfamily arXiv:1312.0941
  [astro-ph.HE]}}.

\bibitem{Rezzolla:2003zy}
L.~Rezzolla, S.~Yoshida, and O.~Zanotti, ``{Oscillations of vertically
  integrated relativistic tori - 1. Axisymmetric modes in a Schwarzschild
  space-time},'' \href{http://dx.doi.org/10.1046/j.1365-8711.2003.07023.x}{{\em
  Mon. Not. Roy. Astron. Soc.} {\bfseries 344} (2003) 978},
  \href{http://arxiv.org/abs/astro-ph/0307488}{{\ttfamily
  arXiv:astro-ph/0307488}}.

\bibitem{Rezzolla:2003zx}
L.~Rezzolla, S.~Yoshida, T.~J. Maccarone, and O.~Zanotti, ``{A New simple model
  for high frequency quasi periodic oscillations in black hole candidates},''
  \href{http://dx.doi.org/10.1046/j.1365-8711.2003.07018.x}{{\em Mon. Not. Roy.
  Astron. Soc.} {\bfseries 344} (2003) L37},
  \href{http://arxiv.org/abs/astro-ph/0307487}{{\ttfamily
  arXiv:astro-ph/0307487}}.

\bibitem{Ganguly:2014pwa}
C.~Ganguly and S.~SenGupta, ``{Penrose process in a charged axion--dilaton
  coupled black hole},''
  \href{http://dx.doi.org/10.1140/epjc/s10052-016-4058-0}{{\em Eur.\ Phys.\ J.\
  C} {\bfseries 76} no.~4, (2016) 213},
  \href{http://arxiv.org/abs/1401.6826}{{\ttfamily arXiv:1401.6826 [hep-th]}}.

\bibitem{Garcia:1995qz}
A.~Garcia, D.~Galtsov, and O.~Kechkin, ``{Class of stationary axisymmetric
  solutions of the Einstein-Maxwell dilaton - axion field equations},''
  \href{http://dx.doi.org/10.1103/PhysRevLett.74.1276}{{\em Phys. Rev. Lett.}
  {\bfseries 74} (1995) 1276--1279}. \url
  {https://doi.org/10.1103/PhysRevLett.74.1276}.

\bibitem{Yazadjiev:1999xq}
S.~Yazadjiev, ``{Exact static solutions in four-dimensional Einstein-Maxwell
  dilaton gravity},'' \href{http://dx.doi.org/10.1142/S0218271899000432}{{\em
  Int. J. Mod. Phys. D} {\bfseries 8} (1999) 635--643},
  \href{http://arxiv.org/abs/gr-qc/9906048}{{\ttfamily arXiv:gr-qc/9906048}}.

\bibitem{2016A&A...586A.130S}
Z.~{Stuchl{\'\i}k} and M.~{Kolo{\v{s}}}, ``{Models of quasi-periodic
  oscillations related to mass and spin of the GRO J1655-40 black hole},''
  \href{http://dx.doi.org/10.1051/0004-6361/201526095}{{\em Astron. Astrophys.}
  {\bfseries 586} (2016) A130},
  \href{http://arxiv.org/abs/1603.07366}{{\ttfamily arXiv:1603.07366
  [astro-ph.HE]}}.

\bibitem{Yagi:2016jml}
K.~Yagi and L.~C. Stein, ``{Black Hole Based Tests of General Relativity},''
  \href{http://dx.doi.org/10.1088/0264-9381/33/5/054001}{{\em Class. Quant.
  Grav.} {\bfseries 33} no.~5, (2016) 054001},
  \href{http://arxiv.org/abs/1602.02413}{{\ttfamily arXiv:1602.02413 [gr-qc]}}.
  \url {https://doi.org/10.1088/0264-9381/33/5/054001}.

\bibitem{Tsang:2008fz}
D.~Tsang and D.~Lai, ``{Corotational Damping of Diskoseismic C-modes in Black
  Hole Accretion Discs},''
  \href{http://dx.doi.org/10.1111/j.1365-2966.2008.14228.x}{{\em Mon. Not. Roy.
  Astron. Soc.} {\bfseries 393} (2009) 992--998},
  \href{http://arxiv.org/abs/0810.1299}{{\ttfamily arXiv:0810.1299
  [astro-ph]}}. \url {https://10.1111/j.1365-2966.2008.14228.x}.

\bibitem{Fu:2008iw}
W.~Fu and D.~Lai, ``{Effects of Magnetic Fields on the Diskoseismic Modes of
  Accreting Black Holes},''
  \href{http://dx.doi.org/10.1088/0004-637X/690/2/1386}{{\em Astrophys. J.}
  {\bfseries 690} (2009) 1386--1392},
  \href{http://arxiv.org/abs/0806.1938}{{\ttfamily arXiv:0806.1938
  [astro-ph]}}. \url {https://10.1088/0004-637X/690/2/1386}.

\bibitem{Fu:2010tf}
W.~Fu and D.~Lai, ``{Corotational Instability, Magnetic Resonances and Global
  Inertial-Acoustic Oscillations in Magnetized Black-Hole Accretion Discs},''
  \href{http://dx.doi.org/10.1111/j.1365-2966.2010.17451.x}{{\em Mon. Not. Roy.
  Astron. Soc.} {\bfseries 410} (2011) 399},
  \href{http://arxiv.org/abs/1006.3763}{{\ttfamily arXiv:1006.3763
  [astro-ph.HE]}}. \url {https://10.1111/j.1365-2966.2010.17451.x}.

\bibitem{Stella:1998mq}
L.~Stella and M.~Vietri, ``{Khz quasi periodic oscillations in low mass x-ray
  binaries as probes of general relativity in the strong field regime},''
  \href{http://dx.doi.org/10.1103/PhysRevLett.82.17}{{\em Phys. Rev. Lett.}
  {\bfseries 82} (1999) 17--20},
  \href{http://arxiv.org/abs/astro-ph/9812124}{{\ttfamily
  arXiv:astro-ph/9812124}}. \url {https://doi.org/10.1103/PhysRevLett.82.17}.

\bibitem{stella1997lense}
L.~Stella and M.~Vietri, ``Lense-thirring precession and quasi-periodic
  oscillations in low-mass x-ray binaries,'' {\em The Astrophysical Journal}
  {\bfseries 492} no.~1, (1997) L59. \url {https://10.1086/311075}.

\bibitem{stella1999correlations}
L.~Stella, M.~Vietri, and S.~M. Morsink, ``Correlations in the quasi-periodic
  oscillation frequencies of low-mass x-ray binaries and the relativistic
  precession model,'' {\em The Astrophysical Journal} {\bfseries 524} no.~1,
  (1999) L63. \url {https://10.1086/312291}.

\bibitem{2001PASJ...53....1K}
S.~{Kato}, ``{Basic Properties of Thin-Disk Oscillations $^{1}$},''
  \href{http://dx.doi.org/10.1093/pasj/53.1.1}{{\em pasj} {\bfseries 53} no.~1,
  (Feb., 2001) 1--24}.

\bibitem{2004PASJ...56..559K}
S.~{Kato}, ``{Wave-Warp Resonant Interactions in Relativistic Disks and kHz
  QPOs},'' \href{http://dx.doi.org/10.1093/pasj/56.3.599}{{\em pasj} {\bfseries
  56} (June, 2004) 559--567}.

\bibitem{2004PASJ...56..905K}
S.~{Kato}, ``{Resonant Excitation of Disk Oscillations by Warps: A Model of kHz
  QPOs},'' \href{http://dx.doi.org/10.1093/pasj/56.5.905}{{\em pasj} {\bfseries
  56} (Oct., 2004) 905--922},
  \href{http://arxiv.org/abs/astro-ph/0409051}{{\ttfamily
  arXiv:astro-ph/0409051 [astro-ph]}}.

\bibitem{2005PASJ...57..699K}
S.~{Kato}, ``{A Vertical Resonance of g-Mode Oscillations in Warped Disks and
  QPOs in Low-Mass X-Ray Binaries},''
  \href{http://dx.doi.org/10.1093/pasj/57.4.699}{{\em pasj} {\bfseries 57}
  (Aug., 2005) 699--703},
  \href{http://arxiv.org/abs/astro-ph/0507234}{{\ttfamily
  arXiv:astro-ph/0507234 [astro-ph]}}.

\bibitem{2008PASJ...60..111K}
S.~{Kato}, ``{Resonant Excitation of Disk Oscillations in Deformed Disks II: A
  Model of High-Frequency QPOs},''
  \href{http://dx.doi.org/10.1093/pasj/60.1.111}{{\em pasj} {\bfseries 60}
  (Feb., 2008) 111}, \href{http://arxiv.org/abs/0709.2467}{{\ttfamily
  arXiv:0709.2467 [astro-ph]}}.

\bibitem{2004ApJ...617L..45B}
M.~{Bursa}, M.~A. {Abramowicz}, V.~{Karas}, and W.~{Klu{\'z}niak}, ``{The Upper
  Kilohertz Quasi-periodic Oscillation: A Gravitationally Lensed Vertical
  Oscillation},'' \href{http://dx.doi.org/10.1086/427167}{{\em apjl} {\bfseries
  617} no.~1, (Dec., 2004) L45--L48},
  \href{http://arxiv.org/abs/astro-ph/0406586}{{\ttfamily
  arXiv:astro-ph/0406586 [astro-ph]}}.

\bibitem{2005AN....326..849B}
M.~{Bursa}, ``{Global oscillations of a fluid torus as a modulation mechanism
  for black-hole high-frequency QPOs},''
  \href{http://dx.doi.org/10.1002/asna.200510426}{{\em Astronomische
  Nachrichten} {\bfseries 326} no.~9, (Nov., 2005) 849--855},
  \href{http://arxiv.org/abs/astro-ph/0510460}{{\ttfamily
  arXiv:astro-ph/0510460 [astro-ph]}}.

\bibitem{2005ragt.meet...39B}
M.~Bursa, ``{High-frequency QPOs in GRO J1655-40: Constraints on resonance
  models by spectral fits},'' in {\em RAGtime 6/7: Workshops on black holes and
  neutron stars}, S.~{Hled{\'\i}k} and Z.~{Stuchl{\'\i}k}, eds., pp.~39--45.
\newblock Dec., 2005.

\bibitem{Beer:2001cg}
M.~E. Beer and P.~Podsiadlowski, ``{The quiescent light curve and evolutionary
  state of gro j1655-40},''
  \href{http://dx.doi.org/10.1046/j.1365-8711.2002.05189.x}{{\em Mon. Not. Roy.
  Astron. Soc.} {\bfseries 331} (2002) 351},
  \href{http://arxiv.org/abs/astro-ph/0109136}{{\ttfamily
  arXiv:astro-ph/0109136}}. \url {
  https://doi.org/10.1046/j.1365-8711.2002.05189.x}.

\bibitem{Motta:2013wga}
S.~E. Motta, T.~M. Belloni, L.~Stella, T.~Mu\~noz Darias, and R.~Fender,
  ``{Precise mass and spin measurements for a stellar-mass black hole through
  X-ray timing: the case of GRO J1655\ensuremath{-}40},''
  \href{http://dx.doi.org/10.1093/mnras/stt2068}{{\em Mon. Not. Roy. Astron.
  Soc.} {\bfseries 437} no.~3, (2014) 2554--2565},
  \href{http://arxiv.org/abs/1309.3652}{{\ttfamily arXiv:1309.3652
  [astro-ph.HE]}}. \url { https://doi.org/.1093/mnras/stt2068}.

\bibitem{Orosz:2011ki}
J.~A. Orosz, J.~F. Steiner, J.~E. McClintock, M.~A.~P. Torres, R.~A. Remillard,
  C.~D. Bailyn, and J.~M. Miller, ``{An Improved Dynamical Model for the
  Microquasar XTE J1550-564},''
  \href{http://dx.doi.org/10.1088/0004-637X/730/2/75}{{\em Astrophys. J.}
  {\bfseries 730} (2011) 75}, \href{http://arxiv.org/abs/1101.2499}{{\ttfamily
  arXiv:1101.2499 [astro-ph.SR]}}. \url {
  https://doi.org/10.1088/0004-637X/730/2/75}.

\bibitem{Reid:2014ywa}
M.~J. Reid, J.~E. McClintock, J.~F. Steiner, D.~Steeghs, R.~A. Remillard,
  V.~Dhawan, and R.~Narayan, ``{A Parallax Distance to the Microquasar GRS
  1915+105 and a Revised Estimate of its Black Hole Mass},''
  \href{http://dx.doi.org/10.1088/0004-637X/796/1/2}{{\em Astrophys. J.}
  {\bfseries 796} (2014) 2}, \href{http://arxiv.org/abs/1409.2453}{{\ttfamily
  arXiv:1409.2453 [astro-ph.GA]}}. \url {
  https://doi.org/10.1088/0004-637X/796/1/2}.

\bibitem{Pei:2016kka}
G.~Pei, S.~Nampalliwar, C.~Bambi, and M.~J. Middleton, ``{Blandford-Znajek
  mechanism in black holes in alternative theories of gravity},''
  \href{http://dx.doi.org/10.1140/epjc/s10052-016-4387-z}{{\em Eur. Phys. J. C}
  {\bfseries 76} no.~10, (2016) 534},
  \href{http://arxiv.org/abs/1606.04643}{{\ttfamily arXiv:1606.04643 [gr-qc]}}.
  \url {https://doi.org/10.1140/epjc/s10052-016-4387-z}.

\bibitem{Bhattacharjee:2019vyy}
A.~Bhattacharjee, I.~Banerjee, A.~Banerjee, D.~Debnath, and S.~K. Chakrabarti,
  ``{The 2004 Outburst of BHC H1743-322: Analysis of spectral and timing
  properties using the TCAF Solution},''
  \href{http://dx.doi.org/10.1093/mnras/stw3117}{{\em Mon. Not. Roy. Astron.
  Soc.} {\bfseries 466} (2017) 1372--1381},
  \href{http://arxiv.org/abs/1901.00810}{{\ttfamily arXiv:1901.00810
  [astro-ph.HE]}}.

\bibitem{Petri:2008jc}
J.~Petri, ``{A new model for QPOs in accreting black holes: application to the
  microquasar GRS 1915+105},''
  \href{http://dx.doi.org/10.1007/s10509-008-9916-2}{{\em Astrophys. Space
  Sci.} {\bfseries 318} (2008) 181--186},
  \href{http://arxiv.org/abs/0809.3115}{{\ttfamily arXiv:0809.3115
  [astro-ph]}}.

\bibitem{Ghez:2008ms}
A.~M. Ghez {\em et~al.}, ``{Measuring Distance and Properties of the Milky
  Way's Central Supermassive Black Hole with Stellar Orbits},''
  \href{http://dx.doi.org/10.1086/592738}{{\em Astrophys. J.} {\bfseries 689}
  (2008) 1044--1062}, \href{http://arxiv.org/abs/0808.2870}{{\ttfamily
  arXiv:0808.2870 [astro-ph]}}. \url {https://doi.org/ 10.1086/592738}.

\bibitem{Gillessen:2008qv}
S.~Gillessen, F.~Eisenhauer, S.~Trippe, T.~Alexander, R.~Genzel, F.~Martins,
  and T.~Ott, ``{Monitoring stellar orbits around the Massive Black Hole in the
  Galactic Center},''
  \href{http://dx.doi.org/10.1088/0004-637X/692/2/1075}{{\em Astrophys. J.}
  {\bfseries 692} (2009) 1075--1109},
  \href{http://arxiv.org/abs/0810.4674}{{\ttfamily arXiv:0810.4674
  [astro-ph]}}. \url {https://10.1088/0004-637X/692/2/1075}.

\bibitem{Stuchlik:2008fy}
Z.~Stuchl\'\i{}k and A.~Kotrlov\'a, ``{Orbital resonances in discs around
  braneworld Kerr black holes},''
  \href{http://dx.doi.org/10.1007/s10714-008-0709-2}{{\em Gen. Rel. Grav.}
  {\bfseries 41} (2009) 1305--1343},
  \href{http://arxiv.org/abs/0812.5066}{{\ttfamily arXiv:0812.5066
  [astro-ph]}}. \url {https://10.1007/s10714-008-0709-2}.

\bibitem{1976ApJ...210..642A}
Y.~Avni, ``{Energy spectra of X-ray clusters of galaxies},''
  \href{http://dx.doi.org/10.1086/154870}{{\em Apj} {\bfseries 210} (1976)
  642--646}.

\bibitem{2001A&A...374L..19A}
M.~A. {Abramowicz} and W.~{Klu{\'z}niak}, ``{A precise determination of black
  hole spin in GRO J1655-40},''
  \href{http://dx.doi.org/10.1051/0004-6361:20010791}{{\em Astron. Astrophys.}
  {\bfseries 374} (2001) L19--L20},
  \href{http://arxiv.org/abs/astro-ph/0105077}{{\ttfamily
  arXiv:astro-ph/0105077 [astro-ph]}}.

\bibitem{2005A&A...436....1T}
G.~{T{\"o}r{\"o}k}, M.~A. {Abramowicz}, W.~{Klu{\'z}niak}, and
  Z.~{Stuchl{\'\i}k}, ``{The orbital resonance model for twin peak kHz quasi
  periodic oscillations in microquasars},''
  \href{http://dx.doi.org/10.1051/0004-6361:20047115}{{\em Astron. Astrophys.}
  {\bfseries 436} no.~1, (2005) 1--8}.

\bibitem{2001AcPPB..32.3605K}
W.~{Kluzniak} and M.~A. {Abramowicz}, ``{Strong-Field Gravity and Orbital
  Resonance in Black Holes and Neutron Stars --- kHz Quasi-Periodic
  Oscillations (QPO)},'' {\em Acta Physica Polonica B} {\bfseries 32} no.~11,
  (Nov., 2001) 3605.

\bibitem{Horak:2004hm}
J.~Horak, M.~Abramowicz, V.~Karas, and W.~Kluzniak, ``{Of NBOs and kHz QPOs: A
  Low-frequency modulation in resonant oscillations of relativistic accretion
  disks},'' \href{http://dx.doi.org/10.1093/pasj/56.5.819}{{\em Publ. Astron.
  Soc. Jap.} {\bfseries 56} (2004) 819--822},
  \href{http://arxiv.org/abs/astro-ph/0408090}{{\ttfamily
  arXiv:astro-ph/0408090}}. \url {https://doi.org/10.1093/pasj/56.5.819}.

\bibitem{torok2005orbital}
G.~T{\"o}r{\"o}k, M.~Abramowicz, W.~Kluzniak, and Z.~Stuchl{\i}k, ``The orbital
  resonance model for twin peak khz quasi periodic oscillations in
  microquasars,'' {\em Astronomy \& Astrophysics} {\bfseries 436} no.~1, (2005)
  1--8.

\bibitem{2004ApJ...603L..93L}
W.~H. {Lee}, M.~A. {Abramowicz}, and W.~{Klu{\'z}niak}, ``{Resonance in Forced
  Oscillations of an Accretion Disk and Kilohertz Quasi-periodic
  Oscillations},'' \href{http://dx.doi.org/10.1086/383245}{{\em apjl}
  {\bfseries 603} no.~2, (Mar., 2004) L93--L96},
  \href{http://arxiv.org/abs/astro-ph/0402084}{{\ttfamily
  arXiv:astro-ph/0402084 [astro-ph]}}.

\bibitem{2001PASJ...53L..37K}
S.~Kato, ``{Trapping of Non-Axisymmetric g-Mode Oscillations in Thin
  Relativistic Disks and kHz QPOs},''
  \href{http://dx.doi.org/10.1093/pasj/53.5.L37}{{\em PASJ} {\bfseries 53}
  no.~5, (Oct., 2001) L37--L39}.

\bibitem{Li:2002yi}
L.-X. Li, J.~Goodman, and R.~Narayan, ``{Non-axisymmetric g-mode and p-mode
  instability in a thin accretion disk},''
  \href{http://dx.doi.org/10.1086/376695}{{\em Astrophys. J.} {\bfseries 593}
  (2003) 980}, \href{http://arxiv.org/abs/astro-ph/0210455}{{\ttfamily
  arXiv:astro-ph/0210455}}. \url {https://doi.org/10.1086/376695}.

\bibitem{2003PASJ...55..257K}
S.~Kato, ``{Damping and Frequencies of Non-Axisymmetric Trapped g-Mode
  Oscillations},'' \href{http://dx.doi.org/10.1093/pasj/55.1.257}{{\em PASJ}
  {\bfseries 55} (Feb., 2003) 257--265}.

\bibitem{2010tbha.book.....A}
M.~A. {Abramowicz}, G.~{Bj{\"o}rnsson}, and J.~E. {Pringle}, {\em {Theory of
  Black Hole Accretion Discs}}.
\newblock 2010.

\bibitem{Torok:2015tpu}
G.~T\"or\"ok, K.~Goluchov\'a, J.~Hor\'ak, E.~\v{S}r\'amkov\'a, M.~Urbanec,
  T.~Pech\'a\v{c}ek, and P.~Bakala, ``{Twin peak quasi-periodic oscillations as
  signature of oscillating cusp torus},''
  \href{http://dx.doi.org/10.1093/mnrasl/slv196}{{\em Mon. Not. Roy. Astron.
  Soc.} {\bfseries 457} no.~1, (2016) L19--L23},
  \href{http://arxiv.org/abs/1512.03841}{{\ttfamily arXiv:1512.03841
  [astro-ph.HE]}}.

\bibitem{Sramkova:2015bha}
E.~Sramkova, G.~Torok, A.~Kotrlova, P.~Bakala, M.~Abramowicz, Z.~Stuchlik,
  K.~Goluchova, and W.~Kluzniak, ``{Black hole spin inferred from 3:2 epicyclic
  resonance model of high-frequency quasi-periodic oscillations},''
  \href{http://dx.doi.org/10.1051/0004-6361/201425241}{{\em Astron. Astrophys.}
  {\bfseries 578} (2015) A90},
  \href{http://arxiv.org/abs/1505.02712}{{\ttfamily arXiv:1505.02712
  [astro-ph.HE]}}.

\bibitem{Horak:2008zg}
J.~Horak, ``{Weak nonlinear coupling between epicyclic modes in slender
  tori},'' \href{http://dx.doi.org/10.1051/0004-6361:20078305}{{\em Astron.
  Astrophys.} {\bfseries 486} (2008) 1},
  \href{http://arxiv.org/abs/0805.2059}{{\ttfamily arXiv:0805.2059
  [astro-ph]}}.

\bibitem{Verde:2009tu}
L.~Verde, ``{Statistical methods in cosmology},''
  \href{http://dx.doi.org/10.1007/978-3-642-10598-2_4}{{\em Lect. Notes Phys.}
  {\bfseries 800} (2010) 147--177},
  \href{http://arxiv.org/abs/0911.3105}{{\ttfamily arXiv:0911.3105
  [astro-ph.CO]}}.

\bibitem{Hazarika:2025axz}
B.~Hazarika, M.~M. Gohain, and P.~Phukon, ``{Signatures of NED on quasi
  periodic oscillations of a magnetically charged black hole},''
  \href{http://dx.doi.org/10.1088/1475-7516/2025/07/035}{{\em JCAP} {\bfseries
  07} (2025) 035}, \href{http://arxiv.org/abs/2504.07821}{{\ttfamily
  arXiv:2504.07821 [gr-qc]}}.

\bibitem{Wang:2021gtd}
Z.~Wang, S.~Chen, and J.~Jing, ``{Constraint on parameters of a rotating black
  hole in Einstein-bumblebee theory by quasi-periodic oscillations},''
  \href{http://dx.doi.org/10.1140/epjc/s10052-022-10475-x}{{\em Eur. Phys. J.
  C} {\bfseries 82} no.~6, (2022) 528},
  \href{http://arxiv.org/abs/2112.02895}{{\ttfamily arXiv:2112.02895 [gr-qc]}}.

\bibitem{Shafee_2005}
R.~Shafee, J.~E. McClintock, R.~Narayan, S.~W. Davis, L.-X. Li, and R.~A.
  Remillard, ``Estimating the spin of stellar-mass black holes by spectral
  fitting of the x-ray continuum,''
  \href{http://dx.doi.org/10.1086/498938}{{\em The Astrophysical Journal}
  {\bfseries 636} no.~2, (Dec, 2005) L113--L116}.
  \url{https://doi.org/10.1086/498938}.

\bibitem{Miller:2009cw}
J.~M. Miller, C.~S. Reynolds, A.~C. Fabian, G.~Miniutti, and L.~C. Gallo,
  ``{Stellar-mass Black Hole Spin Constraints from Disk Reflection and
  Continuum Modeling},''
  \href{http://dx.doi.org/10.1088/0004-637X/697/1/900}{{\em Astrophys. J.}
  {\bfseries 697} (2009) 900--912},
  \href{http://arxiv.org/abs/0902.2840}{{\ttfamily arXiv:0902.2840
  [astro-ph.HE]}}. \url { https://doi.org/10.1088/0004-637X/697/1/900}.

\bibitem{Steiner:2010bt}
J.~F. Steiner, R.~C. Reis, J.~E. McClintock, R.~Narayan, R.~A. Remillard, J.~A.
  Orosz, L.~Gou, A.~C. Fabian, and M.~A.~P. Torres, ``{The Spin of the Black
  Hole Microquasar XTE J1550-564 via the Continuum-Fitting and Fe-Line
  Methods},'' \href{http://dx.doi.org/10.1111/j.1365-2966.2011.19089.x}{{\em
  Mon. Not. Roy. Astron. Soc.} {\bfseries 416} (2011) 941--958},
  \href{http://arxiv.org/abs/1010.1013}{{\ttfamily arXiv:1010.1013
  [astro-ph.HE]}}.

\bibitem{McClintock:2006xd}
J.~E. McClintock, R.~Shafee, R.~Narayan, R.~A. Remillard, S.~W. Davis, and
  L.-X. Li, ``{The Spin of the Near-Extreme Kerr Black Hole GRS 1915+105},''
  \href{http://dx.doi.org/10.1086/508457}{{\em Astrophys. J.} {\bfseries 652}
  (2006) 518--539}, \href{http://arxiv.org/abs/astro-ph/0606076}{{\ttfamily
  arXiv:astro-ph/0606076}}.

\bibitem{2006MNRAS.373.1004M}
M.~{Middleton}, C.~{Done}, M.~{Gierli{\'n}ski}, and S.~W. {Davis}, ``{Black
  hole spin in GRS 1915+105},''
  \href{http://dx.doi.org/10.1111/j.1365-2966.2006.11077.x}{{\em mnras}
  {\bfseries 373} no.~3, (Dec., 2006) 1004--1012},
  \href{http://arxiv.org/abs/astro-ph/0601540}{{\ttfamily
  arXiv:astro-ph/0601540 [astro-ph]}}.

\bibitem{Blum:2009ez}
J.~L. Blum, J.~M. Miller, A.~C. Fabian, M.~C. Miller, J.~Homan, M.~van~der
  Klis, E.~M. Cackett, and R.~C. Reis, ``{Measuring the Spin of GRS 1915+105
  with Relativistic Disk Reflection},''
  \href{http://dx.doi.org/10.1088/0004-637X/706/1/60}{{\em Astrophys. J.}
  {\bfseries 706} (2009) 60--66},
  \href{http://arxiv.org/abs/0909.5383}{{\ttfamily arXiv:0909.5383
  [astro-ph.HE]}}. \url { https://doi.org/10.1088/0004-637X/706/1/60}.

\bibitem{Mills:2021dxs}
B.~S. Mills, S.~W. Davis, and M.~J. Middleton, ``{The Black Hole Spin in GRS
  1915+105, Revisited},''
  \href{http://dx.doi.org/10.3847/1538-4357/abf2b7}{{\em Astrophys. J.}
  {\bfseries 914} no.~1, (2021) 6},
  \href{http://arxiv.org/abs/2101.11655}{{\ttfamily arXiv:2101.11655
  [astro-ph.HE]}}. \url { https://doi.org/10.3847/1538-4357/abf2b7}.

\bibitem{Steiner:2011kd}
J.~F. Steiner, J.~E. McClintock, and M.~J. Reid, ``{The Distance, Inclination,
  and Spin of the Black Hole Microquasar H1743-322},''
  \href{http://dx.doi.org/10.1088/2041-8205/745/1/L7}{{\em Astrophys. J. Lett.}
  {\bfseries 745} (2012) L7}, \href{http://arxiv.org/abs/1111.2388}{{\ttfamily
  arXiv:1111.2388 [astro-ph.HE]}}.

\bibitem{Moscibrodzka:2009gw}
M.~Moscibrodzka, C.~F. Gammie, J.~C. Dolence, H.~Shiokawa, and P.~K. Leung,
  ``{Radiative Models of Sgr A* from GRMHD Simulations},''
  \href{http://dx.doi.org/10.1088/0004-637X/706/1/497}{{\em Astrophys. J.}
  {\bfseries 706} (2009) 497--507},
  \href{http://arxiv.org/abs/0909.5431}{{\ttfamily arXiv:0909.5431
  [astro-ph.HE]}}. \url { https://doi.org/10.1088/0004-637X/706/1/497}.

\bibitem{Shcherbakov:2010ki}
R.~V. Shcherbakov, R.~F. Penna, and J.~C. McKinney, ``{Sagittarius A* Accretion
  Flow and Black Hole Parameters from General Relativistic Dynamical and
  Polarized Radiative Modeling},''
  \href{http://dx.doi.org/10.1088/0004-637X/755/2/133}{{\em Astrophys. J.}
  {\bfseries 755} (2012) 133}, \href{http://arxiv.org/abs/1007.4832}{{\ttfamily
  arXiv:1007.4832 [astro-ph.HE]}}. \url {
  https://doi.org/10.1088/0004-637X/755/2/133}.

\bibitem{Fragione:2020khu}
G.~Fragione and A.~Loeb, ``{An upper limit on the spin of SgrA$^*$ based on
  stellar orbits in its vicinity},''
  \href{http://dx.doi.org/10.3847/2041-8213/abb9b4}{{\em Astrophys. J. Lett.}
  {\bfseries 901} no.~2, (2020) L32},
  \href{http://arxiv.org/abs/2008.11734}{{\ttfamily arXiv:2008.11734
  [astro-ph.GA]}}.

\bibitem{Daly:2023axh}
R.~A. Daly, M.~Donahue, C.~P. O'Dea, B.~Sebastian, D.~Haggard, and A.~Lu,
  ``{New black hole spin values for Sagittarius A* obtained with the outflow
  method},'' \href{http://dx.doi.org/10.1093/mnras/stad3228}{{\em Mon. Not.
  Roy. Astron. Soc.} {\bfseries 527} no.~1, (2023) 428--436},
  \href{http://arxiv.org/abs/2310.12108}{{\ttfamily arXiv:2310.12108
  [astro-ph.GA]}}.

\bibitem{Meyer:2006fd}
L.~Meyer, A.~Eckart, R.~Schoedel, W.~J. Duschl, K.~Muzic, M.~Dovciak, and
  V.~Karas, ``{Near-infrared polarimetry setting constraints on the orbiting
  spot model for Sgr A* flares},''
  \href{http://dx.doi.org/10.1051/0004-6361:20065925}{{\em Astron. Astrophys.}
  {\bfseries 460} (2006) 15},
  \href{http://arxiv.org/abs/astro-ph/0610104}{{\ttfamily
  arXiv:astro-ph/0610104}}.

\bibitem{Genzel:2003as}
R.~Genzel, R.~Schodel, T.~Ott, A.~Eckart, T.~Alexander, F.~Lacombe, D.~Rouan,
  and B.~Aschenbach, ``{Near-infrared flares from accreting gas around the
  supermassive black hole at the galactic centre},''
  \href{http://dx.doi.org/10.1038/nature02065}{{\em Nature} {\bfseries 425}
  (2003) 934--937}, \href{http://arxiv.org/abs/astro-ph/0310821}{{\ttfamily
  arXiv:astro-ph/0310821}}.

\bibitem{Belanger:2006gm}
G.~Belanger, R.~Terrier, O.~C. De~Jager, A.~Goldwurm, and F.~Melia, ``{Periodic
  Modulations in an X-ray Flare from Sagittarius A*},''
  \href{http://dx.doi.org/10.1088/1742-6596/54/1/066}{{\em J. Phys. Conf. Ser.}
  {\bfseries 54} (2006) 420--426},
  \href{http://arxiv.org/abs/astro-ph/0604337}{{\ttfamily
  arXiv:astro-ph/0604337}}.

\bibitem{Bambi:2013fea}
C.~Bambi, ``{Testing the nature of the black hole candidate in GRO J1655-40
  with the relativistic precession model},''
  \href{http://dx.doi.org/10.1140/epjc/s10052-015-3396-7}{{\em Eur. Phys. J. C}
  {\bfseries 75} no.~4, (2015) 162},
  \href{http://arxiv.org/abs/1312.2228}{{\ttfamily arXiv:1312.2228 [gr-qc]}}.

\bibitem{Bambi:2012pa}
C.~Bambi, ``{Probing the space-time geometry around black hole candidates with
  the resonance models for high-frequency QPOs and comparison with the
  continuum-fitting method},''
  \href{http://dx.doi.org/10.1088/1475-7516/2012/09/014}{{\em JCAP} {\bfseries
  09} (2012) 014}, \href{http://arxiv.org/abs/1205.6348}{{\ttfamily
  arXiv:1205.6348 [gr-qc]}}.

\end{thebibliography}\endgroup
\bibliographystyle{./utphys1}

\end{document}